\documentclass[iop,twocolumn]{emulateapj}
\usepackage{natbib}
\pdfoutput=1

\newcommand{\msun}{M$_\odot$}
\newcommand{\kms}{km s$^{-1}$}

\newcommand{\qpah}{$q_{\rm PAH}$}

\newcommand{\xco}{X$_{\rm CO}$}
\newcommand{\aco}{$\alpha_{\rm CO}$}
\newcommand{\sigd}{$\Sigma_{\rm D}$}
\newcommand{\sighi}{$\Sigma_{\rm HI}$}
\newcommand{\sightwo}{$\Sigma_{{\rm H}_{2}}$}
\newcommand{\ico}{$\mathrm{I}_{\rm CO}$}
\newcommand{\hi}{H{\sc I}}
\newcommand{\hii}{H{\sc II}}
\newcommand{\acounit}{\msun\ pc$^{-2}$ (K km s$^{-1}$)$^{-1}$}

\slugcomment{The Astrophysical Journal, in press.}

\shorttitle{\aco\ \& DGR in Nearby Galaxies}
\shortauthors{Sandstrom et al.}

\begin{document}
\maxdeadcycles=1000
%\interfootnotelinepenalty=10000

%%%%%%%%%%%%%%
% TITLE PAGE %
%%%%%%%%%%%%%%

\title{The CO-to-H$_2$ Conversion Factor and Dust-to-Gas Ratio on Kiloparsec Scales in Nearby Galaxies}

\author{K. M. Sandstrom\altaffilmark{1,2}, 
             A. K. Leroy\altaffilmark{3},
             F. Walter\altaffilmark{1},
             A. D. Bolatto\altaffilmark{4},
             K. V. Croxall\altaffilmark{5,6},
             B.~T. Draine\altaffilmark{7},
             C. D. Wilson\altaffilmark{8},
             M. Wolfire\altaffilmark{4},
             D. Calzetti\altaffilmark{9},
             R. C. Kennicutt\altaffilmark{10},
             G. Aniano\altaffilmark{7,11},
             J. Donovan Meyer\altaffilmark{12},
             A. Usero\altaffilmark{13},
             F. Bigiel\altaffilmark{14},
             E. Brinks\altaffilmark{15},
             W. J. G de Blok\altaffilmark{16},
             A. Crocker\altaffilmark{5,9},
             D. Dale\altaffilmark{17},
             C. W. Engelbracht\altaffilmark{18,19},
             M. Galametz\altaffilmark{10},
             B. Groves\altaffilmark{1},
             L. K. Hunt\altaffilmark{20},
             J. Koda\altaffilmark{12},
             K. Kreckel\altaffilmark{1},
             H. Linz\altaffilmark{1},
             S. Meidt\altaffilmark{1},
             E. Pellegrini\altaffilmark{5},
             H.-W. Rix\altaffilmark{1},
             H. Roussel\altaffilmark{21},
             E. Schinnerer\altaffilmark{1},
             A. Schruba\altaffilmark{22},
             K.-F. Schuster\altaffilmark{23},
             R. Skibba\altaffilmark{18,24},
             T. van der Laan\altaffilmark{1}
             P. Appleton\altaffilmark{25},
             L. Armus\altaffilmark{26},
             B. Brandl\altaffilmark{27},
             K. Gordon\altaffilmark{28},
             J. Hinz\altaffilmark{18,29},
             O. Krause\altaffilmark{1},
             E. Montiel\altaffilmark{18,30},
             M. Sauvage\altaffilmark{31},
             A. Schmiedeke\altaffilmark{1,32},
             J. D. T. Smith\altaffilmark{5},
             L. Vigroux\altaffilmark{21}}
        
\affil{$^1$Max Planck Institut f\"{u}r Astronomie, K\"{o}nigstuhl 17, D-69117, Heidelberg Germany}
\affil{$^2$Marie Curie Postdoctoral Fellow}
\affil{$^3$National Radio Astronomy Observatory, 520 Edgemont Road, Charlottesville, VA 22903, USA}
\affil{$^4$Department of Astronomy, University of Maryland, College Park, MD, 20742, USA}
\affil{$^5$Department of Physics \& Astronomy, Mail Drop 111, University of Toledo, 2801 West Bancroft Street, Toledo, OH 43606, USA}
\affil{$^6$Department of Astronomy, The Ohio State University, 140 West 18th Avenue, Columbus, OH 43210, USA}
\affil{$^7$Princeton University Observatory, Peyton Hall, Princeton, NJ 08544-1001, USA}
\affil{$^8$Department of Physics \& Astronomy, McMaster University, Hamilton, Ontario L8S 4M1, Canada}
\affil{$^9$Department of Astronomy, University of Massachusetts, Amherst, MA 01003, USA}
\affil{$^{10}$Institute of Astronomy, University of Cambridge, Madingley Road, Cambridge CB3 0HA, UK}
\affil{$^{11}$Institut d'Astrophysique Spatiale (IAS), b\^{a}timent 121, Universit\'{e} Paris-Sud 11 and CNRS (UMR 8617), F-91405 Orsay, France}
\affil{$^{12}$Department of Physics \& Astronomy, SUNY Stony Brook, Stony Brook, NY 11794-3800, USA}
\affil{$^{13}$Observatorio Astron\'{o}mico Nacional, Alfonso XII, 3, 28014 Madrid, Spain}
\affil{$^{14}$Institut f\"ur theoretische Astrophysik, Zentrum f\"ur Astronomie der Universit\"at Heidelberg, Albert-Ueberle Str. 2, 69120 Heidelberg}
\affil{$^{15}$Centre for Astrophysics Research, University of Hertfordshire, Hatfield AL10 9AB, United Kingdom}
\affil{$^{16}$ASTRON, The Netherlands Institute for Radio Astronomy, Postbus 2, 7990 AA, Dwingeloo, The Netherlands}
\affil{$^{17}$Department of Physics \& Astronomy, University of Wyoming, Laramie, WY 82071, USA}
\affil{$^{18}$Steward Observatory, University of Arizona, Tucson, AZ 85721, USA}
\affil{$^{19}$Raytheon Company, 1151 E. Hermans Road, Tucson, AZ 85756, USA}
\affil{$^{20}$INAF-Osservatorio Astrofisico di Arcetri, Largo E. Fermi 5, I-50125 Firenze, Italy}
\affil{$^{21}$Institut d'Astrophysique de Paris, UMR 7095 CNRS, Universit\'e Pierre et Marie Curie, 75014 Paris, France}
\affil{$^{22}$California Institute for Technology, 1200 E California Blvd, Pasadena, CA 91125 USA}
\affil{$^{23}$IRAM, 300 rue de la Piscine, 38406 St. Martin d'H\'{e}res, France}
\affil{$^{24}$Center for Astrophysics \& Space Sciences, University of California, 9500 Gilman Dr., San Diego, CA 92093, USA}
\affil{$^{25}$NASA Herschel Science Center, IPAC, California Institute of Technology, Pasadena, CA 91125, USA}
\affil{$^{26}$Spitzer Science Center, California Institute of Technology, MC 314-6, Pasadena, CA 91125, USA}
\affil{$^{27}$Leiden Observatory, Leiden University, P.O. Box 9513, 2300 RA Leiden, The Netherlands}
\affil{$^{28}$Space Telescope Science Institute, 3700 San Martin Drive, Baltimore, MD 21218, USA}
\affil{$^{29}$MMT Observatory, Tucson, AZ 85721, USA}
\affil{$^{30}$Department of Physics \& Astronomy, Louisiana State University, Baton Rouge, LA 70803, USA}
\affil{$^{31}$CEA/DSM/DAPNIA/Service dÕAstrophysique, UMR AIM, CE Saclay, F-91191 Gif sur Yvette Cedex, France}
\affil{$^{32}$Universit\"{a}t zu K\"{o}ln, Z\"{u}lpicher Strasse 77, 50937 K\"{o}ln, Germany}

\email{sandstrom@mpia.de}

%%%%%%%%%%%%
% ABSTRACT %
%%%%%%%%%%%%

\begin{abstract}

We present $\sim$kiloparsec (kpc) spatial resolution maps of the CO-to-H$_2$ conversion factor (\aco) and dust-to-gas ratio (DGR) in 26 nearby, star-forming galaxies.  We have simultaneously solved for \aco\ and DGR by assuming that the DGR is approximately constant on kpc scales.  With this assumption, we can combine maps of dust mass surface density, CO integrated intensity and \hi\ column density to solve for both \aco\ and DGR with no assumptions about their value or dependence on metallicity or other parameters.  Such a study has just become possible with the availability of high resolution far-IR maps from the {\em Herschel} key program KINGFISH, $^{12}$CO J=(2$-$1) maps from the IRAM 30m large program HERACLES and \hi\ 21-cm line maps from THINGS.  We use a fixed ratio between the (2$-$1) and (1$-$0) lines to present our \aco\ results on the more typically used $^{12}$CO J=(1$-$0) scale and show using literature measurements that variations in the line ratio do not effect our results.  In total, we derive 782 individual solutions for \aco\ and DGR.  On average, \aco\ = 3.1 \acounit\ for our sample with a standard deviation of 0.3 dex.  Within galaxies we observe a generally flat profile of \aco\ as a function of galactocentric radius. However, most galaxies exhibit a lower \aco\ in the central kpc---a factor of $\sim$2 below the galaxy mean, on average.  In some cases, the central \aco\ value can be factors of 5 to 10 below the standard Milky Way (MW) value of $\alpha_{\rm CO,MW}=4.4$ \acounit.  While for \aco\ we find only weak correlations with metallicity, DGR is well-correlated with metallicity, with an approximately linear slope.  Finally, we present several recommendations for choosing an appropriate \aco\ for studies of nearby galaxies.

\vspace{0.2in}

\end{abstract}

\keywords{dust, extinction --- infrared: ISM --- ISM:  ISM --- molecules: Galaxies --- ISM}

%%%%%%%%
% BODY %
%%%%%%%%

\section{Introduction}\label{sec:intro}

The H$_2$ molecule is difficult to observe in the prevalent interstellar medium (ISM) conditions of a normal star-forming galaxy.  Since it is the primary constituent of molecular gas, inferring the mass of H$_2$ is crucial for studying this  phase of the ISM and the star formation that occurs within it.  The widespread practice is to use the second most abundant molecule, carbon monoxide ($^{12}$CO), as a tracer and convert measured CO integrated intensities into H$_2$ column densities using a ``CO-to-H$_2$'' conversion factor \xco\footnote{We refer to the CO-to-H$_2$ conversion factor in mass units throughout this paper.  Including a factor of 1.36 for helium, \xco$=2\times 10^{20}$ cm$^{-2}$ (K km s$^{-1}$)$^{-1}$ corresponds to \aco$=4.35$ \msun\ pc$^{-2}$  (K km s$^{-1}$)$^{-1}$.  \aco\ can be converted to \xco\ units by multiplying by a factor of $4.6\times10^{19}$.}:
\begin{equation}
\mathrm{N}_{\rm H_2} = {\rm X}_{\rm CO} \mathrm{I}_{\rm CO}.
\end{equation}
In mass surface density units this equation can be rewritten as
\begin{equation}
\Sigma_{\rm H_2} = \alpha_{\rm CO} \mathrm{I}_{\rm CO}
\end{equation}
where $\Sigma_{\rm H_2}$ is the total mass surface density of molecular gas (including a correction for the second most abundant element, He).  A variety of observations have shown that \aco$\approx 4.4$ \acounit\ is characteristic of the local area of the Milky Way \citep{1987ApJ...319..730S,Strong:1996vg,2010ApJ...710..133A}.  Despite uncertainties in the physics behind the conversion factor, the observability of CO ensures that it will remain a widely-used molecular gas tracer, particularly at high redshift.  %For this reason, measurements of \aco\ in nearby galaxies, where we can relate its variations to local environmental conditions, are critical. 

\aco\ is used in a variety of contexts in Galactic and extragalactic studies.  In the following, we define and measure \aco\ on $\sim$kpc scales in nearby galaxies.  At these resolutions, the small scale structure of the ISM is  averaged out and the variation in \aco\ is driven by large scale changes in the galactic environment (e.g. metallicity, galactic dynamics, ISM pressure). In general, extragalactic studies have adopted a single value of \aco\ for entire galaxies.  The new ability to perform systematic studies of \aco\ on sub-galactic scales in nearby galaxies facilitated by high angular resolution maps of gas and dust will let us move beyond this simplistic assumption.  

In addition, studying \aco\ on $\sim$kpc scales has several advantages: 1) because we do not resolve molecular clouds, we avoid issues with sampling the cloud structure (e.g. envelopes of CO-free H$_2$); 2) because our resolution element contains many clouds, we average over cloud evolutionary effects; and 3) we cover large enough fractions of the total molecular gas mass in a galaxy that it becomes reasonable to generalize our results to determine \aco\ appropriate for integrated galaxy measurements.  It is important to note that our definition of \aco\ is distinct from the line-of-sight $\Sigma_{\rm H_2}$/I$_{\rm CO}$ one can measure in small regions of Galactic molecular clouds---in such cases the conversion factor is not well-defined since it does not sample the full structure of the cloud.

In order to measure the conversion factor, one must measure $\Sigma_{\rm H_2}$ (or equivalently, $\Sigma_{\rm gas} -\Sigma_{\rm HI}$) independently of CO and then compare it to observed CO integrated intensities.  This has been done using a variety of techniques, including: measuring total gas masses from $\gamma$-ray emission plus a model for the cosmic ray distribution \citep{Strong:1996vg,2010ApJ...710..133A}, using the observed velocity dispersion and size of the molecular cloud to obtain virial masses \citep{1987ApJ...319..730S,Wilson:1995kc,2008ApJ...686..948B,2012ApJ...744...42D,2012ApJ...750..136W,2012A&A...542A.108G}, and modeling multiple molecular gas lines with varying optical depths and critical densities \citep[e.g.][]{Weiss:2001eq,2009A&A...506..689I,2009A&A...493..525I}.  In general, few of these techniques are effective for constraining \aco\ in galaxies outside the Local Group due to the difficulty of obtaining the necessary observations (e.g. $\gamma$-ray maps or CO mapping at $<100$pc resolution) or doubts about fundamental assumptions (e.g. that CO traces the full extent of molecular gas in the clouds; that the clouds lack contributions to virial balance from magnetic or pressure forces; that simple radiative transfer models can reproduce molecular gas excitation on kpc-scales).  

Another possible technique to measure $\Sigma_{\rm H_2}$ is to use dust as a tracer of total gas column.  Assuming that dust and gas are well mixed, the dust-to-gas ratio (DGR) is not a function of atomic/molecular phase and the fraction of mass in ionized gas is negligible, the observed dust mass surface density can be converted to a gas mass surface density with information on the DGR, i.e. using the following equation:
\begin{equation}
\frac{\Sigma_{\rm D}}{{\rm DGR}} = \Sigma_{\rm HI} + \Sigma_{\rm H_2} = \Sigma_{\rm HI} + \alpha_{\rm CO} \mathrm{I}_{\rm CO}.\label{eq:dgr}
\end{equation}
Here \sigd\ is the mass surface density of dust and $\Sigma_{\rm HI}$ and $\Sigma_{\rm H_2}$ are the mass surface densities of atomic and molecular gas\footnote[2]{In converting from column densities to mass surface densities, we account for helium with a factor of 1.36.  For $\Sigma_{\rm H_2}$, this factor is included in the \aco\ term.  We apply the helium correction to $\Sigma_{\rm HI}$ as well.}, respectively, and DGR is the dust-to-gas mass ratio.  By measuring \sigd\ and $\Sigma_{\rm HI}$ and assuming the DGR (or simultaneously measuring it, as we will discuss shortly), the molecular gas mass can be determined.  While still subject to its own systematic uncertainties (discussed in detail in Section~\ref{sec:acosys}), this technique relies on a different set of assumptions than those mentioned previously.  

Studies of DGR or \aco\ have typically fixed one of the parameters in order to determine the other, so it is difficult to avoid circularity when using a fixed DGR to solve for \aco\ in Equation~\ref{eq:dgr}.  Alternatively, with sufficiently high spatial resolution, DGR can be determined on sight-lines free of molecular gas and extrapolated to regions where CO is detected.  For very nearby galaxies like the Magellanic Clouds, such a technique has been used by \citet{Israel:1997tm} and \citet{2009ApJ...702..352L}, who found that \aco\ determinations from virial masses can be strongly biased by envelopes of ``CO-dark'' molecular gas in low-metallicity systems.  In more distant galaxies, we generally cannot isolate purely atomic lines of sight at the resolution of typical \hi\ and CO observations.  Instead, it is possible to use a technique developed by \citet{2011ApJ...737...12L} to simultaneously measure \aco\ and DGR using the assumption that DGR should be constant over a region of a galaxy.  Since we can now achieve $\sim$kpc spatial resolution in the far-IR with {\em Herschel} and have sensitive, high-resolution CO and \hi\ maps, it is possible to extend this technique, which has thus far only been applied to the Local Group, to more distant galaxies.  

The idea behind this technique (explained in detail in Section~\ref{sec:tech}) is that spatially resolved measurements of \sigd, \sighi\ and \ico\ allow one to solve for \aco\ and DGR in a region when it is smaller than the typical scale over which DGR varies (i.e. it is well represented by one DGR value) and it covers a range of CO/\hi\ ratios.  Having chosen an appropriately sized region, we can determine these two constants, defining the best fit as that which produces the most uniform DGR for the multiple lines-of-sight included in the region.  This technique assumes no prior value of DGR or \aco---it makes use of the (by definition) linear dependence of $\Sigma_{\rm H_2}$ on \aco\ to identify the solution.  The technique is therefore only applicable in regions where CO is detected.  Constraining \aco\ in more extreme conditions would require different techniques \citep[see ][for more discussion]{2012AJ....143..138S}.

One useful aspect of solving simultaneously for \aco\ and DGR is that the absolute normalization of the dust tracer is irrelevant for the determination of \aco\ as long as it is linear with the ``true'' \sigd.  Since the DGR is calibrated from the map itself, any constant term will show up in both DGR and \sigd\ and makes no impact on the assessment of \sightwo.  For example, \citet{2011ApJ...737...12L} showed that the dust optical depth at 160 \micron, determined using an assumed power-law dependence on frequency and an estimate of the dust temperature from the 70/160 ratio, works comparably well as the dust mass surface density.  \citet{Dobashi:2008jq} performed a similar study in the Large Magellanic Cloud using A$_V$ mapping to trace dust mass surface density.  Since we are also interested in the value of DGR itself, we use \sigd\ as our tracer throughout this paper. We note that uncertainties in the absolute value of \sigd, as long as they do not introduce a non-linearity within the region in question, will not affect the determination of \aco.

For this study, we make use of CO J=$(2-1)$ observations from the large program HERACLES on the IRAM 30m \citep{2009AJ....137.4670L}.  It is important to note that we are therefore determining \aco\ appropriate for that CO line.  Since most studies quote \aco\ for the CO J=$(1-0)$ line, throughout the paper we use a line ratio ($R_{21}$) to convert our measurements to the $(1-0)$ scale.  Systematic variations in $R_{21}$ will result in errors in the $(1-0)$ conversion factor, but the $(2-1)$ conversion factor will not be affected since it is what we are directly deriving. Although it is convenient to discuss \aco\ in its typical $(1-0)$ incarnation, we note that the $(2-1)$ conversion factor itself will be important for future studies with ALMA.  For nearby galaxies, at a given angular resolution, mapping in the $(2-1)$ line is more efficient than in the $(1-0)$ line.  For high redshift galaxies, the $(1-0)$ line may be shifted out of ALMA's frequency coverage. Thus, \aco\ for the $(2-1)$ line will be useful regardless of its relationship to $(1-0)$.  

This paper is laid out as follows: Section~\ref{sec:obs} presents the details of the resolved dust and gas observations we use.  In Section~\ref{sec:tech} we describe the technique to simultaneously measure DGR and \aco\ and discuss how the procedure is optimized to deal with more distant galaxies than those in the Local Group (more details on the technique can be found in the Appendix).  We present the results of performing the solution on 26 nearby galaxies in Section~\ref{sec:results} and discuss their implications for our understanding of how DGR and \aco\ vary with metallicity and other ISM properties in Section~\ref{sec:disc}.

\section{Observations}\label{sec:obs}

We make use of observations of dust and gas from a series of surveys of nearby galaxies built upon SINGS \citep[the ``Spitzer Infrared Nearby Galaxies Survey'';][]{2003PASP..115..928K}.  This includes \hi\ from ``The \hi\ Nearby Galaxies Survey'' \citep[THINGS;][]{2008AJ....136.2563W}, $^{12}$CO J=(2$-$1) from the ``HERA CO Line Emission Survey'' \citep[HERACLES;]{2009AJ....137.4670L,Leroy:2013um}and far-infrared dust emission from``Key Insights into Nearby Galaxies: A Far-Infrared Survey with Herschel'' \citep[KINGFISH;][]{2011PASP..123.1347K}.  In addition, several galaxies that were not included in THINGS have \hi\ observations from either archival or new observations.  The sample of galaxies in common between these surveys and for which we have detections of CO emission \citep[see][for details on the HERACLES non-detections]{2012AJ....143..138S} consists of 26 targets listed in Table~\ref{tab:sample}.  

The SINGS and KINGFISH surveys targeted galaxies with a variety of morphologies, located within 30 Mpc.  Due to the requirement of a CO detection for our work, all but one of the viable targets are spiral galaxies, the exception being NGC~3077 which is a starbursting dwarf.  In Table~\ref{tab:sample} we list the positions, distances, orientation parameters and the B-band isophotal radii at 25 mag arcsec$^{-2}$ ($R_{25}$) for our targets.  Throughout the text we define r$_{25}$ = $r/R_{25}$, where $r$ is the galactocentric radius in units of arcminutes.

\begin{deluxetable*}{lcccccccc}
\tablewidth{0pt}
\tabletypesize{\scriptsize}
\tablecolumns{9}
\tablecaption{Galaxy Sample}
\tablehead{ \multicolumn{1}{l}{Galaxy} &
\multicolumn{1}{c}{R.A.} &
\multicolumn{1}{c}{Dec.} &
\multicolumn{1}{c}{Distance} &
\multicolumn{1}{c}{\em{i}} &
\multicolumn{1}{c}{P.A.} &
\multicolumn{1}{c}{Morphology} &
\multicolumn{1}{c}{$R_{25}$} &
\multicolumn{1}{c}{Solution Pixel\tablenotemark{a}} \\
\multicolumn{1}{c}{} &
\multicolumn{1}{c}{(J2000)} &
\multicolumn{1}{c}{(J2000)} &
\multicolumn{1}{c}{(Mpc)} &
\multicolumn{1}{c}{($^{\circ}$)} &
\multicolumn{1}{c}{($^{\circ}$)} &
\multicolumn{1}{c}{} &
\multicolumn{1}{c}{($\arcmin$)} &
\multicolumn{1}{c}{Radius (kpc)} }
\startdata
NGC~0337                    & $00^{\rm h} 59^{\rm m} 50\fs1$ & $-07^\circ 34\arcmin 41\farcs0$ & 19.3 & 51 & 90  & SBd    & \phn1.5  &  3.5 \\
NGC~0628                    & $01^{\rm h} 36^{\rm m} 41\fs8$ & $+15^\circ 47\arcmin 00\farcs0$ & \phn7.2  & 7  & 20  & SAc    & \phn4.9  &  1.3 \\
NGC~0925                    & $02^{\rm h} 27^{\rm m} 16\fs5$ & $+33^\circ 34\arcmin 43\farcs5$ & \phn9.1 & 66 & 287 & SABd   & \phn5.4  &  1.7 \\
NGC~2841                    & $09^{\rm h} 22^{\rm m} 02\fs6$ & $+50^\circ 58\arcmin 35\farcs2$ & 14.1 & 74 & 153 & SAb    & \phn3.5  &  2.6 \\
NGC~2976                    & $09^{\rm h} 47^{\rm m} 15\fs3$ & $+67^\circ 55\arcmin 00\farcs0$ & \phn3.6 & 65 & 335 & SAc    & \phn3.6  &  0.6 \\
NGC~3077                    & $10^{\rm h} 03^{\rm m} 19\fs1$ & $+68^\circ 44\arcmin 02\farcs0$ & \phn3.8 & 46 & 45  & I0pec  & \phn2.7  &  0.7 \\
NGC~3184\tablenotemark{*}   & $10^{\rm h} 18^{\rm m} 17\fs0$ & $+41^\circ 25\arcmin 28\farcs0$ & 11.7 & 16 & 179 & SABcd  & \phn3.7  &  2.1 \\
NGC~3198                    & $10^{\rm h} 19^{\rm m} 55\fs0$ & $+45^\circ 32\arcmin 58\farcs9$ & 14.1 & 72 & 215 & SBc    & \phn3.2  &  2.6 \\
NGC~3351\tablenotemark{*}   & $10^{\rm h} 43^{\rm m} 57\fs7$ & $+11^\circ 42\arcmin 14\farcs0$ & \phn9.3 & 41 & 192 & SBb    & \phn3.6  &  1.7 \\
NGC~3521\tablenotemark{*}   & $11^{\rm h} 05^{\rm m} 48\fs6$ & $-00^\circ 02\arcmin 09\farcs2$ & 11.2 & 73 & 340 & SABbc  & \phn4.2  &  2.0 \\
NGC~3627\tablenotemark{*}   & $11^{\rm h} 20^{\rm m} 15\fs0$ & $+12^\circ 59\arcmin 29\farcs6$ & \phn9.4 & 62 & 173 & SABb   & \phn5.1  &  1.7 \\
NGC~3938                    & $11^{\rm h} 52^{\rm m} 49\fs4$ & $+44^\circ 07\arcmin 14\farcs9$ & 17.9 & 14 & 195 & SAc    & \phn1.8  &  3.3 \\
NGC~4236                    & $12^{\rm h} 16^{\rm m} 42\fs1$ & $+69^\circ 27\arcmin 45\farcs0$ & \phn4.5 & 75 & 162 & SBdm   & 12.0 &  0.8 \\
NGC~4254\tablenotemark{*}   & $12^{\rm h} 18^{\rm m} 49\fs6$ & $+14^\circ 24\arcmin 59\farcs0$ & 14.4 & 32 & 55  & SAc    & \phn2.5  &  2.6 \\
NGC~4321\tablenotemark{*}   & $12^{\rm h} 22^{\rm m} 54\fs9$ & $+15^\circ 49\arcmin 21\farcs0$ & 14.3 & 30 & 153 & SABbc  & \phn3.0  &  2.6 \\
NGC~4536\tablenotemark{*}   & $12^{\rm h} 34^{\rm m} 27\fs1$ & $+02^\circ 11\arcmin 16\farcs0$ & 14.5 & 59 & 299 & SABbc  & \phn3.5  &  2.6 \\
NGC~4569\tablenotemark{*}   & $12^{\rm h} 36^{\rm m} 49\fs8$ & $+13^\circ 09\arcmin 46\farcs0$ & \phn9.9 & 66 & 23  & SABab  & \phn4.6  &  1.8 \\
NGC~4625                    & $12^{\rm h} 41^{\rm m} 52\fs7$ & $+41^\circ 16\arcmin 26\farcs0$ & \phn9.3  & 47 & 330 & SABmp  & \phn0.7  &  1.7 \\
NGC~4631                    & $12^{\rm h} 42^{\rm m} 08\fs0$ & $+32^\circ 32\arcmin 29\farcs0$ & \phn7.6 & 85 & 86  & SBd    & \phn7.3  &  1.4 \\
NGC~4725                    & $12^{\rm h} 50^{\rm m} 26\fs6$ & $+25^\circ 30\arcmin 03\farcs0$ & 11.9 & 54 & 36  & SABab  & \phn4.9  &  2.2 \\
NGC~4736\tablenotemark{*}   & $12^{\rm h} 50^{\rm m} 53\fs0$ & $+41^\circ 07\arcmin 13\farcs2$ & \phn4.7 & 41 & 296 & SAab   & \phn3.9  &  0.8 \\
NGC~5055\tablenotemark{*}   & $13^{\rm h} 15^{\rm m} 49\fs2$ & $+42^\circ 01\arcmin 45\farcs3$ & \phn7.9 & 59 & 102 & SAbc   & \phn5.9  &  1.4 \\
NGC~5457\tablenotemark{b*} & $14^{\rm h} 03^{\rm m} 12\fs6$ & $+54^\circ 20\arcmin 57\farcs0$ & \phn6.7  & 18 & 39  & SABcd  & 12.0 &  1.2 \\
NGC~5713                    & $14^{\rm h} 40^{\rm m} 11\fs5$ & $-00^\circ 17\arcmin 21\farcs0$ & 21.4 & 48 & 11  & SABbcp & \phn1.2  &  3.9 \\
NGC~6946\tablenotemark{*}   & $20^{\rm h} 34^{\rm m} 52\fs2$ & $+60^\circ 09\arcmin 14\farcs4$ & \phn6.8  & 33 & 243 & SABcd  & \phn5.7  &  1.2 \\
NGC~7331                    & $22^{\rm h} 37^{\rm m} 04\fs0$ & $+34^\circ 24\arcmin 56\farcs5$ & 14.5 & 76 & 168 & SAb    & \phn4.6  &  2.6 
\enddata
\label{tab:sample}
\tablenotetext{a}{Solution pixels are defined in Section~\ref{sec:tech}. They are the regions in which we solve for \aco\ and DGR.}
\tablenotetext{b}{M101}
\tablenotetext{*}{CO J$=(1-0)$ maps available from Nobeyama survey of nearby galaxies \citep{Kuno:2007vn}.}
\tablecomments{Distances and morphologies from the compilation of \citet{2011PASP..123.1347K}. Orientation parameters from \cite{2008AJ....136.2563W} where possible, LEDA and NED databases otherwise.}
\end{deluxetable*}

\subsection{Dust Mass Surface Density}\label{sec:dust}

We use dust mass surface density maps derived from pixel-by-pixel modeling of the infrared (IR) spectral energy distribution (SED) observed by {\em Spitzer} and {\em Herschel} with models developed by \citet{2007ApJ...657..810D}.  A detailed description of the modeling for NGC~0628 and NGC~6946 is presented in \citet{2012ApJ...756..138A} and the full sample results are presented in Aniano et al (in prep).  The dust models include a description of the dust properties (size distribution, composition and optical properties of the grains) with a variable fraction of dust in the form of polycyclic aromatic hydrocarbons (PAHs).  The dust is illuminated by a radiation field distribution wherein a fraction of the dust is heated by a minimum radiation field U$_{\rm min}$ while the rest is heated by a power-law distribution of radiation fields extending up to U $=$10$^7$ U$_{\rm MMP}$, where U$_{\rm MMP}$ is the solar neighborhood radiation field from \citet{1983A&A...128..212M}.  The fraction of the dust mass heated by radiation fields where U$>$10$^2$  U$_{\rm MMP}$ is typically very small, so the dust mass surface density is not very sensitive to the exact value of the upper U limit.    

The resolution of the dust mass surface density map is equivalent to that of the lowest resolution IR map included in the modeling.  In order to preserve spatial resolution while still covering the peak of the dust SED, we use the dust modeling at a resolution matched to the  SPIRE 350 \micron\ map (FWHM $\sim25$\arcsec).  This limiting resolution allows us to include the following maps in the dust modeling:  IRAC 3.6, 4.5, 5.8 and 8.0 \micron; MIPS 24 and 70 \micron; PACS 70, 100 and 160 \micron; and SPIRE 250 and 350 \micron.  In theory, we could perform this analysis at even higher resolution using the SPIRE 250 resolution maps, which include all IRAC bands; MIPS 24 \micron; all PACS bands; and SPIRE 250 \micron. However, \citet{2012ApJ...756..138A} found that maps where the limiting resolution exceeds MIPS 70 \micron\ are less reliable due to the comparatively low surface brightness sensitivity of the PACS observations.  At SPIRE 250 \micron\ resolution, they find systematic errors of up to $\sim 30-40$\% in the dust mass (compared to their best estimate, which includes all IRAC, MIPS, PACS and SPIRE bands).  However, when both SPIRE 250 and 350 observations are included, the systematic errors in the dust mass are $\sim 10$\% or less.  We proceed by using the SPIRE 350 resolution dust modeling results.       

Aside from the dust mass surface density, the \citet{2012ApJ...756..138A} modeling also constrains a number of other quantities that we make use of later in interpreting the results.  These include $\overline{U}$, the average radiation field heating the dust; U$_{\rm min}$ the minimum radiation field; $f_{\rm PDR}$ the fraction of the dust luminosity that comes from dust heated by U $>100$ U$_{\rm MMP}$; and $q_{\rm PAH}$ the fraction of the dust mass accounted for by PAHs with fewer than $10^3$ carbon atoms. 

Statistical uncertainties on \sigd\ and the other derived parameters were measured by \citet{2012ApJ...756..138A} with a Monte Carlo approach. In most of the regions we consider the statistical uncertainties on the dust mass surface densities are small, but the S/N of the dust mass maps is generally a function of radius and the outskirts can have S/N $\sim5$ in some cases.  Our error estimates take these uncertainties into account and are described in Section~\ref{sec:unc}.  Possible systematic effects are discussed in Section~\ref{sec:dgrsys}. 

\subsection{HI Surface Density}\label{sec:gas}

To trace the atomic gas surface density in our targets, we use a combination of NRAO\footnote[3]{The National Radio Astronomy Observatory is a facility of the National Science Foundation operated under cooperative agreement by Associated Universities, Inc.} VLA \hi\ observations from THINGS \citep{2008AJ....136.2563W} and supplementary \hi, both new and archival, described in \citet{Leroy:2013um}.  The source and angular resolution of the \hi\ maps for our targets are listed in Table~\ref{tab:hi}.  

\begin{deluxetable}{lcccc}
\tablewidth{0pt}
\tabletypesize{\scriptsize}
\tablecolumns{5}
\tablecaption{HI Observation Summary}
\tablehead{ \multicolumn{1}{c}{Galaxy} &
\multicolumn{1}{c}{Source} &
\multicolumn{3}{c}{FWHM Beam Properties} \\
\cline{3-5}
\multicolumn{1}{c}{} &
\multicolumn{1}{c}{} & 
\multicolumn{1}{c}{Major ($\arcsec$)} &
\multicolumn{1}{c}{Minor ($\arcsec$)} &
\multicolumn{1}{c}{P.A. ($^{\circ}$)}}
\startdata
NGC~0337 &  Archival     &  20.2   & 13.0  & \phantom{$-$}160.3 \\
NGC~0628 &  THINGS    &  11.9   &  \phn9.3   & \phn$-$70.3	 \\
NGC~0925 &  THINGS    &   \phn5.9    &  \phn5.7   &  \phantom{$-0$}30.6	 \\
NGC~2841 &  THINGS    &  11.1   &  \phn9.4   & \phn$-$12.3	 \\
NGC~2976 &  THINGS    &   \phn7.4    &  \phn6.4   &  \phantom{$-0$}71.8	 \\
NGC~3077 &  THINGS    &  14.3   & 13.2  &  \phantom{$-0$}60.5	 \\
NGC~3184 &  THINGS    &   \phn7.5    &  \phn6.9   &  \phantom{$-0$}85.4	 \\
NGC~3198 &  THINGS    &  11.4   &  \phn9.4   & \phn$-$80.4	 \\
NGC~3351 &  THINGS    &   \phn9.9    &  \phn7.1   &  \phantom{$-0$}24.1	 \\
NGC~3521 &  THINGS    &  14.1   & 11.2  & \phn$-$61.7	 \\
NGC~3938 &  Archival     &  18.5   & 18.2  & \phantom{$-0$}48.6 \\
NGC~3627 &  THINGS    &  10.0   &  \phn8.9   & \phn$-$60.9	 \\
NGC~4236 &  New     &  16.7   & 13.9  &  \phantom{$-0$}69.6	 \\
NGC~4254 &  Archival     &  16.9   & 16.2  &  \phantom{$-0$}54.4	 \\
NGC~4321 &  Archival     &  14.7   & 14.1  & \phantom{$-$}163.4	 \\
NGC~4536 &  Archival, New     &  14.7   & 13.8  & \phn$-$11.4	 \\
NGC~4569 &  Archival     &  14.2   & 13.9  &  \phantom{$-0$}32.9	 \\
NGC~4625 &  Archival     &  13.0   & 12.5  & \phn$-$29.2	 \\
NGC~4631 &  Archival     &  14.9   & 13.3  & \phantom{$-$}178.1	 \\
NGC~4725 &  Archival, New     &  18.6   & 17.0  & \phn$-$20.9 \\
NGC~4736 &  THINGS    &  10.2   &  \phn9.1   & \phn$-$23.0	 \\
NGC~5055 &  THINGS    &  10.1   &  \phn8.7   & \phn$-$40.0	 \\
NGC~5457 &  THINGS    &  10.8   & 10.2  & \phn$-$67.1	 \\
NGC~5713 &  Archival     &  15.5   & 14.9  & \phantom{$-$}121.9	 \\
NGC~6946 &  THINGS    &   \phn6.0    &  \phn5.6   &   \phantom{$-00$}6.6	 \\
NGC~7331 &  THINGS    &   \phn6.1    &  \phn5.6   &  \phantom{$-0$}34.3	 
\enddata
\tablecomments{Galaxies with new HI data were observed in VLA project AL735 (P.I. A. Leroy).}\label{tab:hi}
\end{deluxetable}

The \hi\ maps are converted from integrated intensities to column densities as in Equation~5 of \citet{2008AJ....136.2563W}.  We convolve each map with an elliptical Gaussian kernel determined by its individual beam properties to produce a circular Gaussian point spread function (PSF).  We then use kernels created following the procedures in \citet{2011PASP..123.1218A} to convolve the circular Gaussian to match the PSF at SPIRE 350 \micron. For the \hi\, we assume the uncertainties to be the larger of either 0.5 \msun\ pc$^{-2}$ or 10\% of the measured column density.  Systematic uncertainties from \hi\ opacity effects are discussed in Section~\ref{sec:acosys}.  

\subsection{CO Integrated Intensity}
To trace the molecular gas distribution in our targets we use CO J=(2$-$1) mapping from the HERACLES survey \citep{2009AJ....137.4670L}.  Integrated intensity maps were generated from the CO spectral cubes by integrating the spectra over a range in velocity around either 1) the detected CO line in that spectrum or 2) the expected CO velocity predicted from the \hi\ velocity (since \hi\ is detected at high S/N in almost all relevant pixels).  We propagate uncertainties through these masking steps, creating in the end an integrated CO line map and an uncertainty map. The HERACLES maps have PSFs well approximated by a circular Gaussian with FWHM of $13.4\arcsec$.  We convolve the maps with kernels constructed using the techniques described by \citet{2011PASP..123.1218A} to match the resolution of the SPIRE 350 maps.  We have tested the effect of error beams (i.e. extended wings of the PSF or stray light pick-up of the IRAM 30-m) on the HERACLES maps and find that the effect is less than $\sim5$\% for all galaxies and closer to 1\% for most.

Our measurements of \aco\ directly determine the conversion factor appropriate for CO J=$(2-1)$. However, to compare with the standard CO J=$(1-0)$ factor, which is more frequently used, we convert to $(1-0)$ using a fixed value of the line ratio $R_{21}$=$(2-1)/(1-0) = 0.7$. Due to revised telescope efficiencies, the $R_{21}$ we use differs slightly from what was used in \citet{2009AJ....137.4670L}.  The $R_{21}$ we assume was found to be an appropriate average for the HERACLES sample (Rosolowsky et al., in prep; note that we find good agreement with this $R_{21}$ by comparing the HERACLES measurements with published CO J=$(1-0)$ measurements as described in the Appendix).  We discuss the effects of assuming a fixed $R_{21}$ on our results for the $(1-0)$ conversion factor in Section~\ref{sec:acosys}.  To apply the \aco\ we report in the following sections to CO J=$(2-1)$ observations, its value should be divided by 0.7.

\subsection{Ancillary Datasets}\label{sec:ancillary}

\subsubsection{Metallicity}\label{sec:metals}

In the analysis presented in Section~\ref{sec:results} we study the variations of \aco\ and DGR as a function of metallicity.  Wherever possible, we make use of the metallicity measurements from \citet[hereafter M10]{2010ApJS..190..233M} who derived characteristic metallicities as well as radial gradients in oxygen abundance for the SINGS sample.  M10 present results using two different calibrations for the strong line abundances---from \citet[KK04]{2004ApJ...617..240K} and \citet[PT05]{Pilyugin:2005kn}.  Both calibrations are considered in the following analysis.

Our preferred metallicity measurement is a radial gradient from \hii\ region metallicities (M10, Table~8).  For several galaxies no radial gradient measurement is available, so we use a fixed metallicity for the entire galaxy equal to the ``characteristic metallicity'' (M10, Table~9).  In the case of NGC~4236 and 4569, the only metallicity measurements available are from the B-band luminosity-metallicity ($L-Z$) relationship and we use those values from M10 with no gradient.  Finally, two of our galaxies are not in the M10 sample.  For NGC~3077 we use a metallicity of $12+$Log(O/H)$=8.9$ in KK04 with no gradient, from \citet{1994ApJ...429..582C}. To obtain a PT05 measurement for NGC~3077, we subtract 0.6 dex, the average offset between the two calibrations found by M10.  For NGC~5457 (a.k.a. M\,101), a galaxy with a well-known radial metallicity gradient, we use the measurements from \cite{2007ApJ...656..186B}.  These metallicities are from direct methods, so are not on the same scale as either PT05 or KK04.  The metallicities and gradients we use are listed in Table~\ref{tab:metals}.  These are converted to match the r$_{25}$ we adopt in this work, which can be slightly different from that adopted by M10. In order to compare with the local MW, we use the metallicity of the Orion Nebula \hii\ region in the PT05 and KK04 calibrations, i.e. 12+Log(O/H)=8.5 for PT05 and 8.8 for KK04, which is obtained by applying the strong-line metallicity calibrations to the integrated spectrum of Orion from integral field spectroscopy\footnote{Data available at \url{http://www.caha.es/sanchez/orion/}} \citep{Sanchez:2007iy}.

\begin{deluxetable*}{cccccl}
\tablewidth{0pt}
\tabletypesize{\scriptsize}
\tablecolumns{6}
\tablecaption{Adopted Metallicities and Gradients}
\tablehead{\multicolumn{1}{c}{} &
\multicolumn{1}{c}{PT05} &
\multicolumn{1}{c}{PT05} & 
\multicolumn{1}{c}{KK04} &
\multicolumn{1}{c}{KK04} & 
\multicolumn{1}{l}{} \\
\multicolumn{1}{c}{Galaxy} &
\multicolumn{1}{c}{Central Metallicity} &
\multicolumn{1}{c}{Metallicity Gradient} & 
\multicolumn{1}{c}{Central Metallicity} &
\multicolumn{1}{c}{Metallicity Gradient} & 
\multicolumn{1}{c}{Source} \\
\multicolumn{1}{c}{} &
\multicolumn{1}{c}{(12 + Log(O/H))} &
\multicolumn{1}{c}{(dex r$_{25}^{-1}$)} &
\multicolumn{1}{c}{(12 + Log(O/H))} &
\multicolumn{1}{c}{(dex r$_{25}^{-1}$)} &
\multicolumn{1}{c}{} }
\startdata
NGC~0337  &  8.18 $\pm$ 0.07  &  \nodata                    & 8.84 $\pm$ 0.05 &  \nodata                      &  M10 Table 9 \\ 
NGC~0628  &  8.43 $\pm$ 0.02  &  $-$0.25 $\pm$ 0.05 & 9.19 $\pm$ 0.02 & $-$0.54 $\pm$ 0.04   &  M10 Table 8 \\    
NGC~0925  &  8.32 $\pm$ 0.01  &  $-$0.21 $\pm$ 0.03 & 8.91 $\pm$ 0.01 & $-$0.43 $\pm$ 0.02   &  M10 Table 8 \\   
NGC~2841  &  8.72 $\pm$ 0.12  &  $-$0.54 $\pm$ 0.39 & 9.34 $\pm$ 0.07 & $-$0.36 $\pm$ 0.24   &  M10 Table 8 \\   
NGC~2976  &  8.36 $\pm$ 0.06  &  \nodata                    & 8.98 $\pm$ 0.03 &  \nodata             &  M10 Table 9 \\ 
NGC~3077  &  8.30 $\pm$ 0.20  &  \nodata                    & 8.90 $\pm$ 0.20 &  \nodata             &  K11 \\       
NGC~3184  &  8.65 $\pm$ 0.02  &  $-$0.46 $\pm$ 0.06 & 9.30 $\pm$ 0.02 & $-$0.52 $\pm$ 0.05   &  M10 Table 8 \\   
NGC~3198  &  8.49 $\pm$ 0.04  &  $-$0.38 $\pm$ 0.11 & 9.10 $\pm$ 0.03 & $-$0.50 $\pm$ 0.08   &  M10 Table 8 \\   
NGC~3351  &  8.69 $\pm$ 0.01  &  $-$0.27 $\pm$ 0.04 & 9.24 $\pm$ 0.01 & $-$0.15 $\pm$ 0.03   &  M10 Table 8 \\   
NGC~3521  &  8.44 $\pm$ 0.05  &  $-$0.12 $\pm$ 0.25 & 9.20 $\pm$ 0.03 & $-$0.52 $\pm$ 0.15   &  M10 Table 8 \\   
NGC~3627  &  8.34 $\pm$ 0.24  &  \nodata                    & 8.99 $\pm$ 0.10 &  \nodata             &  M10 Table 9 \\ 
NGC~3938  &  8.42 $\pm$ 0.20  &  \nodata                    & 9.06 $\pm$ 0.20 &  \nodata             &  M10 L$-$Z \\   
NGC~4236  &  8.17 $\pm$ 0.20  &  \nodata                    & 8.74 $\pm$ 0.20 &  \nodata             &  M10 L$-$Z \\   
NGC~4254  &  8.56 $\pm$ 0.02  &  $-$0.35 $\pm$ 0.08 & 9.26 $\pm$ 0.02 & $-$0.39 $\pm$ 0.06   &  M10 Table 8  \\  
NGC~4321  &  8.61 $\pm$ 0.07  &  $-$0.31 $\pm$ 0.17 & 9.29 $\pm$ 0.04 & $-$0.28 $\pm$ 0.11   &  M10 Table 8  \\  
NGC~4536  &  8.21 $\pm$ 0.08  &  \nodata                    & 9.00 $\pm$ 0.04 &  \nodata             &  M10 Table 9 \\ 
NGC~4569  &  8.58 $\pm$ 0.20  &  \nodata                    & 9.26 $\pm$ 0.20 &  \nodata             &  M10 L$-$Z \\   
NGC~4625  &  8.35 $\pm$ 0.17  &  \nodata                    & 9.05 $\pm$ 0.07 &  \nodata             &  M10 Table 9 \\ 
NGC~4631  &  8.12 $\pm$ 0.11  &  \nodata                    & 8.75 $\pm$ 0.09 &  \nodata             &  M10 Table 9 \\ 
NGC~4725  &  8.35 $\pm$ 0.13  &  \nodata                    & 9.10 $\pm$ 0.08 &  \nodata             &  M10 Table 9 \\ 
NGC~4736  &  8.40 $\pm$ 0.01  &  $-$0.23 $\pm$ 0.12 & 9.04 $\pm$ 0.01 & $-$0.08 $\pm$ 0.10   &  M10 Table 8 \\   
NGC~5055  &  8.59 $\pm$ 0.07  &  $-$0.59 $\pm$ 0.27 & 9.30 $\pm$ 0.04 & $-$0.51 $\pm$ 0.17   &  M10 Table 8 \\   
NGC~5457  &  8.75 $\pm$ 0.05  &  $-$0.75 $\pm$ 0.06 & 8.75 $\pm$ 0.05 & $-$0.75 $\pm$ 0.07   &  B07 \\   
NGC~5713  &  8.48 $\pm$ 0.10  &  \nodata                    & 9.08 $\pm$ 0.03 &  \nodata             &  M10 Table 9 \\ 
NGC~6946  &  8.45 $\pm$ 0.06  &  $-$0.17 $\pm$ 0.15 & 9.13 $\pm$ 0.04 & $-$0.28 $\pm$ 0.10   &  M10 Table 8 \\   
NGC~7331  &  8.41 $\pm$ 0.06  &  $-$0.21 $\pm$ 0.31 & 9.18 $\pm$ 0.05 & $-$0.49 $\pm$ 0.25   &  M10 Table 8    
\enddata
\label{tab:metals}
\tablerefs{\citet[M10]{2010ApJS..190..233M}, \citet[K11]{2011PASP..123.1347K}, \citet[B07]{2007ApJ...656..186B}.}
\end{deluxetable*}

\subsubsection{Star Formation Rate and Stellar Mass Surface Density Maps}

The star formation rate (SFR) surface densities ($\Sigma_{\rm SFR}$) are calculated from H$\alpha$ and 24 \micron\ maps using the H$\alpha$ maps and procedure described in \citet{2012AJ....144....3L}. The H$\alpha$ maps have been convolved to match the SPIRE 350 \micron\ PSF assuming an initial $\sim1-2$\arcsec\ FWHM Gaussian PSFs, although the large difference between the initial and final PSF makes this choice mostly irrelevant.  We use 24 \micron\ maps from SINGS, processed (background subtracted, convolved and aligned) as described in \citet{2012ApJ...756..138A}.  The \citet{2012ApJ...756..138A} modeling results described in Section~\ref{sec:dust} are also used to remove a non-star-formation related cirrus component from the 24 \micron\ map as described in \citet{2012AJ....144....3L}.  

We calculate the stellar mass surface density ($\Sigma_{*}$) from the IRAC 3.6 \micron\ observations from SINGS, as described in \citet{2008AJ....136.2782L}.  This provides only a rough tracer of stellar mass surface density, since we do not take into account corrections for various contaminants in the 3.6 \micron\ band \citep{Zibetti:2011br,2012ApJ...744...17M}.   

%We proceed as described in Section 8.2 of that paper by using a U$_{\rm cirrus} = 0.5$ U$_{\rm min}$ along with the  measured values of \sigd\ and q$_{\rm PAH}$ from Aniano et al. (2012a, in prep) to determine the cirrus dust emissivity at 24 \micron.  

\subsection{Processing}

After all maps have been convolved to SPIRE 350 \micron\ resolution, we sample them with a hexagonal grid with approximately half-beam spacings (i.e. $12.5\arcsec$).  Uncertainties on the dust mass surface density, CO integrated intensity and \hi\ column density are propagated through the necessary convolutions and samplings.  Surface densities and other quantities have been deprojected using the orientation parameters listed in Table~\ref{tab:sample}.

\section{Solving Simultaneously for \aco\ and DGR}\label{sec:tech}

In order to use the dust mass surface density to trace the total gas mass surface density, we assume 1) that dust and gas are well mixed (i.e. that Equation~\ref{eq:dgr} holds), 2) that the DGR is constant on $\sim$kpc scales in our target galaxies and 3) within a given $\sim$kpc region, the DGR does not change between the atomic and molecular phases.  Then, given multiple measurements of \sigd, \sighi\ and \ico\ that span a range of CO/\hi\ values in a kpc-scale region of a galaxy, we can adjust \aco\ till we find the value that returns the most uniform DGR for the region.  This procedure makes no assumption about the value of the DGR, only that it is constant in that region.  In addition, it makes no assumption about the value of \aco\ or the scales on which it varies.  If \aco\ varies on small scales, our measured \aco\ value will represent the dust (or gas) mass surface density weighted average \aco\ for the region.  We discuss this process in detail in the following sections.  

A simultaneous solution for \aco\ and DGR can only be performed if a range of CO/\hi\ ratios are present in the measurements.  The linear dependence of $\Sigma_{\rm H_2}$ on \aco\ provides leverage to adjust \aco\ in order to best describe all of the points with the same DGR over the range of CO/\hi\ ratios.  An ``incorrect'' \aco\ value will cause a dependence of the measured DGR on CO/HI, increasing the spread in DGR values in the region.  An illustration of this effect is shown in panel c) of Figure~\ref{fig:techillustration}.  Finding a solution or best-fit \aco\ is equivalent to locating a minimum in the DGR scatter at a given value of \aco.    

The basic procedure we use, minimizing the scatter in the DGR values in the region, was suggested in \citet{2011ApJ...737...12L}. There are, however, a variety of other techniques to solve for \aco\ and DGR given multiple measurements, including directly fitting a plane to \ico, \sigd\ and \sighi. It is not clear {\it a priori} which scatter minimization technique is optimal, so we performed a set of simulations, described in the Appendix, to test various techniques and optimize the procedure for our dataset and objectives.  We describe the resulting scatter minimization procedure in more detail below. 

\subsection{Defining the ``Solution Pixel''}

\begin{figure*}
\centering
\epsscale{1.1}
\plotone{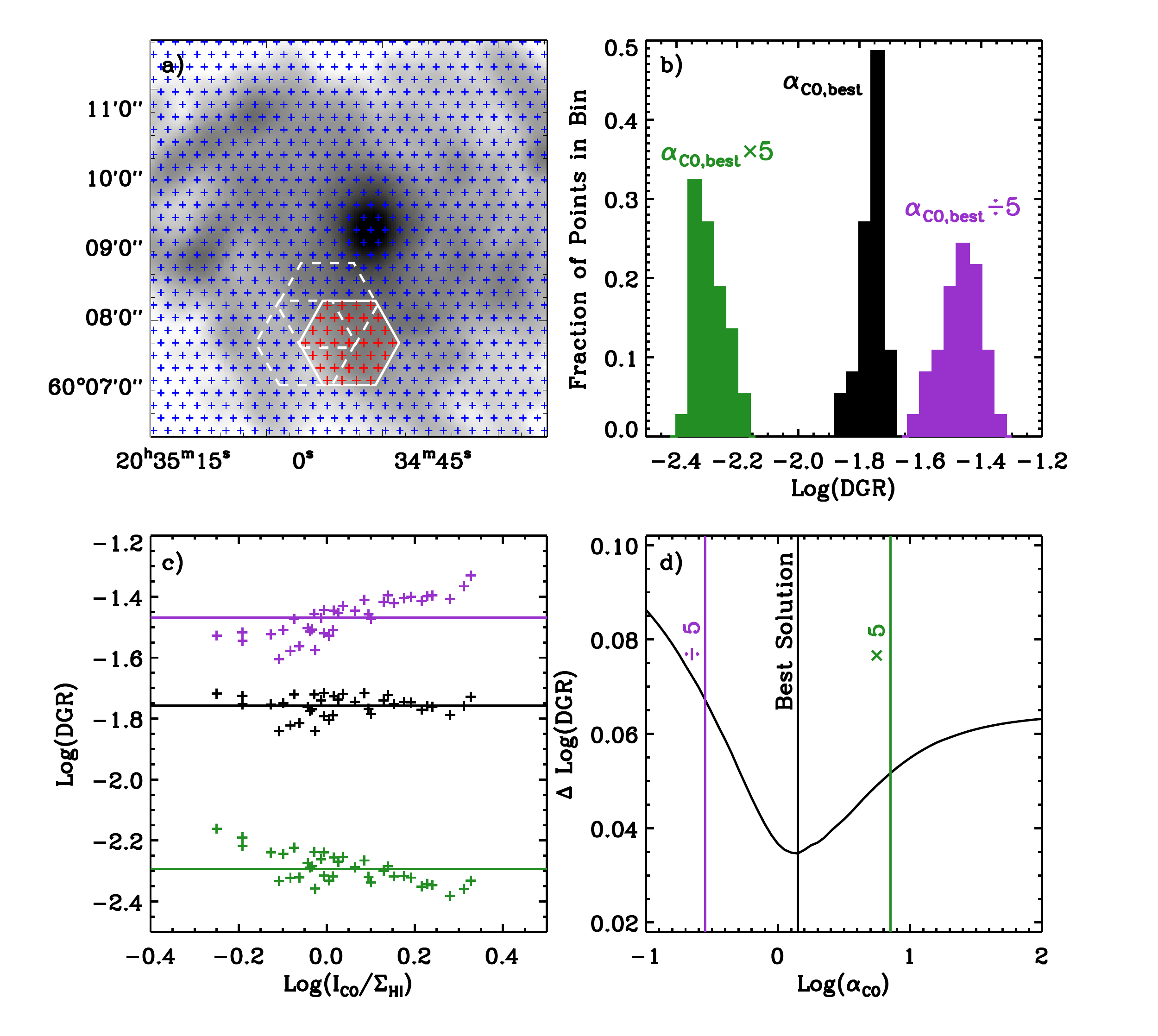}
\caption{Illustration of the technique to determine DGR and \aco\ for a single solution pixel in NGC~6946.  Panel a) shows a portion of the CO map from HERACLES overlaid with the half-beam spaced sampling grid.  The solution pixel in question is shown with a solid white hexagon and the 37 individual sampling points included in the solution are shown in red.  Two neighboring solution pixels are also highlighted with a dashed white line to show how the pixels are arranged and that neighboring solution pixels share $\sim40$\% of their sampling points. Panel b) illustrates that the scatter in the DGR changes as a function of \aco.  Here we have plotted histograms of the measured DGR$_i$ values in this solution pixel at three different values of \aco\ (the optimal value in black and a factor of 5 above and below this value in green and purple, respectively). Panel c) illustrates that this scatter originates in variation of DGR as a function of CO/\hi\ when \aco\ is not at the optimal value.   We show this by plotting the 37 DGR$_i$ values as a function of CO/\hi\ ratio for the same three \aco\ values shown in panel b).  The horizontal lines indicate the mean DGR for each set of points.  The slope in DGR vs CO/\hi\ space is minimized at the optimal \aco.  Finally, in panel d) we show the scatter in the DGR at each value of the full \aco\ grid in black.  The three highlighted \aco\ values are marked with vertical lines. Panel d) highlights the fact that the DGR scatter is minimized at the best fit Log(\aco)$=0.15\pm0.22$ in this region.  The minimization of the scatter in Log(DGR), as shown in panel d), is the technique we have determined to be the most effective for determining \aco\ and DGR, using tests that are described in detail in the Appendix.}
\label{fig:techillustration}
\end{figure*}

To perform the solution, we require multiple measurements of dust and gas tracers in the region.  We also aim, however, to select the smallest possible regions, in order to ensure an approximately constant DGR.  We proceed by dividing each target galaxy into hexagonal regions encompassing 37 of the half-beam spaced sampling points.  We call these regions ``solution pixels'' (see panel a) of Figure~\ref{fig:techillustration} for an example). The 37-point solution pixels are a compromise between small region sizes and having a sufficient number of independent measurements to minimize statistical noise.  The solution pixels tile the galaxy with center-to-center spacing of 37.5$\arcsec$.  Thus, neighboring solution pixels are not independent and share $\sim 40$\% of their sampling points.  The overlap between solution pixels is illustrated in Figure~\ref{fig:techillustration}.  Such a tiling is optimal because it fully samples the data.  To ensure that the final results are not dependent on the placement of the solution pixels we performed a test where solution pixels were distributed randomly throughout the galaxies and compared the resulting radial trends in \aco\ to what we measured using the fixed grid described above.  The radial trends were in agreement, demonstrating that our results are insensitive to the exact placement of the solution pixel grid. The solution pixels correspond to physical scales ranging from $\sim$0.6\,kpc to 3.9\,kpc. We have tiled each galaxy with solution pixels out to the maximum value of r$_{25}$ contained in the HERACLES map.

\subsection{Minimizing the DGR Scatter} 

For each sampling point $i$ in the solution pixel, the measurements of \sigd$_{,i}$, \sighi$_{,i}$\ and \ico$_{,i}$\ along with an assumed \aco\ determine DGR$_i$.  We step through a grid of \aco\ values to find the \aco\ that results in the most uniform DGR for all DGR$_i$ in the solution pixel.  In all solutions presented here, we use an \aco\ grid with 0.05 dex spacing, spanning the range \aco\ $=0.1-100$ \acounit.  

We determine the most ``uniform'' DGR in a solution pixel by minimizing the scatter in the DGR$_i$ values as a function of \aco. The scatter is measured with a robust estimator of the standard deviation\footnote[4]{We use the IDL implementation of Tukey's biweight mean \citep{2002nrc..book.....P} {\tt biweight\_mean.pro}.} of the logarithm of the DGR$_i$ values---this technique appears to work best because outliers have little effect on the measured scatter because of both the logarithmic units and the outlier suppression.  After measuring the scatter in the DGR at every \aco\ value, we find the \aco\ at which the scatter ($\Delta$Log(DGR)) is minimized.  This is taken to be our best fit \aco\ value for the region.  We consider a solution to be found when a minimum has been located in the DGR scatter within the range of our \aco\ grid.  This does not happen in every solution pixel---some do not have sufficient CO/\hi\ contrast, others have too low S/N and for some the minimum is at the edge of the allowable range and the solution is not well-constrained.  The failed solutions are not included in our further analysis.

An illustration of the technique is shown in Figure~\ref{fig:techillustration} for a region in NGC~6946.  In panel a) we show the HERACLES CO J=(2$-$1) map of the galaxy with our half-beam sampling grid, the hexagonal region shows the ``solution pixel'' in question, which includes 37 individual samples from the maps.  Panels b) and c) show how varying \aco\ affects the mean DGR and scatter for the points in the region, illustrating how the scatter increases away from the best \aco\ value. Panel d) shows the scatter as a function of \aco\ for the whole \aco\ grid. A clear minimum exists for this solution pixel at \aco$\sim1.4$ \acounit.  

\subsection{Statistical Uncertainties on \aco\ and DGR}\label{sec:unc}

We judge the uncertainties on the ``best fit'' \aco\ and DGR in several ways.  First, to take into account statistical errors, we perform a Monte Carlo test on the solutions by adding random noise to our measured \sigd, \sighi\ and \ico\ values according to each point's measurement errors.  We repeat the solution with the randomly perturbed data values 100 times and find the standard deviation of the results.  We also perform a ``bootstrapping'' trial, which tests the sensitivity of each solution to individual measurements.  In each bootstrap iteration for a given solution pixel, we randomly select 37 sampling points, with replacement, and derive the solution.  The bootstrap procedure is repeated 100 times for each solution pixel and we measure the resulting standard deviation of the \aco\ values. The standard deviations from the Monte Carlo and bootstrapping iterations are added in quadrature to produce the final quoted error for the \aco\ values we determine.  In addition, to check these uncertainties we also estimate scatter and bias in \aco\ for the given technique from our simulated data trials described in the Appendix, based on the median CO S/N in the solution pixel and the measured minimum of $\Delta$Log(DGR).  The uncertainties from Monte Carlo plus bootstrapping are comparable to what we expect given the simulated data trials.  

\subsection{Systematic Uncertainties on \aco}\label{sec:acosys}

\subsubsection{Uncertainties on \aco\ from $R_{21}$ Variations}\label{sec:r21} 

In the following, we report \aco\ appropriate for the $(1-0)$ line, since it is the canonical CO-to-H$_2$ conversion factor that most observational and theoretical studies utilize.  To do so we have converted between $(2-1)$ (which we have directly measured) and $(1-0)$ using a fixed line ratio $R_{21} = 0.7$.  Deviations from this $R_{21}$ value will result in systematic offsets in the $(1-0)$ conversion factor, while the $(2-1)$ conversion factor will be unaffected since it is what we have directly measured.  To quantify any $(1-0)$ \aco\ offsets, we have investigated the variability of $R_{21}$ in those galaxies with publicly available CO J=$(1-0)$ maps from the Nobeyama survey of nearby spiral galaxies \citep{Kuno:2007vn}.  Galaxies that have Nobeyama maps are marked with an asterisk in Table~\ref{tab:sample}.  The details of this comparison can be found in the Appendix.  We find that deviations from $R_{21}=0.7$ can cause small systematic shifts in \aco\ appropriate for the $(1-0)$ line, but the magnitude of the shifts are generally within the uncertainties on the \aco\ solutions (i.e. typically less than 0.2 dex).  We note that variations of $R_{21}$ within a pixel would introduce additional systematic uncertainties on \aco.

\subsubsection{Variations of \sigd\ Linearity Within Solution Pixels} 

We assume that the dust tracer we employ (\sigd) linearly tracks the true dust mass surface density in a solution pixel.  Because we calibrate DGR based on the values of \sigd\ within each pixel, any multiplicative constant term cancels out in Equation~\ref{eq:dgr} and does not affect the measurement of \aco.  Non-linearities in \sigd\ that are uncorrelated with atomic/molecular phase add scatter to our measurements of \aco\ but do not introduce systematic errors.  In the following we discuss several sources of non-linearity in \sigd\ that are correlated with ISM phase and estimate their systematic error contribution.  

\begin{itemize}
\item {\bf Variation of dust emissivity:} A variety of observations have suggested that dust emissivity increases in molecular gas relative to atomic gas (note, however, that most of the studies use CO to trace molecular gas and may interpret variations in \aco\ as changes in emissivity).  Recent work has suggested that the dust emissivity increases by a factor of $\sim2$ between the atomic and molecular ISM \citep{2011A&A...536A..25P,2012ApJ...751...28M}.  If the dust in molecular regions has a higher emissivity, \sigd\ will overestimate the amount of dust there, causing us to overestimate the amount of gas.  In that case, we would recover a higher \aco\ than actually exists.  As a first approximation, our measured \aco\ would be too high by the change in emissivity between atomic and molecular phases, a factor of $\sim2$ based on the previously mentioned results.     

\item {\bf Variation of DGR:}  Evidence from the depletion of gas phase metals in the Milky Way suggests that DGR increases as a function of the H$_2$ fraction.  To estimate the magnitude of such effects we use the results of \citet{2009ApJ...700.1299J}.  From the minimum level of depletion measured in the Milky Way to complete depletion of all heavy elements, the DGR varies by a factor of 4.  A large fraction of this change in DGR comes from the depletion of oxygen, however, which may not predominantly be incorporated into dust as it is depleted \citep[see discussion in Section 10.1.4 of][]{2009ApJ...700.1299J}. Excluding oxygen, the possible change in DGR is a factor of 2.  Using the correlation between depletion and H$_2$ fraction from Figure 16 of \citet{2009ApJ...700.1299J}, we find that for 10\% to 100\% H$_2$ fractions (as are appropriate for our regions), the possible variation in DGR is a factor of 2 (or less depending on the contribution of oxygen).  As in the case of dust emissivity variations, the effect of DGR increasing in molecular gas would be to artificially increase our measured \aco\ by the same factor as the DGR increases.

\item {\bf Systematic biases in measuring \sigd\ from SED modeling:}  Because warm dust will radiate more strongly per unit mass than cold dust at all wavelengths, the SED will not clearly reflect the presence of cold dust unless it dominates the mass. This means that the SED fitting technique is not sensitive to cold dust contained in GMC interiors (A$_V\gtrsim 1$) at our spatial resolution.  The fraction of the dust mass in these interiors, assuming a spherical cloud with uniform density and total A$_V \approx 8$ mag, is $\sim40\%$ \citep[this estimate agrees well with recent extinction mapping measurements of Milky Way GMCs:][]{2011A&A...536A..48K,2011A&A...535A..16L}.  If we underestimate the mass of dust by missing cold dust in GMC interiors, we would underestimate the amount of molecular gas and adjust \aco\ downwards.  The magnitude of this effect is at the factor of $\sim2$ level and opposite in direction to what we expect for dust emissivity or DGR increases in molecular clouds.
\end{itemize}

To summarize, variations of DGR and dust emissivity between atomic/molecular gas could both bias our \aco\ results towards higher values by factors of $\sim2$.  Systematic biases in accounting for cold dust in the SED modeling act in the opposite direction (i.e. biasing \aco\ towards lower values) also by a factor of $\sim2$. 

\subsubsection{Opaque H~I}

The \hi\ maps we use have not been corrected for any optical depth effects \citep{2008AJ....136.2563W}.  \hi\ observations of M~31 at high spatial and spectral resolution have suggested there may be large local opacity corrections on 50 pc scales \citep{2009ApJ...695..937B}. To estimate the importance of any opaque \hi, we have used the corrected and uncorrected maps of M~31, provided to us by R. Braun.  Convolving to 500 pc spatial resolution, the average resolution element in M~31 has a 20\% correction to the \hi\ column density.  Choosing only regions with N$_H > 10^{21}$ cm$^{-2}$, the average correction is $\sim$30\%.  Essentially all resolution elements have opacity corrections less than a factor of 2. 

If opaque \hi\ exists at the level \citet{2009ApJ...695..937B} have found in M~31, it would have two main effects on our solutions for \aco. First, on average the optically-thin estimate for the atomic gas mass would be too low, resulting in our procedure determining a DGR that is too high (excess dust compared to the amount of gas).  In the molecular regions, then, we will expect too much gas based on that same DGR, and consequently artificially increase \aco.  Second, since the opaque \hi\ features do not appear to be spatially associated with molecular gas \citep[see][section 4.1, for further discussion]{2009ApJ...695..937B}, they will act as a source of intrinsic scatter in the DGR.  In the Appendix, we explore the effect of intrinsic scatter on our solution technique and at the level of opaque \hi\ in M~31, we do not find appreciable bias in the recovered \aco\ due to scatter. We expect the magnitude of the systematic effects due to opaque \hi, if it exists, to be well within the statistical uncertainties we achieve on the \aco\ measurements.

\subsection{Systematic Uncertainties on DGR}\label{sec:dgrsys}

\subsubsection{Absolute Calibration of \sigd}

As we have discussed above, as long as \sigd\ is a linear tracer of true dust mass surface density within a given solution pixel, its absolute calibration has no effect on the \aco\ value we measure.  The same is not true for the DGR value.  Any uncertainties on the calibration of \sigd\ will be directly reflected in the DGR measurement.  The \sigd\ values we used are from fits of the \citet{2007ApJ...657..810D} models to the IR SED using the Milky Way $R_V = 3.1$ grain model.  The extent to which the appropriate dust emissivity $\kappa_{\nu}$ deviates from the value used by this model represents a systematic uncertainty on the DGR values we derive.  Our knowledge of $\kappa_{\nu}$ in different environments is limited, but there are constraints from observations of dust extinction curves and depletions in the Large and Small Magellanic Clouds \citep[c.f.][]{2001ApJ...548..296W} where measured $R_V$ can deviate significantly from the canonical 3.1 value.  \citet{2007ApJ...663..866D} demonstrated that the \sigd\ values decreased by a factor of $\sim1.2$ when the LMC or SMC dust model was used instead of the MW $R_V=3.1$ model.  Given that our sample is largely dominated by spiral galaxies and hence does not probe environments with metallicity comparable to the SMC (due to the faintness of CO in such regions and our S/N limitations), we expect that the systematic uncertainties on our DGR when comparing with other results from \citet{2007ApJ...657..810D} model fits is small. It is important to note, however, that different dust models, even fit to the same $R_V=3.1$ extinction curve have systematic offsets in their dust mass predictions due to different grain size distributions, grain composition, etc. Therefore, comparison of our DGR values to results from studies not using the \citet{2007ApJ...657..810D} models will have systematic offsets. 
 
\section{Results}\label{sec:results}

\subsection{NGC 0628 Results Example}

We divided each of the 26 galaxies in our sample into solution pixels and performed the simultaneous solution for DGR and \aco\ in each pixel.  As an example, we present the results for NGC~0628 in the following section.  The results for all solution pixels in all galaxies can be seen in the Appendix.

Figure~\ref{fig:ngc0628_panel1} shows, from left to right, the \hi, CO and \sigd\ maps used in our analysis.  The circles overlaid on the map represent the centers of the solution pixels we have defined.  Figure~\ref{fig:ngc0628_panel2} shows the same circles representing the solution pixel centers, now filled in with a color representing the best \aco\ solution for that pixel on the left, a gray scale showing the uncertainty on that \aco\ solution in the middle, and the DGR on the right.  Finally, in Figure~\ref{fig:ngc0628_panel3} we show these measured \aco\ values as a function of galactocentric radius (r$_{25}$).  For comparison, Figure~\ref{fig:ngc0628_panel3} also shows the local Milky Way \aco$=4.4$ \acounit\ with a solid horizontal line (dotted lines show a factor of 2 above and below, see Section~\ref{sec:mwaco} for details on how the Milky Way value has been measured).  We note that the Milky Way may show a gradient of \aco\ with radius (also discussed in Section~\ref{sec:mwaco}), but for purposes of comparison with the most widely-used conversion factor we use a constant \aco\ on all plots.  

\begin{figure*}
\centering
\epsscale{1.1}
\plotone{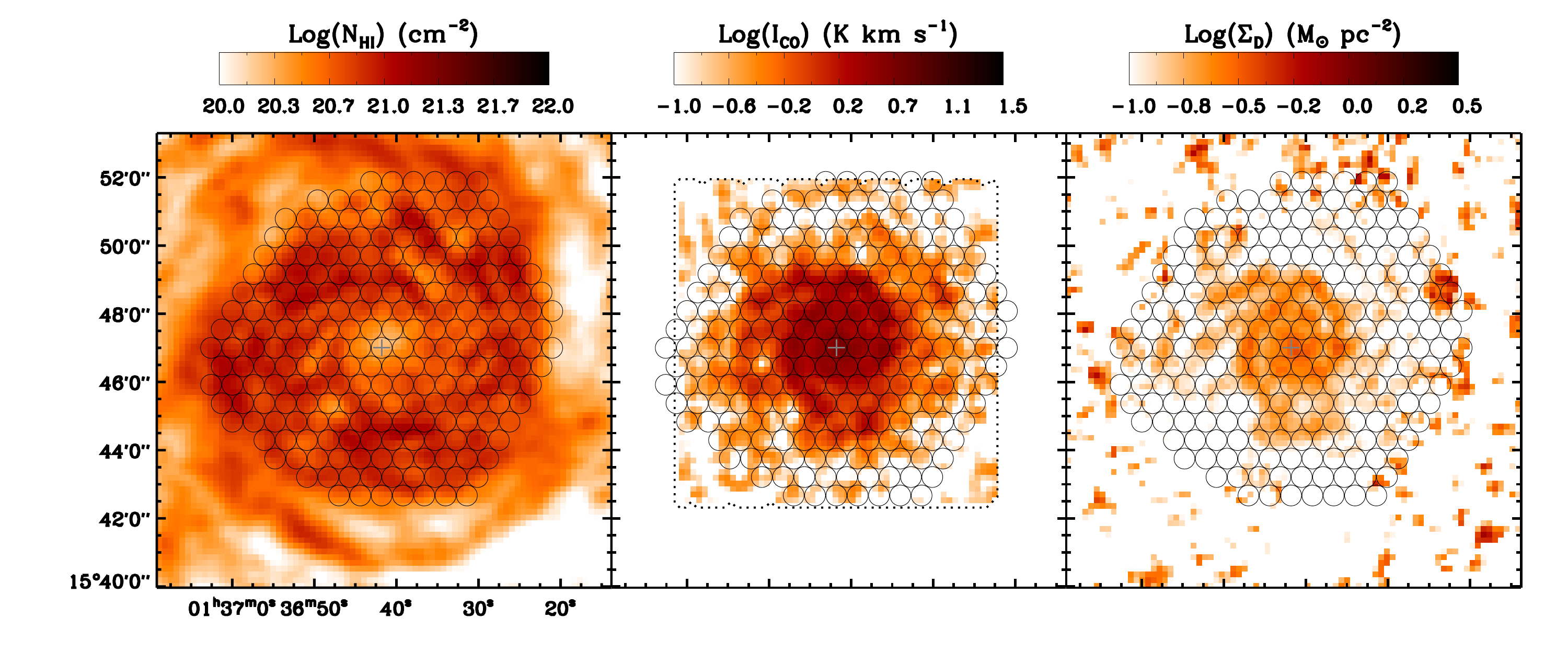}
\caption{The \hi, CO and \sigd\ maps for NGC~0628, from left to right (D = 7.2 Mpc, 1\arcsec = 35 pc).  The centers of the pixels in which we perform the simultaneous \aco\ and DGR solutions are shown as circles overlaid on the images.  The gray cross shows the central solution pixel for the galaxy.  On the middle panel, the coverage of the HERACLES CO map is shown with a dotted line. Similar plots for all galaxies in the sample can be found in the Appendix.}
\label{fig:ngc0628_panel1}
\end{figure*}

\begin{figure*}
\centering
\epsscale{1.1}
\plotone{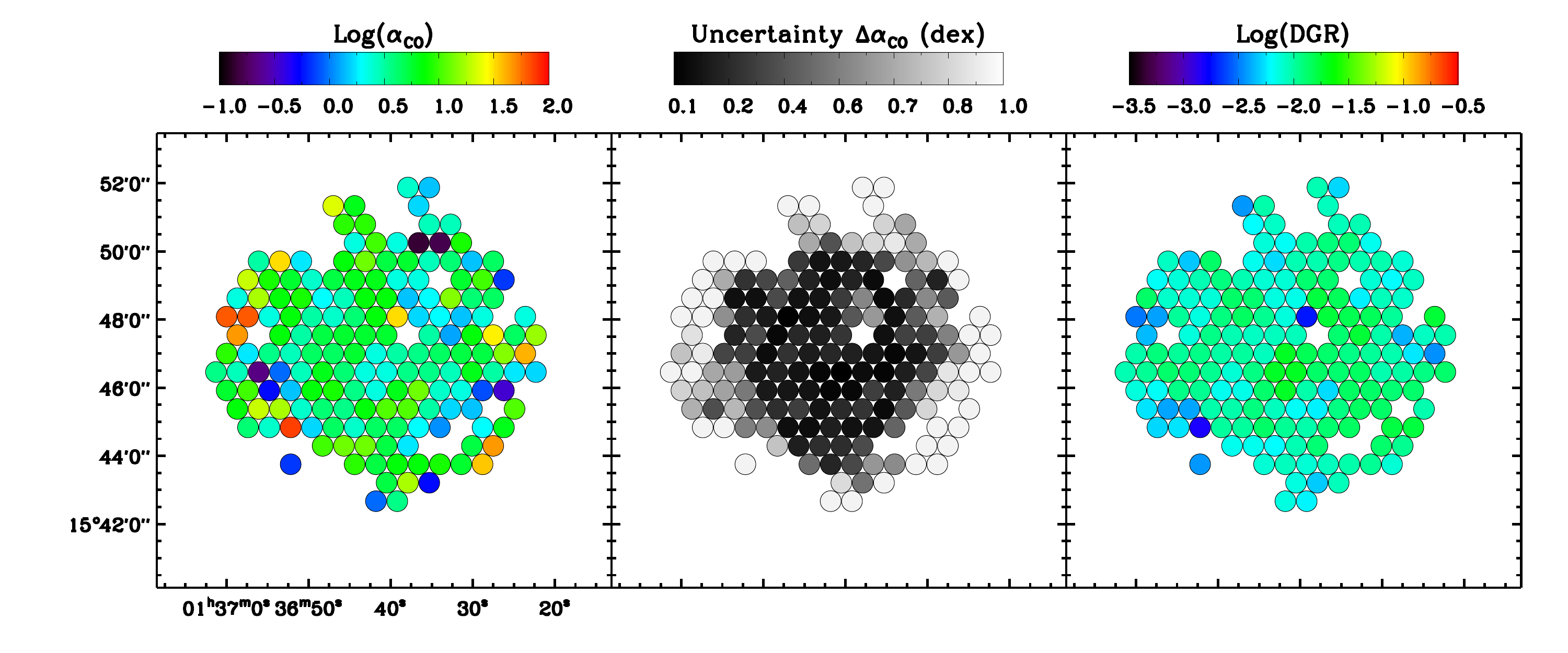}
\caption{Results of the simultaneous \aco\ and DGR solutions for NGC~0628.  The centers of the solution pixels are represented with circles as shown on Figure~\ref{fig:ngc0628_panel1}.  The left panel shows the resulting \aco, the middle shows the uncertainty on that value and the right shows the DGR.  Solution pixels where the technique failed do not appear.  For NGC~0628 it is clear that where there are good solutions (as judged by the uncertainty on \aco\ in the middle panel) most of the values are close to the MW Log(\aco) $=$ 0.64.  In pixels with good solutions, the DGR varies smoothly across the galaxy, which shows that our assumption of a single DGR in each solution pixel is self-consistent.}
\label{fig:ngc0628_panel2}
\end{figure*}

\begin{figure*}
\centering
\epsscale{0.9}
\plotone{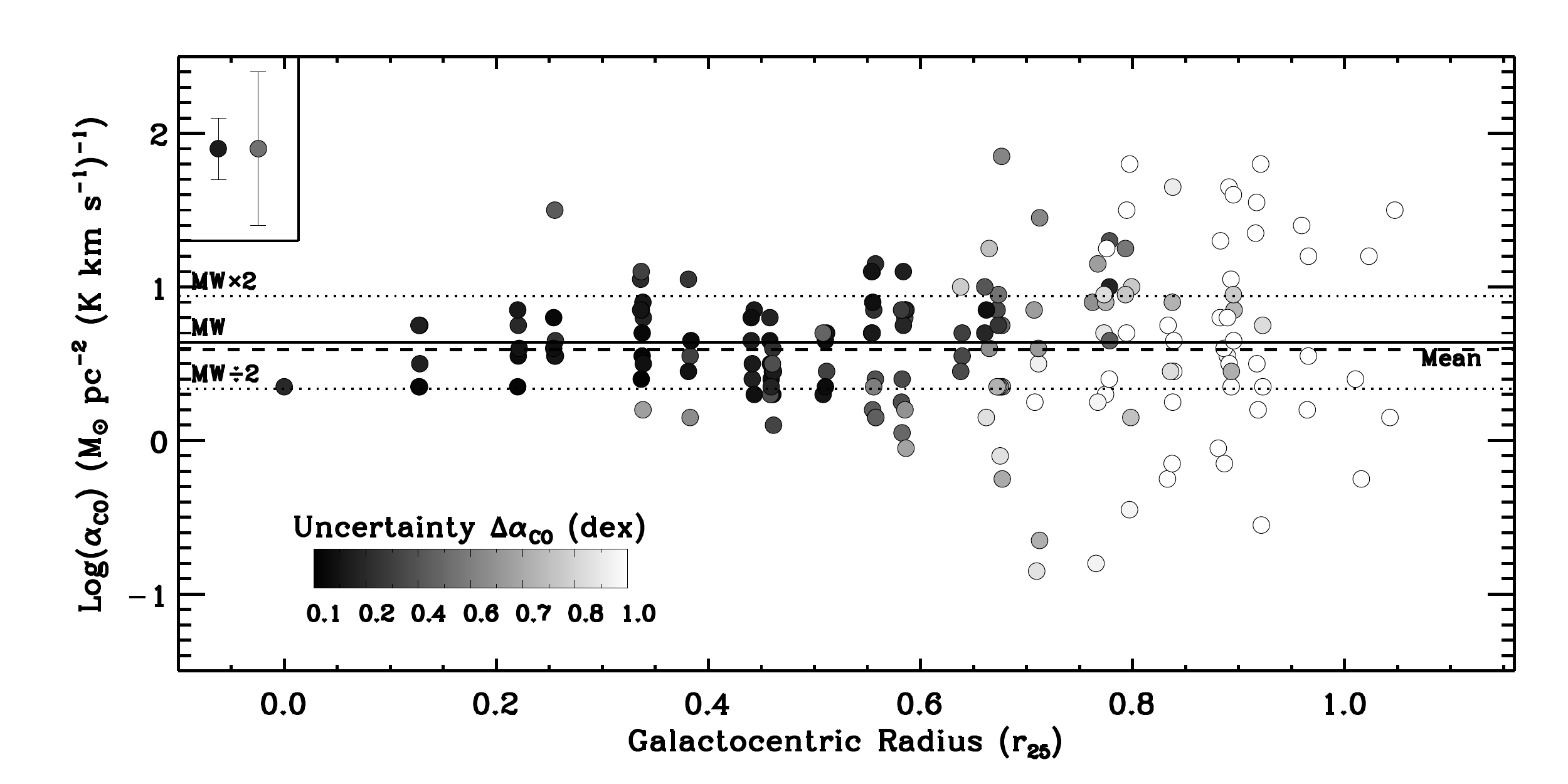}
\caption{\aco\ solutions for NGC~0628 as a function of galactocentric radius (in units of r$_{25}$).  The solid horizontal line shows the MW \aco=4.4 \acounit\ (note that the MW could possibly have a gradient in \aco\ with radius that we do not show here).  The dotted lines show a factor of 2 above and below the MW value.  The dashed horizontal line shows the average value for NGC~0628.  The gray scale color of the points represents the uncertainty on \aco\ as  shown in Figure~\ref{fig:ngc0628_panel2}---darker points have lower uncertainties.  For comparison with the gray color table, two representative error bars for the \aco\ solutions are shown in the top left corner of the plot.  For NGC~0628, almost all of the high confidence \aco\ solutions are within a factor of 2 of the MW value.}
\label{fig:ngc0628_panel3}
\end{figure*}

In general, NGC~0628 shows \aco\ consistent with the Milky Way value within a factor of 2 at all galactocentric radii.  Outside a radius of $r_{25}\sim0.6$ we find few good solutions.  There is a weak trend for lower \aco\ at smaller radii, with the central solution pixel having a conversion factor \aco$=2.2$ \acounit. Recent work by \citet{2013ApJ...764..117B} found consistent results for \aco\ by inverting the star-formation rate surface density map using a fixed molecular gas depletion time.    

\subsection{Completeness of Solutions}\label{sec:complete}

The technique we have used to solve for DGR and \aco\ simultaneously only works if there is sufficient S/N in the CO map and a range of CO/\hi\ ratios in each solution pixel.  These two constraints impose limits on where in the galaxies we can achieve solutions. In order to perform statistical tests on our sample of \aco\ and DGR values, we need to understand what biases these limits introduce into our results.  For example, the failure of the technique in regions with low CO S/N generally limits our good solutions to the inner parts of galaxies, where metallicity tends to be higher.  In order to judge the existence of trends in \aco\ versus metallicity, we therefore need to understand where we have achieved good solutions.

\begin{figure}
\centering
\epsscale{1.1}
\plotone{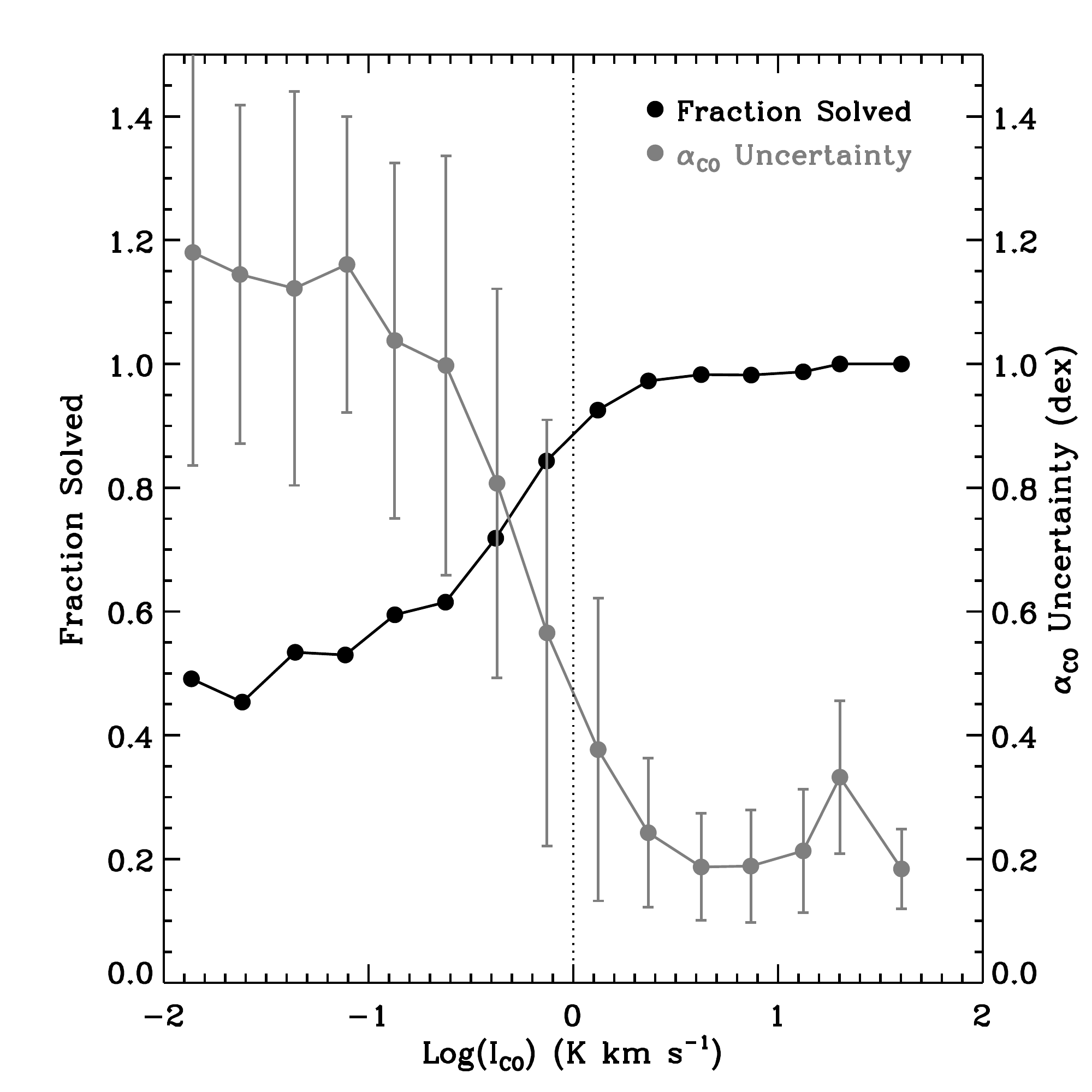}
\caption{The completeness and quality of our \aco\ solutions as a function of the mean \ico\ in a solution pixel.  The black points show the fraction of solution pixels where a solution was found.  The gray points show the mean uncertainty on \aco\ with error bars showing the standard deviation.  For \ico$>1$ K km s$^{-1}$, $>90$\% of solution pixels have solutions and their typical uncertainty is $<0.5$ dex.  We use this 90\% completeness cut to select a well-defined sample of \aco\ measurements for statistical analysis in the following work.}
\label{fig:biasvis}
\end{figure}

To investigate these effects, we have examined the fraction of solution pixels that have solutions and the uncertainty on the derived \aco\ as a function of the range of CO/\hi\ ratios and mean \ico\ in a pixel.  In general, the \hi\ in our target galaxies has a quite flat radial profile, while the CO drops off approximately exponentially \citep{2011AJ....142...37S}.  Because of these trends, wherever there is sufficient signal in CO to achieve a good solution, the CO/\hi\ range is adequate as well.  Thus, we find that the mean \ico\ of a solution pixel is the best predictor for the existence and quality of a solution.  Figure~\ref{fig:biasvis} shows the fraction of pixels with solutions and the average uncertainty on the solutions as a function of the mean \ico.  We identify a cut-off at \ico $>1$ K km s$^{-1}$ above which we obtain solutions $>$90\% of the time and those solutions have an average uncertainty of $<0.5$ dex.  For any statistical analysis that follows we use only pixels above this cut-off.

\subsection{Properties of Full-Sample \aco\ Solutions}

In Tables~\ref{tab:galav} and~\ref{tab:sampav} we list the average \aco\ derived galaxy by galaxy and for the full sample.  These averages only include solution pixels where \ico$>$ 1 K \kms.  The first two columns of Table~\ref{tab:sampav} list the mean and standard deviation of the individual solution pixel measurements.  This approach treats each solution pixel equally, regardless of how strongly it contributes to the total molecular gas mass.  The third column instead lists the mean \aco\ derived from the total \sightwo\ and the total \ico\ for the galaxy (above \ico=1 K km s$^{-1}$), which is equivalent to a mean where the pixels are weighted by their \ico.  If a single \aco\ value were to be applied to the data, this would be the optimal value to use. For most galaxies, the CO weighted mean is higher than the straight mean of the solution pixels. This is due to the fact that the molecular gas is not evenly distributed across the galaxy and the area-weighted average is different then the molecular gas weighted average. 

\begin{deluxetable*}{lcccc}
\tablewidth{0pt}
\tabletypesize{\scriptsize}
\tablecolumns{5}
\tablecaption{Galaxy Average \aco}
\tablehead{ 
\multicolumn{1}{c}{Galaxy} &
\multicolumn{1}{c}{Mean \aco} &
\multicolumn{1}{c}{Std. Dev.} &
\multicolumn{1}{c}{CO Weighted} &
\multicolumn{1}{c}{\# Meas.} \\
\multicolumn{1}{c}{} &
\multicolumn{1}{c}{} &
\multicolumn{1}{c}{(dex)} &
\multicolumn{1}{c}{Mean \aco} &
\multicolumn{1}{c}{}
} 
\startdata
NGC~0337 & 22.4           & 16.2$-$31.0 (0.14) & 21.8           &   2 \\
NGC~0628 & \phantom{0}3.9 &  2.1$-$7.4  (0.28) & \phantom{0}5.1 &  67 \\
NGC~0925 & 10.0           &  6.8$-$14.7 (0.17) & 10.0           &   1 \\
NGC~2841 & \phantom{0}5.0 &  2.5$-$10.3 (0.31) & \phantom{0}5.7 &  19 \\
NGC~2976 & \phantom{0}3.3 &  0.9$-$12.5 (0.58) & \phantom{0}4.7 &  18 \\
NGC~3077 & \phantom{0}4.6 &  2.3$-$9.4  (0.31) & \phantom{0}5.4 &   7 \\
NGC~3184 & \phantom{0}5.3 &  2.9$-$9.5  (0.26) & \phantom{0}6.3 &  43 \\
NGC~3198 & 11.0           &  6.4$-$18.9 (0.24) & 11.9           &  11 \\
NGC~3351 & \phantom{0}2.7 &  1.0$-$6.9  (0.41) & \phantom{0}2.9 &  28 \\
NGC~3521 & \phantom{0}7.6 &  4.9$-$11.8 (0.19) & \phantom{0}7.3 &  42 \\
NGC~3627 & \phantom{0}1.2 &  0.4$-$3.3  (0.44) & \phantom{0}1.8 &  43 \\
NGC~3938 & \phantom{0}5.5 &  4.1$-$7.5  (0.13) & \phantom{0}5.8 &  19 \\
NGC~4236 & \nodata        &  \nodata     & \nodata        &   0 \\
NGC~4254 & \phantom{0}3.4 &  2.1$-$5.7  (0.22) & \phantom{0}4.7 &  46 \\
NGC~4321 & \phantom{0}2.2 &  1.1$-$4.6  (0.32) & \phantom{0}2.2 &  57 \\
NGC~4536 & \phantom{0}2.6 &  1.0$-$6.7  (0.41) & \phantom{0}2.6 &  13 \\
NGC~4569 & \phantom{0}1.1 &  0.3$-$4.1  (0.57) & \phantom{0}1.2 &  14 \\
NGC~4625 & \nodata        &  \nodata     & \nodata        &   0 \\
NGC~4631 & 10.8           &  5.6$-$19.5 (0.26) & \phantom{0}9.8 &  40 \\
NGC~4725 & \phantom{0}1.2 &  0.4$-$3.2  (0.44) & \phantom{0}1.8 &   7 \\
NGC~4736 & \phantom{0}1.0 &  0.5$-$2.0  (0.29) & \phantom{0}1.1 &  33 \\
NGC~5055 & \phantom{0}3.7 &  1.9$-$7.4  (0.30) & \phantom{0}4.0 &  86 \\
NGC~5457 & \phantom{0}2.3 &  1.1$-$4.8  (0.32) & \phantom{0}2.9 & 142 \\
NGC~5713 & \phantom{0}4.6 &  1.7$-$12.6 (0.44) & \phantom{0}5.4 &  13 \\
NGC~6946 & \phantom{0}2.0 &  0.9$-$4.4  (0.35) & \phantom{0}1.8 & 158 \\
NGC~7331 & \phantom{0}9.8 &  6.2$-$15.3 (0.20) & 10.7           &  32 
\enddata
\tablecomments{Averages include only solution pixels with \ico $>1$ K km s$^{-1}$. The number of measurements meeting this criterion are shown in the last column of the table.}\label{tab:galav}
\end{deluxetable*}

\begin{deluxetable*}{lcccccc}
\tablewidth{0pt}
\tabletypesize{\scriptsize}
\tablecolumns{7}
\tablecaption{Sample Average \aco}
\tablehead{ 
\multicolumn{1}{c}{Sample} &
\multicolumn{1}{c}{Mean \aco} &
\multicolumn{1}{c}{Std. Dev.} &
\multicolumn{1}{c}{CO Weighted} &
\multicolumn{1}{c}{Gal. Weighted} &
\multicolumn{1}{c}{Gal \& CO Weighted} \\
\multicolumn{1}{c}{} &
\multicolumn{1}{c}{all L.O.S.} &
\multicolumn{1}{c}{(dex)} &
\multicolumn{1}{c}{Mean \aco} &
\multicolumn{1}{c}{Mean \aco} &
\multicolumn{1}{c}{Mean \aco}
} 
\startdata
%All Galaxies      & 3.1  & 1.2$-$8.0 (0.42)  & 3.9  & 3.9  & 4.3 \\
Incl $<65^\circ$  & 2.6  & 1.0$-$6.6 (0.41)  & 2.9  & 3.1  & 3.5 \\
Incl $>65^\circ$  & 7.2  & 3.0$-$17.6 (0.39) & 8.2  & 6.5  & 6.7
\enddata
\tablecomments{Averages include only solution pixels with \ico $>1$ K km s$^{-1}$.}\label{tab:sampav}
\end{deluxetable*}

\begin{figure*}
\centering
\epsscale{0.9}
\plotone{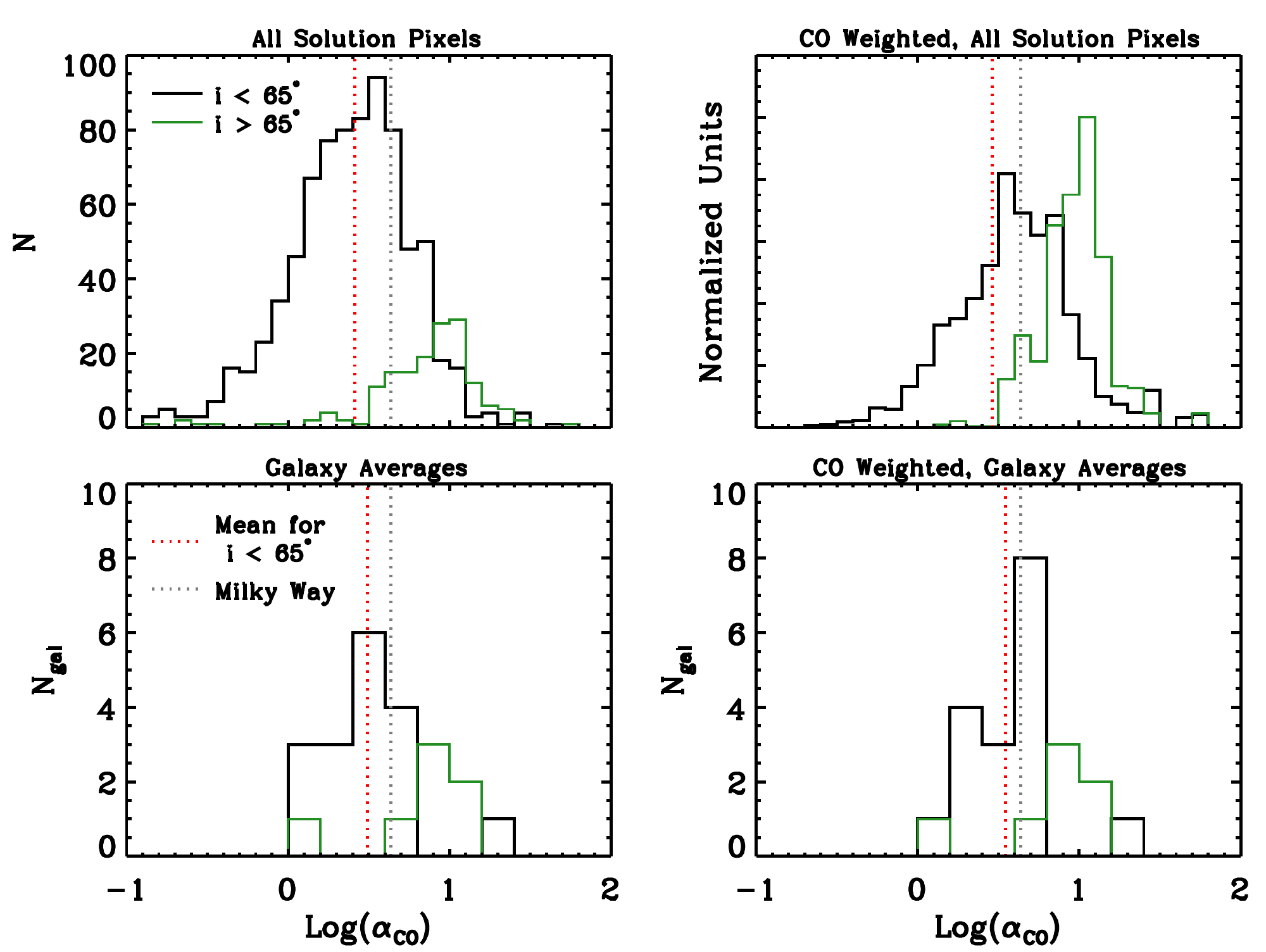}
\caption{Histograms showing the distribution of \aco\ results with various weighting schemes.  The averages are reported in Table~\ref{tab:sampav}.  The panels show histograms of galaxies with $i<65^{\circ}$ in black and galaxies with $i>65^{\circ}$ in green.  In each panel we highlight the mean value of \aco\ for galaxies with $i<65^{\circ}$ with a dotted vertical red line and the Milky Way \aco\ in gray.  These histograms include solution pixels above the \ico\ $>$ 1 K km s$^{-1}$ cutoff.  The top left panel shows the histogram of the solution pixels.  The top right panel shows those same values weighted by their \ico---this makes pixels contribute to this histogram in proportion to their molecular gas mass.  The bottom panels show histograms of the average \aco\ from each galaxy as listed in Table~\ref{tab:galav}. Galaxies with high inclinations tend to show a higher than average \aco, because the solution pixels include gas and dust at a range of radii.  On average, we find \aco\ slightly lower than the MW value in our sample.}
\label{fig:samphist}
\end{figure*}

Figure~\ref{fig:samphist} shows several histograms illustrating the distribution of our measured \aco\ values.  The mean, derived with several different weighting schemes, is listed in Table~\ref{tab:sampav}.  The top panels of Figure~\ref{fig:samphist} show histograms of all solution pixels while the bottom panels show histograms of the galaxy averages from Table~\ref{tab:galav}.  On these histograms we highlight galaxies with high inclinations in green ($i > 65^{\circ}$).  It is clear that the high inclination galaxies tend to have higher \aco\ on average than the more face-on galaxies.  In the highest inclination galaxies the pixel will include contributions from gas at larger radii which tends to have lower DGR and less molecular gas. This is equivalent to the challenge faced by \citet{2011ApJ...737...12L} in the SMC, where some of the \hi\ along the line of sight originates in an essentially dust-free envelope, and in M~31, where regions on the minor axis of the galaxy have contributions from gas and dust at a variety of radii.  In addition, optical depth effects for \hi\ may be accentuated for highly inclined galaxies.  All galaxies with $i>65^{\circ}$ show average \aco\ above the mean \citep[except for NGC~0925, which has a somewhat uncertain inclination;][]{2008AJ....136.2648D}.    We thus eliminate all galaxies with $i>65^{\circ}$, leaving 782 total \aco\ measurements.

In Table~\ref{tab:sampav} we list the average values of \aco\ for the full galaxy sample.  Excluding the high inclination galaxies, we find an average \aco$=2.6$ \acounit\ for the individual solution pixel results.  Weighted by \ico, the average value is slightly higher, \aco$=2.9$ \acounit.  To avoid highly resolved galaxies like NGC~5457 and 6946 contributing more points to the average, we also calculate averages where each galaxy contributes uniformly.  These are listed in the last two columns of Table~\ref{tab:sampav}.  The average value for our sample, \aco$=3.1$ \acounit, is only slightly lower than what is found in the Milky Way disk.                  

The standard deviation in \aco\ is 0.38 dex, treating all lines of sight equally (with our \ico\ and inclination cut-offs).  A key question we would like to answer is to what degree this scatter represents 1) true scatter within each galaxy, 2) galaxy-to-galaxy offsets or 3) variation of \aco\ as a function of local parameters.  If the scatter comes from variations with local environmental parameters, we may be able to generate a prescription for \aco\ as a function of other observables.  In the following sections we explore the variations of \aco\ within and among our galaxies to understand if and why it varies.

\subsection{Radial Variations in \aco}

\begin{figure*}
\centering
\epsscale{1.2}
\plotone{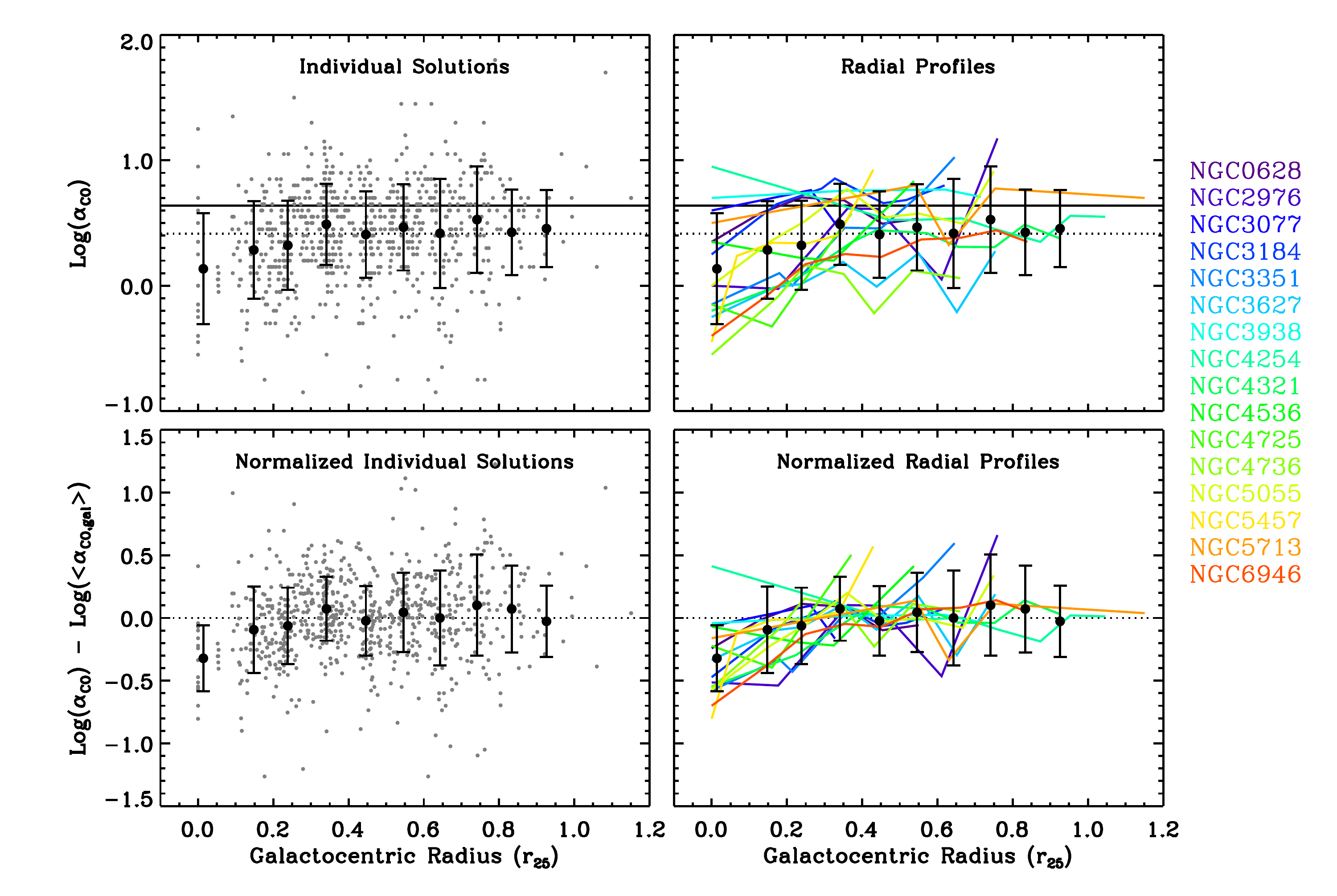}
\caption{Solutions for \aco\ plotted versus their galactocentric radius (r$_{25}$).  The top left panel shows the \aco\ solutions and the bottom left panel shows the same values normalized by each galaxy's average \aco\ (listed in Table~\ref{tab:galav}).  All individual solutions are shown as gray points.  The mean and standard deviation of all of the solutions in 0.1 r$_{25}$ bins are shown with black symbols (this mean treats all lines of sight equally).  The solid black horizontal line in the top panel shows \aco\ for the Milky Way.  The dotted black horizontal line shows the average, with no weighting, of all solution pixels: \aco=2.6 \acounit.  In the right panels, the radial profile for each galaxy is shown with a solid colored line overlaid with the same binned average from the left-hand panels. The average \aco\ radial profile of our galaxies is mostly flat as a function of r$_{25}$ with a decrease in \aco\ towards the center of the galaxy.  On average, the central \aco\ value is $\sim$0.3 dex lower than the rest of the galaxy.}
\label{fig:acovsr25}
\end{figure*}

In Figure~\ref{fig:acovsr25} we present a summary of the \aco\ values we find as a function of galactocentric radius. Plots of \aco\ from each individual galaxy as a function of r$_{25}$ are presented in the Appendix.  Each individual solution that makes our \ico\ and inclination cut is shown in Figure~\ref{fig:acovsr25} with gray circles.  We also display the mean and standard deviation of the \aco\ values in 0.1 r$_{25}$ bins for each galaxy (this mean treats all solution pixels equally).  The top panel of Figure~\ref{fig:acovsr25} shows the measured \aco\ and the bottom panel shows those same values normalized by each galaxy's average \aco\ from Table~\ref{tab:galav}.     

The average radial profile of our galaxies is mostly flat as a function of r$_{25}$, but almost all galaxies show a decrease in \aco\ in the inner $\sim0.2$ r$_{25}$ compared to their average value.  The mean central decrease is $\sim0.3-0.4$ dex, but it can be as much as 0.8 dex.  For several galaxies this decrease leads to a central \aco\ value that is an order-of-magnitude lower than the MW \aco\ (e.g. NGC~4736, 5457 and 6946).  The central depression persists even after correcting for differences in $R_{21}$ (see the following Section for further discussion).

Normalizing each galaxy by its average \aco\ and making each galaxy contribute equally to the average leads to much smaller scatter at r$_{25} > 0.3$ (this can be seen in the bottom panel of Figure~\ref{fig:acovsr25} where the radial profiles show much less scatter than the individual measurements, which is not the case in the top, unnormalized panel).  Inside that radius, scatter in the normalized \aco\ profiles is reduced, but not by as much outside that radius, indicating that individual galaxies show different central profiles of \aco.  The overall conclusions from examination of Figure~\ref{fig:acovsr25} are: 1) outside r$_{25}\sim0.3$ most of the scatter in our \aco\ measurements can be explained by galaxy to galaxy differences, 2) inside that galactocentric radius, galaxy-specific trends in \aco\ dominate the scatter and 3) in general the inner region of galaxies shows \aco\ lower than the rest of the disk.   

\subsection{\aco\ in Galaxy Centers}

In Table~\ref{tab:centers} we list the central \aco\ values and compare them to the MW value and the mean for the galaxy.  Almost all of the central solution pixels have \aco\ lower than the galaxy mean.  The galaxies NGC~3351, 3627, 4321, 4736, 5457 and 6946 show \aco\ values a factor of 5 or more lower than the MW value with 3$\sigma$ confidence.  For 4736, 5457 and 6946, in particular, the central solution pixel has \aco\ an order-of-magnitude below the MW value.  Only one galaxy, NGC~0337, shows a central \aco\ higher than the MW value at 3$\sigma$.  In Table~\ref{tab:centers}, we also list the CO $(2-1)/(1-0)$ ratio measured as described in Section~\ref{sec:r21}.  While the central solution pixels do tend to have higher $R_{21}$ than the value we have adopted, the difference is less than a factor of 2 in all cases, and does not alter our main conclusions about the low $(1-0)$ conversion factor in these regions. In particular, for the 10 galaxies with measured $R_{21}$ and $i<65^{\circ}$, the central \aco\ corrected for $R_{21}$ is still on average 0.3 dex lower (i.e. a factor of 2) than the galaxy average.

\begin{deluxetable*}{lccccccc}
\tablewidth{0pt}
\tabletypesize{\scriptsize}
\tablecolumns{8}
\tablecaption{Central \aco\ Measurements}
\tablehead{ 
\multicolumn{1}{c}{Galaxy} &
\multicolumn{1}{c}{Log$(\alpha_{\rm CO,Cen}$)} &
\multicolumn{1}{c}{$\Delta_{MW}$ =} &
\multicolumn{1}{c}{$\Delta_{MW}/\sigma$} &
\multicolumn{1}{c}{$\Delta_{\rm mean}$ =} &
\multicolumn{1}{c}{$\Delta_{\rm mean}/\sigma$} &
\multicolumn{1}{c}{R$_{21}$} &
\multicolumn{1}{c}{Log$(\mathrm{R}_{21}/0.7)$} \\
\multicolumn{1}{c}{} & 
\multicolumn{1}{c}{$\pm\sigma$} &
\multicolumn{1}{c}{Log$(\alpha_{\rm CO}/\alpha_{\rm CO,MW}$)} &
\multicolumn{1}{c}{} &
\multicolumn{1}{c}{Log$(\alpha_{\rm CO}/\langle\alpha_{\rm CO}\rangle)$} &
\multicolumn{1}{c}{} &
\multicolumn{1}{c}{} &
\multicolumn{1}{c}{}
} 
\startdata
NGC~0337 & $+$1.25 $\pm$ 0.17 & $+$0.61 & 3.6 & $-$0.10 &  0.6    & \nodata         & \nodata  \\
NGC~0628 & $+$0.35 $\pm$ 0.24 & $-$0.29 & 1.2 & $-$0.24 &  1.0    & \nodata         & \nodata  \\
%NGC~0925 & $+$0.65 $\pm$ 0.24 & $+$0.01 & 0.0 & $-$0.35 &  1.4    & \nodata         & \nodata  \\
%NGC~2841 & $+$0.40 $\pm$ 0.61 & $-$0.24 & 0.4 & $-$0.30 &  0.5    & \nodata         & \nodata  \\
NGC~2976 & $+$0.00 $\pm$ 0.49 & $-$0.64 & 1.3 & $-$0.51 &  1.0    & \nodata         & \nodata  \\
NGC~3077 & $+$0.60 $\pm$ 0.17 & $-$0.04 & 0.2 & $-$0.06 &  0.4    & \nodata         & \nodata  \\ 
NGC~3184 & $+$0.25 $\pm$ 0.16 & $-$0.39 & 2.5 & $-$0.47 &  3.0    & 0.77 $\pm$ 0.05 & $+$0.04  \\
%NGC~3198 & $+$0.60 $\pm$ 0.31 & $-$0.04 & 0.1 & $-$0.44 &  1.4    & \nodata         & \nodata  \\
NGC~3351 & $-$0.15 $\pm$ 0.14 & $-$0.79 & 5.6 & $-$0.58 &  4.1    & 1.10 $\pm$ 0.24 & $+$0.19  \\
%NGC~3521 & $+$0.70 $\pm$ 0.36 & $+$0.06 & 0.2 & $-$0.18 &  0.5    & 1.02 $\pm$ 0.06 & $+$0.16  \\
NGC~3627 & $-$0.25 $\pm$ 0.14 & $-$0.89 & 6.3 & $-$0.34 &  2.4    & 0.54 $\pm$ 0.05 & $-$0.12  \\ 
NGC~3938 & $+$0.70 $\pm$ 0.19 & $+$0.06 & 0.3 & $-$0.04 &  0.2    & \nodata         & \nodata  \\
%NGC~4236 & $+$1.90 $\pm$ 0.80 & $+$1.26 & 1.6 & \nodata & \nodata & \nodata         & \nodata  \\
NGC~4254 & $+$0.95 $\pm$ 0.82 & $+$0.31 & 0.4 & $+$0.41 &  0.5    & 1.03 $\pm$ 0.06 & $+$0.17  \\
NGC~4321 & $-$0.20 $\pm$ 0.17 & $-$0.84 & 4.8 & $-$0.55 &  3.1    & 1.25 $\pm$ 0.16 & $+$0.25  \\
NGC~4536 & $+$0.35 $\pm$ 0.13 & $-$0.29 & 2.2 & $-$0.07 &  0.5    & 1.26 $\pm$ 0.28 & $+$0.26  \\
%NGC~4569 & $-$0.85 $\pm$ 0.51 & $-$1.49 & 2.9 & $-$0.89 &  1.8    & 1.14 $\pm$ 0.18 & $+$0.21  \\
NGC~4625 & $+$1.05 $\pm$ 0.41 & $+$0.41 & 1.0 & \nodata & \nodata & \nodata         & \nodata  \\
%NGC~4631 & $+$0.60 $\pm$ 0.35 & $-$0.04 & 0.1 & $-$0.43 &  1.3    & \nodata         & \nodata  \\
NGC~4725 & $-$0.15 $\pm$ 0.71 & $-$0.79 & 1.1 & $-$0.22 &  0.3    & \nodata         & \nodata  \\
NGC~4736 & $-$0.55 $\pm$ 0.17 & $-$1.19 & 6.9 & $-$0.56 &  3.2    & 1.35 $\pm$ 0.09 & $+$0.29  \\
NGC~5055 & $+$0.00 $\pm$ 0.25 & $-$0.64 & 2.6 & $-$0.57 &  2.3    & 1.10 $\pm$ 0.08 & $+$0.20  \\
NGC~5457 & $-$0.45 $\pm$ 0.20 & $-$1.09 & 5.5 & $-$0.80 &  4.1    & 0.90 $\pm$ 0.08 & $+$0.11  \\
NGC~5713 & $+$0.50 $\pm$ 0.25 & $-$0.14 & 0.6 & $-$0.16 &  0.6    & \nodata         & \nodata  \\
NGC~6946 & $-$0.40 $\pm$ 0.31 & $-$1.04 & 3.4 & $-$0.70 &  2.3    & 1.07 $\pm$ 0.14 & $+$0.19  
%NGC~7331 & $+$1.15 $\pm$ 0.44 & $+$0.51 & 1.2 & $+$0.16 &  0.4    & \nodata         & \nodata  
\enddata
\tablecomments{Galaxies with $i>65^{\circ}$ have been omitted from the Table.}
\label{tab:centers}
\end{deluxetable*}

The central solution pixels can be outliers from the rest of the pixels in our sample in the sense of having high CO/\hi\ ratios.  Since there are relatively few solution pixels with these conditions, they may not have been well represented in the simulated data we used to test the accuracy of the solution technique.  High CO/\hi\ ratios could bias the results:  if the S/N of the \hi\ maps is almost always higher than the S/N of CO, the DGR scatter could be reduced in these conditions purely by decreasing the importance of CO in assessing the gas mass surface density.  Since this bias would move the \aco\ results towards lower values in the centers, we performed a test to judge whether our low central \aco\ measurements could be due to this effect. The details of the test are described in the Appendix.  In brief, we generated simulated datasets with known \aco\ values where the N$_{\rm HI}$, \ico\ and \sigd\ S/N is matched to the observations in each central solution pixel and performed Monte Carlo trials to see how well the known input \aco\ was recovered.  In all cases, we found no evidence for a bias in the central solution pixel. 

Taking into account the offset from $R_{21}$ variations, our results show that the central \aco\ in  NGC~3351, 3627, 4321, 4736, 5457 and 6946 is lower than the MW value by a factor of $4-10$.  Many of these galaxies show \aco\ closer to the Milky Way value at larger radii, in line with the general trend seen in Figure~\ref{fig:acovsr25}.  For NGC~3627, the mean \aco\ for the galaxy is low as well, suggesting that the central region is not distinct from the rest of the disk (this galaxy is an interacting member of the Leo Triplet, so it is unique in our sample).  Conversely, examination of the individual galaxy \aco\ measurements as a function of radius (shown in the Appendix) for NGC~3351, 4321, 4736, 5457 and 6946 demonstrates that these galaxies show an unresolved region of lower than average \aco\ in their centers (note that our solution pixel grid oversamples the data, so an unresolved central depression affects the r$_{25}=0$ point and the six adjacent solution pixels).   The closest galaxy that shows a central depression is NGC~4736 at D$=$4.66 Mpc.  At this distance, our solution pixel covers a region of radius 0.8 kpc.  Even for this nearby example, we do not resolve the central depression.  We tested whether the depressions were resolved using a grid of independent solution pixels (i.e. not overlapping) and found that the depression only affected the central pixel, consistent with it being unresolved.  We note that our ability to detect any central depression may be a function of the galaxy's distance due to the increased size of the central solution pixel.  Even at D$=14.3$ Mpc, however, we detect a clear central depression in NGC~4321.

The type of nuclear activity in each galaxy is not a good predictor for whether or not it displays a central \aco\ depression.  Of the galaxies that show the clearest central depression NGC~3351, 5457 and 6946 are classified as star-formation or \hii\ region dominated, while NGC~4321 and 4736 have signatures of active galactic nucleus (AGN) or low-ionization nuclear emission-line region (LINER) activity \citep[for details on the nuclear classifications, see][]{2011PASP..123.1347K}.  Several galaxies with flat radial \aco\ profiles do show evidence for AGN activity: NGC~3627, 4254, and 4725, for example.  Enhanced AGN activity could affect molecular gas properties in the nuclei \citep[e.g.][]{2008ApJ...677..262K}, but the relatively weak AGN present in the KINGFISH galaxies may not dominate on the kpc scales we study here.

To summarize, we find that several galaxies in our sample show central \aco\ values that are substantially lower than the Milky Way value and also well below the galaxy average.  This appears to take the form of depression in the central region that is unresolved by our solution pixel grid.  We discuss these central regions and the physical conditions that may lead to low \aco\ further in Section~\ref{sec:centralaco}.

\subsection{Correlations of \aco\ with Environmental Parameters}\label{sec:acocorr}

Variations in \aco\ may be related to variations in the local environmental conditions including metallicity, ISM pressure, interstellar radiation field strength, gas temperature, dust properties or other variables.  Correlations between \aco\ and quantities that trace these environmental conditions may allow us to identify the drivers of \aco\ variations.  They may also provide tools to predict the appropriate \aco\ in a given environment.  In the following, we examine the correlations of our measured \aco\ with several observables that trace ISM conditions.  For each tracer, we use the average value in each solution pixel.  The physical interpretation of these correlations or lack thereof is discussed in Section~\ref{sec:drivers}.

\begin{figure*}
\centering
\epsscale{1.1}
\plotone{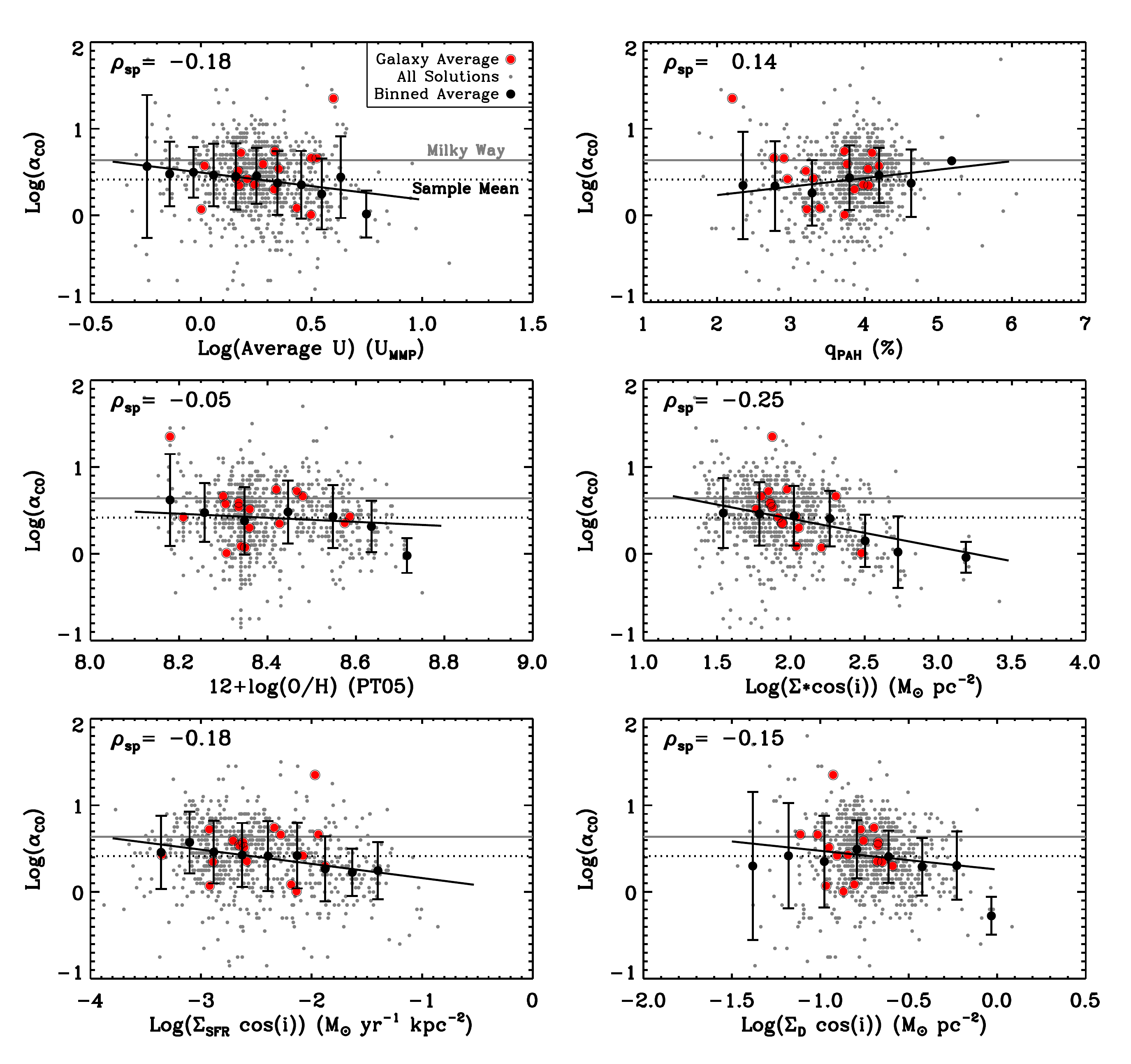}
\caption{\aco\ as a function of environmental parameters.  The panels show \aco\ plotted versus average radiation field intensity ($\overline{U}$; top-left) and PAH fraction (q$_{\rm PAH}$; top-right) from the \citet{2012ApJ...756..138A} fits, metallicity (12$+$Log(O/H); middle-left), stellar mass surface density ($\Sigma_{*}$; middle-right), star-formation rate surface density ($\Sigma_{\rm SFR}$; lower-left) and dust mass surface density (\sigd; lower-right).  The solid gray line shows the Milky Way \aco\ and the dotted black line shows the sample average \aco=2.6 \acounit, treating all solution pixels equally.  Individual \aco\ solutions above our \ico\ and inclination cuts are shown with gray symbols.  The mean and standard deviation of those values in bins are shown with black circles and error bars.  A linear fit to the measurements is shown with a black line.  The mean values for all galaxies are shown with red circles.  The Spearman rank correlation coefficient for each panel is listed in the top left.  All variables except for metallicity show a statistically significant correlation with \aco. None of these correlations allow \aco\ to be predicted with significantly better precision than the scaling with r$_{25}$.  However, the extreme values of $\Sigma_{\rm SFR}$ ($>0.5$ \msun\ yr$^{-1}$ kpc$^{-2}$) and and $\Sigma_{*}$ ($>1000$ \msun\ pc$^{-2}$) are always associated with low \aco\ and are located in galaxy centers.}
\label{fig:drivers}
\end{figure*}

In Figure~\ref{fig:drivers} we plot our measured \aco\ values as a function of the average radiation field intensity ($\overline{U}$), the PAH fraction (\qpah), metallicity (12$+$Log(O/H)), stellar mass surface density ($\Sigma_{*}$), star-formation rate surface density ($\Sigma_{\rm SFR}$) and the dust mass surface density (\sigd).  Each panel shows the individual measurements with gray symbols and the average for the galaxies with red symbols.  The Milky Way \aco\ is highlighted with a horizontal gray line and the mean \aco\ treating all solution pixels equally (\aco$=2.6$ \acounit) is shown with a dashed horizontal line.  Overlaid on each panel is the binned mean and standard deviation and a linear fit to the gray points.  In Table~\ref{tab:corr} we list the linear fit results as well as the rank correlation coefficient for each panel and its significance (in standard deviations away from the null hypothesis).  The rank correlation coefficients suggest there are significant correlations between \aco\ and all of the variables except metallicity.  Due to possible inconsistencies between various metallicity measurements and incomplete knowledge of metallicity gradients in some galaxies, we examine the trends with metallicity separately in Section~\ref{sec:acovsz}.  

\begin{deluxetable}{lcccc}
\tablewidth{0pt}
\tabletypesize{\scriptsize}
\tablecolumns{5}
\tablecaption{\aco\ Correlation Properties}
\tablehead{ 
\multicolumn{1}{c}{Log(\aco) vs.} &
\multicolumn{1}{c}{Correlation\tablenotemark{a}} &
\multicolumn{1}{c}{$\sigma$ from} &
\multicolumn{2}{c}{Linear Fit\tablenotemark{b}} \\
\cline{4-5}
\multicolumn{1}{c}{Variable} &
\multicolumn{1}{c}{Coeff.} & 
\multicolumn{1}{c}{Null} &
\multicolumn{1}{c}{Offset} &
\multicolumn{1}{c}{Slope} 
}
\startdata
r$_{25}$                & $+$0.19 & 5.2 & 0.29 & $+$0.29 \\
Log($\overline{U}$)     & $-$0.18 & 5.0 & 0.50 & $-$0.31 \\
q$_{\rm PAH}$ (\%)      & $+$0.14 & 3.9 & 0.11 & $+$0.10 \\
12$+$Log(O/H)           & $-$0.05 & 1.4 & 2.39 & $-$0.24 \\
Log($\Sigma_{*}$)       & $-$0.26 & 7.1 & 1.05 & $-$0.33 \\
Log($\Sigma_{\rm D}$) & $-$0.15 & 4.1 & 0.26 & $-$0.22 \\
Log($\Sigma_{\rm SFR})$ & $-$0.18 & 5.1 & 0.00 & $-$0.16 
\enddata
\tablenotetext{a}{Spearman rank correlation coefficient.}
\tablenotetext{b}{Correlation with metallicity using a homogeneous sample of metallicity measurements is discussed in Section~\ref{sec:acovsz}.}
\tablecomments{These correlations are shown in
Figure~\ref{fig:drivers}.}\label{tab:corr}
\end{deluxetable}

Several other radiation field properties are measured from the dust SED modeling.  These include U$_{\rm min}$, the minimum radiation field heating the dust, and f$_{\rm PDR}$ the fraction of the dust luminosity that arises in ``PDR-like'' regions where U$>$100 U$_{\rm MMP}$.  We have investigated the dependence of \aco\ on these quantities and find weak trends with low significance.  

\begin{figure*}
\centering
\epsscale{1.1}
\plotone{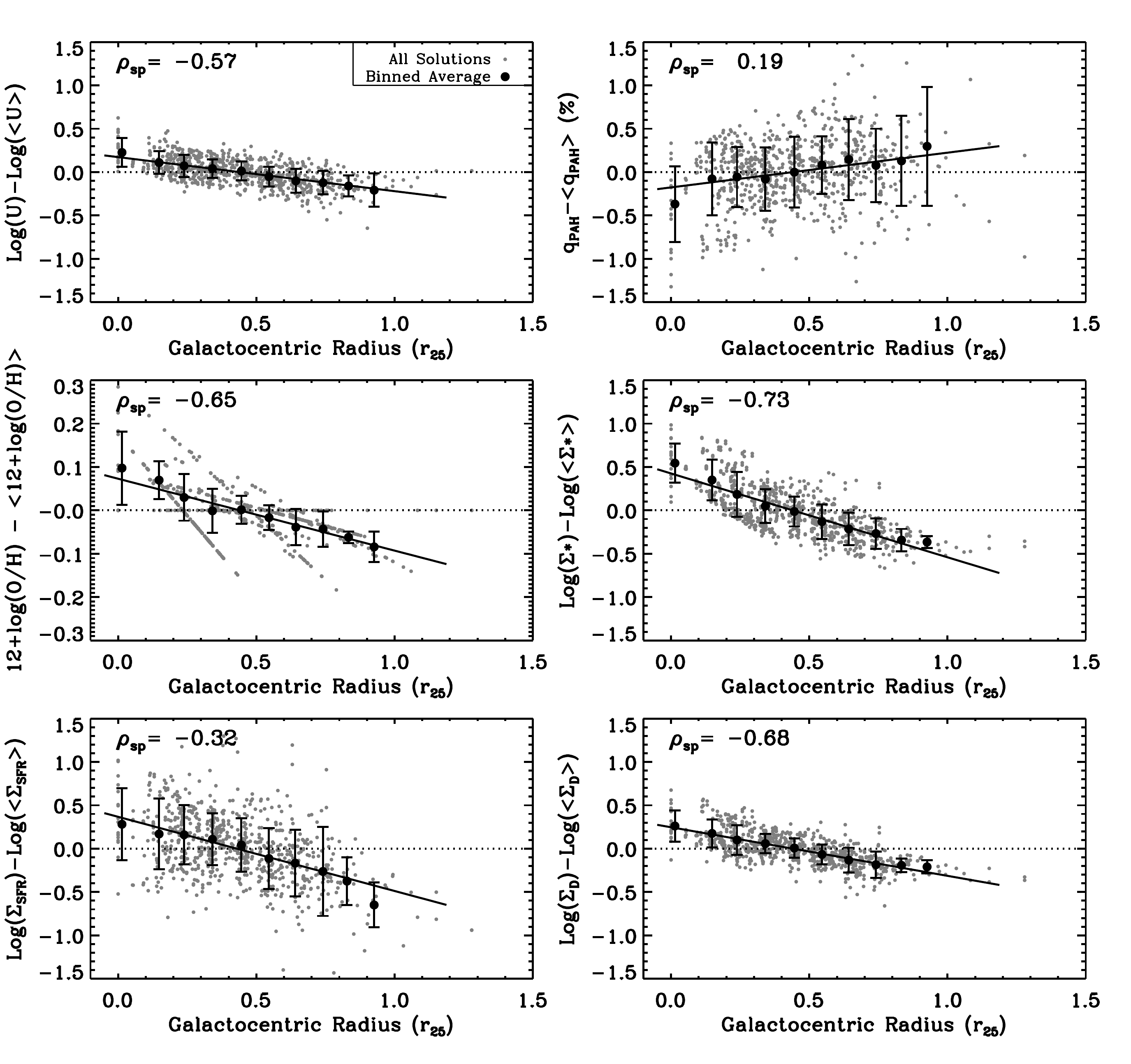}
\caption{Normalized environmental parameters as a function of galactocentric radius r$_{25}$.  In each panel we show the average value of each quantity in the solution pixel, normalized by the mean value for the galaxy, with gray circles.  The mean and standard deviation of the binned, normalized values are shown with black circles and error bars.  A linear fit is shown with a solid black line.  The Spearman rank correlation coefficient for each panel is listed in the top left. All parameters (radiation field strength, PAH fraction, metallicity, stellar mass surface density, star-formation rate surface density, dust mass surface density and \aco) show significant radial correlations.  Since we have assumed radial gradients in the metallicity for many galaxies, normalized metallicity directly reflects the slopes of these gradients.  In general, these radial trends are much stronger than the correlations between the variables and \aco, presented in Figure~\ref{fig:drivers}.}
\label{fig:driversvsr25}
\end{figure*}

The existence of a correlation between \aco\ and environmental parameters does not directly identify the cause of the variations in \aco, particularly since all of the parameters change radially to first order.  In Figure~\ref{fig:driversvsr25} we illustrate the radial variations of the same variables by plotting them normalized by their average value as a function of r$_{25}$.  All of the variables show significant correlations with radius. Table~\ref{tab:r25corr} lists the correlation coefficients and linear fits to the normalized radial profiles of the parameters.  In all cases, the correlation of the variables with r$_{25}$ is more significant than the correlation of \aco\ with those variables.

\begin{deluxetable}{lcccc}
\tablewidth{0pt}
\tabletypesize{\scriptsize}
\tablecolumns{5}
\tablecaption{Correlation Properties vs. r$_{25}$}
\tablehead{ 
\multicolumn{1}{c}{Variable vs.} &
\multicolumn{1}{c}{Correlation\tablenotemark{a}} &
\multicolumn{1}{c}{$\sigma$ from} &
\multicolumn{2}{c}{Linear Fit} \\
\cline{4-5}
\multicolumn{1}{c}{r$_{25}$} &
\multicolumn{1}{c}{Coeff.} & 
\multicolumn{1}{c}{Null} &
\multicolumn{1}{c}{Offset} &
\multicolumn{1}{c}{Slope} 
}
\startdata
Log(\aco$/\langle$\aco$\rangle_{\rm gal}$)                   & $+$0.18 & \phn5.0 &  $-$0.10 & $+$0.23  \\
Log($\overline{U}/\langle\overline{U}\rangle_{\rm gal}$) & $-$0.57 &    15.9 & \phs0.17 & $-$0.39  \\
q$_{\rm PAH}/\langle$q$_{\rm PAH}\rangle_{\rm gal}$      & $+$0.19 & \phn5.3 &  $-$0.17 & $+$0.40  \\
Log((O/H)$/\langle$O/H$\rangle_{\rm gal}$)           & $-$0.65 &    18.2 & \phs0.07 & $-$0.17  \\
Log($\Sigma_{*}/\langle\Sigma_{*}\rangle_{\rm gal}$)       & $-$0.73 &    20.5 & \phs0.43 & $-$0.97  \\
Log($\Sigma_{\rm D}/\langle\Sigma_{\rm D}\rangle_{\rm gal}$) & $-0.68$ & 19.1 & \phs0.25 & $-$0.56 \\
Log($\Sigma_{\rm SFR}/\langle\Sigma_{\rm SFR}\rangle_{\rm gal}$) & $-$0.31 & \phn8.8 & \phs0.37 & $-$0.86  
\enddata
\tablenotetext{a}{Spearman rank correlation coefficient.}
\tablecomments{These correlations are shown in Figure~\ref{fig:driversvsr25}.}\label{tab:r25corr}
\end{deluxetable}

If all of the variations of these parameters were primarily related to radius, we would not expect to find stronger correlations between \aco\ and any parameter than between \aco\ and radius.  Our results suggest, however, that the correlation between \aco\ and $\Sigma_{*}$ is stronger ($r_s = -0.25$) and more significant ($7.1\sigma$) than the correlation with radius r$_{25}$ ($r_s = +0.19$, $5.2\sigma$).  One possible explanation is that our $\Sigma_{*}$ measurement may trace the stellar profile better than r$_{25}$ itself.  Other explanations for the correlation of \aco\ with $\Sigma_{*}$ will be discussed in Section~\ref{sec:drivers}.    

Figure~\ref{fig:acovsr25} illustrates that the normalized radial profile of \aco\ shows a factor of $\sim2$ standard deviation in a typical radial bin.  This uncertainty can be explained primarily by the uncertainty on our \aco\ solutions themselves, which is typically close to a factor of $\sim2$.  It is therefore unlikely that we could predict \aco\ as a function of other variables to better precision than a factor of $\sim2$ unless the galaxy average \aco\ is highly correlated with that variable.  In general, the galaxy averages shown in Figure~\ref{fig:drivers} do not appear to be more tightly correlated with the environmental parameters.  Figure~\ref{fig:drivers} thus illustrates that aside from outliers at the extremes of these plots, the minimum standard deviation of \aco\ in these bins is a factor of 2 or more and correlations with environmental parameters do not allow us to predict the behavior of \aco\ within the galaxies better than our normalized radial profile.  

As previously discussed, the average profile of \aco\ versus radius is mostly flat with a central depression in some galaxies.  Many of the environmental parameters we plot in Figures~\ref{fig:drivers} and~\ref{fig:driversvsr25} show radial trends extending over the entire range we cover, although the normalized trends generally span less than an order of magnitude. This range does not provide the leverage to separate the dominantly radial correlations of multiple variables.  Future studies attempting to associate changes in \aco\ with environmental parameters will need either higher precision measurements of \aco\ or to span a greater range of environments.  

One possible way to overcome the limitations of separating various radial trends is to normalize all of the variables by their mean in radial bins and then search for residual trends between them.  We have investigated such non-radial variations in 0.1 r$_{25}$ bins and in general, find only very weak trends.  NGC~4254 is one of few galaxies that shows a marginally significant correlation---the normalized \aco\ measurements correlate with normalized $\overline{U}$ and $\Sigma_{*}$ at the level of 4-5$\sigma$ from the Spearman rank correlation coefficient.  In both cases, the correlation is positive, such that \aco\ increases in region with higher $\overline{U}$ and $\Sigma_{*}$ compared to the average in that radial bin.  NGC~6946 also shows a correlation at 5$\sigma$ between normalized \aco\ and f$_{\rm PDR}$.  In this case, the correlation is negative, meaning \aco\ decreases in regions with high f$_{\rm PDR}$.

NGC~6946 is unique in that the non-radial structure is clearly visible in the maps of \aco\ shown in the Appendix.  The low \aco\ values in that galaxy appear to track the spiral arm structure seen in the \ico\ and \sigd\ maps.  To make this observation quantitative, we have investigated the correlation between the radially normalized \aco, \ico, \sigd\ and \sighi.  These correlations are weak (all are $<5\sigma$ signficance in the Spearman rank correlation) but more widespread.  NGC 3627, 4254, 4321, 5457 and 6946 all show some degree of correlation at $>3\sigma$ with either radially normalized \ico, \sighi\ or \sigd.  For all galaxies aside from NGC~6946, the correlations are positive, in the sense that the conversion factor is higher where there is more \hi\ and dust or more CO emission.  NGC~6946 shows negative correlations with all of these parameters.  

\subsubsection{\aco\ versus Metallicity}\label{sec:acovsz}

Metallicity has been suggested by several theoretical and observational studies to be an important driver for \aco\ variations.  Unfortunately, metallicity measurements in nearby galaxies are often very uncertain and subject to systematic errors from different calibrations and techniques.  In Figure~\ref{fig:drivers} we show our \aco\ measurements as a function of metallicity from the PT05 calibration (middle-left panel).  Our measurements span $\sim0.5$ dex in metallicity with no statistically significant trend in \aco.  One possible reason for the lack of a clear trend in this figure is that we have combined metallicity measurements obtained with different techniques.  In addition, many of the galaxies in our sample lack constraints on possible gradients.  

\begin{figure*}
\centering
\epsscale{1.1}
\plottwo{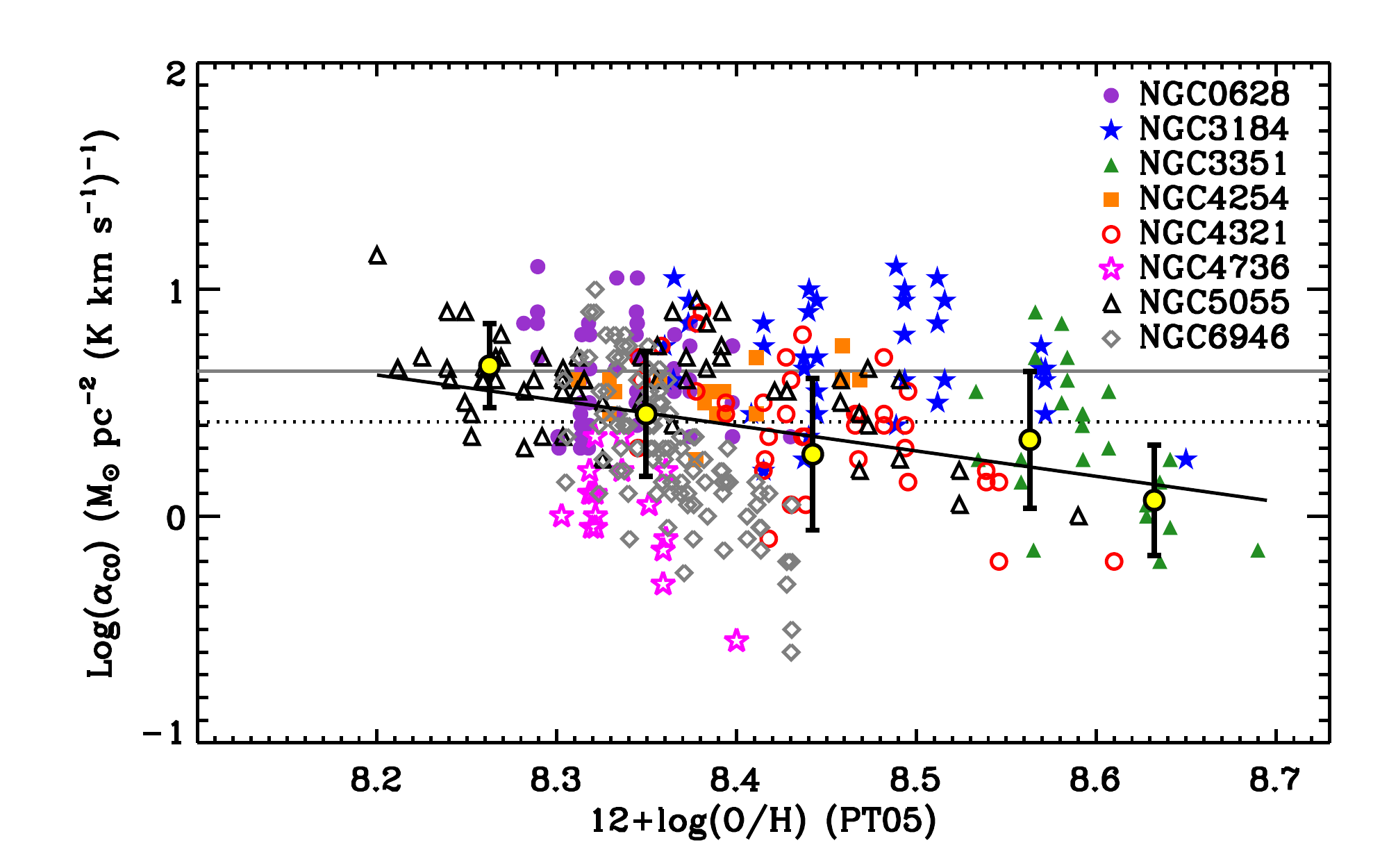}{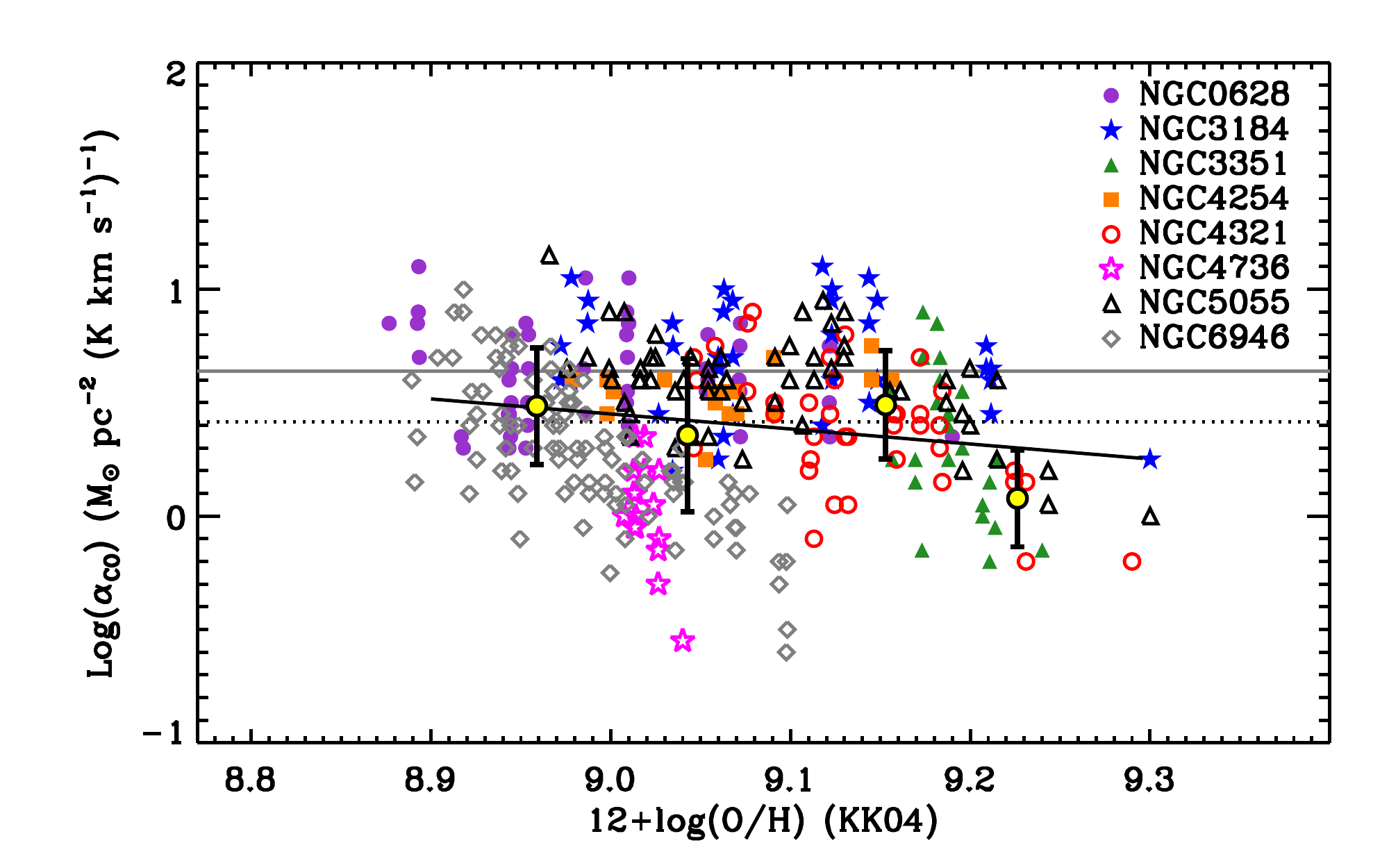}
\caption{Measured \aco\ values as a function of metallicity for galaxies with \hii\ region abundance gradients from \citet{2010ApJS..190..233M} in the PT05 (left) and KK04 (right) calibrations.  Only measurements with uncertainties less than $<0.3$ dex are shown.  The mean and standard deviation in 0.1 dex bins of metallicity are shown with yellow circles and black error bars.  A linear fit to \aco$(Z)$ is overlaid with a solid black line.  The correlation of \aco\ with metallicity is weak for both calibrations.  In particular it is clear that galaxies with shallow metallicity gradients like NGC~6946, can still span a wide range of \aco\ values, whereas those with large gradients, like NGC~3184, may have essentially constant \aco.}
\label{fig:m10metal}
\end{figure*}

To explore any metallicity trends that may be washed out due to systematic effects when combining metallicities from different sources, we isolate a sample of galaxies with uniformly determined metallicities from \citet{2010ApJS..190..233M} and plot those separately in Figure~\ref{fig:m10metal}.  We have eliminated galaxies whose metallicities have been determined from integrated spectrophotometry (those listed as ``M10 Table 9'' in Table~\ref{tab:metals}) because the drift scan observations were not always along angles that allow a robust gradient measurement.  We also isolate our highest confidence \aco\ measurements by showing only those with uncertainties less than 0.3 dex (a factor of $\sim2$).  We show metallicities in both the PT05 and KK04 calibrations for comparison.  

In general, there is a slight negative correlation of \aco\ with metallicity for this subsample.  In the PT05 calibration, we find a rank correlation coefficient of $r=-0.2$, which is 3.8$\sigma$ from the null result given the number of measurements.  Using KK04 metallicities, we find $r=-0.1$ at 2.0$\sigma$ from null.  If we remove the S/N cut on the \aco\ measurements, the correlations in both PT05 and KK04 essentially disappear, yielding results that are less than 1$\sigma$ from the null result.   

The weakness of the correlation between \aco\ and metallicity suggest that in the regions of the galaxies we are studying, metallicity may not be the primary driver of \aco\ variations.  Along these lines, it is interesting to note the contrast between a galaxy like NGC~6946, which has a relatively shallow metallicity gradient, and galaxies like NGC~0628 and 3184 which are thought to have much steeper gradients.  NGC~6946 has \aco\ spanning a range of an order of magnitude, while NGC~0628 and 3184 have much smaller ranges---the opposite of what we would expect if their metallicity gradient was the dominant factor controlling \aco.

\subsection{Properties of Full-Sample DGR Solutions}

In the Appendix we show the measured DGR for all galaxies and solution pixels in figures similar to Figure~\ref{fig:ngc0628_panel2}.  The DGR we report is the mean value for all of the individual sampling points in a solution pixel.  Generally, the scatter in the DGR$_i$ values within a solution pixel is small ($<0.1$ dex).  This means that the uncertainty on the \aco\ value generally dominates the uncertainty on DGR as well.  Therefore, it is important to note that the errors in the DGR and \aco\ values are highly correlated.  We assign a representative uncertainty on the DGR using the $\pm1\sigma$ bounds on \aco.  In general, when good solutions are obtained, the DGR measurements vary smoothly across the galaxy.  This supports our key assumption that DGR varies on scales larger than our solution pixels.  The smoothness of the DGR maps can be seen by inspecting the figures in the Appendix and the results for NGC~0628 shown in Figure~\ref{fig:ngc0628_panel2}. 

\begin{figure*}
\centering
\epsscale{0.8}
\plotone{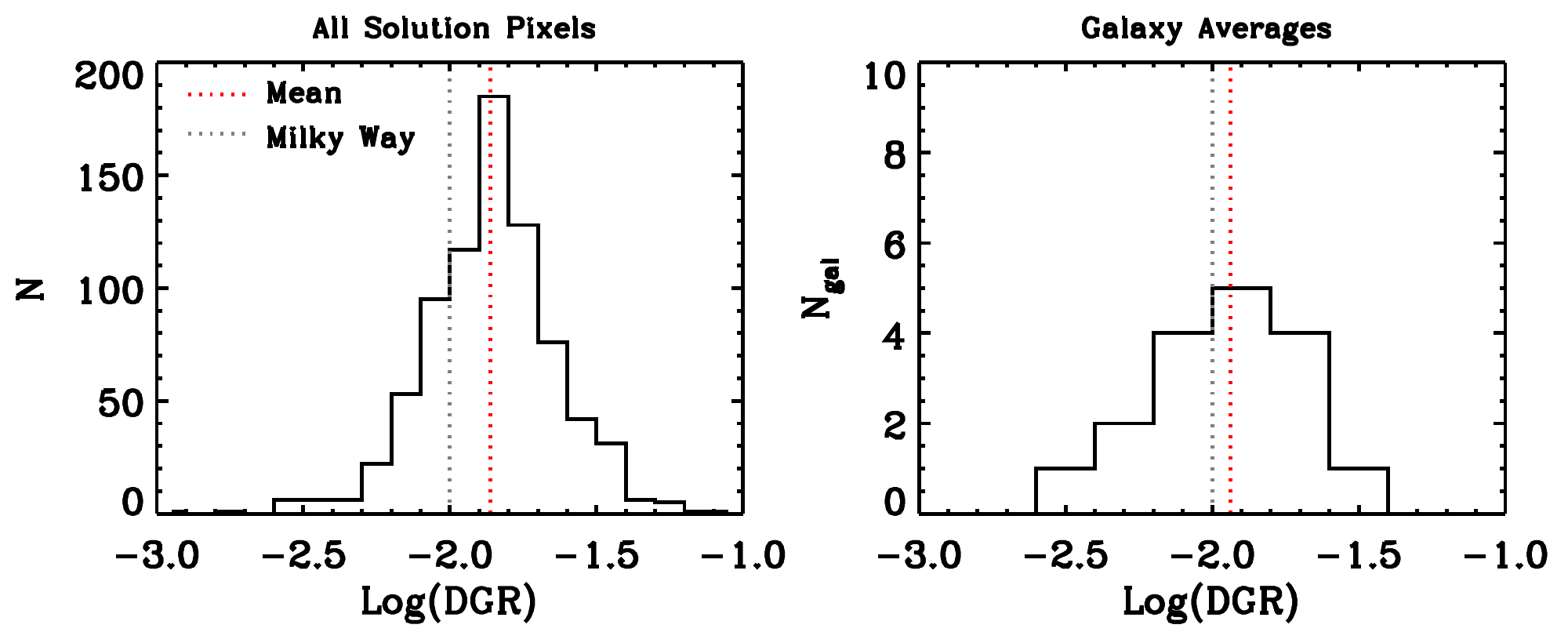}
\caption{Histograms of the measured DGR values for all solution pixels (left) and galaxies (right) in our sample.  The dotted red line shows the mean value for the sample and the dotted gray line shows the local MW DGR of 0.01 for comparison.  The average DGR for our sample is slightly higher than in the MW.  The standard deviation of the DGR measurements is 0.22 dex, which is smaller than the scatter in the \aco\ values for these same pixels.}
\label{fig:dgrhist}
\end{figure*}

In Figure~\ref{fig:dgrhist} we show a histogram of all of the measured DGR values for our galaxies that comply with our cuts on inclination and \ico.  The average of our measurements treating all lines of sight equally is Log(DGR)$=-1.86$ with a standard deviation of 0.22 dex.  This is slightly higher than the value typically adopted for the solar neighborhood of Log(DGR$_{\rm MW}$)$=-2.0$.  Forcing all galaxies to contribute equally to the average despite differing numbers of solution pixels gives Log(DGR)$=-1.96$.  The lower scatter in the DGR measurements for all solution pixels compared to the \aco\ measurements, despite the uncertainty on \aco\ dominating the uncertainty on DGR, is due to the contribution of \hi\ to the total gas mass surface density.  Because \hi\ makes up some fraction of the gas mass the DGR has a smaller possible range over which to vary compared to \aco.

\subsection{Correlations of DGR with Environmental Parameters and Radius}\label{sec:dgrvsz}

\begin{figure*}
\centering
\epsscale{1.1}
\plotone{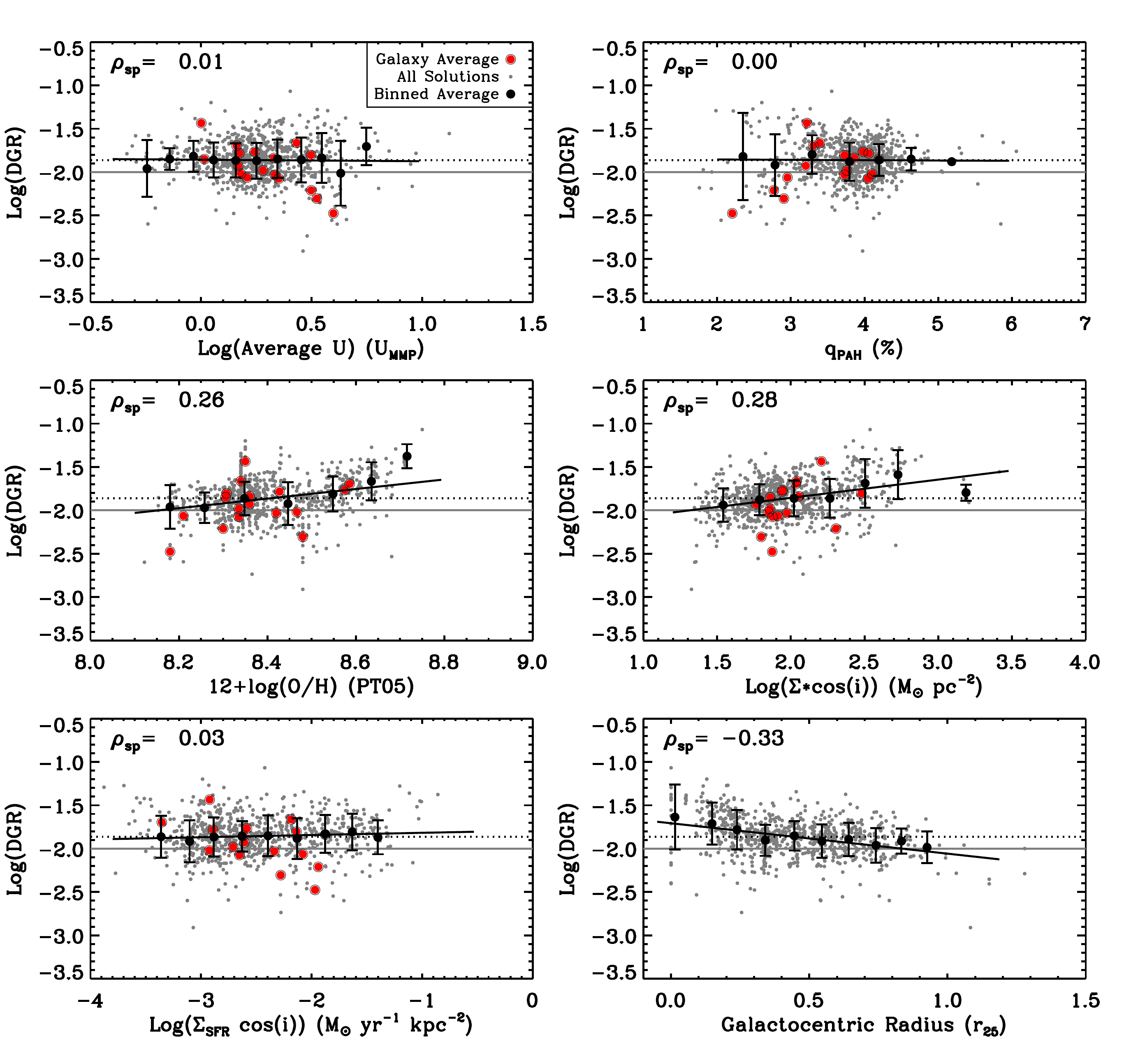}
\caption{DGR as a function of environmental parameters.  The panels show DGR plotted versus average radiation field intensity ($\overline{U}$; top-left) and PAH fraction (q$_{\rm PAH}$; top-right) from the \citet{2012ApJ...756..138A} fits, metallicity (12$+$Log(O/H); middle-left), stellar mass surface density ($\Sigma_{*}$; middle-right), star-formation rate surface density ($\Sigma_{\rm SFR}$; lower-left) and galactocentric radius (r$_{25}$; lower-right).  The solid gray line shows the Milky Way DGR and the dotted black line shows the sample average.  Each individual solution above our \ico\ and inclination cuts are shown with gray symbols.  The mean and standard deviation of those values in bins are shown with black circles and error bars.  A linear fit to the measurements is shown with a black line.  The mean values for all galaxies are shown with red circles.  The Spearman rank correlation coefficient for each panel is listed in the top left.  DGR shows statistically significant correlations with r$_{25}$, 12$+$Log(O/H) and $\Sigma_{*}$.}
\label{fig:driversdgr}
\end{figure*}

DGR may also vary as a function of environmental parameters.  To explore any such trends we show our measured DGR values as a function of the same tracers we have previously studied in Figure~\ref{fig:driversdgr}.  We list the correlation coefficients and linear fit parameters in Table~\ref{tab:corr_dgr}.  

\begin{deluxetable}{lcccc}
\tablewidth{0pt}
\tabletypesize{\scriptsize}
\tablecolumns{5}
\tablecaption{DGR Correlation Properties}
\tablehead{ 
\multicolumn{1}{c}{Log(DGR) vs.} &
\multicolumn{1}{c}{Correlation\tablenotemark{a}} &
\multicolumn{1}{c}{$\sigma$ from} &
\multicolumn{2}{c}{Linear Fit\tablenotemark{b}} \\
\cline{4-5}
\multicolumn{1}{c}{Variable} &
\multicolumn{1}{c}{Coeff.} & 
\multicolumn{1}{c}{Null} &
\multicolumn{1}{c}{Offset} &
\multicolumn{1}{c}{Slope} 
}
\startdata
r$_{25}$                & $-$0.33 & 9.2 & $-$1.71 &  $-$0.35  \\
Log($\overline{U}$)     & $+$0.01 & 0.3 & $-$1.87 &  $+$0.02  \\
q$_{\rm PAH}$ (\%)      & $+$0.00 & 0.1 & $-$1.85 &  $+$0.00  \\
12$+$Log(O/H)           & $+$0.26 & 7.3 & $-$6.50 &  $+$0.55  \\
Log($\Sigma_{*}$)       & $+$0.28 & 7.9 & $-$2.27 &  $+$0.21  \\
Log($\Sigma_{\rm SFR})$ & $+$0.03 & 0.8 & $-$1.79 &  $-$0.03  
\enddata
\tablenotetext{a}{Spearman rank correlation coefficient.}
\tablenotetext{b}{Correlation with metallicity using a homogeneous sample of metallicity measurements is discussed in Section~\ref{sec:dgrvsz}.}
\tablecomments{These correlations are shown in Figure~\ref{fig:drivers}.}\label{tab:corr_dgr}
\end{deluxetable}

DGR shows significant correlations with galactocentric radius, metallicity and $\Sigma_{*}$.  Even with heterogeneously determined metallicities, it shows a much clearer trend with 12$+$Log(O/H) than \aco\ does over the same range.  A correlation of DGR with metallicity is expected---given the high depletions of elements such as Mg, Si, Ca and Ti in the local area of the Milky Way, to first order the mass of dust should be proportional to the amount of heavy elements.
 
\begin{figure*}
\centering
\epsscale{1.1}
\plottwo{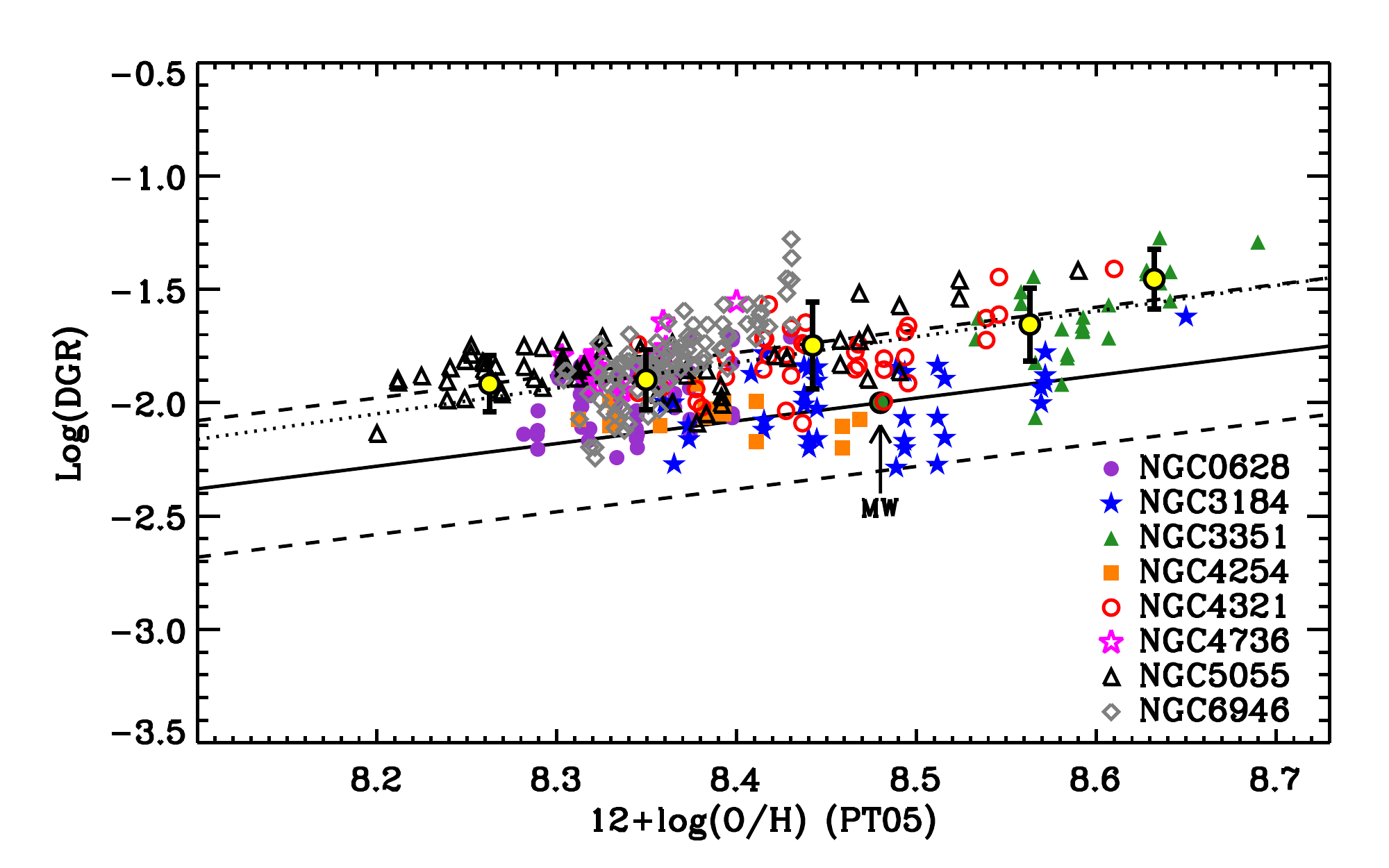}{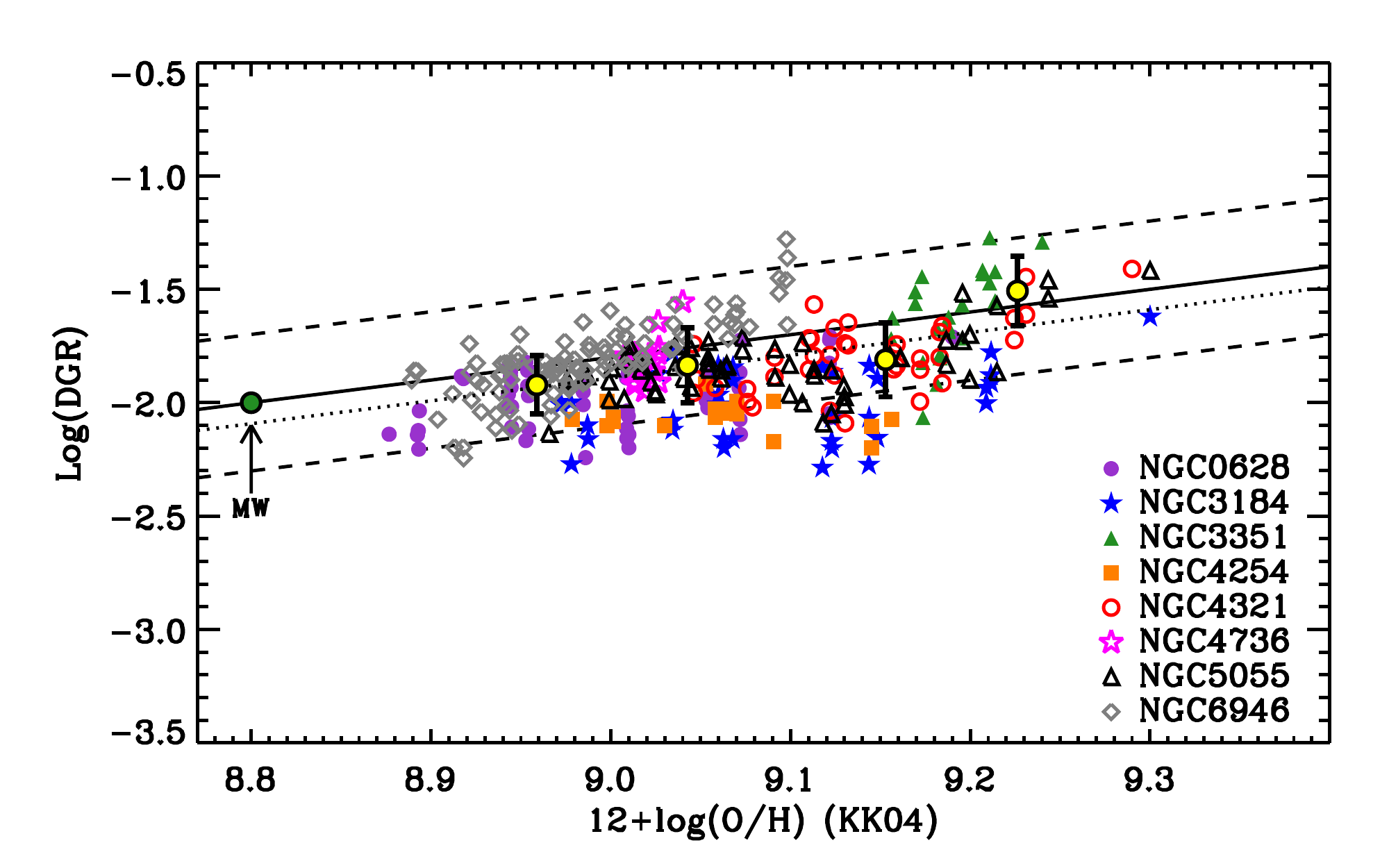}
\caption{Measured DGR values as a function of metallicity for galaxies with \hii\ region abundance gradients from \citet{2010ApJS..190..233M} in the PT05 (left) and KK04 (right) calibrations.  Only solutions where the uncertainty in \aco\ is less than 0.3 dex are shown, although the significance of the correlation between DGR and Z is relatively unchanged if all points are included.  A linear fit to DGR$(Z)$ is overlaid with a dotted black line. There is a clear correlation of DGR with Z.  The best fit slope is slightly below linear, but if NGC~3184 is removed the correlation is consistent with linear. The yellow circles show the mean DGR in bins of 0.1 dex in metallicity with error bars representing the standard deviation in those bins. The solid black line shows a linear scaling of the MW DGR and metallicity, with dashed lines above and below showing a factor of 2 difference from MW.  On the PT05 scale, our sample appears to have higher DGR for a given metallicity compared to the MW, while the opposite is true on the KK04 scale.}
\label{fig:dgrvsz}
\end{figure*}

In Figure~\ref{fig:dgrvsz}, we show DGR as a function of metallicity for the same sample of galaxies from Figure~\ref{fig:m10metal}.  In addition to the best linear fit, shown with a dotted black line, we also plot a prediction for scaling the local MW DGR linearly with metallicity (solid black line) and a factor of 2 higher and lower DGR with dashed lines.  The scaling of the plot is such that it covers three orders of magnitude, the same range covered in Figure~\ref{fig:m10metal}.  It is clear that the DGR values have much less scatter than the \aco\ values over the same range of metallicity---a product of the limited allowable range of DGR set by the fact that some dust is associated with \hi.  

The DGR values show a stronger correlation with metallicity than the \aco\ values.  In the PT05 calibration we find a rank correlation coefficient of 0.35 which is significant at 6.4$\sigma$.  For KK04, the correlation coefficient is $0.39$ at 6.9$\sigma$.  In a given metallicity bin, the standard deviation of the DGR values is $\sim0.15$ dex for PT05 and $\sim0.18$ for KK04.  This suggests that, on average, we should be able to predict DGR to better than a factor of 2 given the metallicity in one of these calibrations. In addition, we do not see evidence for major galaxy-to-galaxy offsets in the DGR versus metallicity plots, which distinguishes the DGR and \aco\ behavior, and most galaxies appear to have similar slopes in the plots.  

Neither the KK04 nor PT05 calibration values fall directly along the line of scaled MW DGR.  On the PT05 scale, almost all of our sample has higher DGR for a given metallicity than the Milky Way scaling, while the opposite is true on the KK04 scale.  For purposes of scaling the local Milky Way metallicity, we have used strong-line abundance measurements calculated from the integrated spectrum of the Orion Nebula.  

The best linear fit to the data is shown with a dotted line in both the PT05 and KK04 panels.  This fit is parameterized by the following equation where uncertainties have been determined using bootstrapping:  
\begin{equation}
\mathrm{Log(DGR)} = a + b(12+\mathrm{Log(O/H)}-c).
\end{equation}
For the PT05 measurements we find the following constants: 
\begin{eqnarray}
a = -1.86\pm0.01\nonumber\\
b = 0.85\pm0.11\nonumber\\
c = 8.39\nonumber
\end{eqnarray}
For the KK04 measurements the constants are:
\begin{eqnarray}
a = -1.86\pm0.01\nonumber\\
b = 0.87\pm0.11\nonumber\\
c = 9.05 \nonumber
\end{eqnarray}
In these equations, the constant $c$ is the mean metallicity of the sample of points.  We use this parameterization in order to avoid covariance in the offset ($a$) and slope ($b$). In both cases the slope is below unity.  It can be seen in Figure~\ref{fig:dgrvsz} that the flatter than unity slope is mainly due to NGC~3184 (blue stars).  Leaving this galaxy out of the fits produces slopes of 1.13$\pm$0.10 and 1.01$\pm$0.11 for PT05 and KK04, respectively.  It also increases the significance of both correlations by 1-2$\sigma$.  

We note that it is necessary to use the same strong-line metallicity calibration that we have (i.e. KK04 or PT05) in order to predict DGR (or \aco) using our fits.  To the extent that applying our DGR(Z)  is merely an interpolation over a given metallicity range there is relatively little uncertainty introduced in the process.  Any systematic errors in the metallicity scale itself are minimized if the same scale is used in the calibration and the application.

\section{Comparison to Literature}\label{sec:acolit}

\subsection{Measurements of \aco\ in the Milky Way and Nearby Galaxies}\label{sec:mwaco}

The local region of the Milky Way is seen to have a conversion factor close to \aco$=4.4$ \acounit.  This standard value has been recovered to within a factor of 2 by various techniques: $\gamma$-ray measurements \citep{1996ApJ...463..609D,2010ApJ...710..133A}; virial masses \citep{1987ApJ...319..730S}; and dust \citep{2001ApJ...547..792D,2008ApJ...679..481P,2011A&A...536A..19P}.  A number of studies of the Galactic Center, however, have found very different answers.  Using $\gamma$-ray observations, \citet{Strong:2004ds} found evidence that \aco\ in the Galactic Center was factors of 5$-$10 lower than in the disk.  Similar results using dust as a tracer for total gas mass have been found  by \citet{1995ApJ...452..262S} who suggest that \aco\ is lower by factors of 3$-$10.  Modeling of multiple $^{12}$CO, $^{13}$CO and C$^{18}$O lines by \citet{Dahmen:1998ta} found that the standard conversion factor from the disk overestimated molecular gas mass in the Galactic Center by an order-of-magnitude.  Thus, several independent measurements suggest that the Milky Way has \aco\ near its center between 3$-$10 times lower than in the solar neighborhood.  The properties of the transition from the standard disk value to this lower number are not well constrained although $\gamma$-rays provide some hint of a gradient \citep{Strong:2004ds}.

The agreement among the various techniques in the Galactic Center provides confidence in the resulting low \aco.  On their own, each technique is subject to a number of important systematic uncertainties.  For $\gamma$-ray modeling, a key limitation is our knowledge of the cosmic ray distribution.  In the case of multi-line modeling approaches, it is not clear that the approximations involved adequately represent the variety of excitation conditions in the molecular gas \citep{2000A&A...358..433M,Bayet:2006jx}.  Virial mass based techniques rely on the assumption that the clouds are in virial equilibrium, with gravity opposed primarily by turbulent motions. ÊIf magnetic stresses and thermal pressures are significant, or if GMCs are actually not self-gravitating, the virial technique may fail to reflect the true conversion factor, particularly in regions like the Galactic Center \citep[e.g.][]{Dahmen:1998ta}.

Measurements of \aco\ with various techniques in nearby galaxies have provided somewhat contradictory results.  Virial mass studies of GMCs in nearby galaxies tend to find \aco\ very similar to the Milky Way disk \citep{Wilson:1995kc,2008ApJ...686..948B}, even in galaxy centers \citep{2012ApJ...744...42D} and in low metallicity galaxies like the SMC \citep{2008ApJ...686..948B}.  In contrast, a number of studies using dust-based techniques have found very high \aco\ in low-metallicity galaxies, suggesting substantial amounts of ``dark molecular gas'' or ``CO-free H$_{2}$'' around GMCs \citep{Israel:1997tm,Dobashi:2008jq,2009ApJ...702..352L,2011ApJ...737...12L}. \citet{2011ApJ...737...12L} used dust to trace the total amount of gas and determine \aco\ for the major galaxies of the Local Group.  Their results suggested that 1) at low metallicity, \aco\ can be well above the standard MW value, 2) in the central few kpc of M~31, the conversion factor was slightly lower than the standard value, and 3) the 10 kpc ring of M~31, M~33 and the LMC (which collectively span $\sim0.5$ dex in metallicity) all showed \aco\ in agreement with the MW value. Studies of the centers of several nearby galaxies with ``large velocity gradient'' (LVG) or related modeling have found evidence for \aco\ up to an order-of-magnitude lower than the Milky Way value \citep[e.g.][]{Weiss:2001eq,2009A&A...506..689I,2009A&A...493..525I}.    

The study presented here is the first to create maps of \aco\ in galaxies outside the Local Group.  Therefore, we are able to study \aco\ across a range of environments and connect the various trends that have been found in previous work.  We find that for \ico$>1$ K km s$^{-1}$, \aco\ is relatively constant for a range of radii and metallicities in galaxies.  In the central $\sim$kpc, we observe that galaxies often show a lower than average \aco, sometimes by up to factors of 10. These low \aco\ values we find in several galaxy centers are not unprecedented, given the results from multi-line modeling studies and the evidence that the Galactic Center shows a similar depression.  Our lowest metallicities are still above the transition abundance seen by \citet{2011ApJ...737...12L} where galaxies in the Local Group began to display much higher \aco\ (i.e. 12$+$Log(O/H) $\lesssim$ 8.2).  The picture that emerges from the synthesis of these literature results is one where \aco\ is generally around the standard MW value with a factor of $\sim2$ variability in the disks of galaxies, in the regime where CO is bright and metallicity is greater than $\sim1/2$ Z$_{\odot}$, with some galaxies showing lower \aco\ in their central kpc.

\subsection{Direct Comparison of \aco\ with Virial Mass Measurements}

\begin{figure}
\centering
\epsscale{1.2}
\plotone{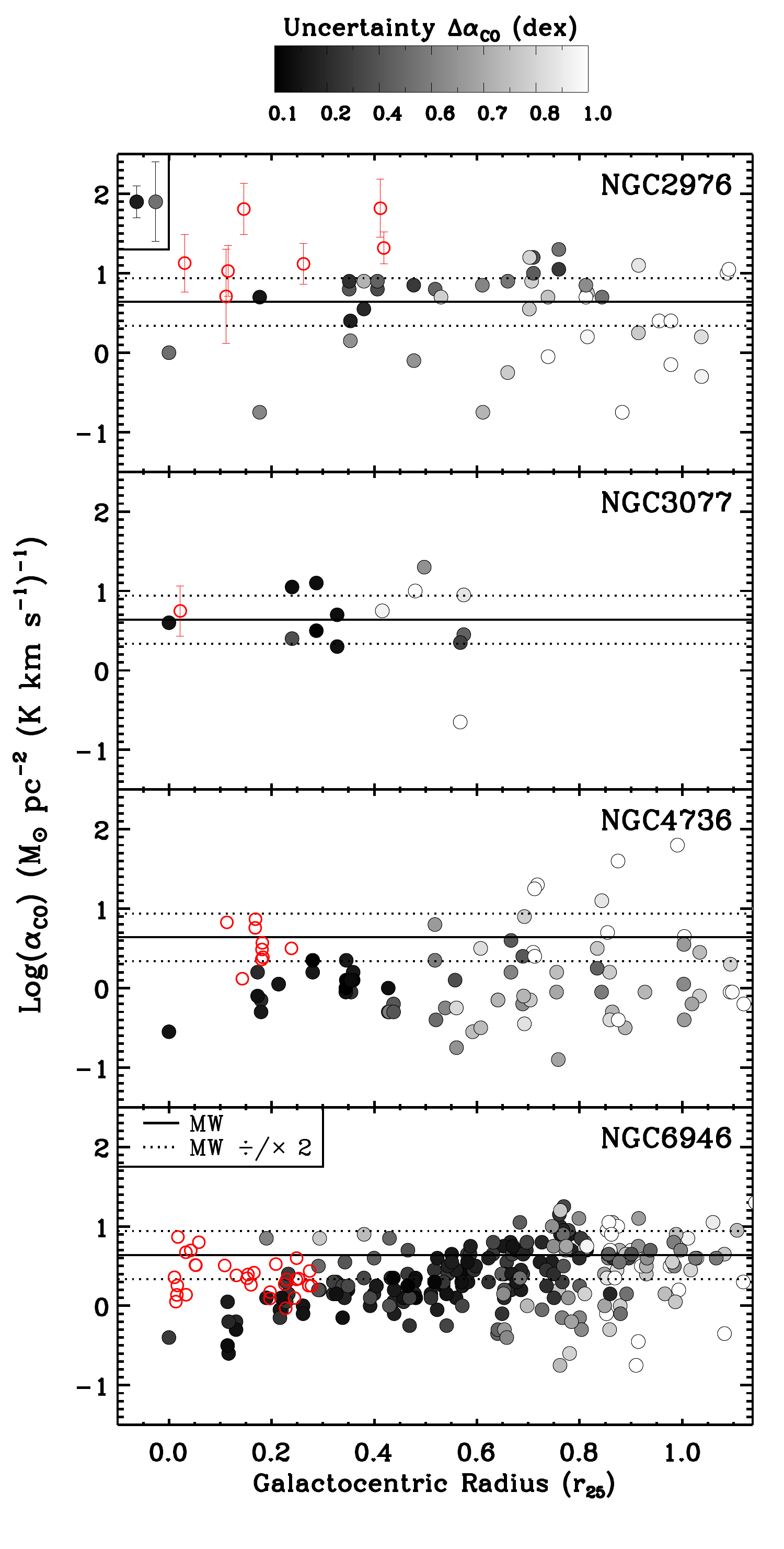}
\caption{A comparison between virial mass-based measurements of \aco\ from the literature (shown with red circles) and the results of this study.  The gray-scale color table for the points from this study shows the uncertainty on each \aco\ measurement, as in Figure~\ref{fig:ngc0628_panel3}.  Sample error bars for comparison with the gray-scale color table are shown in the top left corner of the plot.  Virial mass \aco\ measurements for NGC~2976 and 3077 are taken from \cite{2008ApJ...686..948B}.  Measurements for NGC~4736 and 6946 are from \citet{Meyer:2013tf} and \cite{2012ApJ...744...42D}, respectively.  For all of the targets, except NGC~3077, there is a discrepancy between the virial mass and dust based \aco.  At larger radii, the two measurements are in better agreement in NGC~6946.}
\label{fig:virialcomp}
\end{figure}

In several cases, our \aco\ measurements overlap with previous $^{12}$CO J=(1$-$0) virial mass based studies.  \citet{2008ApJ...686..948B} derived GMC properties and virial masses for NGC~2976 and 3077. Their results for NGC~3077 are in good agreement with previous studies by \citet{Meier:2001cl} and \citet{Walter:2002iy}.  Recently, \citet{2012ApJ...744...42D} and \citet{Meyer:2013tf} have made \aco\ determinations in the central regions of NGC~6946 and 4736, respectively, using virial masses.  In Table~\ref{tab:lit} we provide a summary of the galaxies in our sample with existing literature measurements.  Figure~\ref{fig:virialcomp} shows a comparison between these literature virial mass based \aco\ measurements with what is derived in this study.  In almost all cases we find a lower \aco\ using the dust-based technique than the virial mass technique.  

\begin{deluxetable*}{lcccc}
\tablewidth{0pt}
\tabletypesize{\scriptsize}
\tablecolumns{5}
\tablecaption{Literature \aco\ Measurements}
\tablehead{ \multicolumn{1}{c}{Galaxy} &
\multicolumn{1}{c}{Regan et al. (2001)} &
\multicolumn{1}{c}{3$\sigma$ Below MW} &
\multicolumn{2}{c}{Literature \aco\ Measurements} \\ 
\cline{4-5} 
\multicolumn{1}{c}{} & 
\multicolumn{1}{c}{ Central Excess?} &
\multicolumn{1}{c}{ \aco (this study)} &
\multicolumn{1}{c}{Type} &
\multicolumn{1}{c}{Reference} }
\startdata
NGC~0628 &  N  & N          &  & \\
NGC~2976 &  \nodata  & N & VM & B08 \\
NGC~3077 &  \nodata  & N & VM & B08 \\
NGC~3351 &  Y  & Y          &  & \\
NGC~3521 &  N  & N        &  & \\
NGC~3627 &  Y   & Y        &  &  \\
NGC~4321 &  Y  & Y           &  & \\
NGC~4631 &  \nodata & N & ML & I09\\
NGC~4736 &  Y    & Y        & VM & DM13 \\
NGC~5055 &  Y    & N\tablenotemark{a}        &   &  \\
NGC~6946 &  Y   & Y         & VM, ML & DM12a, IB01, MT04, W02  \\
NGC~7331 &  N  & N         & ML & IB99
\enddata
\label{tab:lit}
\tablenotetext{a}{2.6$\sigma$ below MW \aco.}
\tablerefs{B08 - \cite{2008ApJ...686..948B}; DM12a - \cite{2012ApJ...744...42D}; DM13 - \citet{Meyer:2013tf}; I09 - \cite{2009A&A...506..689I}; IB99 - \cite{Israel:1999va}; IB01 - \cite{2001A&A...371..433I}; M01 - \cite{Meier:2001cl}; MT04 - \cite{2004AJ....127.2069M};  R01 - \cite{2001ApJ...561..218R}; W02 - \cite{Walsh:2002bd}}
\tablecomments{VM - virial mass; ML - multi-line modeling.}
\end{deluxetable*}

For NGC~2976, \citet{2008ApJ...686..948B} found conversion factors $\sim 4$ times larger than the Milky Way value on average.  Almost all of our measurements for NGC~2976 have \aco\ consistent with the Milky Way value within their uncertainties.  The central solution pixel has \aco\ a factor of  4$-$5 below the Milky Way value.  In contrast, the virial mass and dust based techniques are in agreement for NGC~3077, both returning a conversion factor consistent with the Milky Way.  The center of NGC~6946 stands out in our study as places with particularly low \aco.  However, virial mass measurements find \aco\ more consistent with the Milky Way value.  Figure~\ref{fig:virialcomp} illustrates that the discrepancy decreases with radius for NGC 6946---many of the points with galactocentric radii closer to $r_{25} \sim 0.2$ agree with our dust-based measurements.  We will discuss the discrepancy between the virial and dust-based \aco\ values in the galaxy centers further in Section~\ref{sec:centralaco}.  It is interesting to note that the normalcy of NGC~3077, currently undergoing a starburst, compared to the also highly star-forming centers of 4736 and 6946, suggests that conditions other than just high star-formation surface density must contribute to determining \aco. 

\subsection{Direct Comparison with Multi-Line Modeling Techniques}\label{sec:mllit}

The center of NGC~6946 has been observed extensively in molecular gas lines and modeled in a variety of ways.  \citet{2001A&A...371..433I} modeled $^{12}$CO and $^{13}$CO lines plus [C~I] (492 GHz) observations and found a conversion factor $\sim 10$ times lower than the Milky Way value in the center. Using LVG modeling of multiple CO isotopomers including $^{12}$C$^{18}$O, \cite{Walsh:2002bd} found a conversion factor $4-5$ times below the Milky Way value.  Similarly, \citet{2004AJ....127.2069M} used an LVG analysis of C$^{18}$O observations, in addition to multiple CO lines, to argue that the central region had a conversion factor 4 times below the Milky Way value.  All of these studies suggest a central \aco\ between $4-10$ times lower than the Milky Way value, in good agreement with what we observe with dust, but in contrast with the virial mass results.  

Several other galaxies in our sample have been modeled with multi-line techniques, however they are at high inclination and therefore do not have reliable \aco\ measurements from our work.  \citet{2009A&A...506..689I} used multiple $^{12}$CO and $^{13}$CO lines plus [C~I] (492 GHz) observations towards the central region of NGC~4631 to constrain \aco\ and found a value 6 times lower than the standard Milky Way value.  Similarly, \cite{Israel:1999va} argued for \aco\ $\sim 5$ times lower than the Milky Way value in the center of NGC~7331.  In both of these highly inclined galaxies we find \aco\ a factor of 2 or more higher than the MW \aco.  This is due to the failure of our technique at high inclinations.

\subsection{Comparison with CO Exponential Disk Profiles}

\citet{2001ApJ...561..218R} found that approximately half of their sample of galaxies observed in the BIMA SONG survey show excess CO emission in their centers compared to the extrapolation of their best fit exponential disk profile.  They argued that this could be due to enhanced reservoirs of molecular gas in the centers or a change in the conversion factor. Table~\ref{tab:centers} lists the central excess classification for the galaxies in our sample studied by \citet{2001ApJ...561..218R}.  Indeed, we find that in every case where they found excess CO above the exponential profile, we find a conversion factor significantly below the Milky Way value.  Conversely, for the three galaxies we have observed and that they found to {\em not} have an excess, we find conversion factors consistent with the Milky Way value or slightly higher.  Similar results are found when comparing with the radial profiles determined from the HERACLES CO maps we utilize in this study \citep[see radial profiles in ][]{2011AJ....142...37S}.  In general, the offset between the central \aco\ and galaxy average \aco\ from our measurements is similar in magnitude to the excess observed at the centers compared to the extrapolated exponential disk profile.      

\section{Discussion}\label{sec:disc}

\subsection{Drivers of \aco\ Variations}\label{sec:drivers}

In the following, we explore the correlations or lack thereof between \aco\ and parameters that may influence it.  Since our resolution elements are large compared to an individual GMC, our \aco\ value is an average over a population of GMCs, including any molecular gas that may be in a more diffuse phase and not in self-gravitating clouds.  

For an individual, self-gravitating, turbulence-supported molecular cloud, there are several key parameters that influence \aco: the density and temperature of the gas and the fraction of the mass that exists in the ``CO-dark'' phase (where H$_2$ self-shields and CO is photodissociated).  The general dependence of \aco\ on density and temperature can be illustrated using a simple model---a spherical, homogeneous cloud where self-gravity is opposed primarily by turbulence and $^{12}$CO J=(1$-$0) is optically thick. Following the derivation in \citet{2011piim.book.....D},    
\begin{equation}
\alpha_{\rm CO} = 3.4 n_3^{0.5} (e^{5.5/T_{ex}}-1)\ \text{\msun}\ \mathrm{pc}^{-2} (\mathrm{K~km~s}^{-1})^{-1}.
\end{equation}
Here, $n_3$ is the H nucleon density $n_H=10^3 n_3$ cm$^{-3}$ and $T_{ex}$ is the excitation temperature.  Real molecular clouds are certainly not spherical and homogenous, but this model allows us to understand the basic dependence of \aco\ on density and temperature \citep[note that simulated molecular clouds with non-spherical, non-homogenous structure show similar basic scalings;][]{Shetty:2011ef}.  Molecular clouds with increased density will have higher \aco\ while clouds with increased temperature have lower \aco.  

Because CO is optically thick at solar metallicity, the abundance of C or O does not directly influence \aco.  However, metallicity may indirectly influence \aco\ by altering the density or temperature structure of the gas, since the heating and cooling rates may be affected by metallicity or correlate with metallicity.  At lower metallicities, the transition between ionized and neutral carbon to CO may shift relative to the \hi/H$_2$ transition due to a lack of dust shielding, leading to layers of ``CO-dark'' H$_2$ \citep{1985ApJ...291..722T,1988ApJ...334..771V,1993ApJ...402..195W,1999ApJ...527..795K,Bell:2006ie}.  \aco\ will increase when there are significant amounts of gas in these layers. Recent work by \citet{2010ApJ...716.1191W} used PDR modeling and a spherical model for a turbulent molecular cloud to study variations in the conversion factor.  They find that the ``dark molecular gas'' fraction (i.e. the fraction of the molecular cloud mass where H$_2$ exists but CO is photodissociated) depends primarily on the extinction through the cloud or DGR. Studies of simulated molecular clouds have also found a dependence of \aco\ on extinction through the cloud \citep{2011MNRAS.412..337G}.

The simple model we have described above assumes that the cloud's self-gravity is balanced by turbulence.  If there are other forces that contribute to the virial equation, \aco\ may be altered.  For a self-gravitating cloud, an increase in the pressure on the surface of the cloud will lead to a larger velocity dispersion.  Because CO is optically thick, the amount of emission that ``escapes'' in the CO line is directly tied to the velocity dispersion of the cloud and increasing the velocity dispersion increases the CO emission for the same mass of H$_2$.  Therefore, if a self-gravitating cloud were subject to significant external pressure we would expect a lower \aco.  Conversely, if magnetic fields played an important role in supporting the cloud, this would lower the velocity dispersion and raise \aco.  In ultra-luminous infrared galaxies (ULIRGs), the molecular gas may not be in individual self-gravitating clouds, but instead in a large-scale molecular medium where stellar mass contributes to the gravitational potential. \citep[e.g.][]{1998ApJ...507..615D}.  In this case, the velocity dispersion is larger than what would be expected from the gas mass alone and the CO emission is enhanced for a given gas mass (resulting in a lower \aco).

In a $\sim$kpc region of a galaxy, many individual molecular clouds will be averaged together in our measurement.  Therefore, another factor which could contribute to the variation of our measured \aco\ is changes in the cloud populations.  In addition, any significant component of diffuse CO emission would be included as well. Several studies have suggested that CO emission from diffuse gas may not be negligible, even in the local area of the Milky Way \citep{1998A&A...339..561L,2008ApJ...680..428G}.  Interestingly, it appears that locally, \aco\ is similar in diffuse molecular gas and self-gravitating GMCs \citep{2010A&A...518A..45L}.  Recent work by \citet{2012A&A...541A..58L} has shown that under differing environmental conditions, the conversion factor appropriate for diffuse gas can vary substantially. Therefore, depending on the amount of diffuse molecular gas and the local conditions in a galaxy, \aco\ may be very different from what is observed in the local area of the Milky Way.  The contribution of diffuse molecular gas with a different \aco\ has been cited previously as a cause for discrepancies between different techniques for measuring \aco\ in M~51 \citep{2010ApJ...719.1588S}.   

To summarize our theoretical expectations, \aco\ measured in $\sim$kpc regions of nearby galaxies can vary as a function of environment for many reasons: changes in the excitation temperature and density of the gas or contributions from pressure or magnetic support in self-gravitating clouds; envelopes of ``CO-dark'' gas; changes in the molecular cloud population; or contributions from diffuse CO emission.  Unfortunately, directly measuring temperature, density and velocity dispersion in large samples of extragalactic GMCs is challenging due to the need to resolve individual clouds in multiple molecular gas emission lines.  Therefore, we are left with more indirect tracers of  changes to the GMC properties.  In Section~\ref{sec:acocorr} we examined the correlation of \aco\ with $\overline{U}$, q$_{PAH}$, metallicity, $\Sigma_{*}$, $\Sigma_{\rm SFR}$, \sigd\ and galactocentric radius.  

The average radiation field $\overline{U}$, measured from the dust SED, could influence \aco\ through the gas excitation temperature.  If photoelectric heating dominates over other heat sources (e.g. cosmic rays) at the $\tau_{\rm CO} \sim 1$ surface, then the intensity of the radiation field may play a role in determining T$_{ex}$ \citep{1993ApJ...402..195W}.  Likewise, q$_{\rm PAH}$ and metallicity could both influence the efficiency of the photoelectric effect \citep{1994ApJ...427..822B,Rollig:2006fx}, thereby changing the heating rate.  We expect that $\Sigma_{\rm SFR}$ should be responsible for higher $\overline{U}$ in many regions.  Enhanced SFR leading to higher radiation field intensities has been suggested as the explanation for high observed \aco\ in several outer-disk molecular clouds in M~33 \citep{2010ApJ...725.1159B}.  If the gas excitation temperature were affected by the radiation field or photoelectric heating, we would expect negative correlations (i.e. lower \aco\ at higher $\overline{U}$, $\Sigma_{SFR}$, q$_{\rm PAH}$, etc).  The correlations we observe between \aco\ and $\overline{U}$, q$_{\rm PAH}$ and $\Sigma_{\rm SFR}$ are generally weak.  $\overline{U}$ and \aco\ show the strongest association and the slope of the trend is negative, with lower \aco\ at higher $\overline{U}$.  This is consistent with the expectation of higher radiation field intensities leading to warmer molecular gas temperatures, but the correlation is weak and other variables may play a more important role.

\begin{figure*}
\centering
\epsscale{1.1}
\plottwo{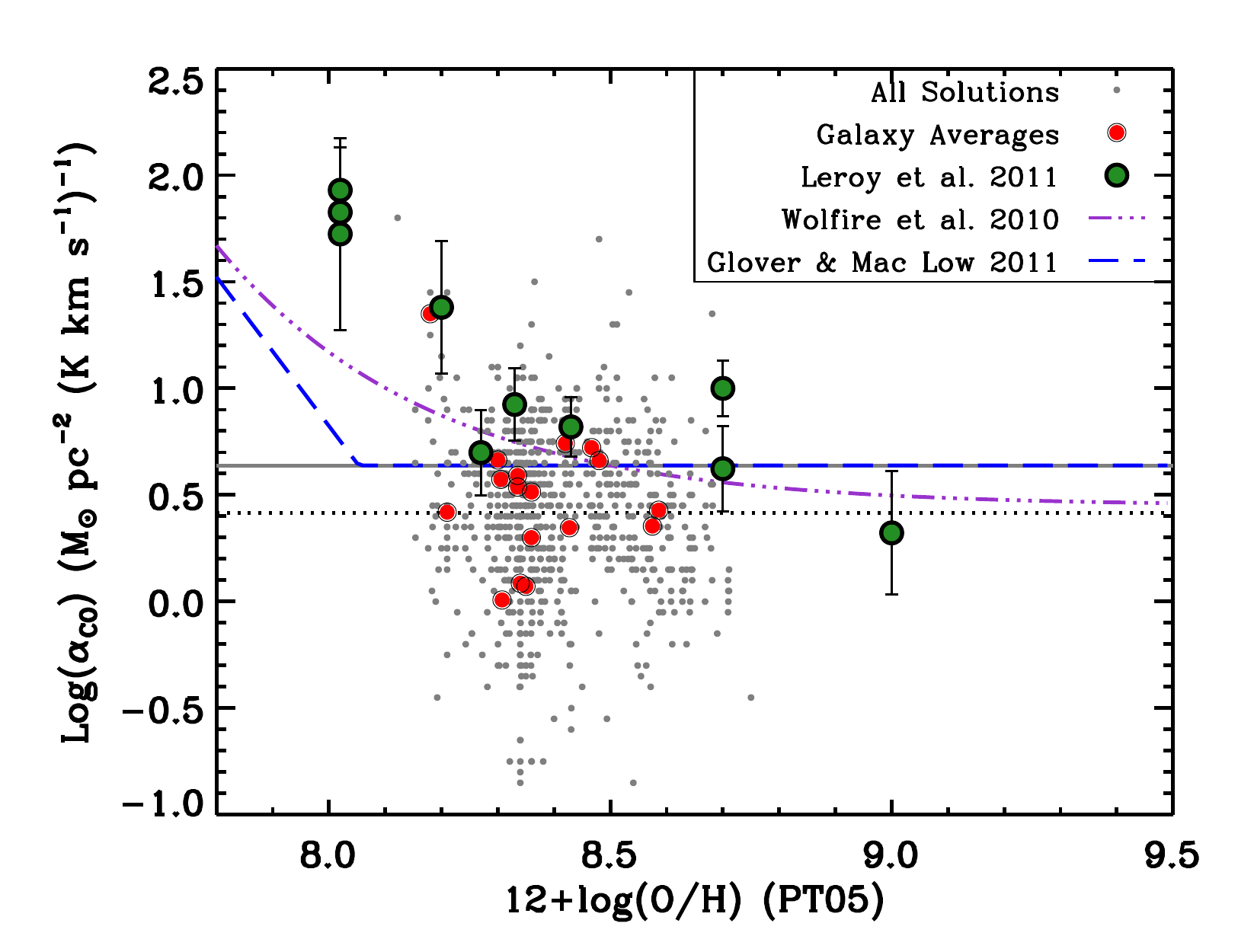}{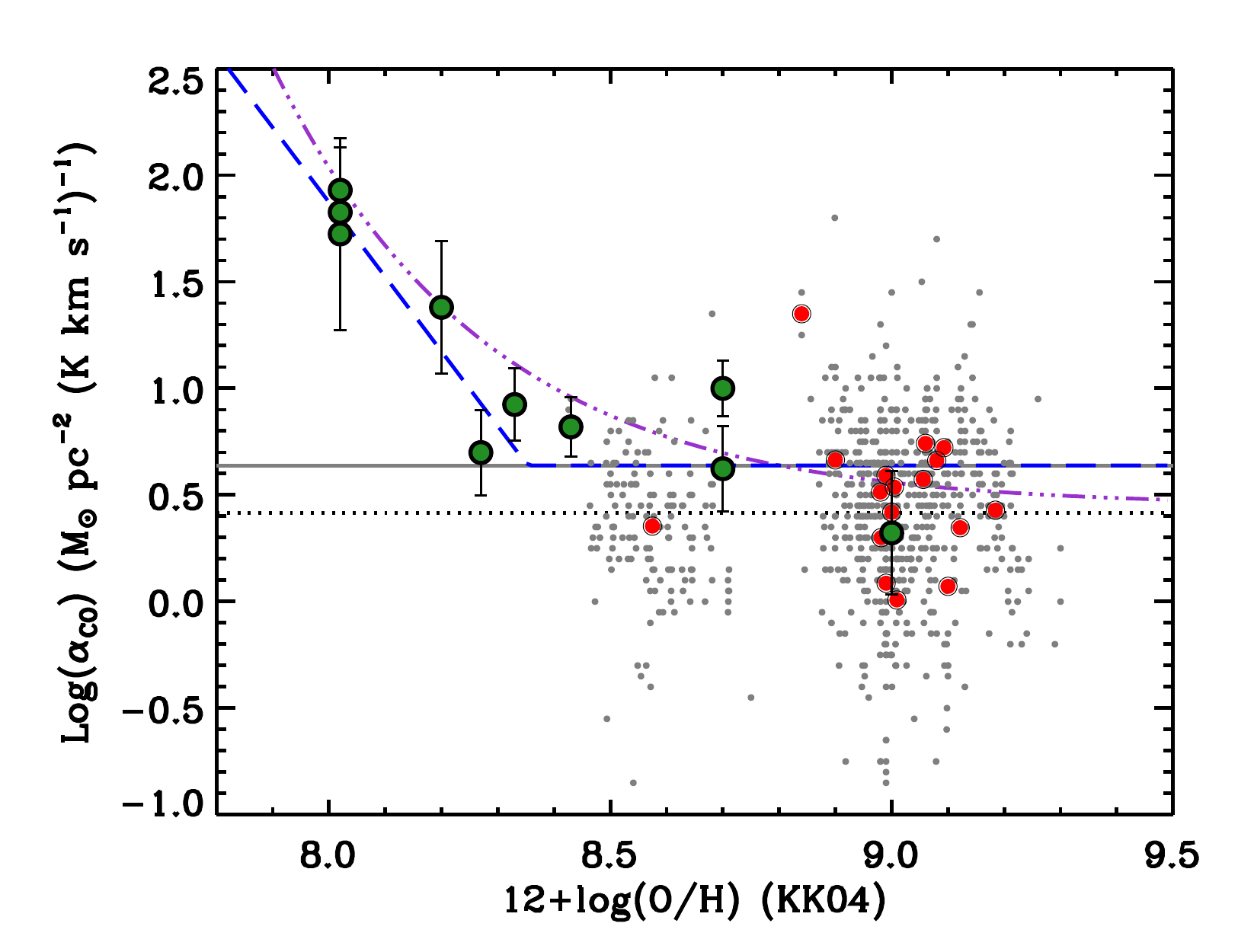}
\caption{\aco\ as a function of metallicity compared to previous measurements and models. The left panel shows measurements in the PT05 calibration while the right panel shows KK04.  Measurements from this paper are shown with gray points (individual solutions) and red circles (galaxy averages).  We show all of the solution pixels where \ico $>1$ \kms\ and $i>65^{\circ}$, regardless of the source of the metallicity measurement (i.e. without the requirement of having measured \hii\ region metallicities from M10).  Measurements of \aco\ in Local Group galaxies from \citet{2011ApJ...737...12L} are shown with green circles.  The MW \aco\ is shown with a gray line and the average of our solution pixels with no weighting is shown with a dotted black line.  Predictions based on the model of \citet{2010ApJ...716.1191W} are shown with a purple dot-dashed line and those based on \citet{2011MNRAS.412..337G} are shown with a dashed blue line.  These predictions assume a linear dependence of DGR on metallicity and a fixed gas mass surface density for molecular clouds of $\Sigma_{\rm GMC}=100$ \msun\ pc$^{-2}$.  The model predictions are normalized to have \aco $=4.4$ \acounit\ at the metallicity adopted for the Milky Way in each calibration.  Note that the metallicities we use for  NGC~5457 are not from strong line calibrations, so it appears in the same position both plots.  Regardless of the metallicity calibration, our measurements do not extend to low enough metallicities to constrain the effects of ``CO-dark'' H$_2$.}
\label{fig:acocomp}
\end{figure*}

We see a weak trend of \aco\ with metallicity.  Since our observations are limited to regions with metallicities similar to or higher than the Milky Way, we may not expect to see a strong correlation between \aco\ and metallicity.  The fraction of gas mass in the layer where CO is photodissociated is predicted to be $\sim 30$\% at Milky Way metallicity/DGR \citep{2010ApJ...716.1191W}.  Thus, increasing the DGR should only have a minimal effect on the conversion factor at MW metallicity and above.  To illustrate this point, in Figure~\ref{fig:acocomp} we plot our \aco\ measurements as a function of metallicity and overlay the \citet{2011ApJ...737...12L} Local Group measurements and the models of \citet{2010ApJ...716.1191W} and \citet{2011MNRAS.412..337G}.  The model predictions are normalized to have \aco $=4.4$ \acounit\ at the MW metallicity (i.e. 12$+$Log(O/H) = 8.5 in PT05 and 8.8 in KK04; from the Orion Nebula).  Both models assume a linear scaling of DGR with metallicity and a fixed gas mass surface density for molecular clouds.  The model predictions are discussed in detail in \citet{Leroy:2013um}.  In both calibrations, our measurements do not extend into the regime where ``CO-dark'' H$_2$ dominates \aco.  We note that an investigation of the  \aco-DGR relationship using our results would require taking into account the strong correlation of the uncertainties in these parameters.   

The strongest correlation we observe for \aco\ is that with $\Sigma_{*}$.  This correlation is the only one that is stronger than the correlation of \aco\ with galactocentric radius r$_{25}$.  In essence, our definition of r$_{25}$ is based on a scale length of the stellar disk, defined by a B-band surface brightness, so it is possible the stronger correlation of \aco\ with $\Sigma_{*}$ is due to it being a better proxy for the disk scale length than r$_{25}$.  In addition, of the environmental parameters we have available for our analysis, $\Sigma_{*}$ is the most straightforward to measure and may show the smallest systematic uncertainties.  It may be the strongest correlation simply because the other parameters are more uncertain.  The question remains, however, as to why \aco\ would correlate with the stellar mass surface density.  There are a number of possibilities: pressure contributions to GMC virial balance, enhanced fractions of diffuse CO emission correlated with ISM pressure, changes to the population of GMCs, or other effects.  We do not speculate here about what causes this correlation, only note that observations of multiple molecular gas lines at GMC resolution will help understand these trends and are possible with ALMA.

Although we see some evidence for weak correlations between \aco\ and environmental variables, it is not possible with this dataset to distinguish the cause of these variations.  Our primary result is that in regions with \ico $>1$ \kms\ in these galaxies, \aco\ is mostly insensitive to environment except in the central regions.  This may be due to our limited S/N for each individual \aco\ measurement plus the relatively small range of environmental conditions we probe.  Another possibility is that in the disks of most galaxies, the properties of molecular gas are insensitive to environment because most of the gas resides in molecular clouds that are not under significant external pressure or subject to enhanced turbulence or magnetic fields.  The conversion factor appropriate for such clouds should be similar to the standard Milky Way value. 

\subsection{Comparison with Galaxy Simulations}

\begin{figure}
\centering
\epsscale{1.2}
\plotone{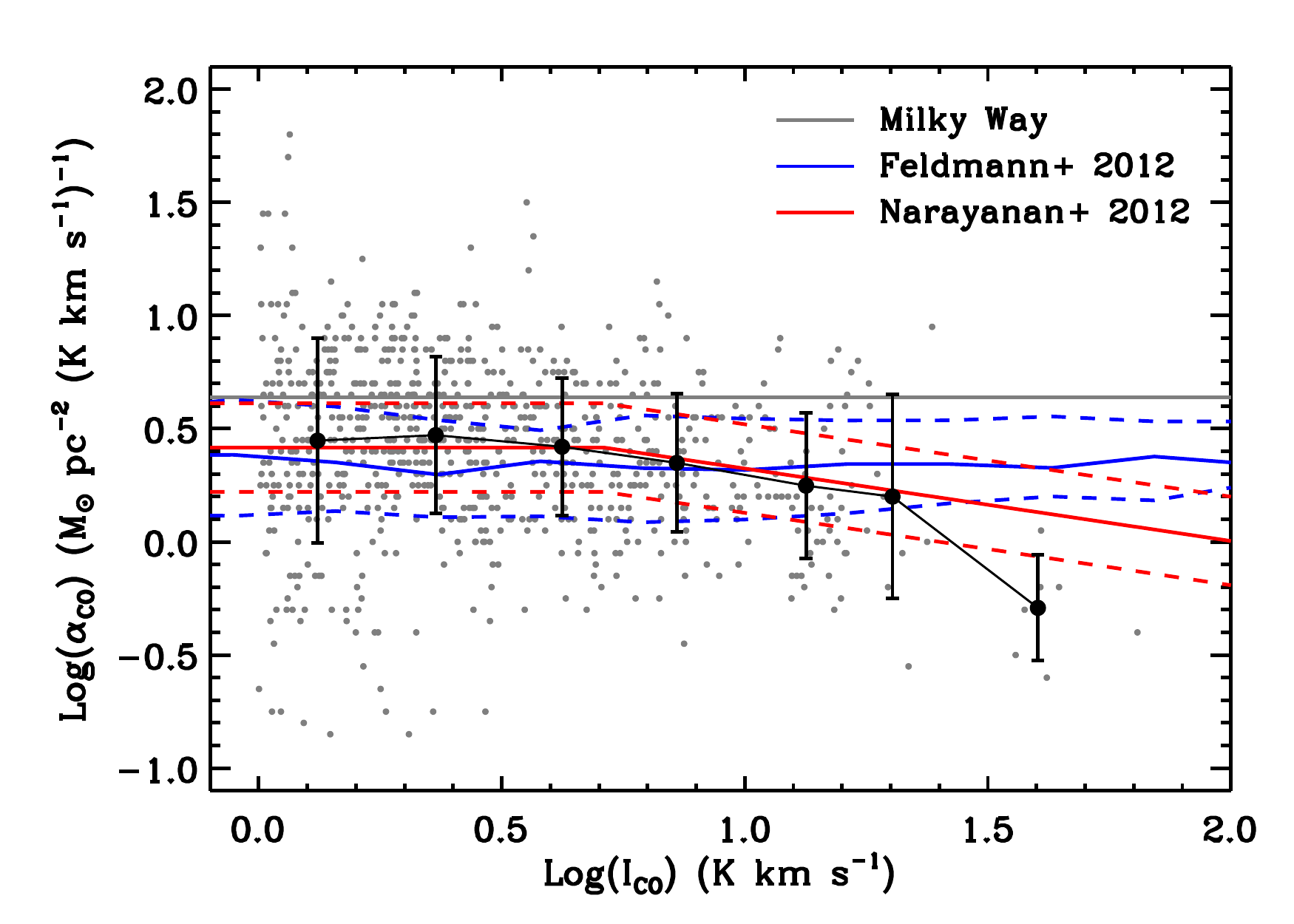}
\caption{A comparison between our \aco\ measurements and the predictions of simulations by \citet{2012MNRAS.tmp.2537N} and \citet{2012ApJ...747..124F}.  On the x-axis we show the average \ico\ in a given solution pixel.  All solutions above our \ico\ and inclination cuts are shown as gray circles.  The simulation predictions are shown at three possible metallicities: $Z=Z_{\odot}$ in a solid line and a factor of 2 above and below that with dashed lines (for \citet{2012ApJ...747..124F} we show $Z=Z_{\odot}$, $3\times Z_{\odot}$ and $0.3\times Z_{\odot}$).  The mean and standard deviation in 0.25 K km s$^{-1}$ bins are shown with black circles and error bars.  The predictions from the simulations have been divided by a filling factor of 3 to correct for our different resolutions. In general, the average behavior of our \aco\ measurements agrees well with the predictions of these models except at the highest \ico.}
\label{fig:comptheory}
\end{figure}

Because of the resolution of our observations, our measurements average over a population of molecular clouds.  This means that environmental variations in the cloud population can also be responsible for changes in \aco.  Therefore, it is interesting to compare our results to what has been found with galaxy simulations, despite the fact that these simulations are forced to adopt a sub-grid model that prescribes \aco\ for unresolved clouds. In Figure~\ref{fig:comptheory} we show our measured \aco\ values as a function of the average \ico\ in each solution pixel.  Due to our completeness cut, we have no points below 1 K km s$^{-1}$.  The mean and standard deviation of the \aco\ measurements in bins of 0.25 K km s$^{-1}$ are shown with black circles and error bars.

Overlaid on Figure~\ref{fig:comptheory} we show the predictions of two recent simulations.  \citet[F12]{2012ApJ...747..124F} couples full galaxy simulations with the \citet{2011MNRAS.412..337G} cloud-based conversion factor predictions assuming a fixed gas temperature and either a constant line width of 3 \kms\ or a virial scaling.  The \citet[N12]{2012MNRAS.tmp.2537N} predictions include sub-grid semi-analytic models for the chemical and thermal state of the gas and dust.  In the N12 models, the temperature of the molecular gas is not fixed and depends on heating by cosmic rays and, in dense regions, on energy transfer with dust.  For both models it is necessary to adopt a filling factor or clumping correction due to the difference in resolution between the simulations and our observations.  F12 and N12 both have $\sim60-70$pc resolution, so we apply the same correction to both.  For purposes of comparing with the trend of our observed \aco\ as a function of CO integrated intensity, we have adopted a filling factor correction of 3.  This correction factor is larger than purely the ratio of the resolutions because our beam will contain more than one molecular cloud.  Verifying that this absolute scaling is correct would require a more detailed investigation that is beyond the scope of this paper.

We find that the predicted trends of \aco\ with CO integrated intensity from N12 and F12 match the average observed behavior well, although both underpredict \aco\ in the central regions of some galaxies.  There is considerable scatter around this trend, however.  F12 argue that on $\sim$kpc scales metallicity is the primary driver of the conversion factor variations.  This metallicity dependence reflects, to first order, the \citet{2011MNRAS.412..337G} dependence between \aco\ and extinction through the cloud with the addition that F12 assume a metallicity dependent DGR. Our \aco\ results do not show a strong dependence on metallicity, but may not extend to low enough metallicities to clearly distinguish such variations.  N12 do see a decrease in \aco\ in regions with high star-formation rate surface densities due to enhanced gas temperatures and velocity dispersions, similar to what we have found in some galaxy centers.  

\subsection{Low Central \aco\ and Discrepancies with Virial Mass Based Measurements}\label{sec:centralaco}

For some galaxy centers in our sample we measure \aco\ lower than the typical Milky Way value by factors of 5$-$10.  Interestingly, the virial mass based \aco\ measurements for several of these regions do not agree with the values found here, instead recovering \aco\ values close to that of the Milky Way. From Figure~\ref{fig:virialcomp} we found that the discrepancy between virial mass and dust-based \aco\ decreases with radius for NGC~6946---suggesting that this is a property of the galaxy centers rather than an issue with the resolution of the virial mass techniques (which tends to be comparable to the size of the GMCs). For NGC~6946, results of multi-line modeling studies are generally in much better agreement with our results.  In the following we discuss processes that could lower \aco\ measured from dust and multi-line modeling techniques, while leaving virial mass based \aco\ similar to the MW.

For self-gravitating GMCs, all three techniques should derive a lower \aco\ if the gas excitation temperature were changing independently of the other variables.  In contrast, if the velocity dispersion of the cloud changed due to external pressure or additional sources of turbulence, the virial and dust techniques could arrive at different results.  In such a situation, the measured line-width would suggest a larger virial mass and the cloud's CO emissivity would also increase.  Thus, one could derive a conversion factor similar to the MW value from the comparison of CO to virial mass since both have increased. \citet{1998ApJ...493..730O} have suggested that external pressure in the Galactic Center can account for up to an order-of-magnitude change in the conversion factor.  However, \citet{2012ApJ...744...42D} found that GMCs in NGC~6946 did not generally show enhanced velocity dispersion at a given size or luminosity, in contrast to the Galactic Center clouds identified by \citet{1998ApJS..118..455O}. At the resolution of our CO maps, we cannot distinguish between changes in GMCs internal velocity dispersion (which would affect \aco) and changes in the velocity dispersion within the population of clouds, so we cannot test whether there is enhanced velocity dispersion in GMCs with our measurements. 

It is possible that molecular gas in the centers of some galaxies is not in self-gravitating GMCs.  If that is the case, the velocity dispersion may reflect other hydrodynamical processes and the virial mass estimate will be incorrect.  For instance, if the molecular gas is not in GMCs but rather in a larger-scale molecular medium that is gravitationally bound to the stars and gas, the velocity dispersion will be larger and more CO emission will be produced for a given amount of gas.  This explanation has been suggested by \citet{1993ApJ...414L..13D} and \citet{1998ApJ...507..615D} to explain the properties of molecular gas in the nuclear disks of ultra-luminous infrared galaxies (ULIRGs) \citep[see also recent work by][]{2012ApJ...751...10P}.  Another option is a contribution to the CO emission from diffuse molecular gas, which could potentially have a lower \aco\ than GMCs \citep{2010A&A...518A..45L}.     

The dust-based and multi-line modeling \aco\ results may agree better than the virial results in galaxy centers because they do not assume a relationship between the velocity dispersion and the mass of molecular gas, which would cause issues in the cases outlined above.  At the moment, only NGC~6946 (out of our $i<65^{\circ}$ galaxies) has both virial and multi-line modeling results.  In the other galaxies, we may be able to find some indication of how optical depth, excitation and velocity dispersion affect \aco\ by examining the ratios of $^{12}$CO and $^{13}$CO lines.  Usero et al. (in prep) presents measurements of $^{13}$CO and $^{12}$CO lines towards regions of several HERACLES galaxies.  Selecting all of their pointings towards the disks of galaxies in our sample (in regions above our \ico\ and inclination cuts), the uncertainty weighted mean of $R_{12/13}= ^{12}$CO J=$(1-0)/^{13}$CO J=$(1-0)$ is 8.24$\pm$0.11.  The only galaxy centers with $^{13}$CO measurements are NGC~0628, 3184, 5055 and 6946.  NGC~0628, 3184 and 5055 show $R_{12/13}$ consistent with or slightly lower than the average disk value.  NGC~6946, on the other hand, has $R_{12/13} = 18.53\pm0.81$. The high $R_{12/13}$ for the center of NGC~6946 has been noted in several previous studies \citep{2001ApJS..135..183P,2001A&A...371..433I,2004AJ....127.2069M}.

\subsection{Dust-to-Gas Ratio}\label{sec:dgrdisc}

As shown in Figures~\ref{fig:driversdgr} and~\ref{fig:dgrvsz}, we observe a good correlation of DGR with metallicity.  Unlike most previous studies of DGR resolved within galaxies, we made no assumption about \aco\ in determining these values.  Although we do not probe a wide range of metallicity due to the limitations on CO S/N, it is clear in Figure~\ref{fig:dgrvsz} in particular, that DGR is correlated with metallicity with less than a factor of 2 scatter over 0.5 dex in metallicity.  A linear dependence of DGR on metallicity suggests a constant fraction of heavy elements are locked up in dust grains.  Chemical evolution and dust life-cycle models have varying predictions for the dependence of DGR on metallicity \citep{1998ApJ...501..643D,1998ApJ...496..145L,Hirashita:2002bh}.  Most of the dramatic differences between the models, however, occur at lower metallicities and observations of dwarf galaxies may provide more leverage on distinguishing between models \citep[e.g.][]{2012ApJ...752..112H}.  

\subsection{Recommendations on Choosing \aco}\label{sec:recs}

Due to the requirements on CO S/N to apply the technique we have used, our measurements only tell us about \aco\ above \ico$=$1 K km s$^{-1}$ in the galaxies we have targeted.  Generally, this limits us to the inner parts of galaxies ($<$r$_{25}$) where metallicities are comparable to the MW disk.  However, these are the regions where H$_2$ contributes most significantly to the total gas mass \citep{2011AJ....142...37S}, so applying our conversion factor results to unresolved galaxies should work reasonably well for recovering the total H$_2$ mass.  A forthcoming paper extending our study to the \hi\ dominated regions of the KINGFISH galaxies, will test this assertion.  Note that our recommendations are only applicable to spiral galaxies with metallicity similar to that of the MW.

For an unresolved galaxy, we recommend adopting our galaxy-based average \aco\ of 3.1 \acounit\ with uncertainty of 0.3 dex (a factor of $\sim2$).  This is the mean and standard deviation of the average value for all galaxies with inclination less than 65$^{\circ}$.

In dealing with resolved galaxies, we recommend adopting a flat radial \aco\ profile for regions with \ico\ $>1$ K km s$^{-1}$ except in the central $\sim 0.1$r$_{25}$, where the average \aco\ is a factor of 2 lower.  When dealing with a single galaxy, our average radial profile from Figure~\ref{fig:acovsr25} suggest that the \aco\ values have $\sim0.2$ dex of scatter in each radial bin (0.3 dex for the central bin). Therefore, for studies of molecular gas within a single galaxy, the relative values of \aco\ adopted from this profile have a factor of $\sim 1.5$ uncertainty. When comparing resolved galaxies, it is necessary to further recognize that the absolute normalization of the radial profiles, i.e. the galaxy average value, has an additional 0.3 dex of uncertainty.

In the case where additional information about the galaxy is available, such as profiles of stellar mass surface density, star formation rate surface density or line ratios of $^{12}$CO $(2-1)/(1-0)$ or $^{12}$CO/$^{13}$CO, we suggest using these maps to pick out regions that may have \aco\ very different from the mean.  This includes regions with $\Sigma_{*} \gtrsim 1000$ \msun\ pc$^{-2}$, $\Sigma_{\rm SFR} \gtrsim 0.1$ \msun\ yr$^{-1}$ kpc$^{-2}$, or R$_{21} \gtrsim 1$ at our working resolution of $\sim1$ kpc.  In these regions, \aco\ can be 5$-$10 times lower than the MW \aco, and are often systematically lower than the average radial profile.

\section{Summary \& Conclusions}\label{sec:conclusions}

The availability of high angular resolution far-infrared maps from the {\em Herschel} Space Observatory and sensitive $^{12}$CO J=(2$-$1) and \hi\ maps has recently allowed us to trace dust and gas mass surface densities in nearby galaxies on $\sim$kpc scales.  On these scales, we expect (and have verified) that the DGR is approximately constant.  Therefore, we are able to combine the dust mass surface density maps with matched resolution CO and \hi\ to solve simultaneously for \aco\ and DGR.  The solution technique finds the \aco\ that best minimizes the scatter in DGR values in $\sim$kpc regions across the galaxies.  We have performed a thorough investigation into the efficacy of this technique, using data from the KINGFISH key program on {\em Herschel}, the large IRAM 30m survey HERACLES and the THINGS \hi\ survey with the VLA.  Our tests show that above \ico\ $\sim1$ K km s$^{-1}$ we reliably achieve accurate solutions for \aco.  We have used a fixed ratio between the (2$-$1) and (1$-$0) lines to present our \aco\ results on the more typically used $^{12}$CO J=(1$-$0) scale and have shown using literature measurements that variations in the line ratio do not effect our results in these galaxies.

We find that the average \aco\ for the galaxies in our sample is 3.1 \acounit.  This value is slightly lower than the Milky Way \aco\ $=4.4$ \acounit, but the MW value is well within the standard deviation of our measurements. Treating all 782 solutions independently (instead of weighting each galaxy equally), we find \aco\ $=2.6$ \acounit\ with 0.4 dex standard deviation.  In all averages, we have removed galaxies with inclinations higher than 65$^{\circ}$ since our solution pixel samples gas along the line-of-sight with a range of DGR and therefore does not conform to our main assumption---that the DGR is constant in the solution pixels.    

Within the galaxies we observe a relatively flat profile of \aco\ as a function of galactocentric radius aside from in the galaxy centers.  Normalizing each galaxy by its average \aco\, the average galaxy shows a factor of $\sim2$ lower \aco\ in its center.  In several galaxies, this central value can be factors of 3 or more lower than the galaxy average.  In several notable cases, the central \aco\ value is factors of 5 to 10 lower than the standard MW \aco$=4.4$ \acounit.  

We have investigated the correlations between \aco\ and environmental parameters in an attempt to isolate factors that drive variations in \aco.  The strongest correlation we find is between \aco\ and stellar mass surface density ($\Sigma_{*}$).  If the strength of this correlation relative to the other parameters is real and not due to issues with measuring the other parameters, there are a number of possible explanations for why \aco\ should depend on stellar mass surface density, including the influence of ISM pressure on both molecular clouds and a more diffuse molecular medium.  Distinguishing among these possibilities will require ALMA observations that resolve individual GMCs in multiple molecular gas lines across a range of extragalactic environments.  

We do not observe a strong correlation between \aco\ and metallicity in our galaxies.  In the range of metallicities we have sampled, this conclusion is not unexpected.  The abundance of C or O does not play a major role in determining \aco\ at these metallicities since CO is optically thick in molecular clouds.  The metallicity primarily influences \aco\ through the effects of dust shielding (due to the dependence of DGR on metallicity).  Models and simulations have suggested that dust shielding can influence the amount of gas in ``CO-dark'' layers of GMCs, but at MW metallicity, only $\sim30$\% of the gas is in this layer.  At lower metallicities, a strong dependence of \aco\ on metallicity has been observed using the same technique we employ \citep{2011ApJ...737...12L}.

In the centers of several galaxies, we find \aco\ values 5$-$10 times lower than the MW \aco.  These regions are also significantly below the average \aco\ for their galaxy.  Comparison of our measured \aco\ with values from the literature shows good agreement between our dust-based results and multi-line modeling results.  In contrast, our \aco\ values in these regions can be much lower than \aco\ determined from virial mass based techniques.  The discrepancy with virial mass measurements becomes smaller at larger galactocentric radii, suggesting this is a particular property of the gas in galaxy centers.  At the resolution of our observations, the central region with low \aco\ is unresolved ($\lesssim$ kpc).  We suggest several explanations for the low \aco\ value and the fact that it is not reflected in virial masses including ISM pressure contributions to GMC virial balance, increases in molecular gas temperature and/or a more diffuse molecular medium similar to what is found in ULIRGs.  With the limited amount of $^{13}$CO and other molecular line observations in this region, it is not possible to distinguish between these scenarios, but NGC~6946 at least shows some evidence for lower CO optical depth towards its center.  As is explored in \citet{Leroy:2013um}, these galaxy centers show enhanced star-formation efficiencies when we apply our \aco, close to what is seen along the ``starburst sequence'' in ultra-luminous infrared galaxies \citep[e.g.][]{2010ApJ...714L.118D}. 

In addition to \aco\ we also simultaneously measure DGR in all of our regions.  On average, we find Log(DGR)$=-1.86$ treating all solution pixels equally, with a standard deviation of 0.22 dex. When we force each galaxy to contribute equally, the average is essentially indistinguishable from the MW DGR.  Unlike \aco, DGR is well-correlated with metallicity, with a slope slightly shallower than linear (although this slope is mainly due to NGC~3184 being offset from the main trend, removing it from the sample gives a slope consistent with linear).  The approximately linear dependence of DGR on metallicity agrees with the predictions of dust evolution models, but our measurements do not cover a wide enough metallicity range to distinguish between them.

The results presented here suggest a picture where \aco\ is slightly lower than the typical value of 4.4 \acounit\ in the disks of most galaxies, and mainly constant as a function of radius despite changes in metallicity, radiation field intensity and star-formation rate surface density.  Galaxy centers appear to be a different regime, where external pressure or changes in the character of molecular gas (i.e. mostly confined to self-gravitating GMCs versus a more diffuse molecular medium) may bring about large changes in \aco.  Through the galaxy, however, DGR appears to be an approximately linear function of metallicity.  The simple behavior of DGR provides a unique tool to study the ISM in nearby galaxies, if we can obtain measurements of metallicity gradients with trustworthy calibration.

\acknowledgements

We thank the referee for useful comments that helped to improve the quality of the manuscript.  K.S. thanks R. Shetty, S. Glover, R. Klessen, D. Narayanan, A. Stutz for helpful conversations.  The authors thank R. Feldmann and N. Gnedin for helpful comments regarding the comparison of our measurements to their simulations.  K.S. is supported by a Marie Curie International Incoming Fellowship.  A.D.B. wishes to acknowledge partial support from a CAREER grant NSF-AST0955836, NASA-JPL1373858, and from a Research Corporation for Science Advancement Cottrell Scholar award.

This work is based on observations made with Herschel. Herschel is an ESA space observatory with science instruments provided by European-led Principal Investigator consortia and with important participation from NASA. PACS has been developed by a consortium of institutes led by MPE (Germany) and including UVIE (Austria); KU Leuven, CSL, IMEC (Belgium); CEA, LAM (France); MPIA (Germany); INAF-IFSI/OAA/OAP/OAT, LENS, SISSA (Italy); IAC (Spain). This development has been supported by the funding agencies BMVIT (Austria), ESA-PRODEX (Belgium), CEA/CNES (France), DLR (Germany), ASI/INAF (Italy), and CICYT/MCYT (Spain).  SPIRE has been developed by a consortium of institutes led by Cardiff University (UK) and including Univ. Lethbridge (Canada); NAOC (China); CEA, LAM (France); IFSI, Univ. Padua (Italy); IAC (Spain); Stockholm Observatory (Sweden); Imperial College London, RAL, UCL-MSSL, UKATC, Univ. Sussex (UK); and Caltech, JPL, NHSC, Univ. Colorado (USA). This development has been supported by national funding agencies: CSA (Canada); NAOC (China); CEA, CNES, CNRS (France); ASI (Italy); MCINN (Spain); SNSB (Sweden); STFC (UK); and NASA (USA).  HIPE is a joint development by the Herschel Science Ground Segment Consortium, consisting of ESA, the NASA Herschel Science Center, and the HIFI, PACS and SPIRE consortia.  This research has made use of the NASA/IPAC Extragalactic Database (NED) which is operated by the Jet Propulsion Laboratory, California Institute of Technology, under contract with the National Aeronautics and Space Administration. This research has made use of NASA's Astrophysics Data System Bibliographic Services.

%%%%%%%%%%%%%%
% REFERENCES %
%%%%%%%%%%%%%%

\bibliographystyle{apj}
%\bibliography{/Users/karin/full}

%%%%%%%%%%%%%%
% APPENDIX %
%%%%%%%%%%%%%%
\clearpage

\appendix

\section{Technique for Simultaneously Determining DGR and \aco}\label{sec:technique}

Prior to choosing the technique described in Section~\ref{sec:tech} we have explored a variety of possible ways to simultaneously constrain DGR and \aco.    The various techniques represent, in essence, different ways to judge how uniform the DGR is in the region in question, at any given value of \aco.  After defining these variations on the essential technique, we construct simulated datasets with known \aco\ and DGR and realistic statistical errors and intrinsic scatter to test the solution techniques.  These simulated data tests allow us to optimize the performance of the technique given the properties of our dataset.  

\subsection{Potential Techniques}

The general procedure for finding the most ``uniform'' DGR in the solution pixel given the spatially resolved measurements of \sighi, \sigd, and \ico\ is to step through a grid of \aco\ values, calculate DGR at each sampling point in the solution pixel using that \aco, measure some statistic that can be used to judge how uniform the DGR values are across the region and finally, either minimize or maximize that statistic as a function of \aco.  Below we describe the various statistics we have tested to judge the uniformity of the DGR values in a given region.  The reason we investigate all of these possible statistics is because each weights outliers differently, and it is not obvious {\em a priori} which will work best to recover the underlying \aco.  For techniques that involve minimizing scatter in the DGR, we use three different statistics to measure scatter: the standard deviation (RMS), the median absolute deviation (MAD) and a robust estimator of the standard deviation from Tukey's biweight (BIW) \citep{2002nrc..book.....P}. 

\begin{enumerate}

\item Minimum Fractional Scatter in DGR (FS): At each value of \aco\ we calculate the DGR for all points in the region in question.  We then divide each DGR value by the mean DGR in the region and measure the scatter of the resulting values (with RMS, MAD and BIW statistics).  

\item Minimum Logarithmic Scatter in DGR (LS): At each \aco\ we calculate the scatter of the logarithm of the DGR values (with RMS, MAD and BIW).  This is similar to the minimization of fractional scatter, but with a different weighting for the individual points.  This also throws away data points with negative values, since their logarithm is undefined.

\item Maximum Correlation Coefficient of \sigd\ and $\Sigma_{\rm Gas}$ (LC or RC): At each value of \aco\ we compute the linear and rank correlation coefficients between \sigd\ and $\Sigma_{\rm gas} =$ \sighi\ $+$ \aco \ico.  The best \aco\ value will maximize the correlation between these two quantities.

\item Minimum $\chi^2$ of Best-Fit Plane to \ico, N$_{\rm HI}$ and \sigd\ (PF): The \sigd, \sighi\ and $\Sigma_{\rm H_2}$ values describe a plane with two free parameters.  At each value of \aco\ we fit a plane to the measurements to determine DGR.  The best value of \aco\ will be the one that minimizes the $\chi^2$ of the best-fit plane.

\item Minimum Correlation Coefficient of DGR vs. \ico/N$_{\rm HI}$ (CHC): In this technique we search for the \aco\ value that minimizes the correlation of DGR with \ico/N$_{\rm HI}$.  This technique weights the points differently depending on how much local $\Sigma_{\rm gas}$ depends on \aco.

\end{enumerate}

\subsection{Definition of the ``Solution Pixel'' Size and Location}

A key aspect of the technique we use to constrain DGR and \aco\ is defining the region over which we assume there to be one value each of DGR.  Under ideal circumstances, the regions would be as small as possible given the resolution of the maps.  Realistically, however, there are several considerations that must be taken into account when defining the regions including: covering an adequate range of CO/\hi\ ratios, having enough points to measure scatter accurately, and keeping the region to $\sim$kpc scales over which it is reasonable to assume that the DGR is indeed constant.. 

To complement the hexagonal, half-beam spaced grid with which we sampled the original data (i.e. our ``sampling points'') we again use hexagonal spacings to define the ``solution pixels'' as the individual samples can be divided naturally into concentric hexagons.  We tested a variety of solution pixel sizes ranging from 19-point to 271-point hexagons.  

We found that we 37-point solution pixels was a good compromise between size and solution quality.  Because of the underlying half-beam spaced sampling grid, the 37 point solution pixel contains $\sim 9$ independent measurements.  For the next smallest possible hexagon (19-points), there are not sufficient numbers of independent measurements to reliably judge the scatter in the DGR.

In order to fully sample the map, the center of each hexagonal pixel is offset by 1/2 of the spacing from its neighbor.  This results in the ``solution pixels'' not being independent--each neighboring pixel shares $\sim 40$\% of the same data (the oversampling can be exactly described as $n^2/(3n^2 - 3n + 1)$ for concentric hexagons).

\subsection{Test of the Solution Technique with Simulated Data}\label{sec:simgal}

We have investigated the efficacy of the various techniques using simulated data modeled on our observations.  Our goal is to identify the technique or techniques that return the most accurate values of \aco\ and DGR over the range of conditions and signal-to-noise levels in our dataset.  We also aim to characterize any biases in the recovered \aco\ as a function of input \aco\ and S/N.  In performing these tests, it is important to model the simulated data on our real observations.  This is because the ability of the techniques to judge the DGR uniformity depend not only on the S/N of the measurements but also on the range of CO/\hi\ ratios.  Since CO and \hi\ can be correlated, anti-correlated or independent of each other on kpc scales, depending on the state of the ISM, it is difficult to generate a reasonable set of simulated data from scratch that encompasses the range of CO/\hi\ behaviors in the observations.  Therefore, we expect the best test of the technique will come from simulated data that are closely modeled on the observations.  In the following, we describe how the simulated data are generated and discuss the results of the test.

The simulated data are generated via a Monte Carlo procedure in which we randomly choose a galaxy from our sample (listed in Table~\ref{tab:sample}) at SPIRE 350 resolution, choose a solution pixel and create simulated data based on the selected CO, \hi\ and \sigd\ samples.  For each trial we randomly sample a range of \aco\ values between $0.5-50$  \msun\ pc$^{-2}$ (K km s$^{-1}$)$^{-1}$ and intrinsic scatter per axis (evenly divided among the \ico, \sighi\ and \sigd\ axes, for lack of better knowledge of the sources of intrinsic scatter) between $0.0-0.4$ dex.  We note that because of the properties of lognormal distributions, the scatter in the DGR is not simply $\sqrt{3}\times$ the input intrinsic scatter per axis, so we approximate it after-the-fact using the input scatter per axis and the mean H$_2$/\hi\ ratio for the simulated measurements.

Our procedure is the following: 1) select the observations with the sampling described above, 2) use the \ico\ and \sighi\ measurements along with the random \aco\ value to create a simulated $\Sigma_{\rm gas}$ vector, 3) calculate the DGR value that preserves the average S/N level of the \sigd\ measurements and create a simulated, noise-free \sigd\ vector by multiplying the $\Sigma_{\rm gas}$ from step 2 by this DGR, 4) apply the randomly-selected intrinsic scatter to each axis and 5) add simulated measurement errors according to the observed uncertainties in the \ico, \sighi\ and \sigd\ maps.  These simulated measurements are then used as input to solve for \aco\ and DGR using the techniques described above.

\begin{figure*}[h]
\centering
\epsscale{1.1}
\plottwo{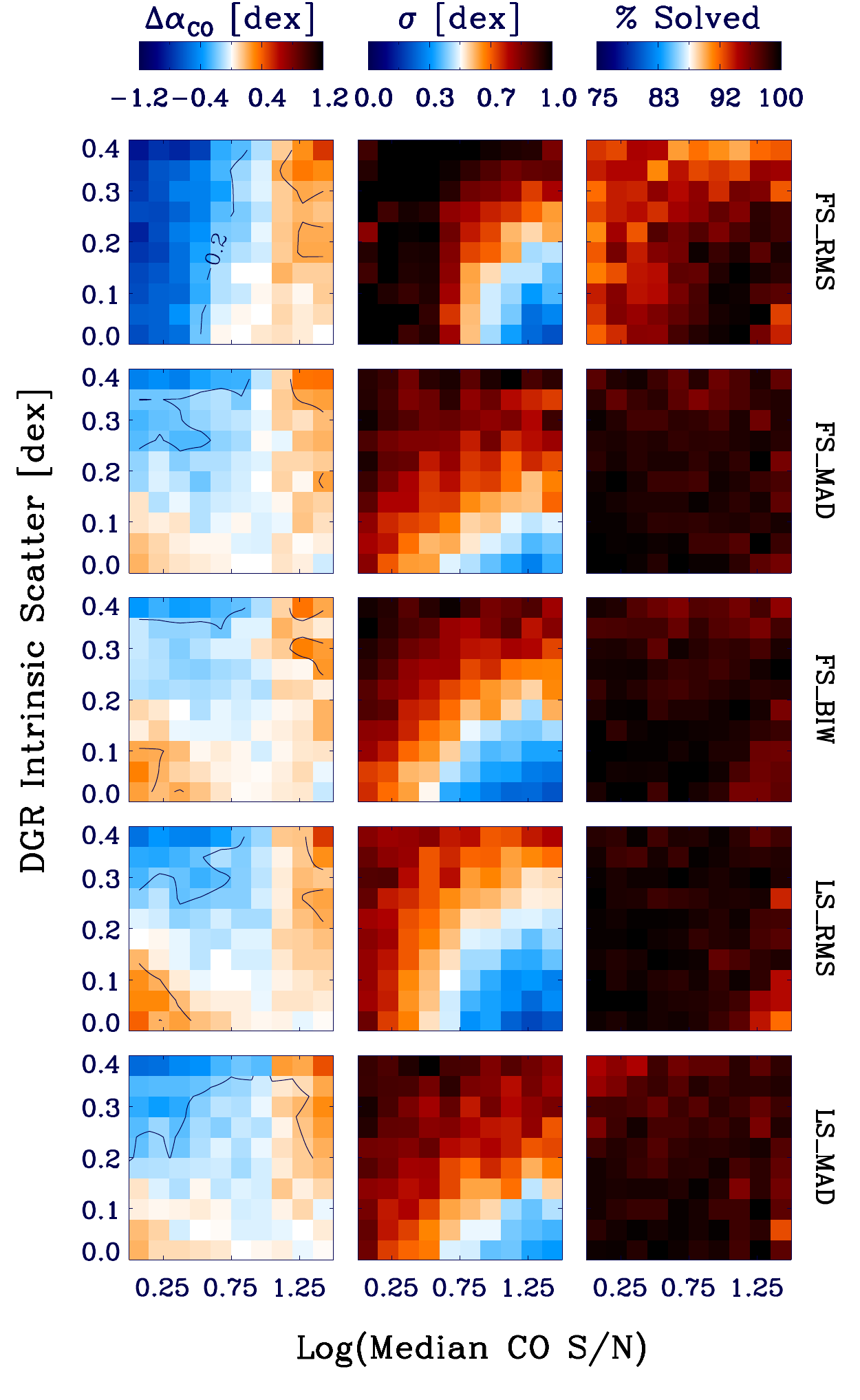}{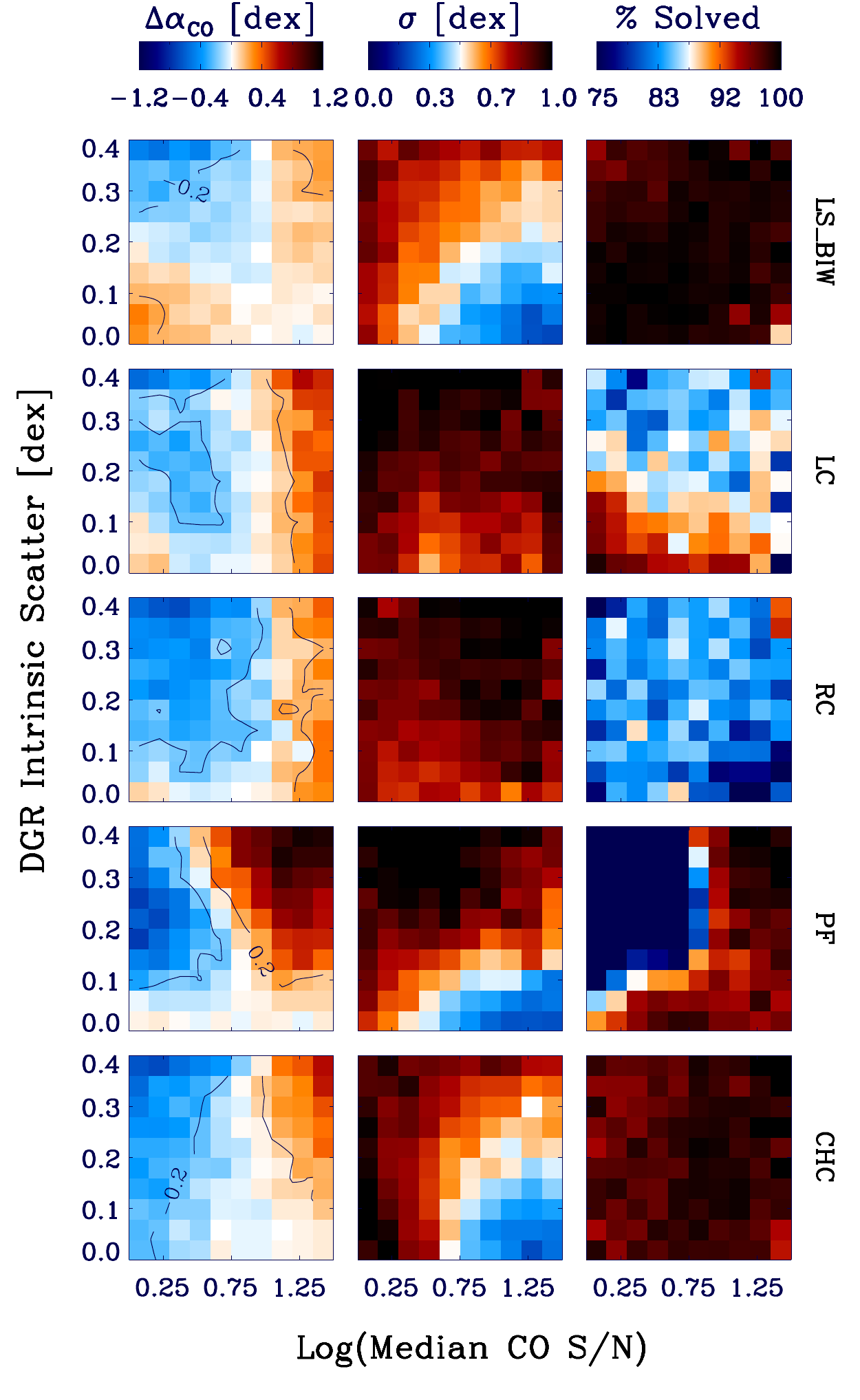}
\caption{Results of simulated data tests to optimize the solution technique.  Columns correspond to: left, the difference between the known input \aco\ and the recovered best-solution \aco\ ($\Delta$\aco); middle, the standard deviation of the difference between input and recovered \aco\ ($\sigma$); and right, the fraction of trials in each bin where a solution is achieved.  The results are presented as two-dimensional histograms, binned in CO S/N and the simulated intrinsic scatter added to the data.  The rows show results for the various solution techniques described previously.  From top to bottom in the left panel these are: minimization of the fractional scatter using the mean (FS\_RMS), median (FS\_MAD) and robust mean (FS\_BIW) and the minimization of the logarithmic scatter using the mean (LS\_RMS) and median (LS\_MAD).  In the right panel we show minimization of the logarithmic scatter using the robust mean (LS\_BIW), minimizing the linear correlation coefficient (LC), minimizing the rank correlation coefficient (RC), minimizing the $\chi^2$ of the best-fit plane (PF) and minimizing the correlation coefficient between DGR and \ico/\sighi\ (CHC).  The contours in the $\Delta$\aco\ column show the regions of parameter space where the difference in the input and recovered \aco\ is less than 0.2 dex. The LS\_BIW technique minimizes the difference between input and recovered \aco\ and achieves a small $\sigma$ over the largest area, although the distinction from FS\_BIW is not large.}
\label{fig:mctest2}
\end{figure*}

From this Monte Carlo simulation, we can then choose the technique that is the most robust (i.e. most frequently achieves a solution) and accurate (i.e. to get closest to the known input \aco).  In Figure~\ref{fig:mctest2} we show the results of this simulation for the techniques we have outlined.  The panels show (left) the mean difference between the input and recovered \aco\ value, (middle) the standard deviation of the difference of the input and recovered \aco, and (right) the percent of the Monte Carlo trials that achieve a solution.  These values are binned in a two-dimensional space defined by the intrinsic scatter added to the simulated data and the median S/N of the simulated CO measurements. 

It is clear that the FS and LS techniques provide the least bias in the recovered \aco\ over the relevant range of intrinsic scatter and S/N (except for the FS\_RMS which is significantly biased at low CO S/N).  The correlation coefficient (LC and RC), planefit (PF) and CO/\hi\ correlation (CHC) techniques all provide much more limited regions with low bias in the recovered \aco\ as well as having a lower fraction of solutions.  In general the MAD based techniques have slightly less bias towards the low CO S/N region, but show a larger scatter between the input and recovered \aco.  Overall, we find that the LS\_BIW (robust biweight mean of the logarithmic scatter) is the technique that shows the best performance in accuracy, precision and robustness over the range of characteristics of our dataset.  

\begin{figure*}[h]
\centering
\epsscale{0.5}
\plotone{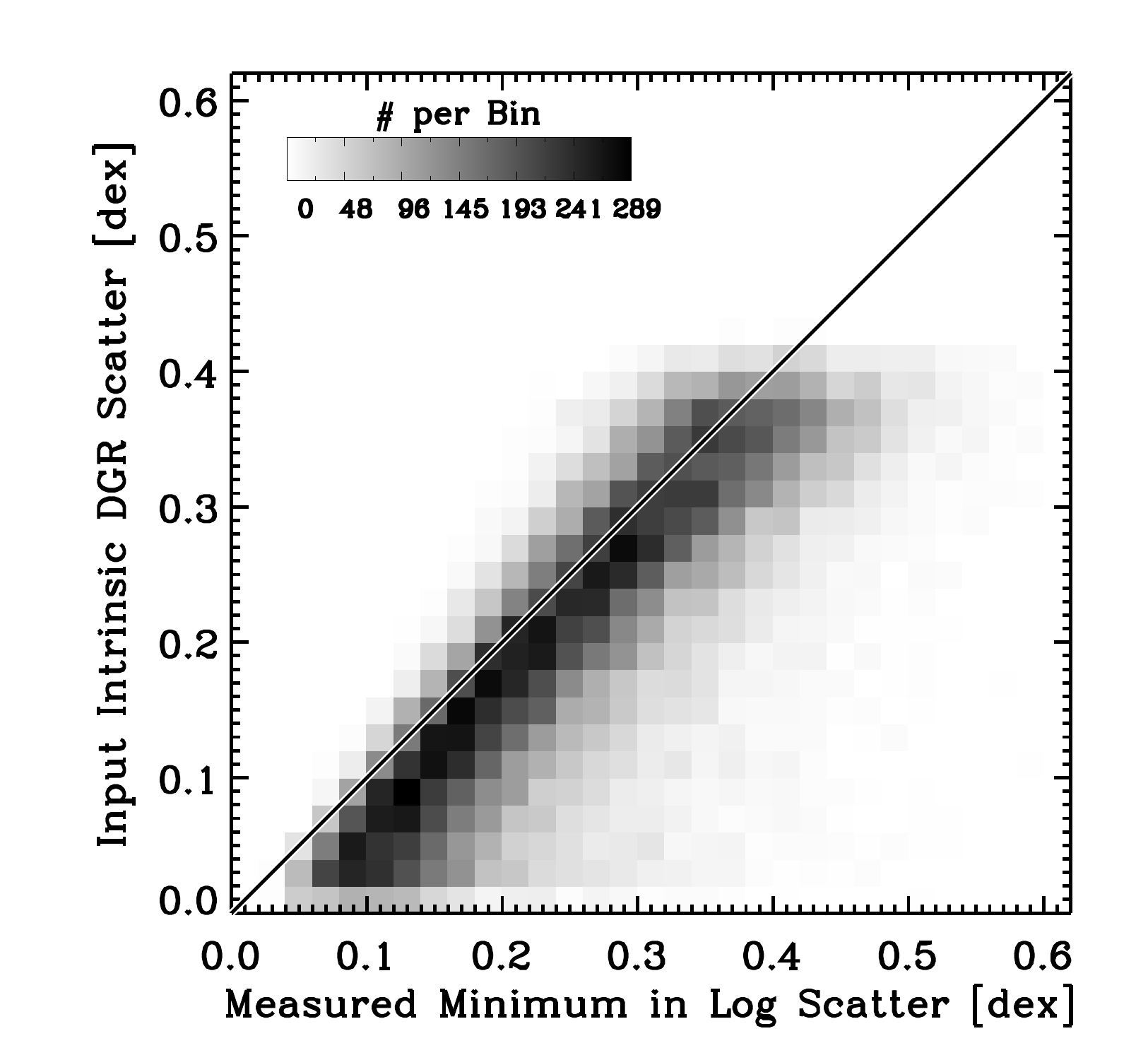}
\caption{Results of simulated data tests comparing the known input intrinsic scatter in DGR with the recovered minimum in the DGR scatter.  The gray scale represents the number of trials falling in a given bin.  We have uniformly sampled input intrinsic scatter values between $0.0-0.4$ dex.  The line shows a one-to-one relationship.}
\label{fig:recover_intsc}
\end{figure*}

In addition to helping us choose the best technique, the tests with simulated data also allow us to investigate how the measured minimum in the logarithmic scatter corresponds to the known intrinsic scatter added to the simulated data.  In addition to being an interesting quantity scientifically, the measured scatter can also help us to identify which region of the parameter space a given solution falls in, allowing us to judge how biased we expect it to be given our knowledge from these simulated data tests.  Figure~\ref{fig:recover_intsc} shows a comparison of the known input intrinsic scatter with the measured minimum in the logarithmic scatter for the LS\_BIW technique.  Most of the outliers come from regions with low CO S/N.  At very low levels of intrinsic scatter, the measured scatter is larger due to the influence of observational uncertainties.  In general, however, the measured minimum log scatter is a very good proxy for the intrinsic scatter in the DGR. 

Based on these results with simulated data, we can use the measured median CO S/N and the measured minimum of the scatter in the DGR values to predict how biased and how uncertain a given determination of \aco\ for our dataset.  To do this we use the measured minimum in the DGR scatter as an estimate of the intrinsic DGR scatter (it is biased slightly higher, but this is a minor effect and represents a conservative estimate of the intrinsic scatter).  Then using this value and the median CO S/N in the solution pixel we interpolate in the grid presented in Figure~\ref{fig:mctest2} for ``LS\_BIW'' to predict the bias inherent in the technique.  

\subsection{Testing for Bias at High CO S/N and Low \sighi}

One of the key results of our work is the measurement of low \aco\ values in the centers of some galaxies (see Section~\ref{sec:centralaco}).  These locations are outliers from the average solution pixel in the sense of having high CO S/N and often a relative deficiency in \sighi.  Since they make up a small subset of all solution pixels, our Monte Carlo simulation from the previous Section may not adequately judge the bias of the technique in these regions, since they are not sampled frequently enough (only 26 out of $\sim 900$ solution pixels).  We therefore perform a separate Monte Carlo simulation for the galaxy centers in order to verify that these low \aco\ values are not due to biases in the technique (we have demonstrated that the average pixel in the sample is not biased using the experiment in the previous section).

The results of the technique are biased if it returns an \aco\ systematically different from the true \aco.  For this test, we will define several different \aco\ values.  First, \aco$_{\rm ,meas}$ is the measured value for a given solution pixel, we do not know if this measured value is biased, it is only what we have measured for that pixel given the true data.  Second, \aco$_{\rm ,in}$ is a known \aco\ value we have used to generate simulated data.  Finally, \aco$_{\rm ,out}$ is the \aco\ value that we recover for \aco$_{\rm ,in}$ using our solution technique.

The essence of this test is that we take a range of \aco$_{\rm ,in}$ values and map them to their \aco$_{\rm ,out}$ values, by generating simulated data given \aco$_{\rm ,in}$ and running it through our solution technique.  This simulated data generation proceeds as described in the previous section and is designed to preserve the CO/\hi\ ratio and S/N of the measurements for a given pixel.  In addition to observational noise, we also add intrinsic scatter between 0.01$-$0.1 dex to the simulated data, which reflects the level of intrinsic scatter we have found in the real solutions.  For each solution pixel we generate 10$^4$ random \aco$_{\rm ,in}$ values that span our range of \aco$=0.1-100$.  

With this simulation, we can determine for any given value of \aco$_{\rm ,out}$, which values of \aco$_{\rm ,in}$ ended up there.  If a solution is biased, the correspondence between \aco$_{\rm ,in}$ and \aco$_{\rm ,out}$ will not be one-to-one.  For example, a bias may make us consistently recover \aco$_{\rm ,out}=1$ \acounit\ when \aco$_{\rm ,in}=0.1$ \acounit.

\begin{figure}
\centering
\epsscale{0.5}
\plotone{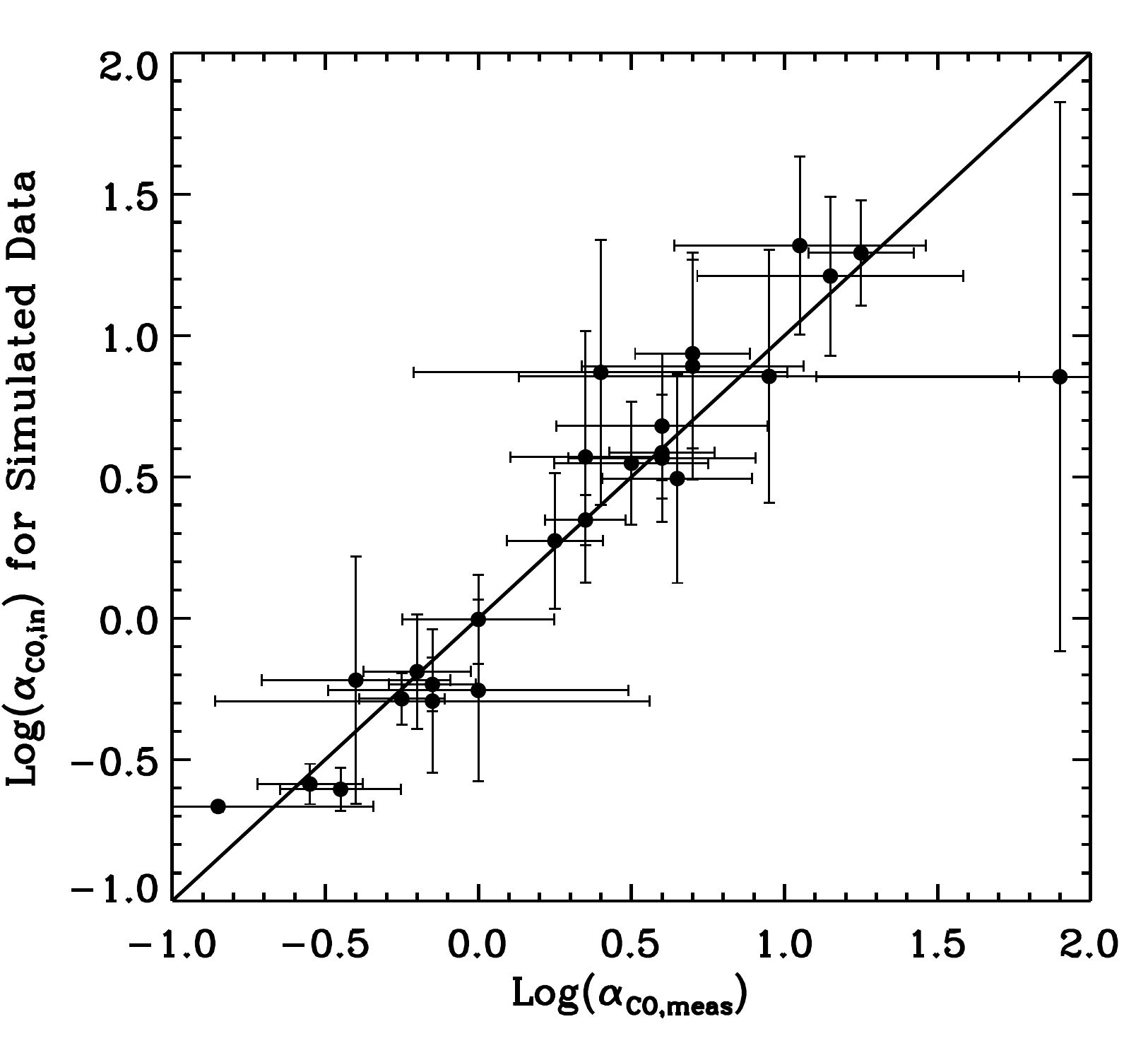}
\caption{This figure shows the results of simulated data tests focused on the galaxy centers, where high CO S/N and high CO/\hi\ ratios could bias the recovered \aco.  The $x$-axis shows the measured \aco$_{\rm ,meas}$ and uncertainty for each of the 26 central points in our sample.  The $y$-axis shows the mean and standard deviation of the {\em known, input} \aco$_{\rm ,in}$ values from the simulations where we recovered \aco$_{\rm ,out}$ equal to the value we have measured.  This test shows that the \aco\ values we measured in the galaxy centers are not biased within their uncertainties.}
\label{fig:centersbias}
\end{figure}

In judging the biases of the actual solutions for our central solution pixels, we then select all of the \aco$_{\rm ,out}$ measurements equal to \aco$_{\rm ,meas}$ and find the mean and standard deviation of the \aco$_{\rm ,in}$ values that gave those results.  In Figure~\ref{fig:centersbias} we show a plot of the mean and standard deviation of the \aco$_{\rm ,in}$ values versus the \aco$_{\rm ,meas}$ plus its uncertainty for all galaxy centers in the sample.  This figure shows that within the uncertainties of \aco$_{\rm meas}$, none of the central pixel measurements show a bias. 

\begin{figure}
\centering
\epsscale{0.45}
\plotone{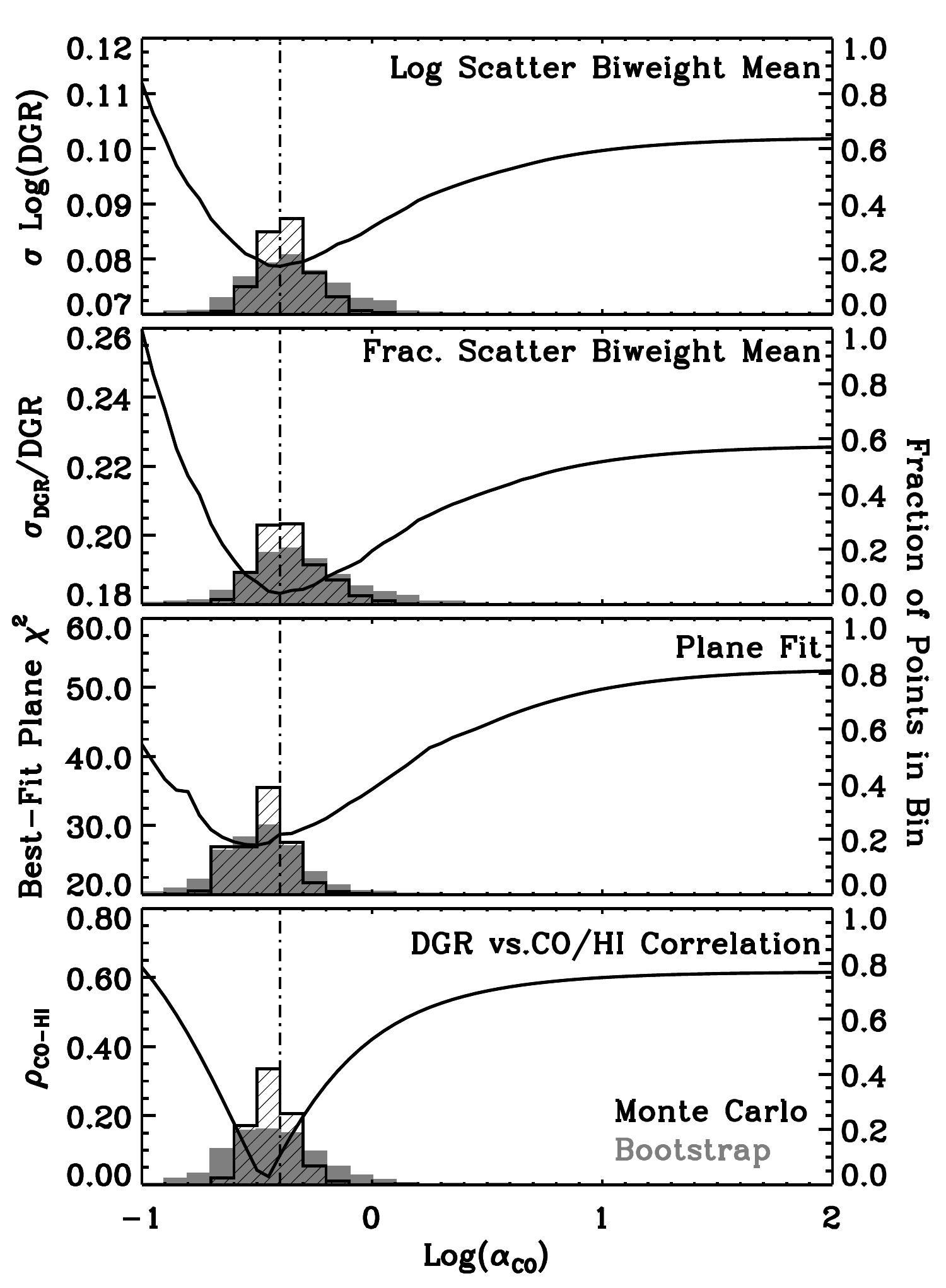}
\caption{Solution for the central region of NGC~6946.  The panels show different statistics we use to measure the uniformity of the DGR in a given solution pixel, all showing a clear minimum at \aco$=0.4$\acounit.  A thorough discussion of the various statistics is presented in the Appendix.  From top to bottom we show: the scatter in Log(DGR) measured using the robust biweight mean (LS\_BIW); the fractional scatter DGR determined with the biweight mean (FS\_BIW); the $\chi^2$ of the best-fit plane to \ico, \sighi\ and \sigd\ (PF); and the correlation coefficient between DGR and the CO/\hi\ ratio (CHC).  The dot-dash vertical line shows the solution for \aco\ from the LS\_BIW technique, which is what we adopt in the rest of the study.  The agreement between different techniques (i.e. minimizing scatter in DGR, plane-fit and minimizing the correlation between DGR and CO/HI) illustrates the high confidence of this solution and shows that it is not an artifact of the solution technique.}
\label{fig:ngc6946curve}
\end{figure}

In addition, a second check on these low central \aco, at least for NGC~6946, is shown in Figure~\ref{fig:ngc6946curve} where we show the minimization curve for several techniques towards its central solution pixel.  All techniques agree within their errors, highlighting the minimum at \aco$\sim -0.4$.  This emphasizes that the low \aco\ found in the center is not an artifact of the particular technique we are using, but can be recovered via minimization of scatter in the DGR, minimization of the correlation coefficient between CO/\hi\ and DGR and minimization of the $\chi^2$ of the best-fit plane to CO, \hi\ and \sigd.  

\subsection{Summary of Investigations into the Solution Technique}\label{sec:appsummary}

To summarize, we have tested a number of variations on the basic technique of finding the most ``uniform'' DGR in a given region by stepping through a grid of \aco\ values.  We created simulated data with known \aco\ and DGR intrinsic scatter matched to the range of S/N in our dataset.  Via a Monte Carlo simulation, we found the robust mean of the logarithmic scatter to be the technique that spans the range of CO S/N and possible intrinsic DGR scatter with the highest accuracy and  rate of success.  Given knowledge of the CO S/N for a region and the measured minimum of the logarithmic scatter, we can use the simulated data tests to constrain how biased our \aco\ measurement can be.  We have also specifically checked that the central regions of galaxies, which only comprise 26 pixels out of the sample, are not biased due to their sometimes unusual CO/\hi\ ratios.  We found no significant bias for these solutions and show in addition that a variety of techniques can reproduce the low central \aco\ solutions, not just the minimization of the DGR scatter.  

\subsection{Variations in $R_{21}$}

\begin{figure}
\centering
\epsscale{0.6}
\plotone{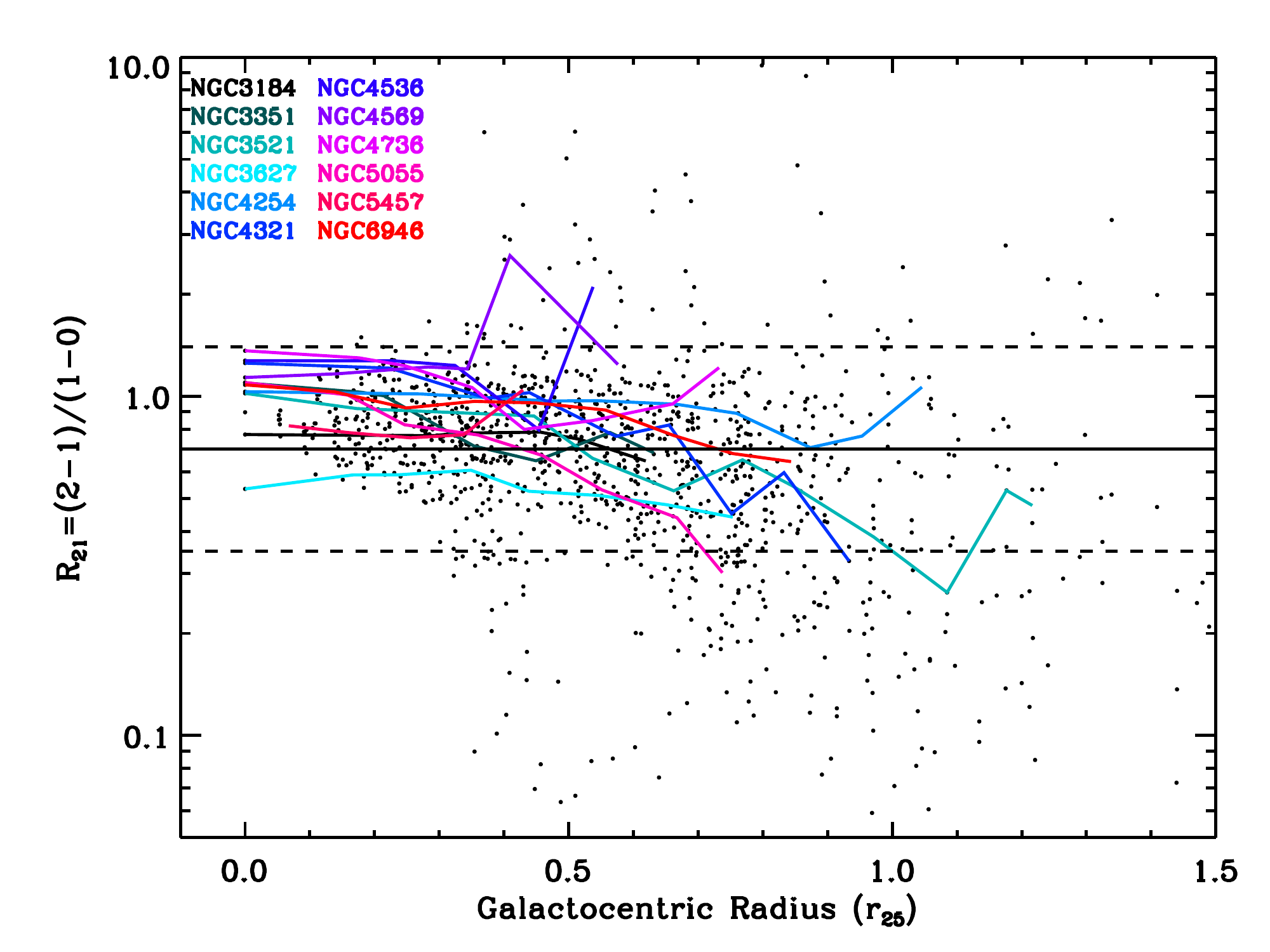}
\caption{The ratio of the CO J$=(2-1)$ and $(1-0)$ intensities measured from HERACLES and Nobeyama observations, respectively.  The black points show all of the individual measurements from the solution pixels and the colored lines show radial profiles, binned by $0.1$r$_{25}$.  The value we have assumed for converting our $(2-1)$ \aco\ measurements to the more commonly used $(1-0)$ scale is shown with a solid black line (R$_{21}=0.7$), and the dashed lines show factors of 2 above and below that value. The variations of $R_{21}$ are generally less than a factor of 2, which is less than the uncertainty on most of our \aco\ measurements.  Note that variations in $R_{21}$ only affect the $(1-0)$ \aco, since we are directly measuring the conversion factor appropriate for the $(2-1)$ line.}
\label{fig:r21}
\end{figure}

In Figure~\ref{fig:r21} we show the measured CO $(2-1)$/$(1-0)$ line ratio in the solution pixels we have defined for each galaxy (the black points) and in radial profiles (colored lines).  The assumed $R_{21}$ is shown with a horizontal solid, black line and a factor of 2 above and below that value are marked with black dashed lines.  There are variations in $R_{21}$, but in regions with good CO S/N (the only regions in which we will achieve good solutions) these are generally less than a factor of two away from the $R_{21}$ we have assumed.  This deviation is within the uncertainties on the \aco\ solutions, which typically have 0.2 dex uncertainty, at minimum.

\clearpage

\appendix
\section{Atlas of Galaxies and Results}
\begin{figure*}
\centering
\epsscale{2.2}
\plottwo{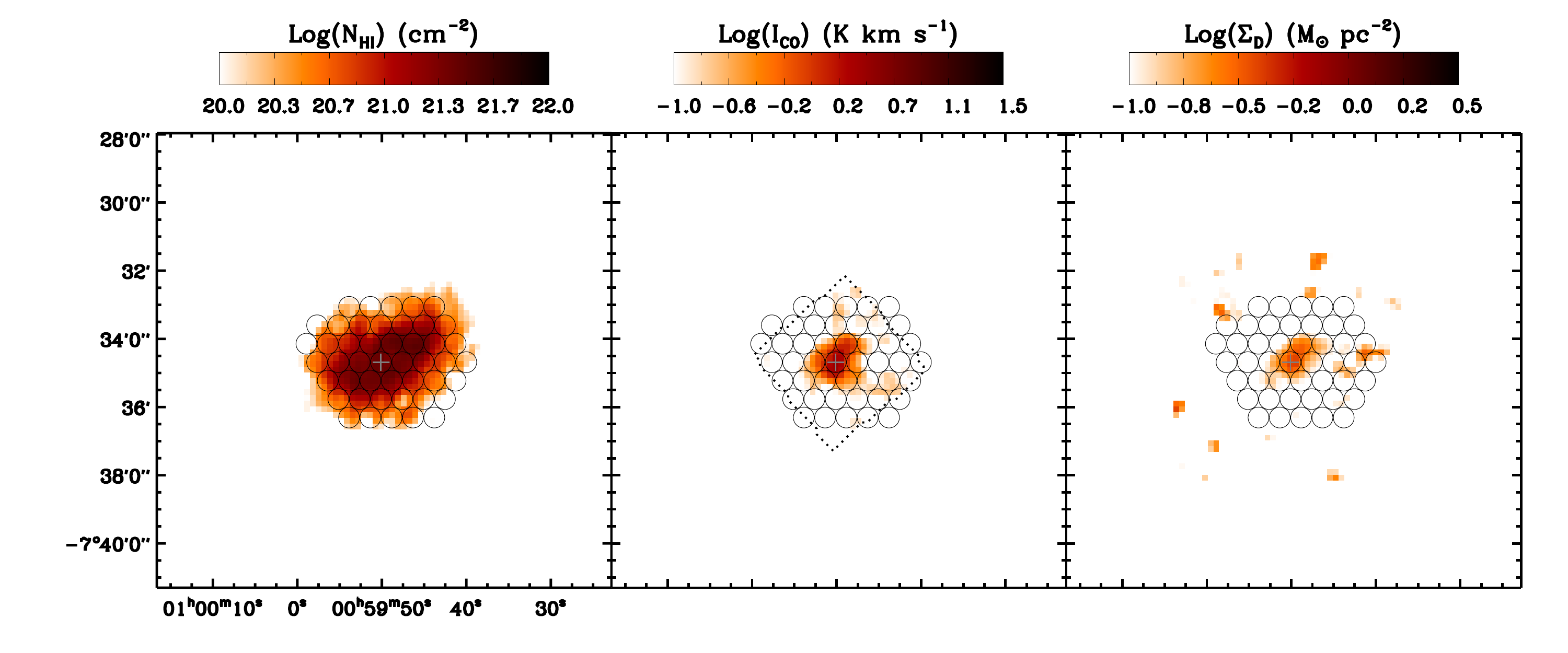}{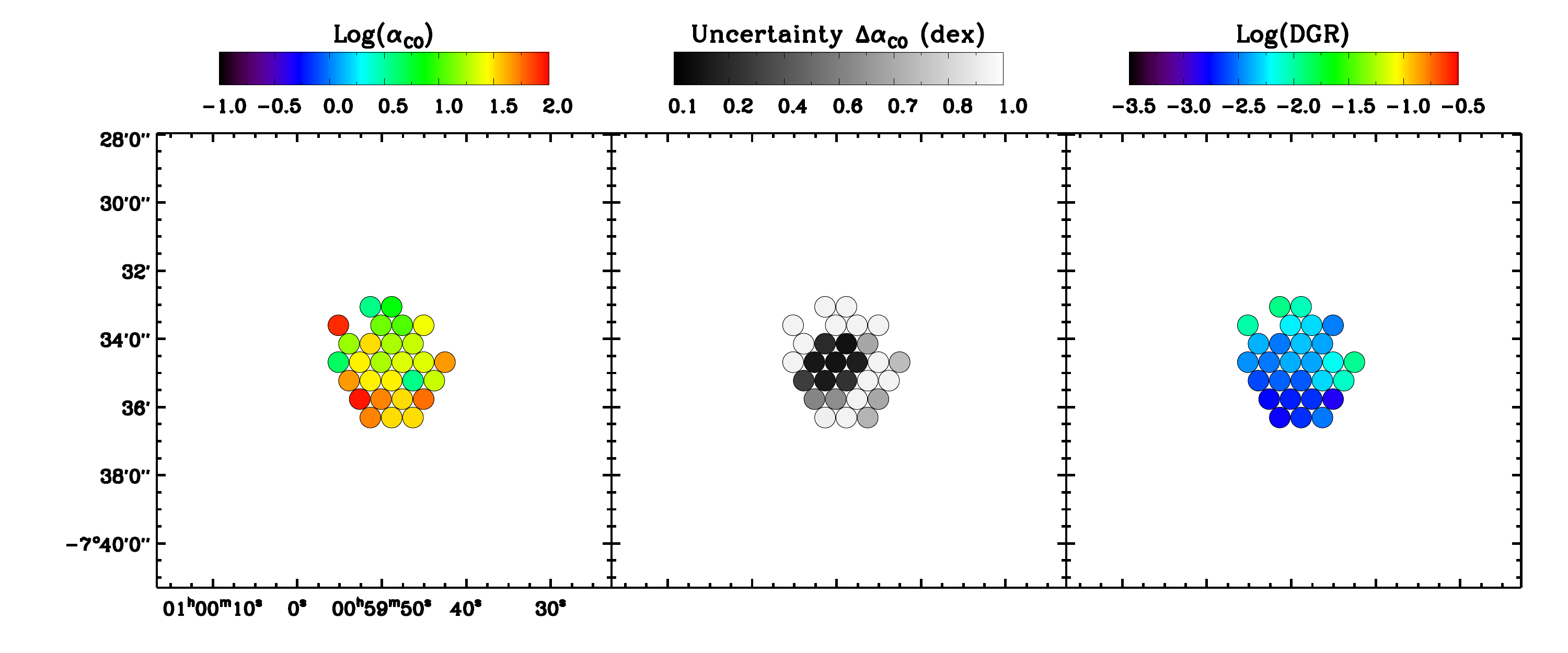}
\epsscale{1.0}
\plotone{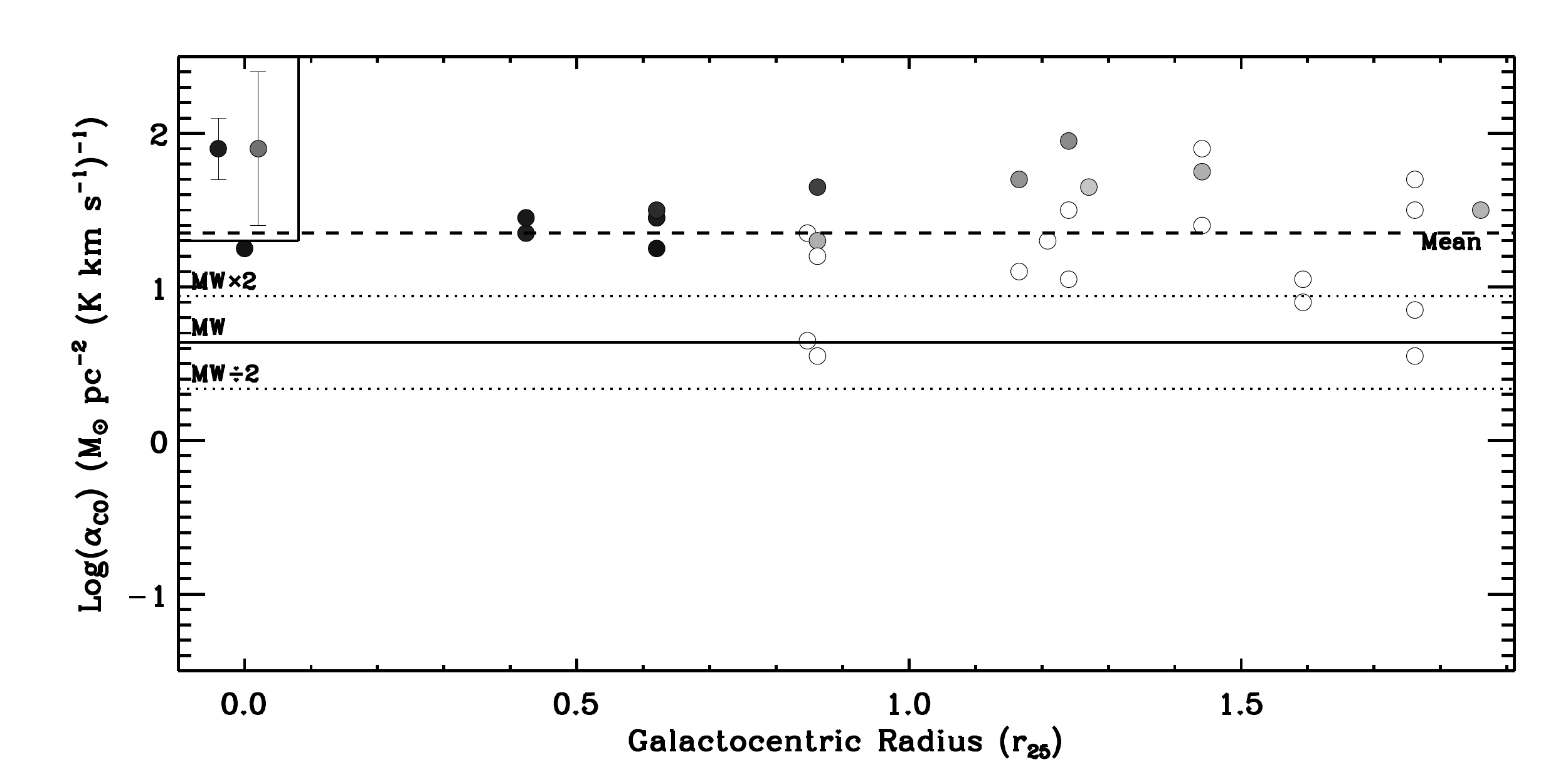}
\caption{Results for NGC 0337 (D = 19.3 Mpc; 1\arcsec = 94 pc).}
\label{fig:ngc0337_panel3}
\end{figure*}

\begin{figure*}
\centering
\epsscale{2.2}
\plottwo{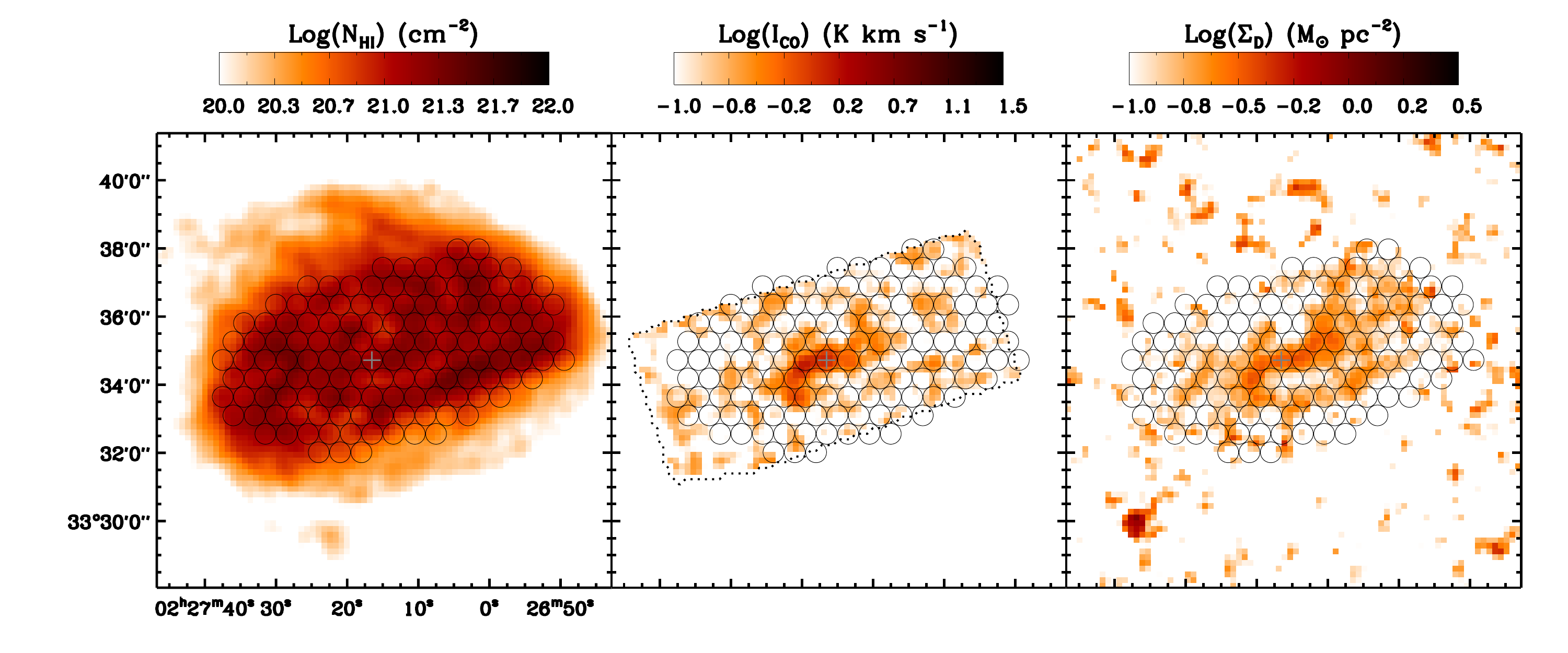}{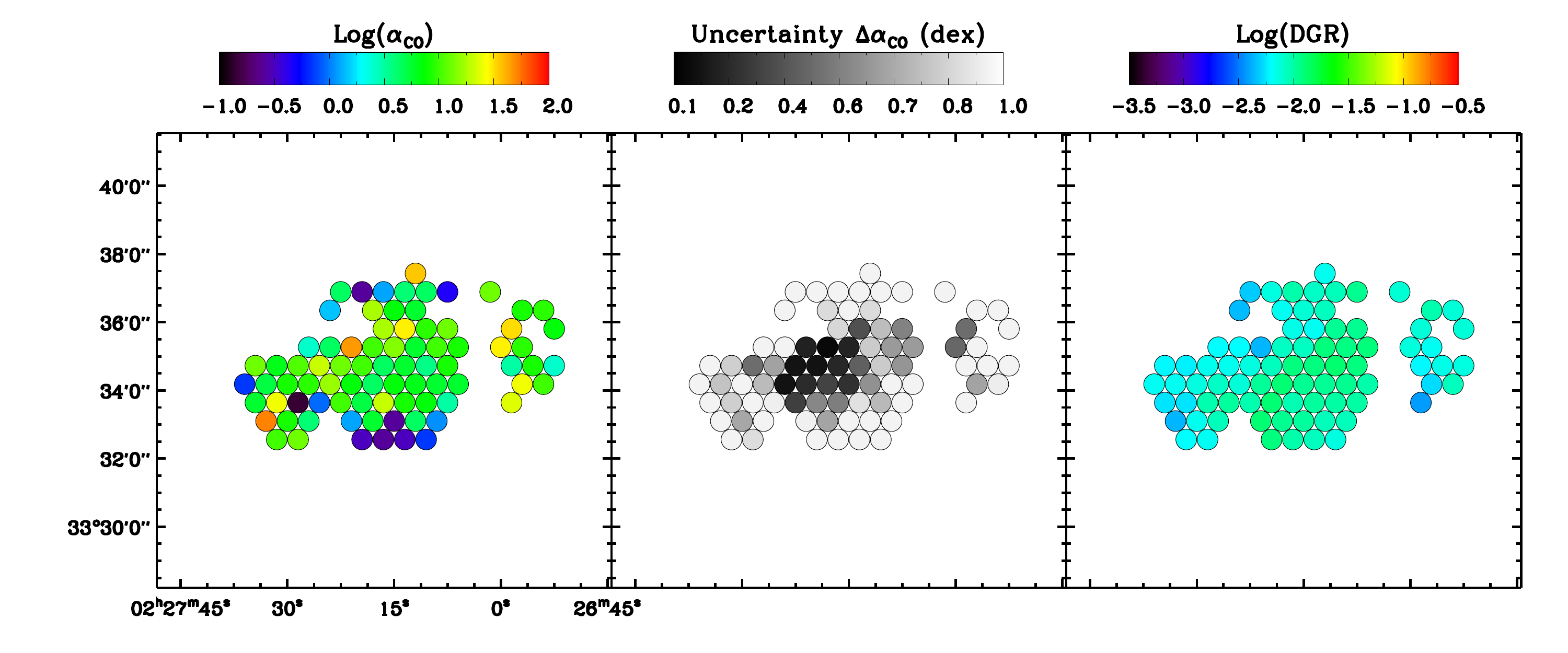}
\epsscale{1.0}
\plotone{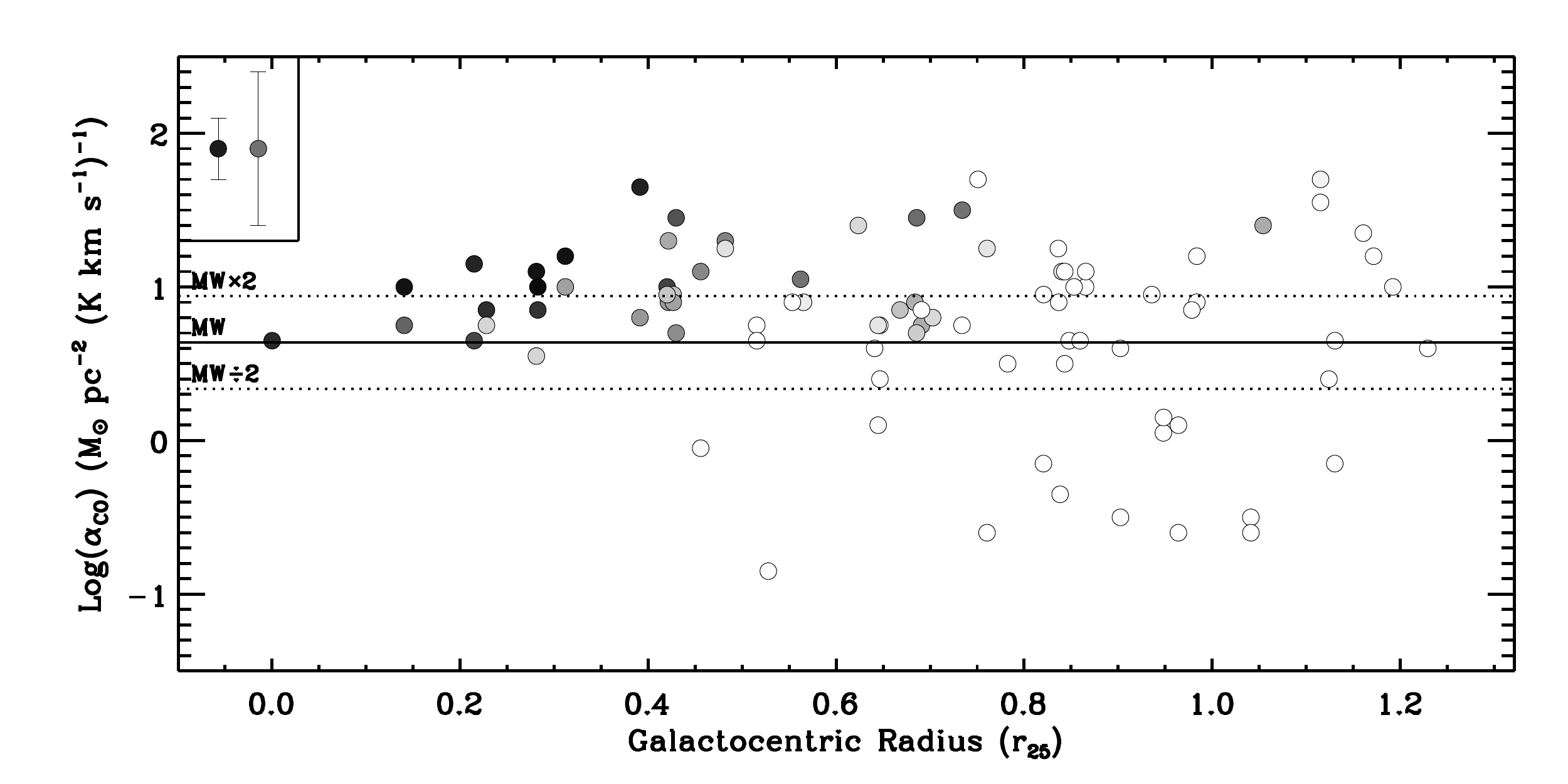}
\caption{Results for NGC 0925 (D = 9.1 Mpc; 1\arcsec = 44 pc).}
\label{fig:ngc0925_panel3}
\end{figure*}

\begin{figure*}
\centering
\epsscale{2.2}
\plottwo{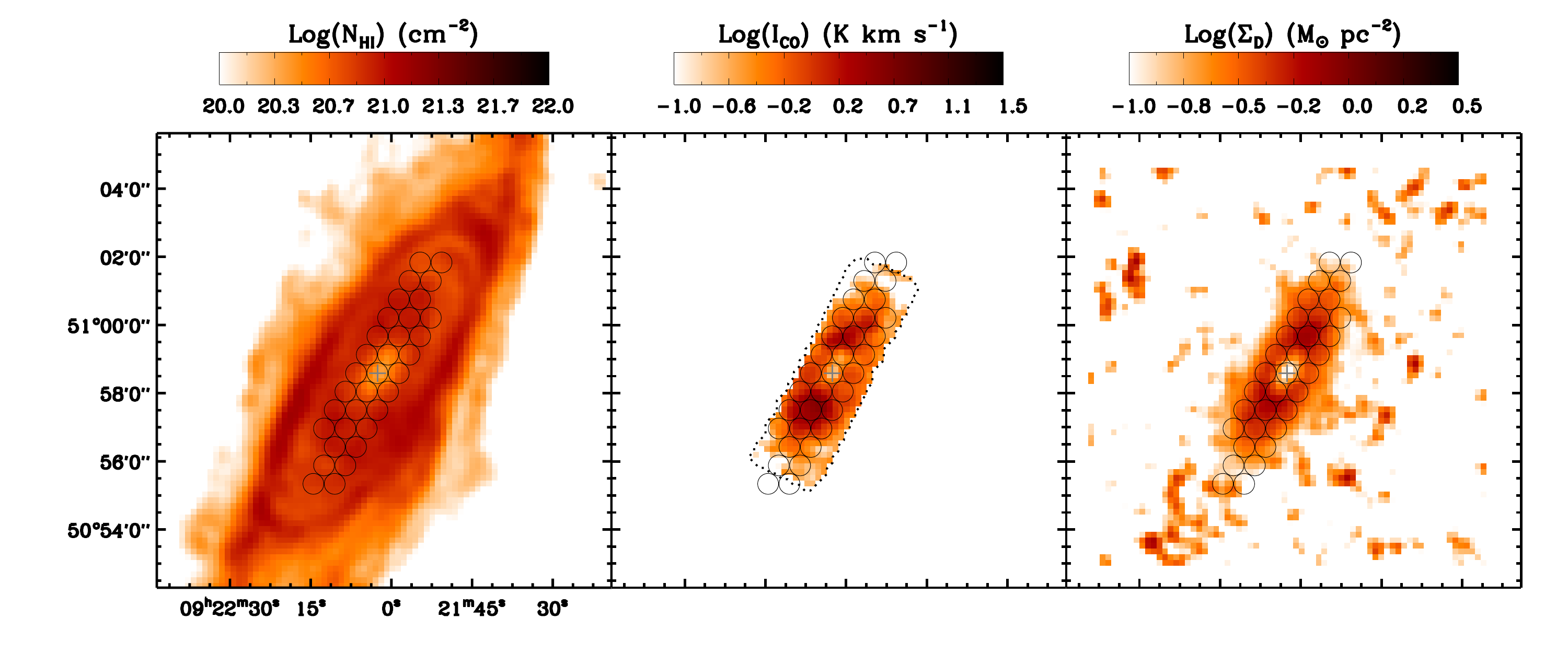}{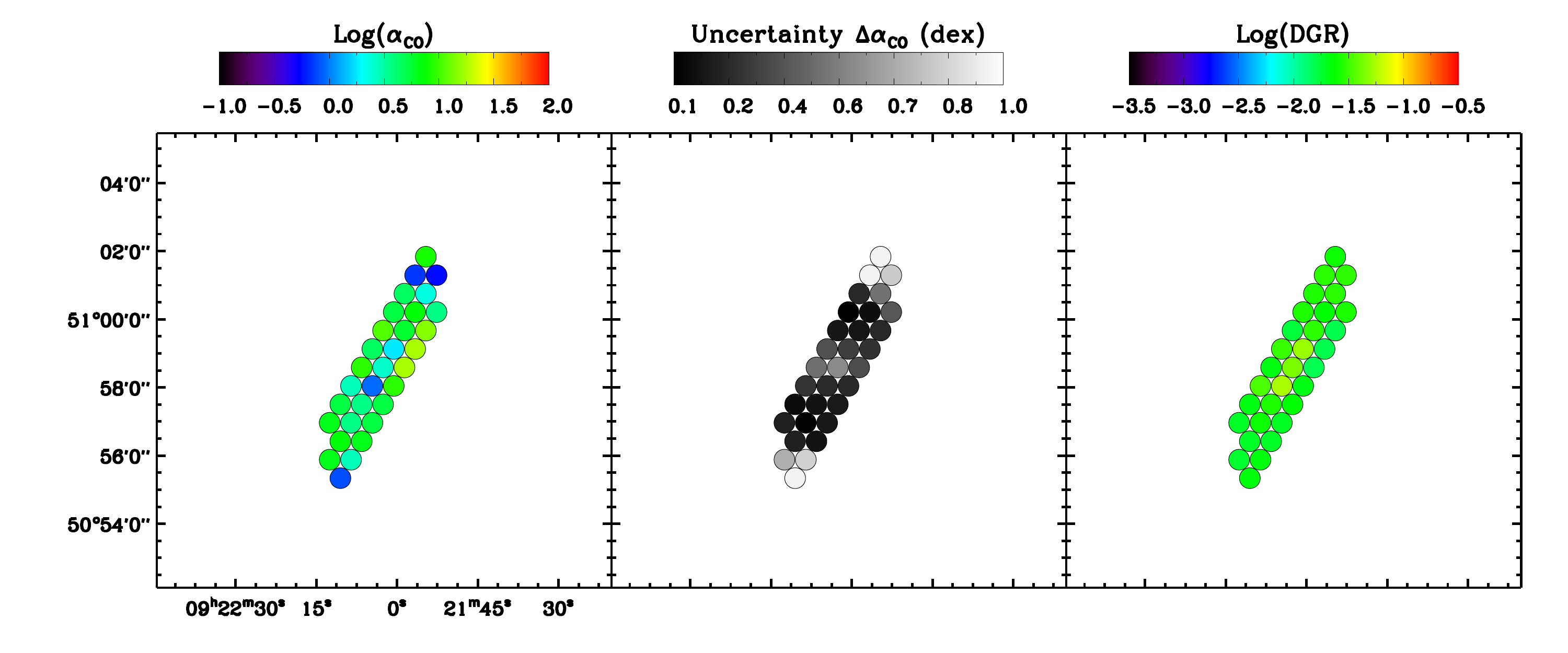}
\epsscale{1.0}
\plotone{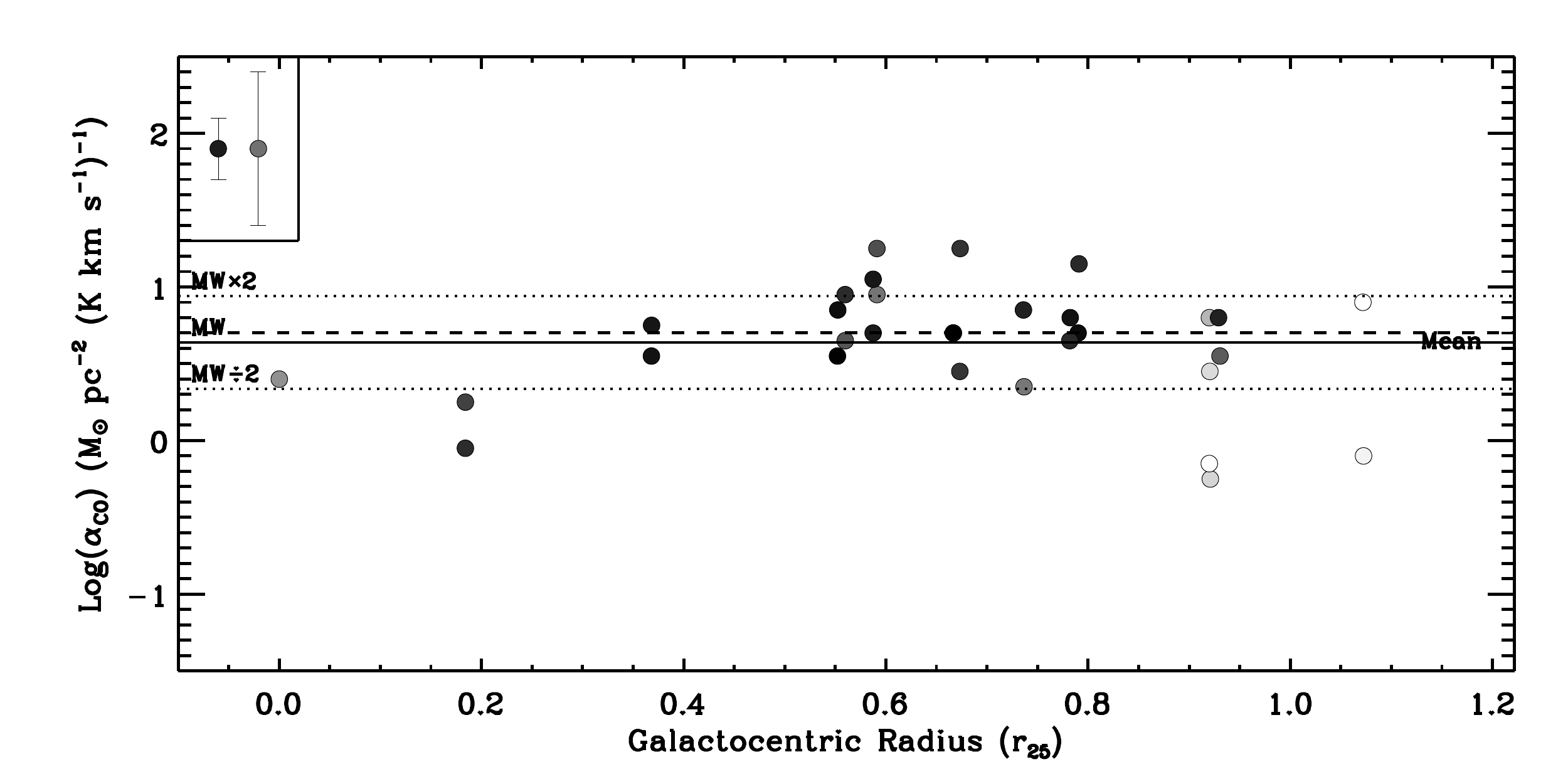}
\caption{Results for NGC 2841 (D = 14.1 Mpc; 1\arcsec = 68 pc).}
\label{fig:ngc2841_panel3}
\end{figure*}

\begin{figure*}
\centering
\epsscale{2.2}
\plottwo{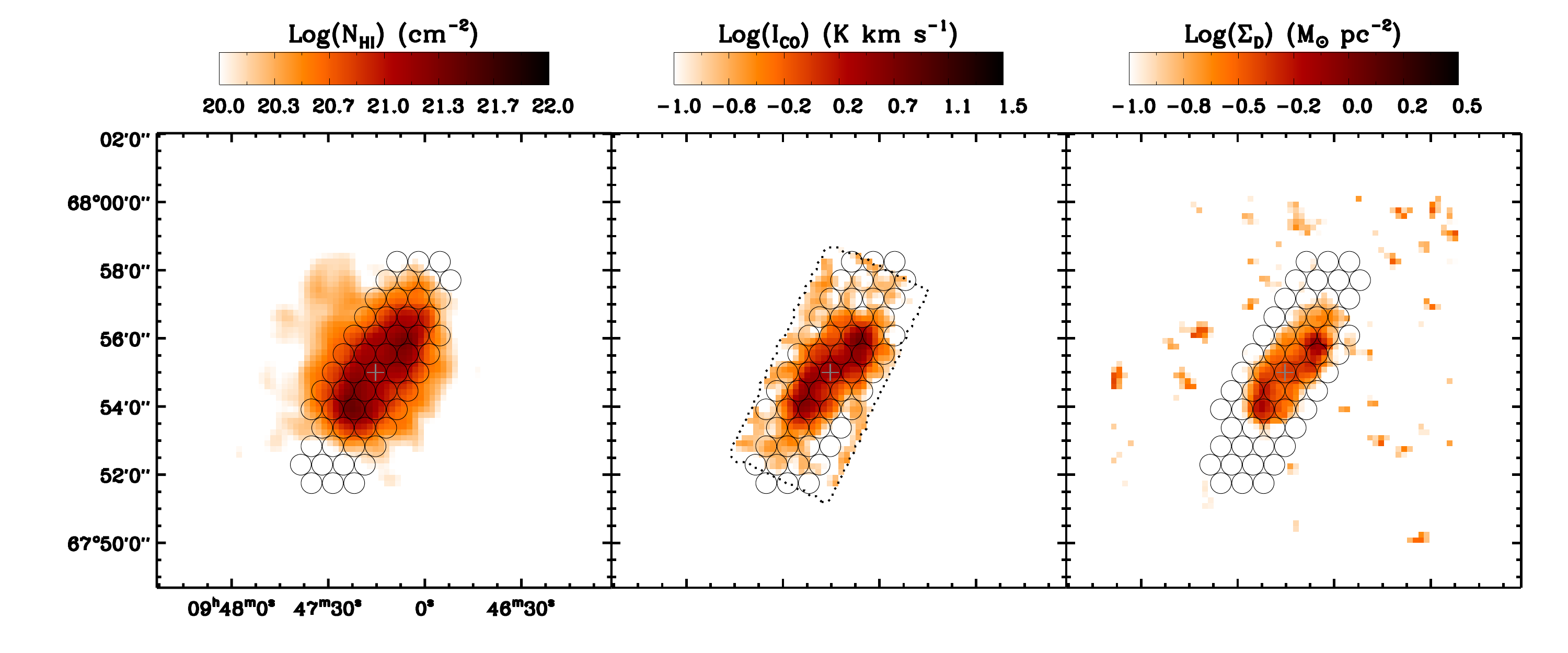}{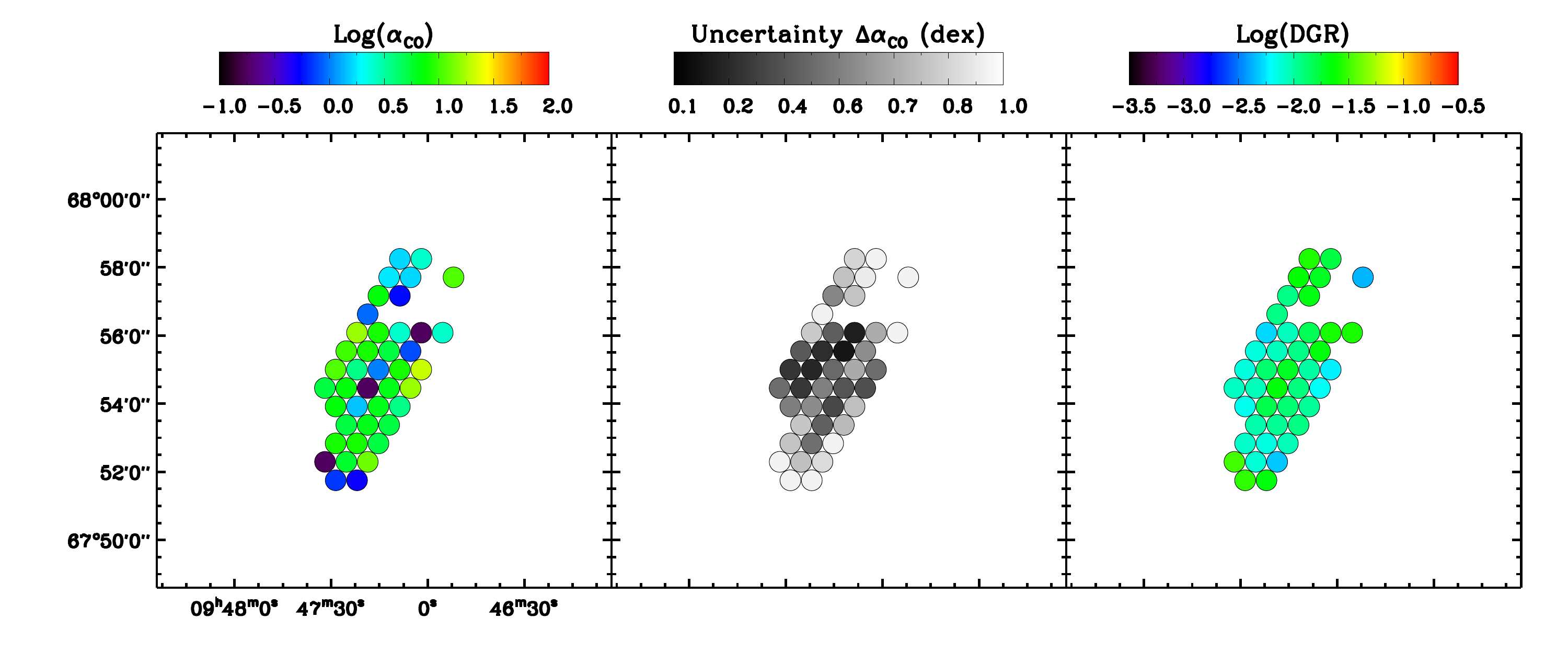}
\epsscale{1.0}
\plotone{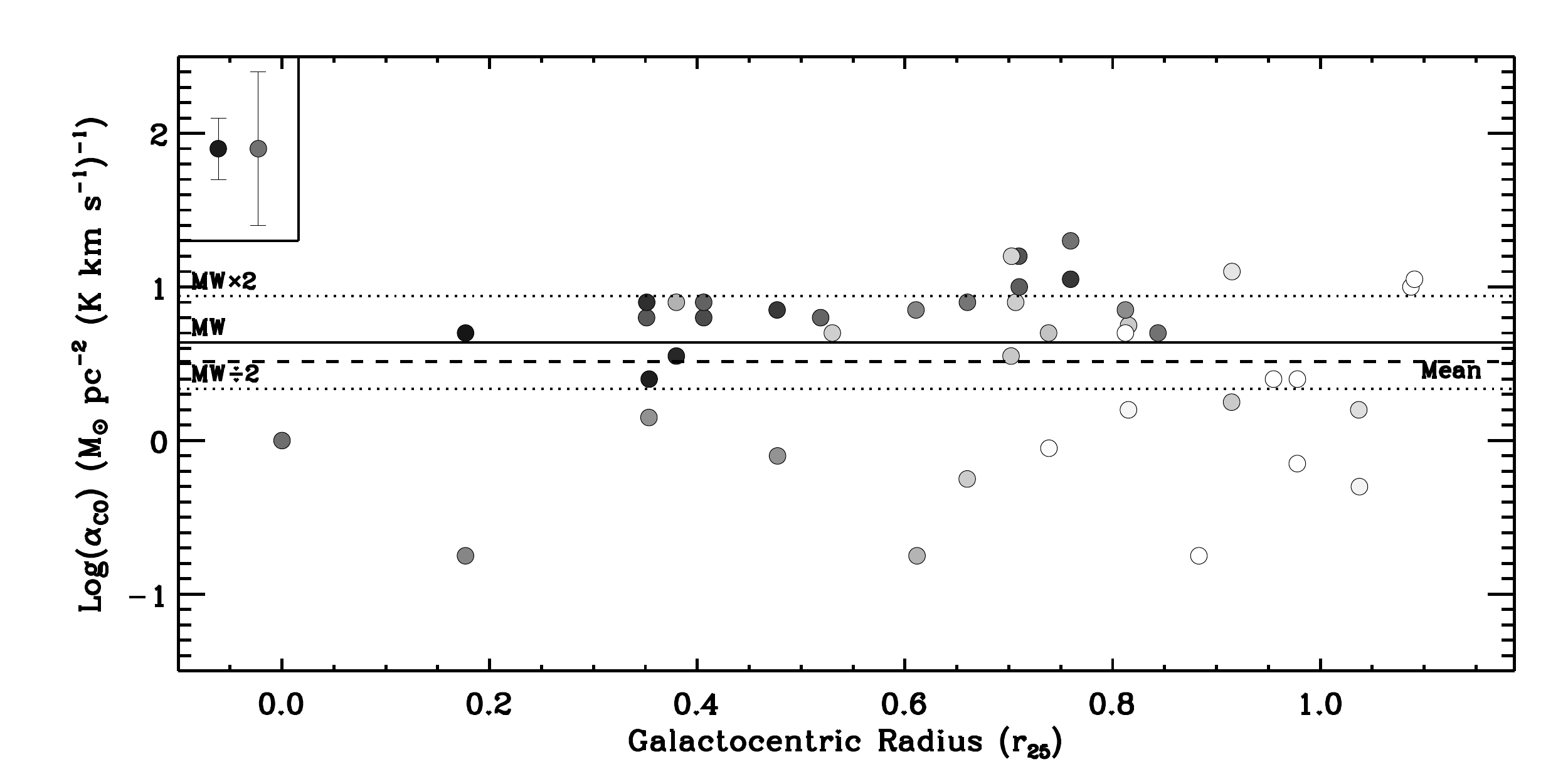}
\caption{Results for NGC 2976 (D = 3.6 Mpc; 1\arcsec = 17 pc).}
\label{fig:ngc2976_panel3}
\end{figure*}

\begin{figure*}
\centering
\epsscale{2.2}
\plottwo{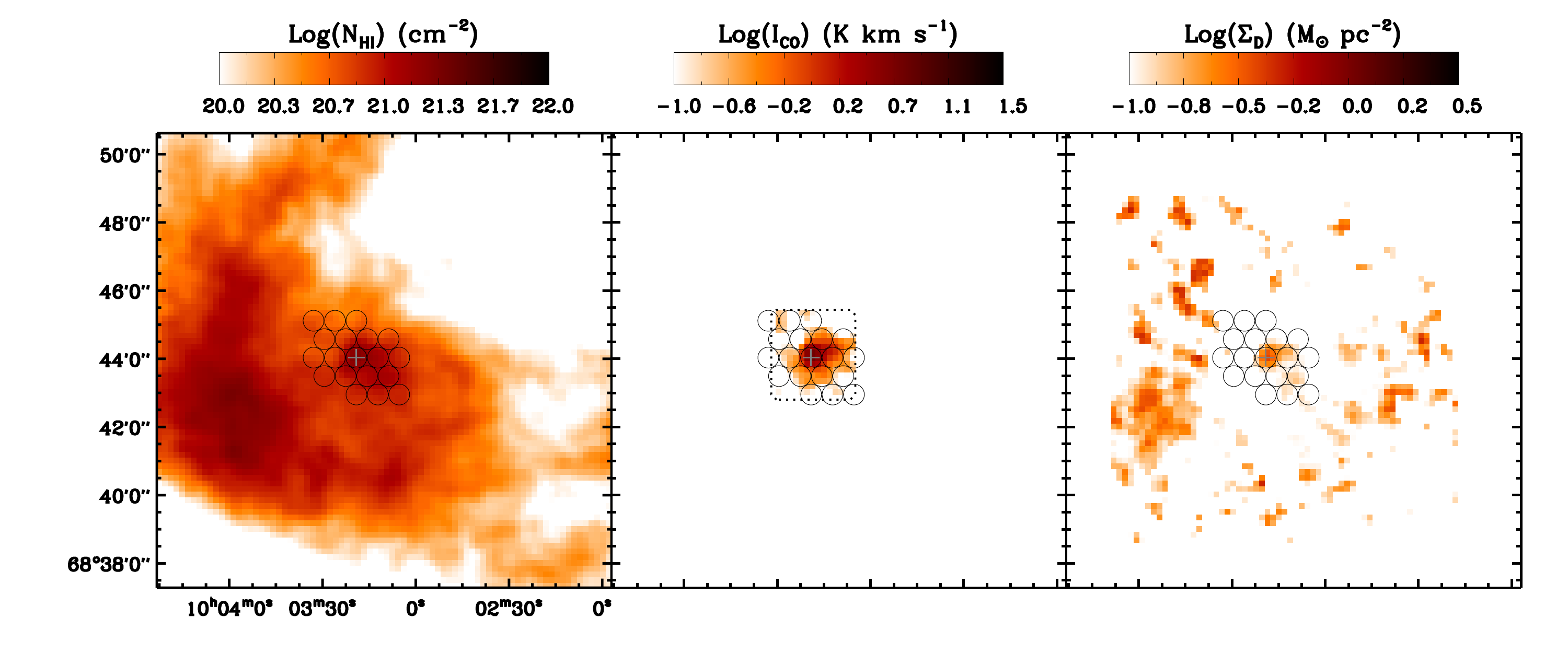}{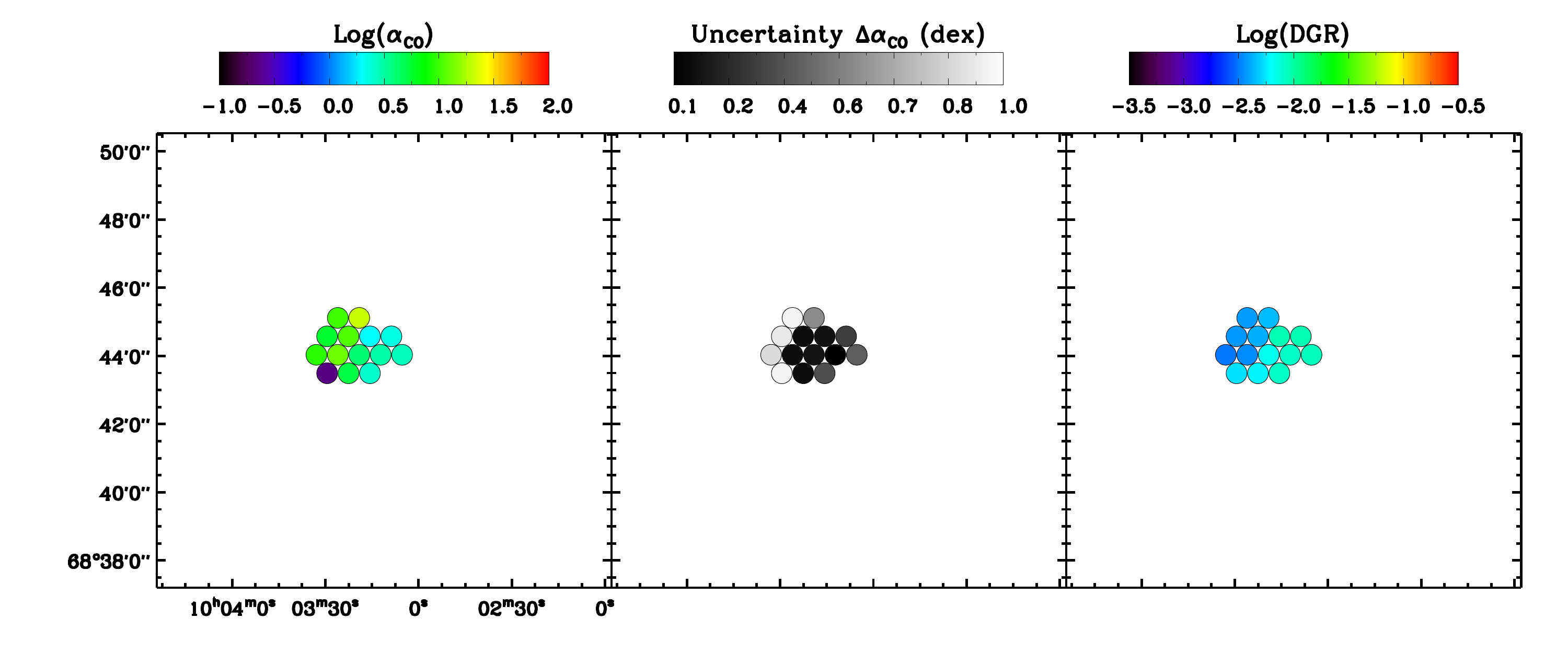}
\epsscale{1.0}
\plotone{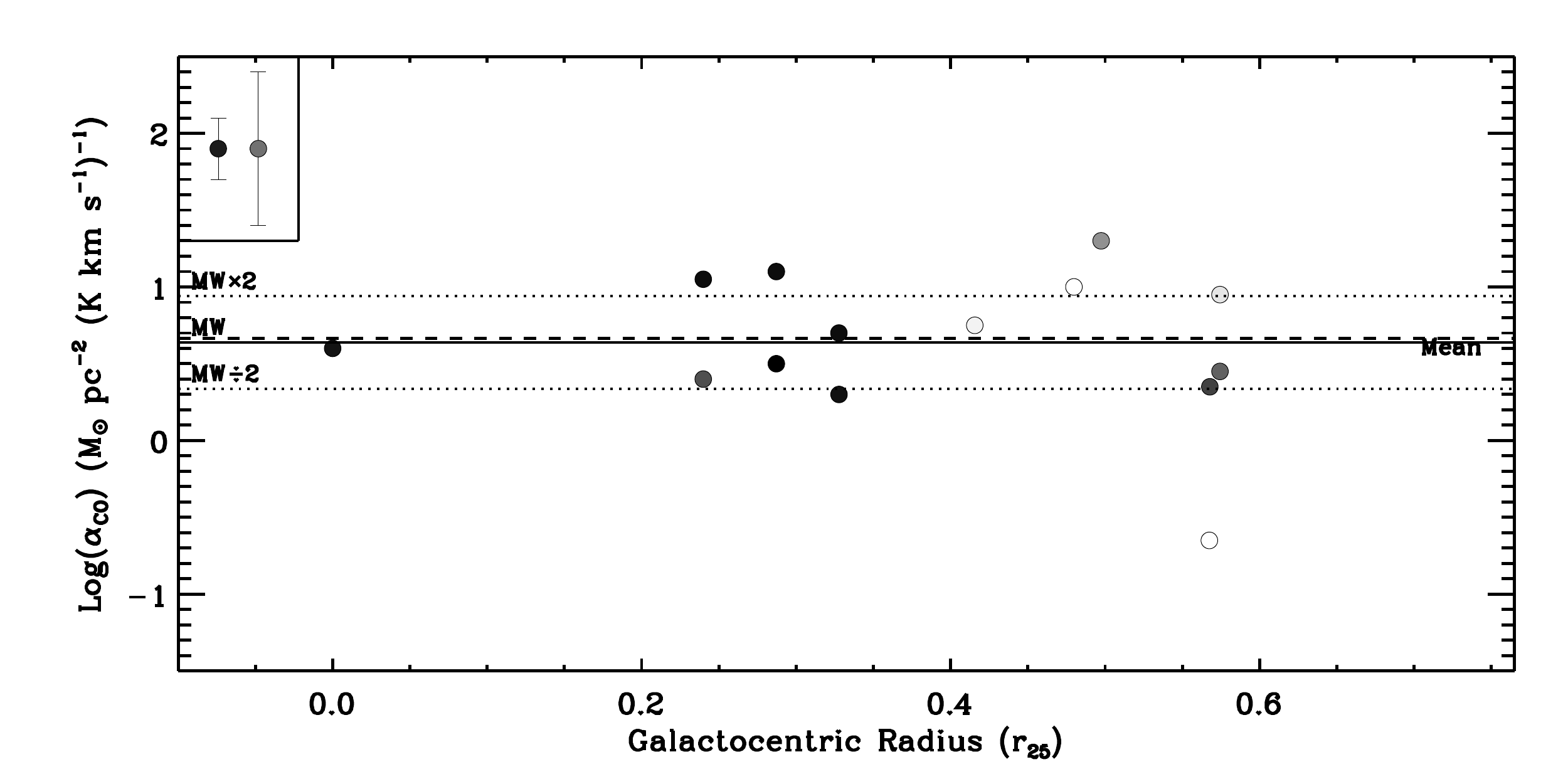}
\caption{Results for NGC 3077 (D = 3.8 Mpc; 1\arcsec = 19 pc).}
\label{fig:ngc3077_panel3}
\end{figure*}

\begin{figure*}
\centering
\epsscale{2.2}
\plottwo{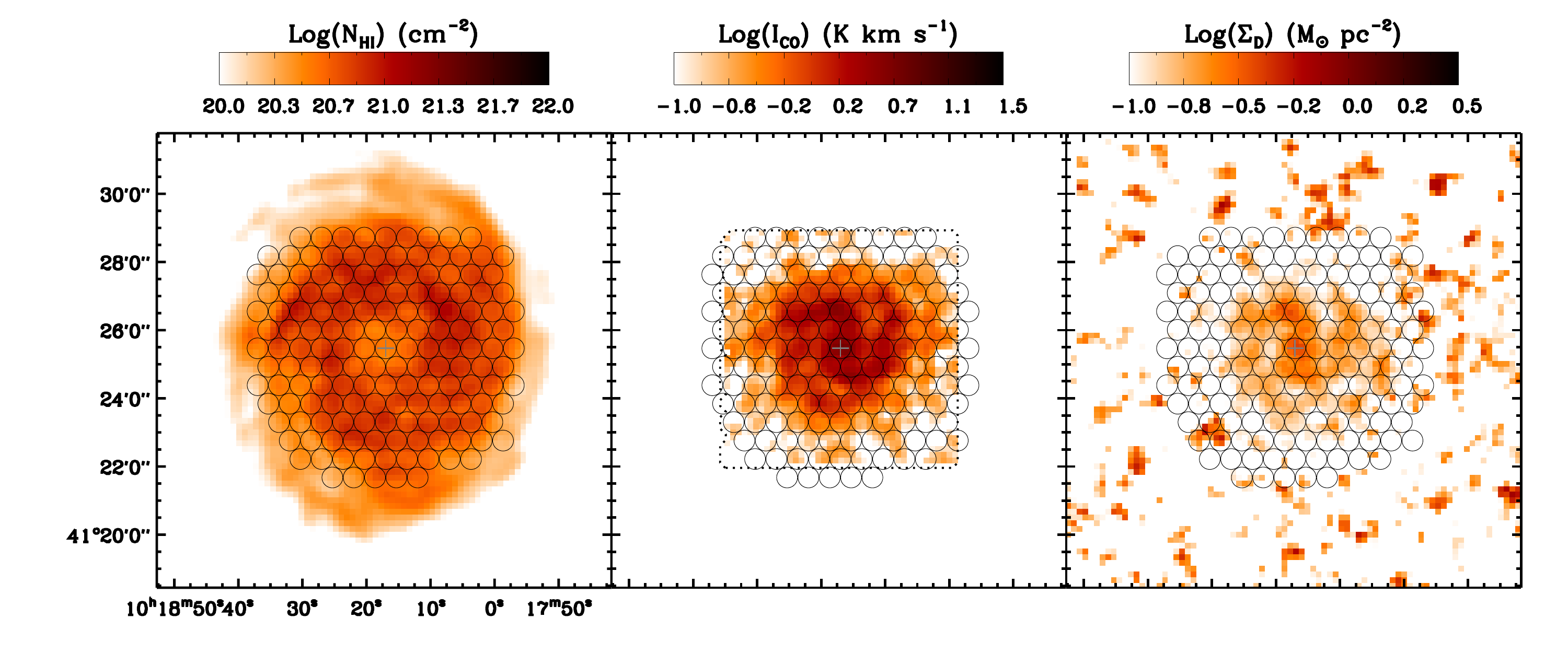}{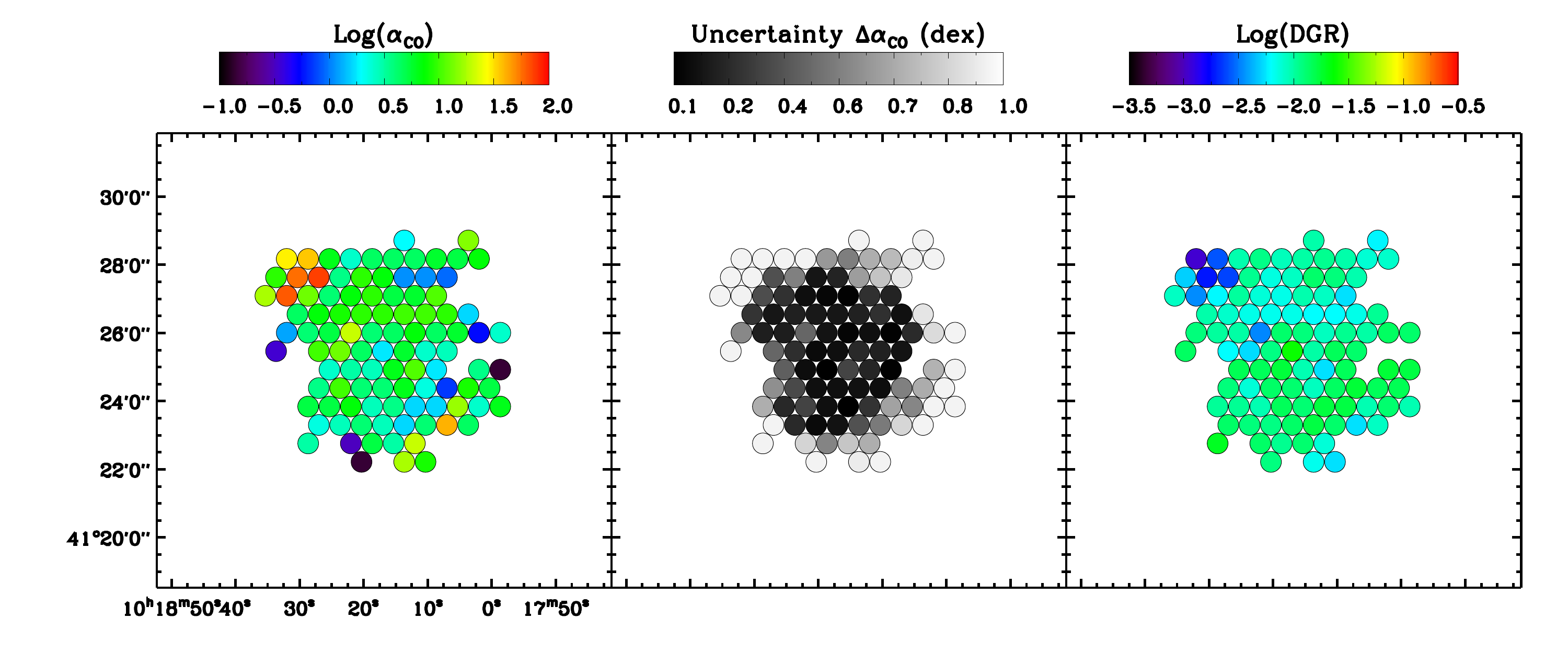}
\epsscale{1.0}
\plotone{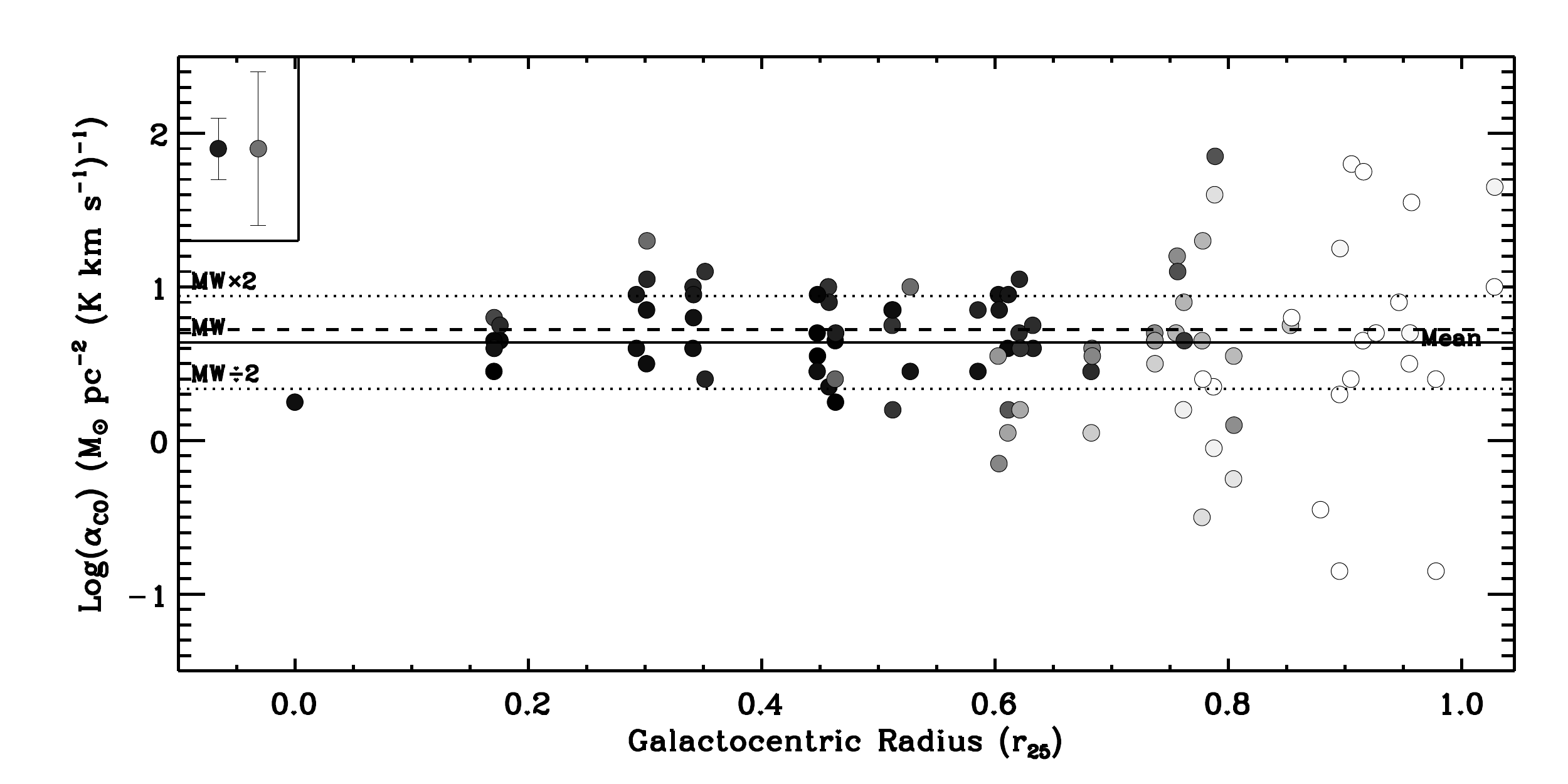}
\caption{Results for NGC 3184 (D = 11.7 Mpc; 1\arcsec = 57 pc).}
\label{fig:ngc3184_panel3}
\end{figure*}

\begin{figure*}
\centering
\epsscale{2.2}
\plottwo{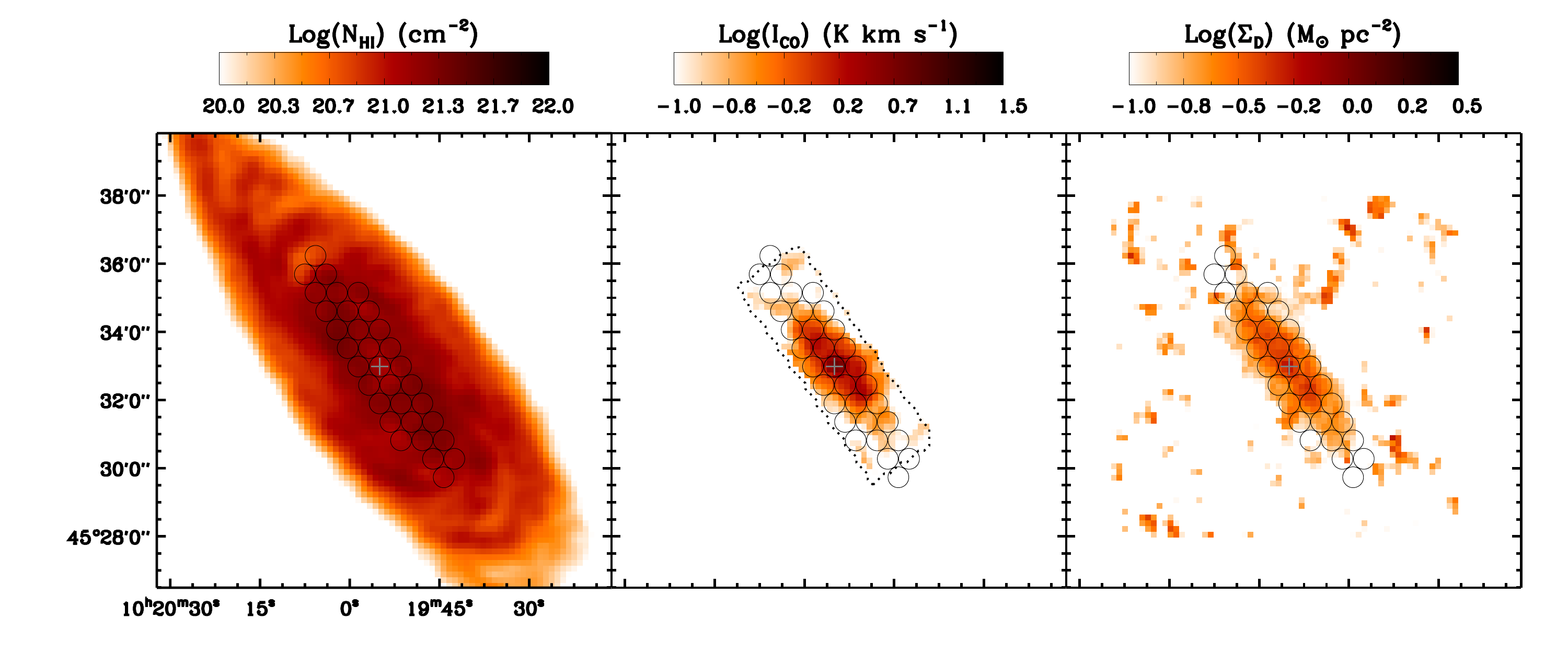}{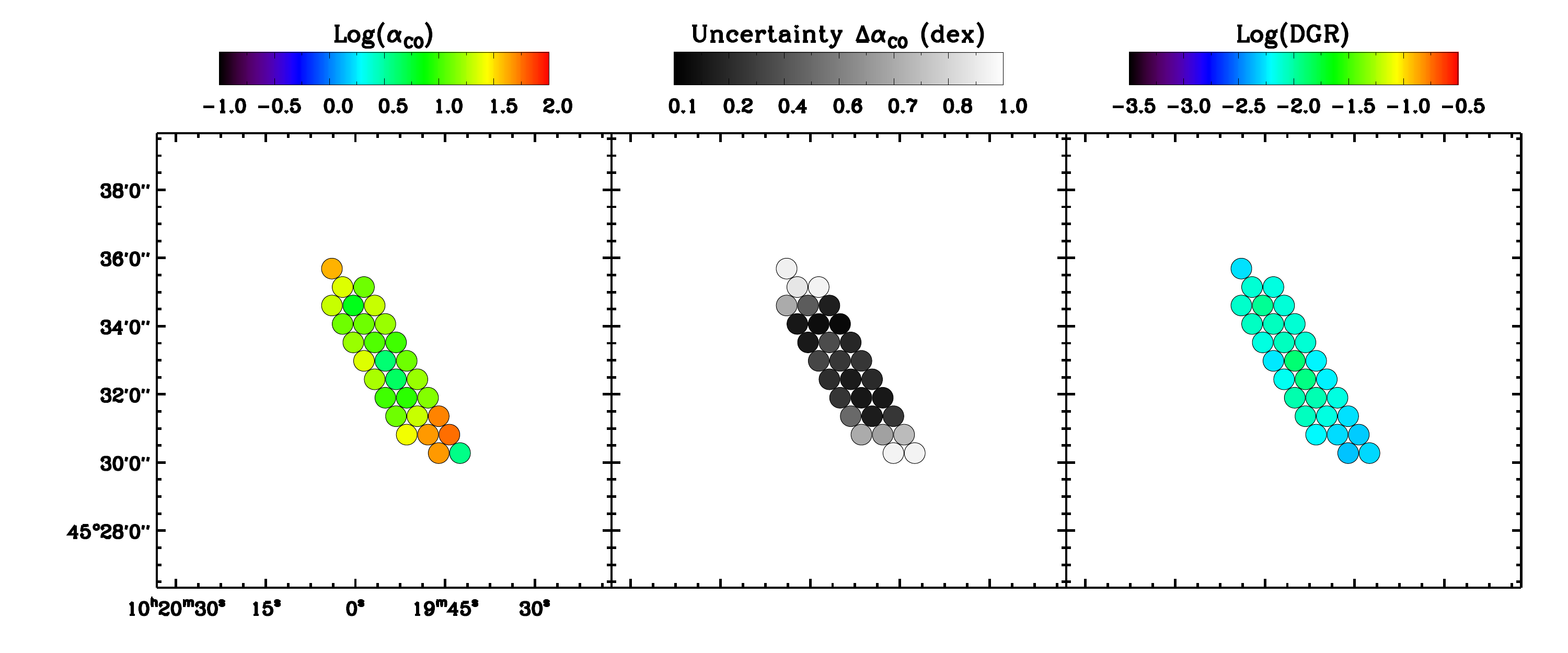}
\epsscale{1.0}
\plotone{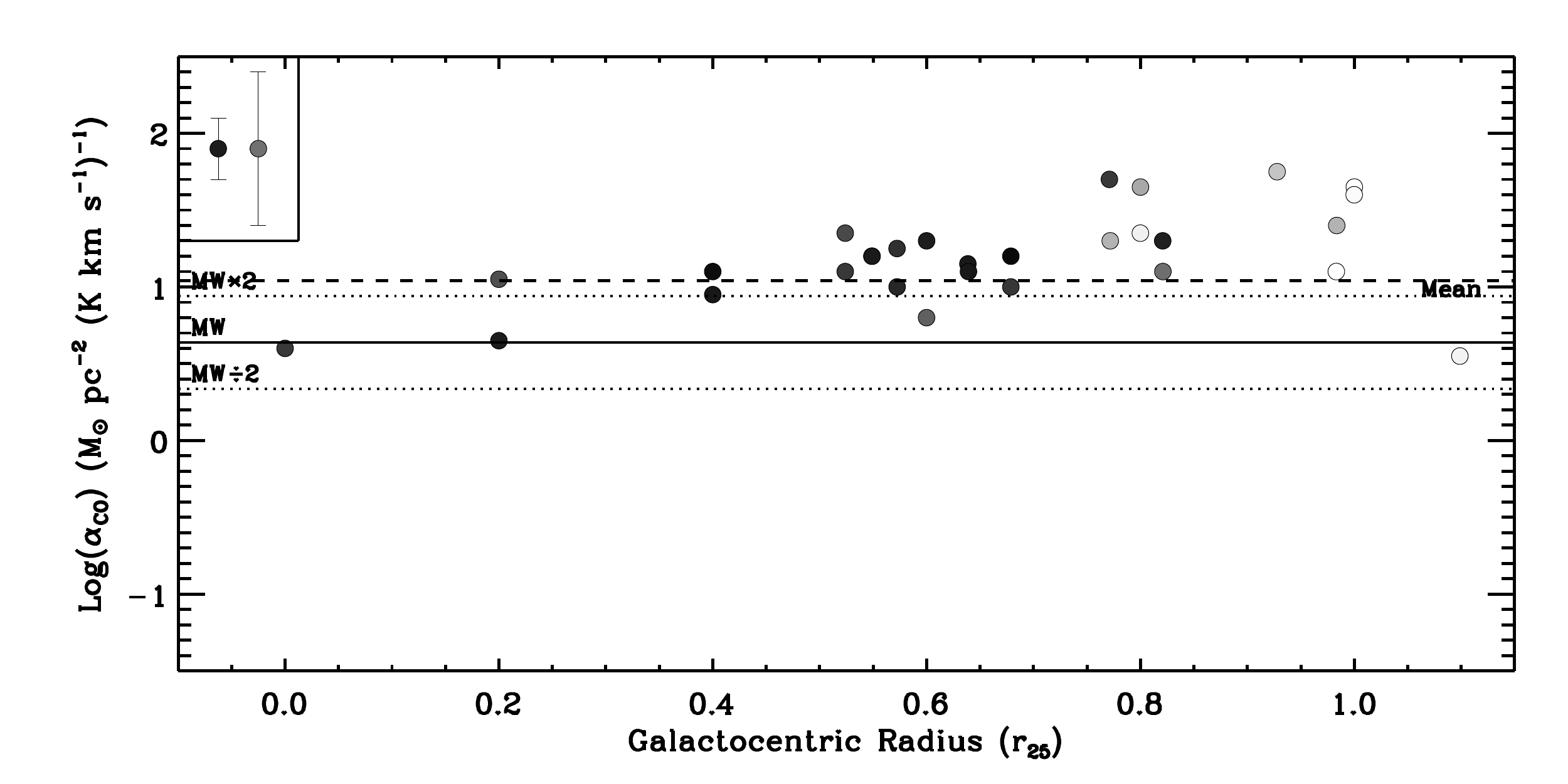}
\caption{Results for NGC 3198 (D = 14.1 Mpc; 1\arcsec = 68 pc).}
\label{fig:ngc3198_panel3}
\end{figure*}

\begin{figure*}
\centering
\epsscale{2.2}
\plottwo{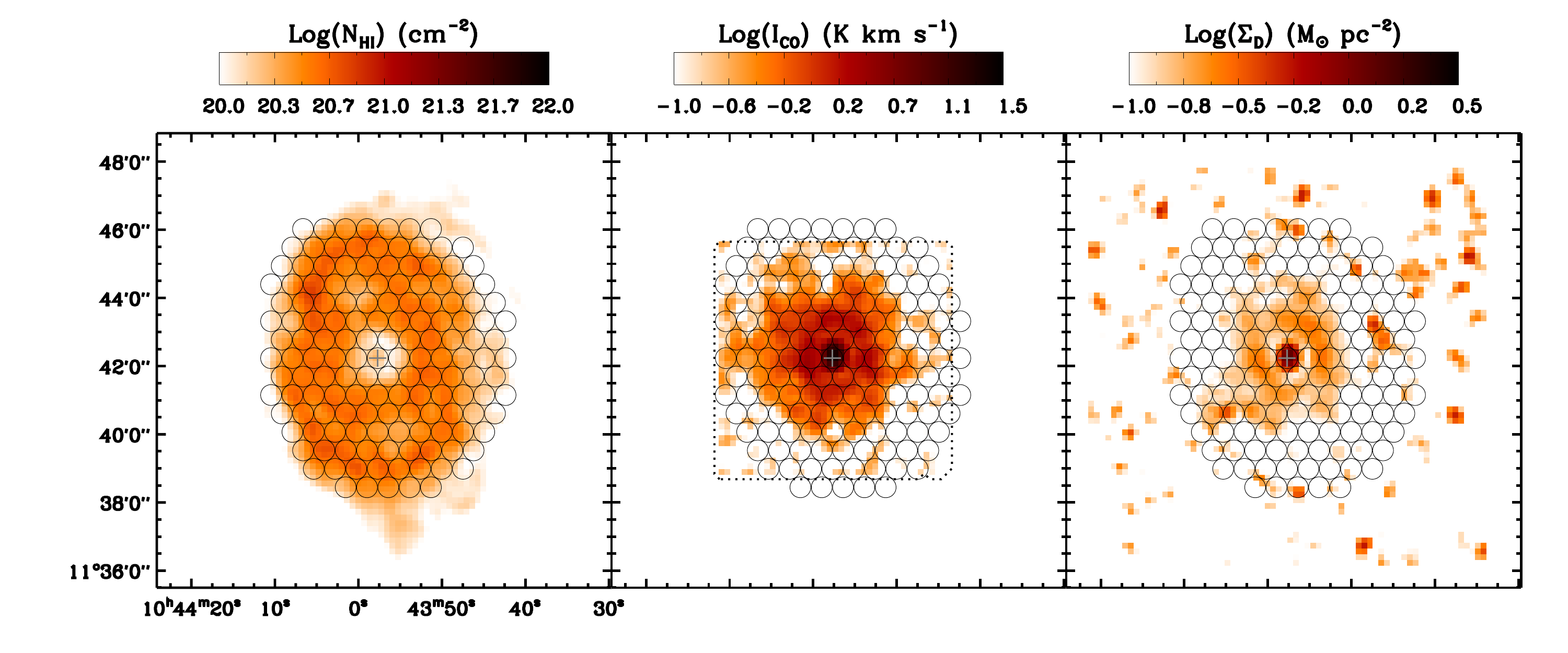}{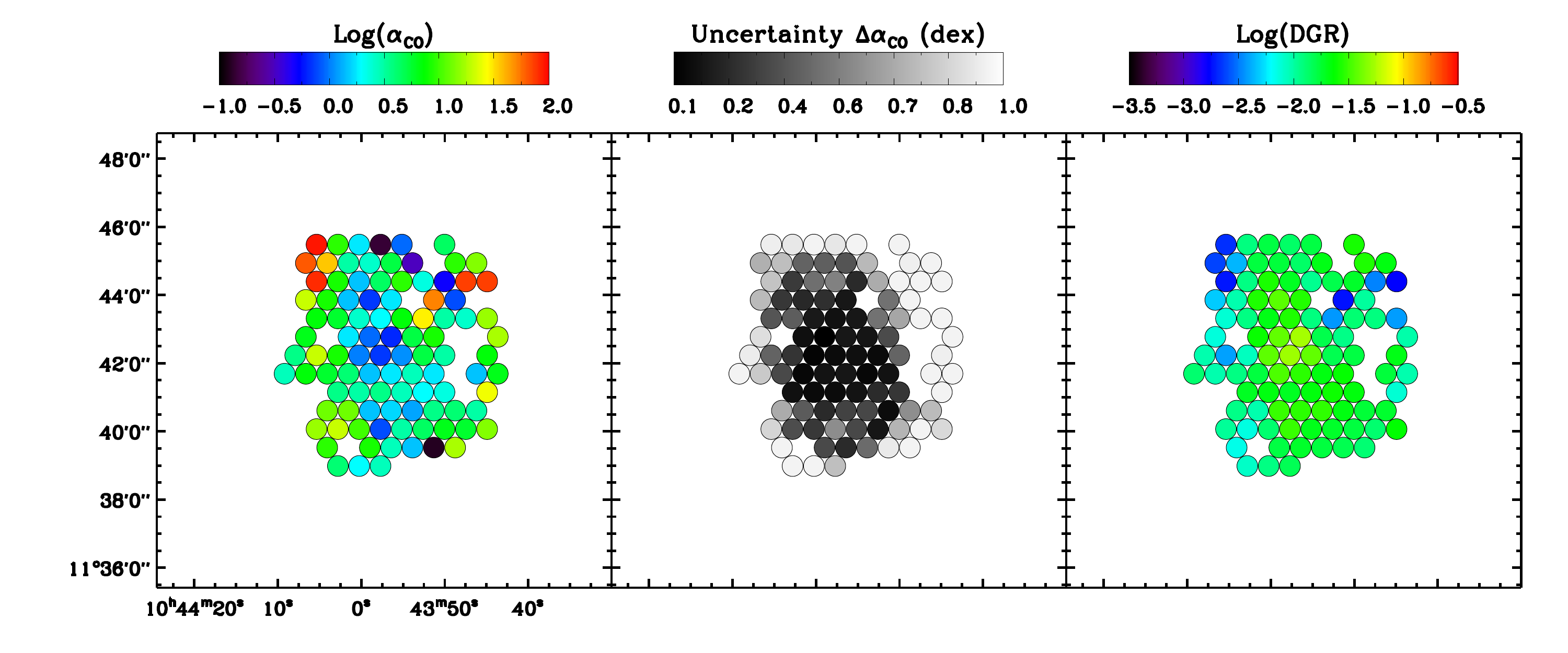}
\epsscale{1.0}
\plotone{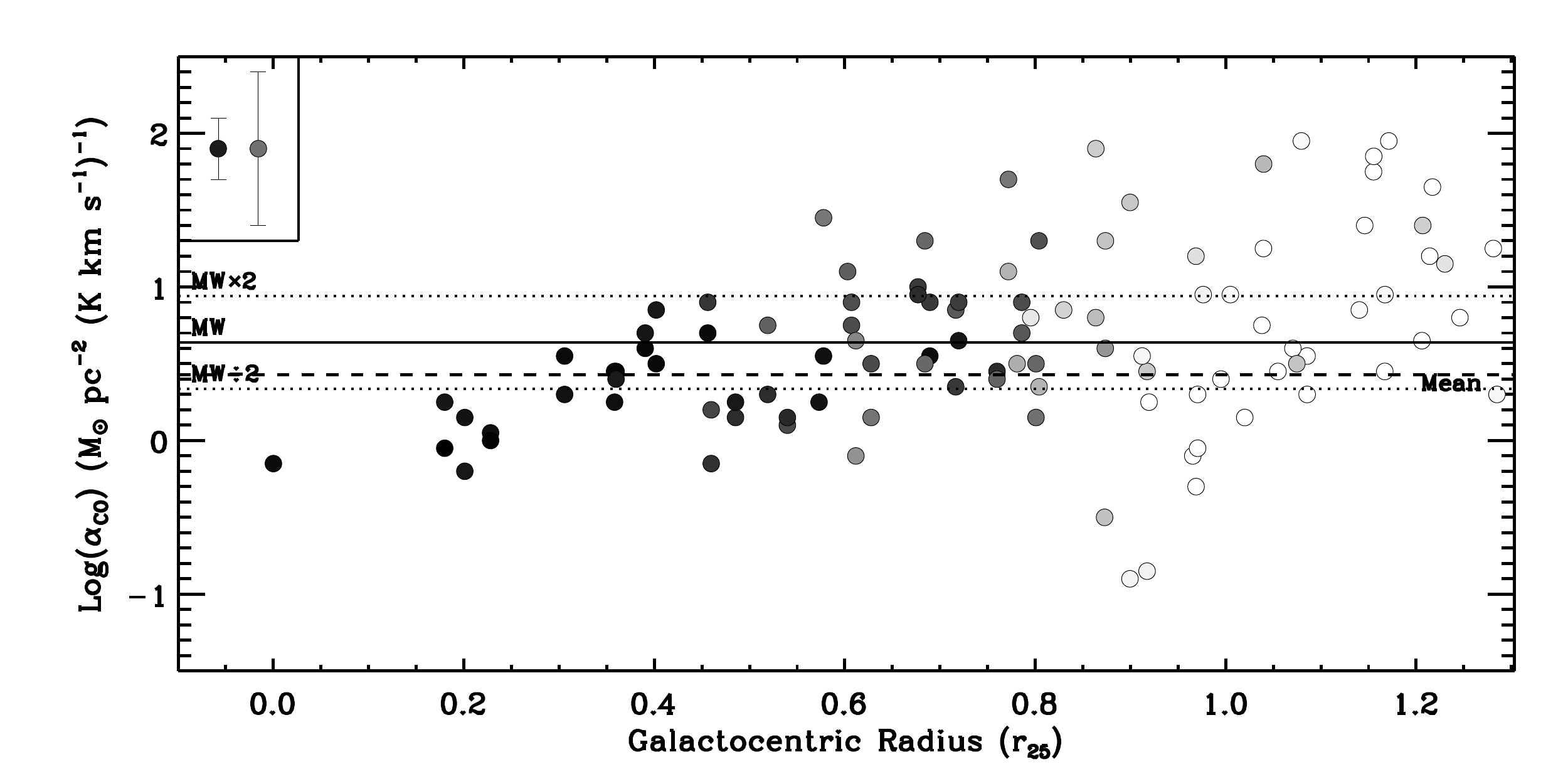}
\caption{Results for NGC 3351 (D = 9.3 Mpc; 1\arcsec = 45 pc).}
\label{fig:ngc3351_panel3}
\end{figure*}

\begin{figure*}
\centering
\epsscale{2.2}
\plottwo{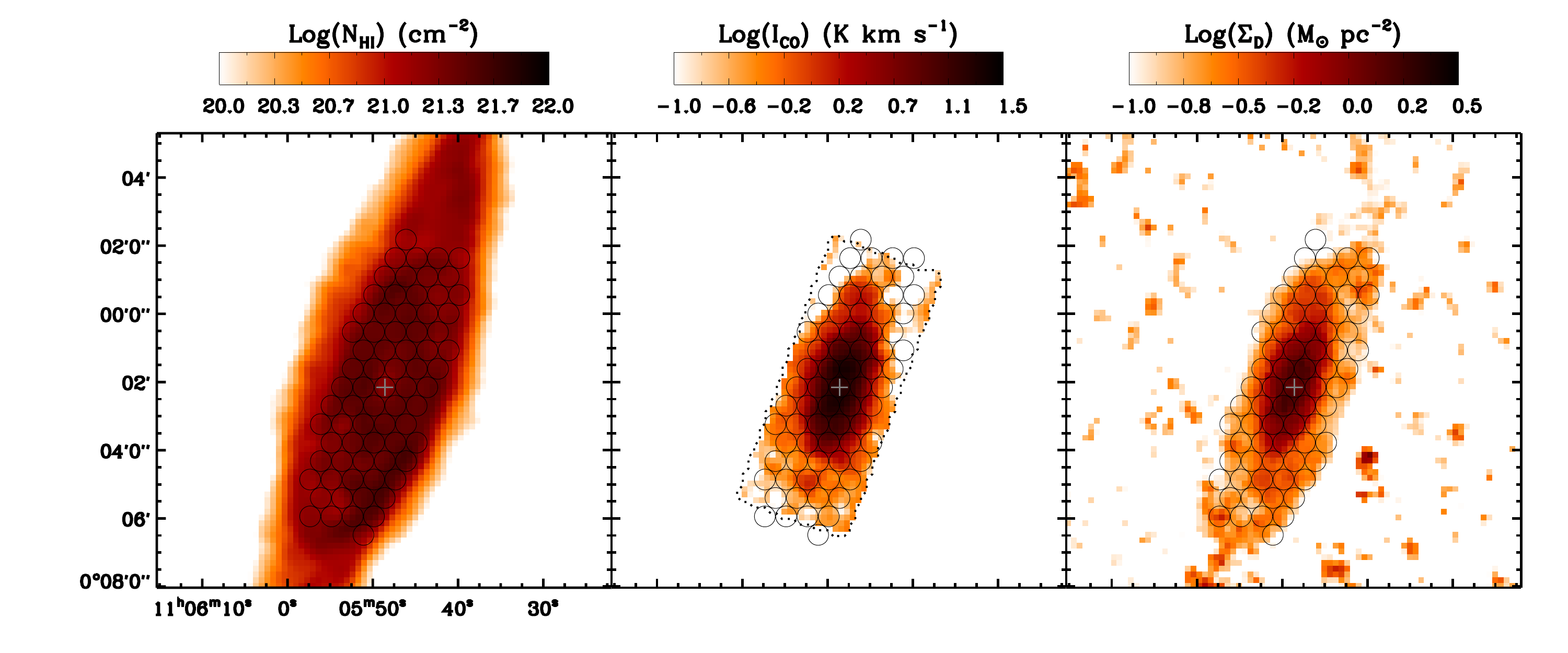}{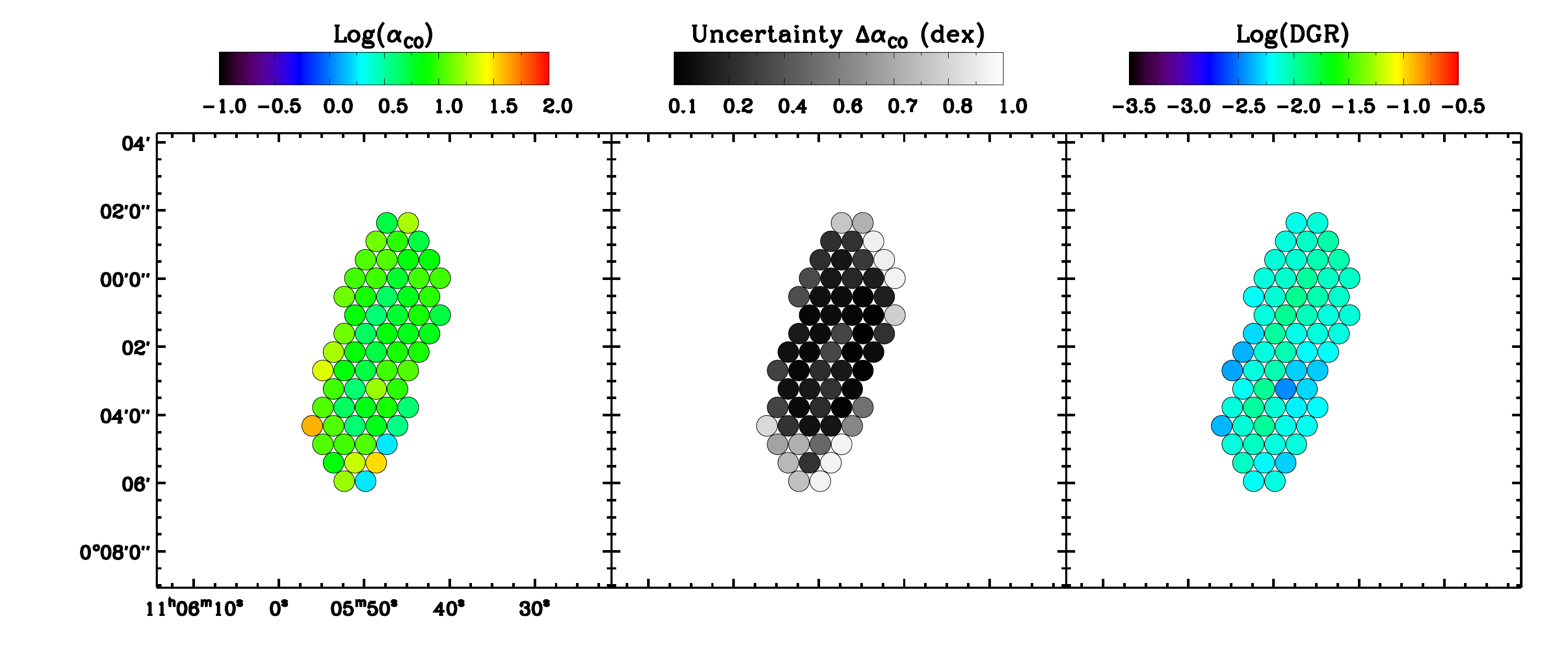}
\epsscale{1.0}
\plotone{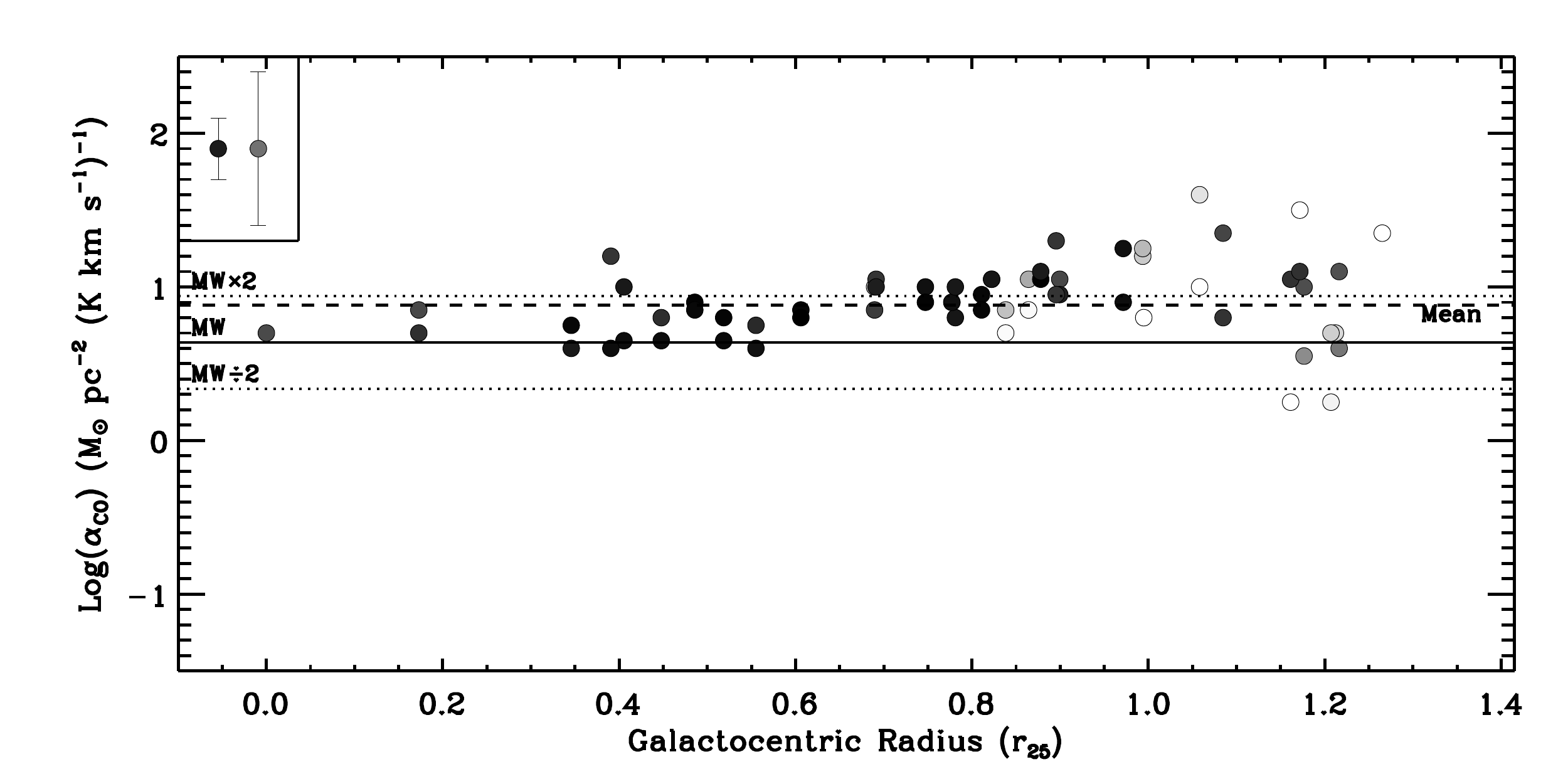}
\caption{Results for NGC 3521 (D = 11.2 Mpc; 1\arcsec = 54 pc).}
\label{fig:ngc3521_panel3}
\end{figure*}

\begin{figure*}
\centering
\epsscale{2.2}
\plottwo{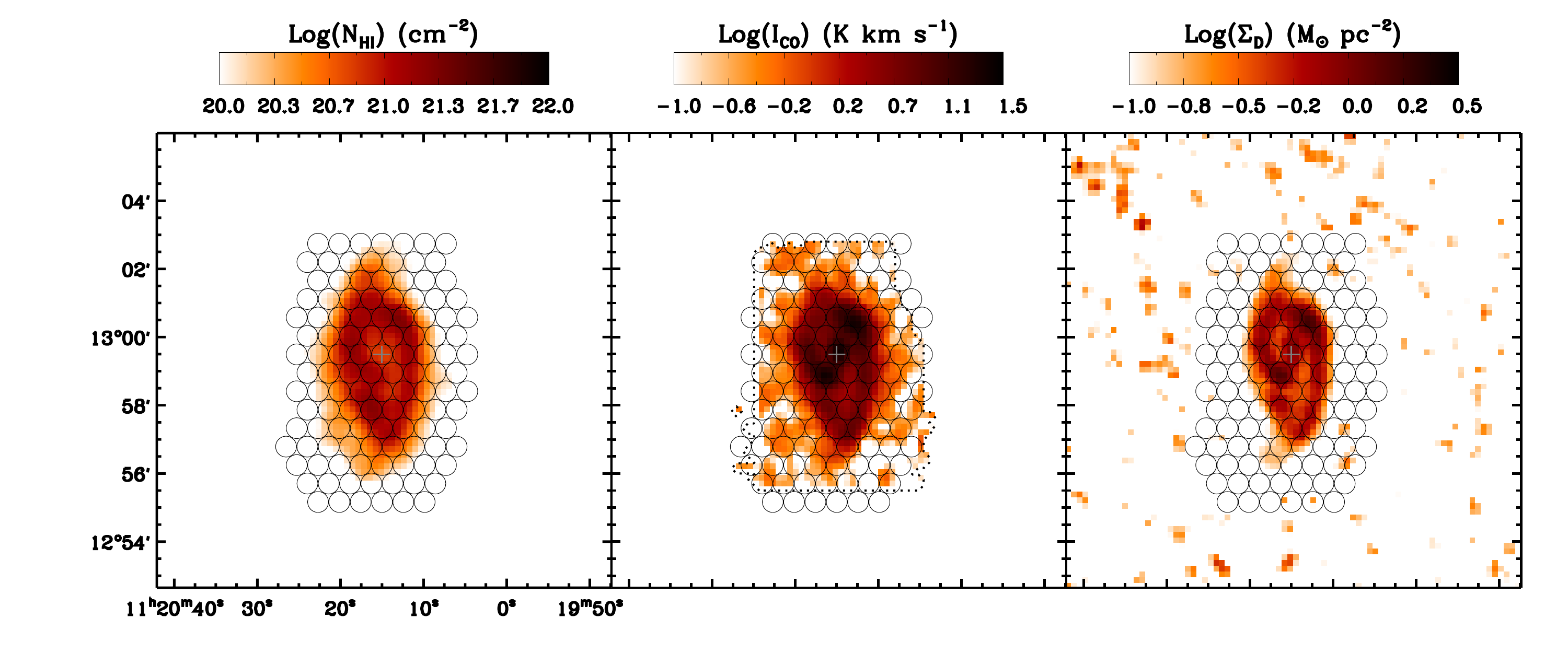}{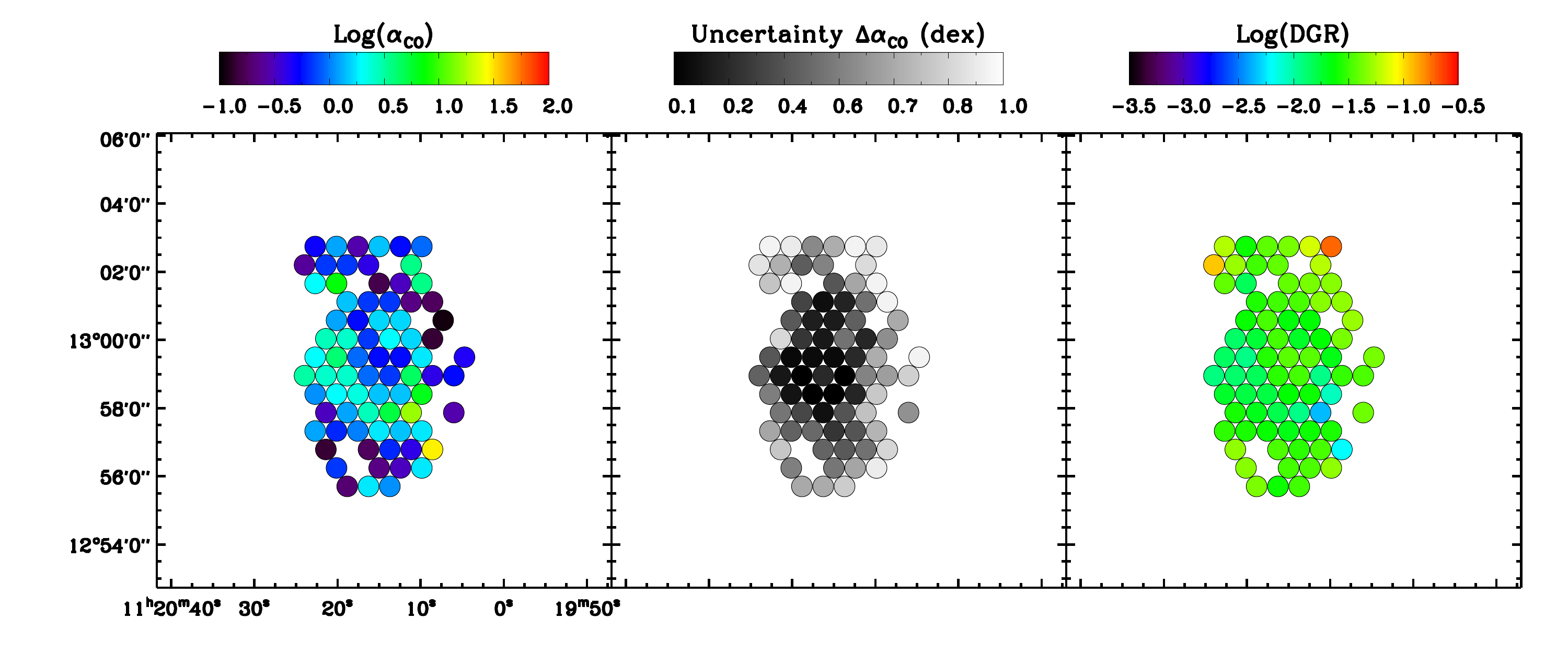}
\epsscale{1.0}
\plotone{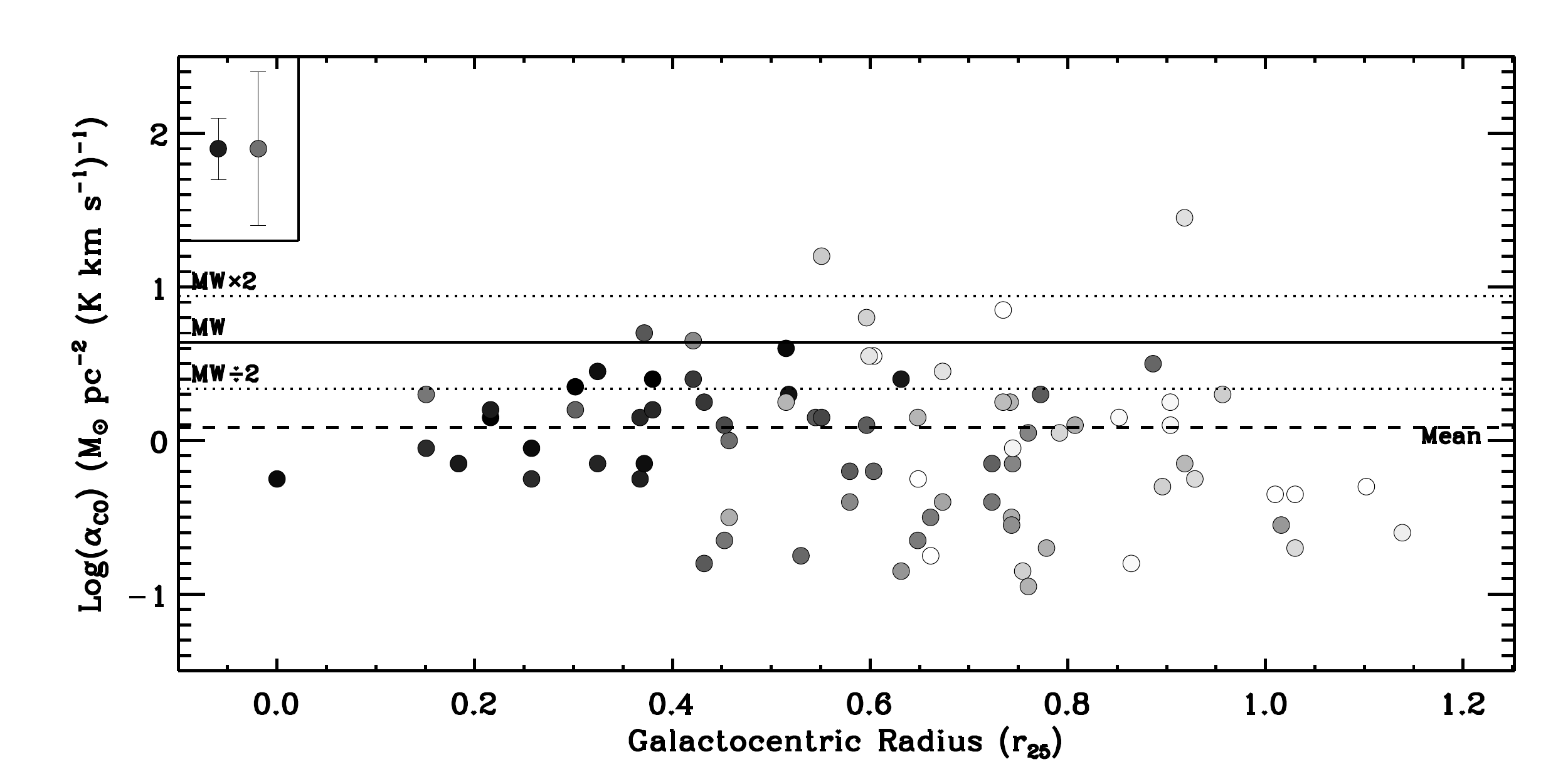}
\caption{Results for NGC 3627 (D = 9.4 Mpc; 1\arcsec = 45 pc).}
\label{fig:ngc3627_panel3}
\end{figure*}

\begin{figure*}
\centering
\epsscale{2.2}
\plottwo{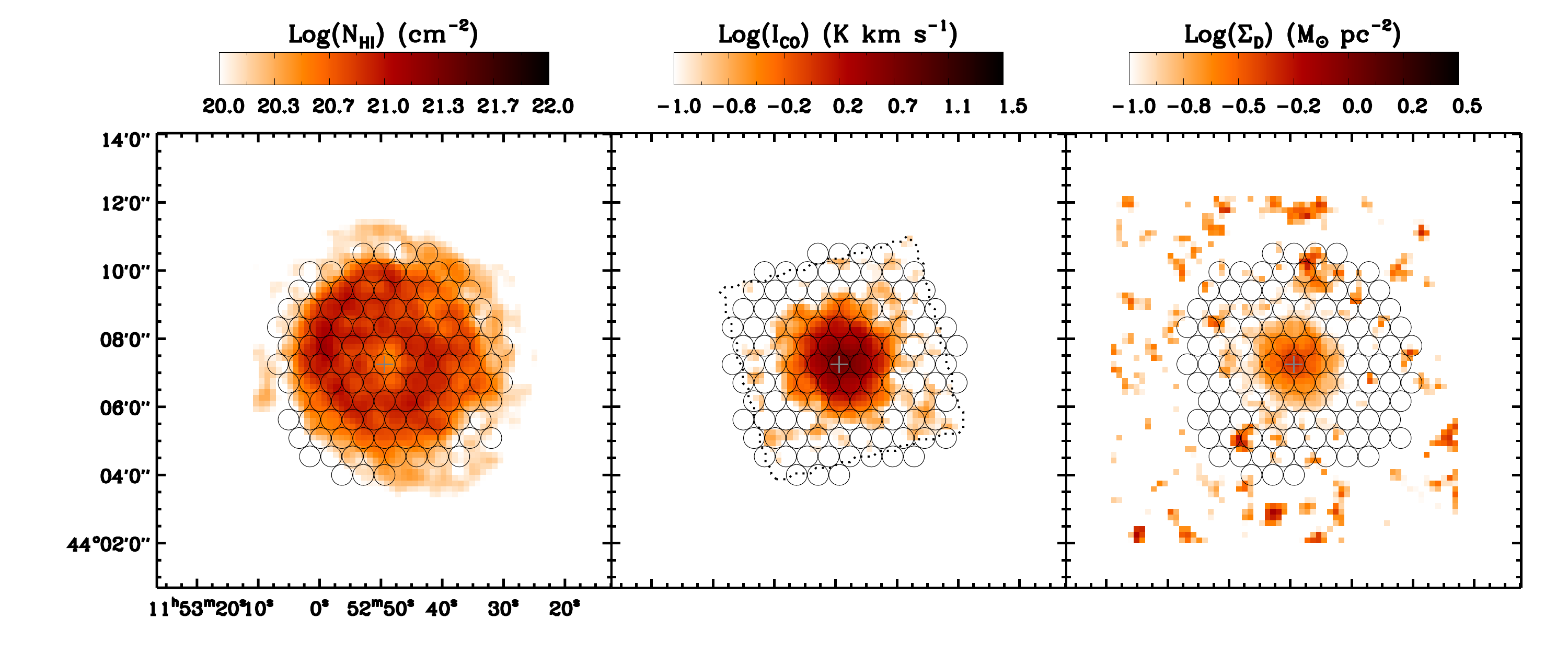}{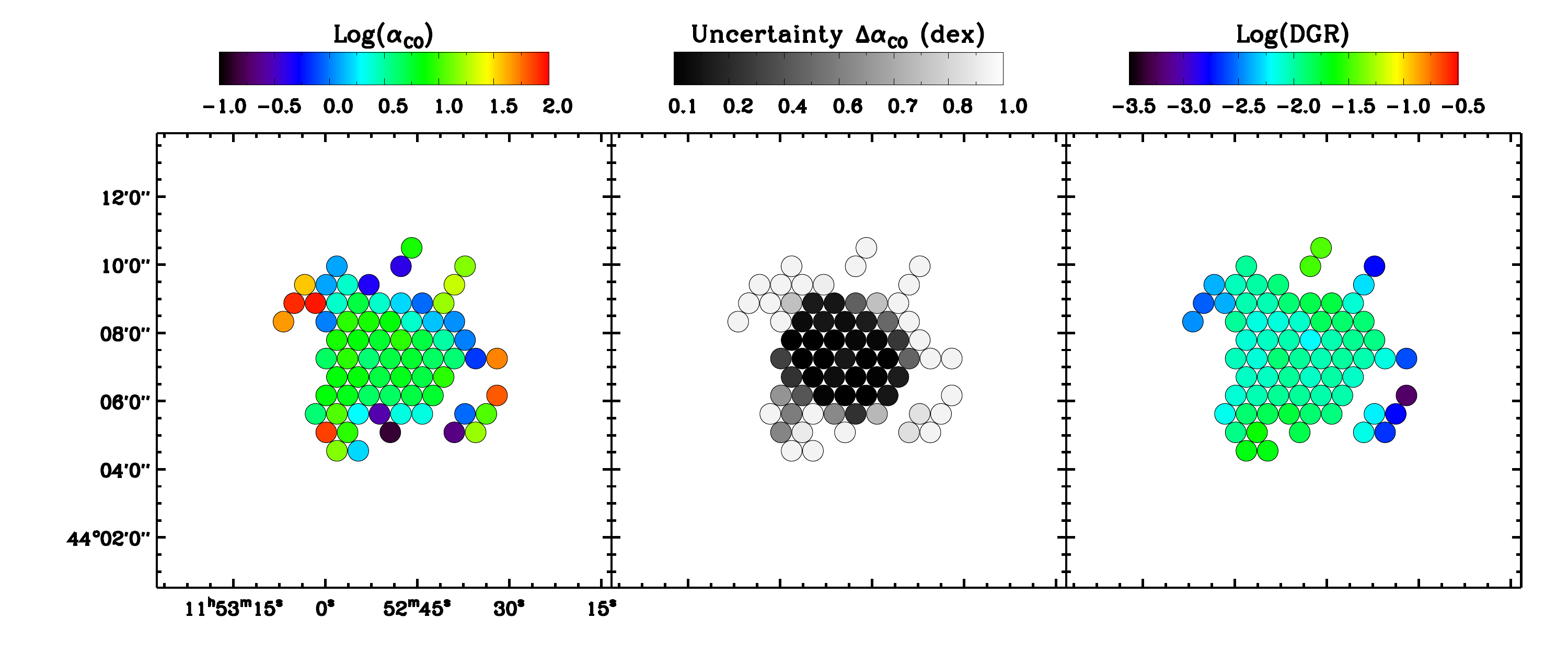}
\epsscale{1.0}
\plotone{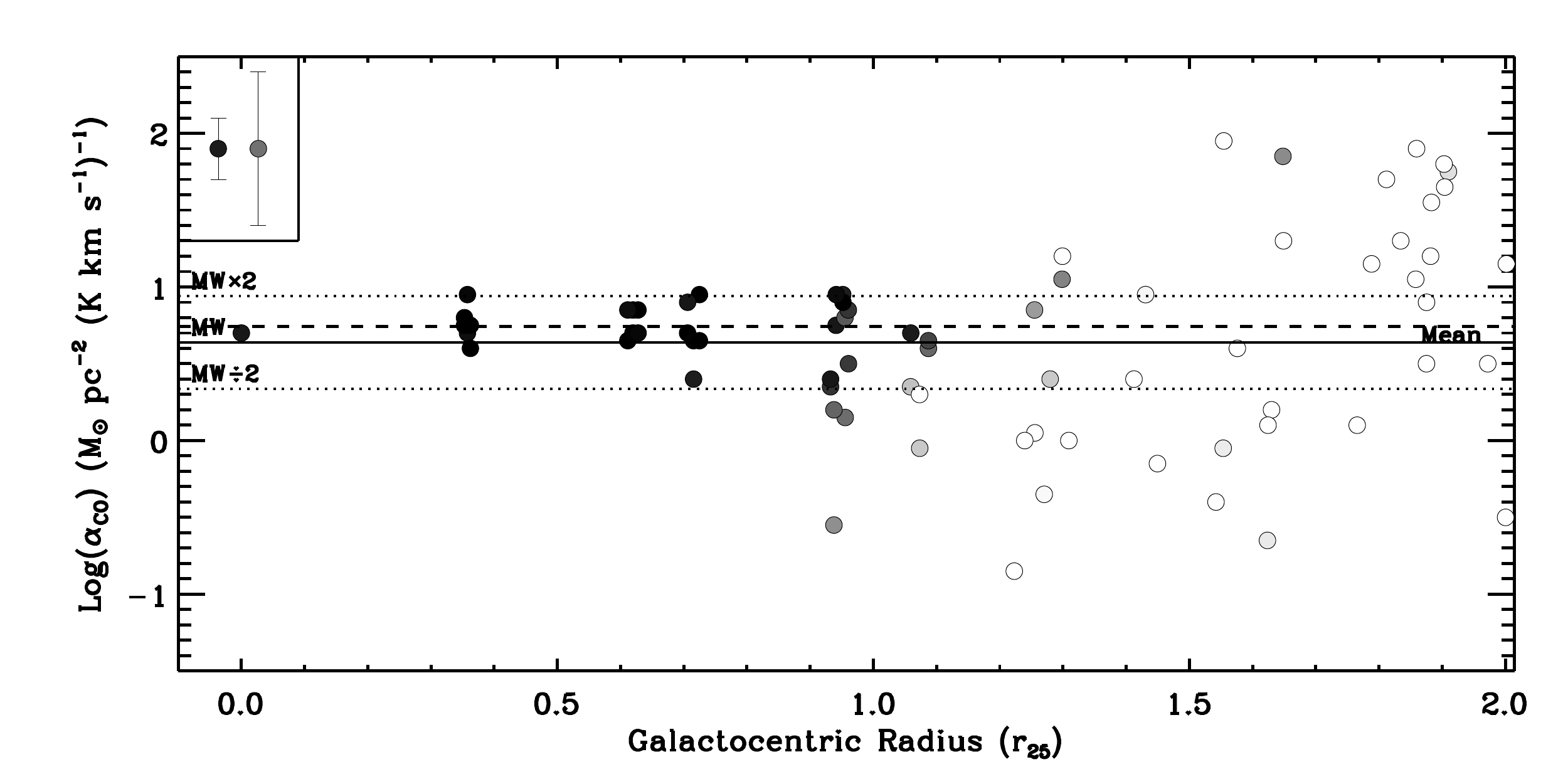}
\caption{Results for NGC 3938 (D = 17.9 Mpc; 1\arcsec = 87 pc).}
\label{fig:ngc3938_panel3}
\end{figure*}

\begin{figure*}
\centering
\epsscale{2.2}
\plottwo{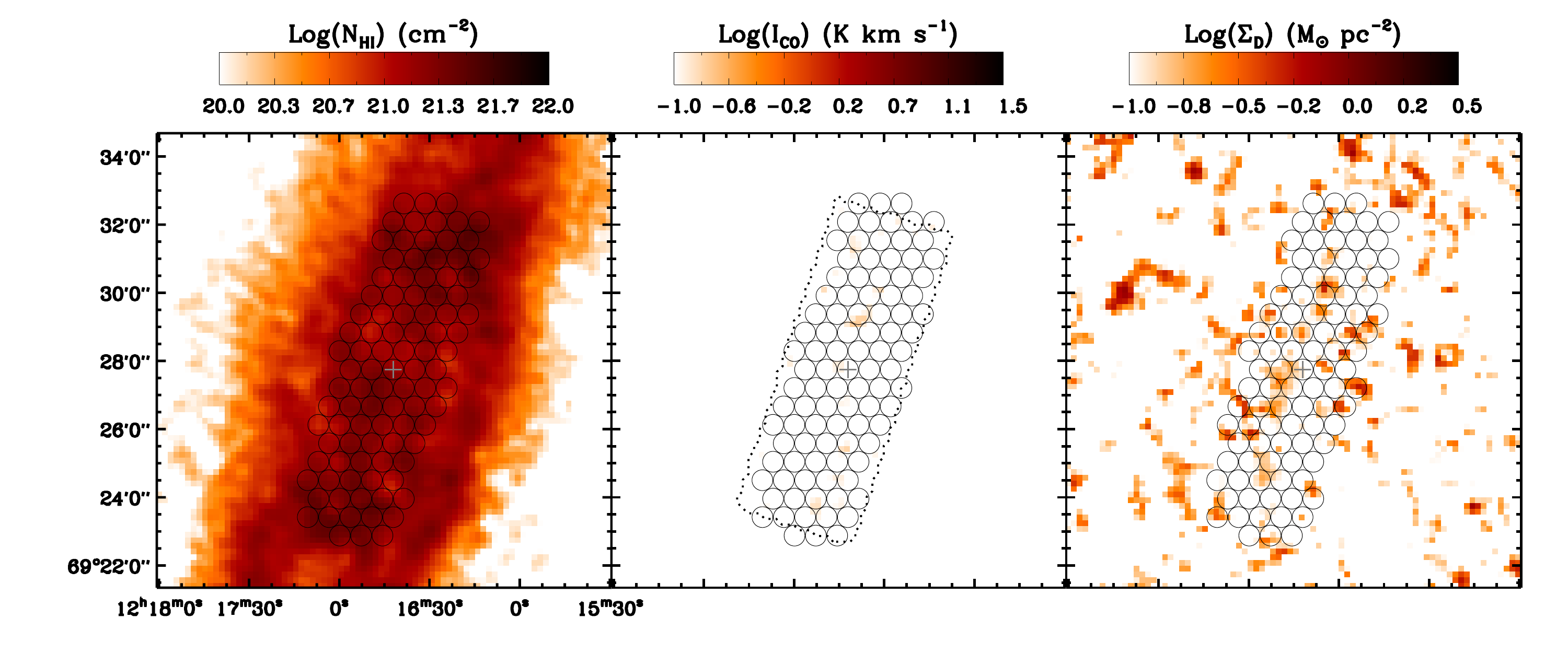}{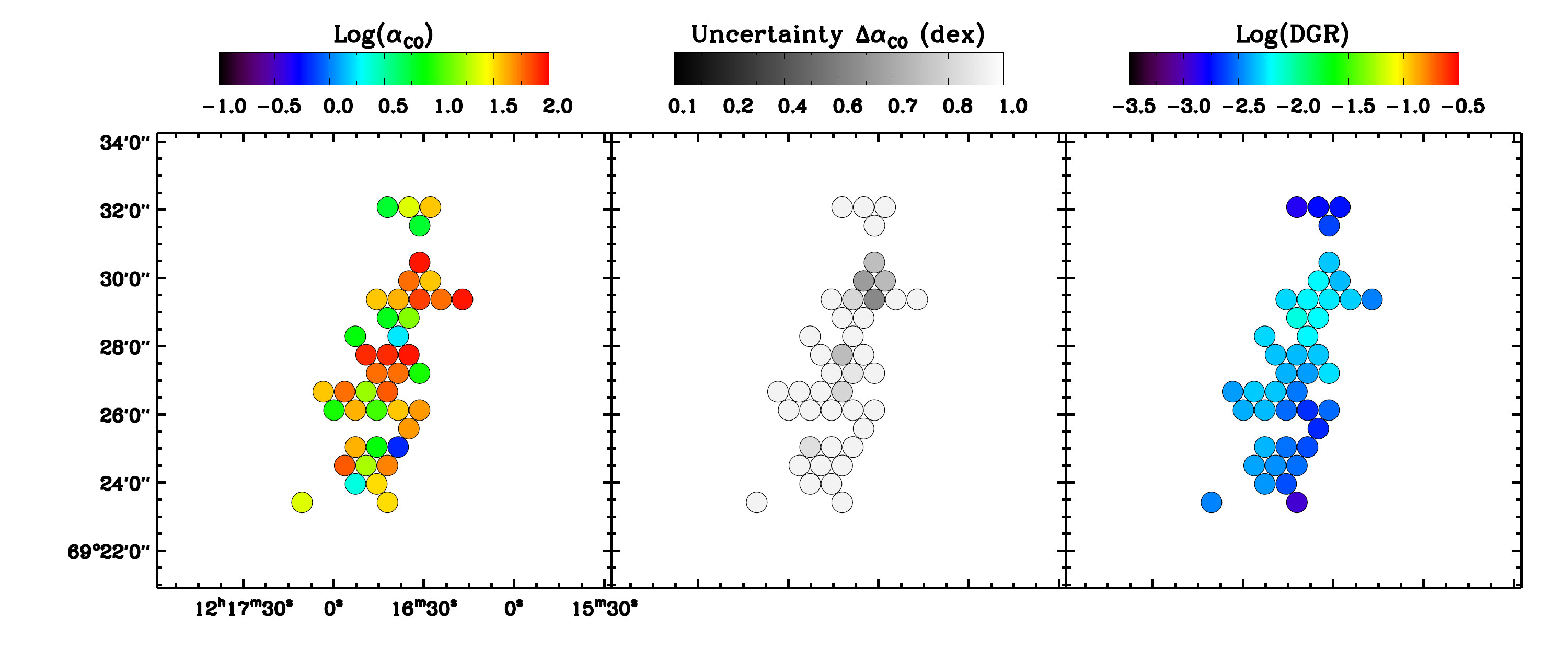}
\epsscale{1.0}
\plotone{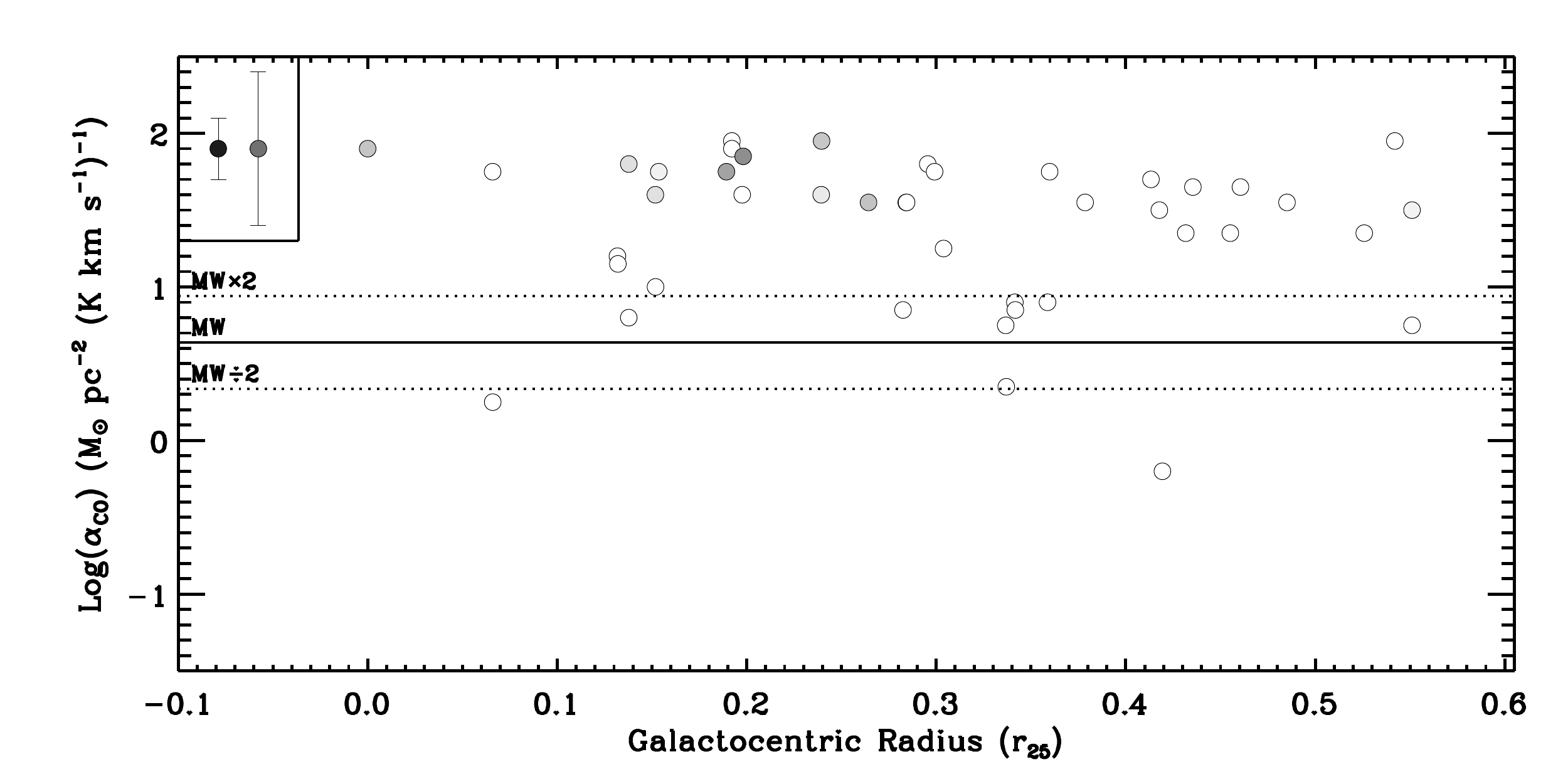}
\caption{Results for NGC 4236 (D = 4.5 Mpc; 1\arcsec = 22 pc).}
\label{fig:ngc4236_panel3}
\end{figure*}

\begin{figure*}
\centering
\epsscale{2.2}
\plottwo{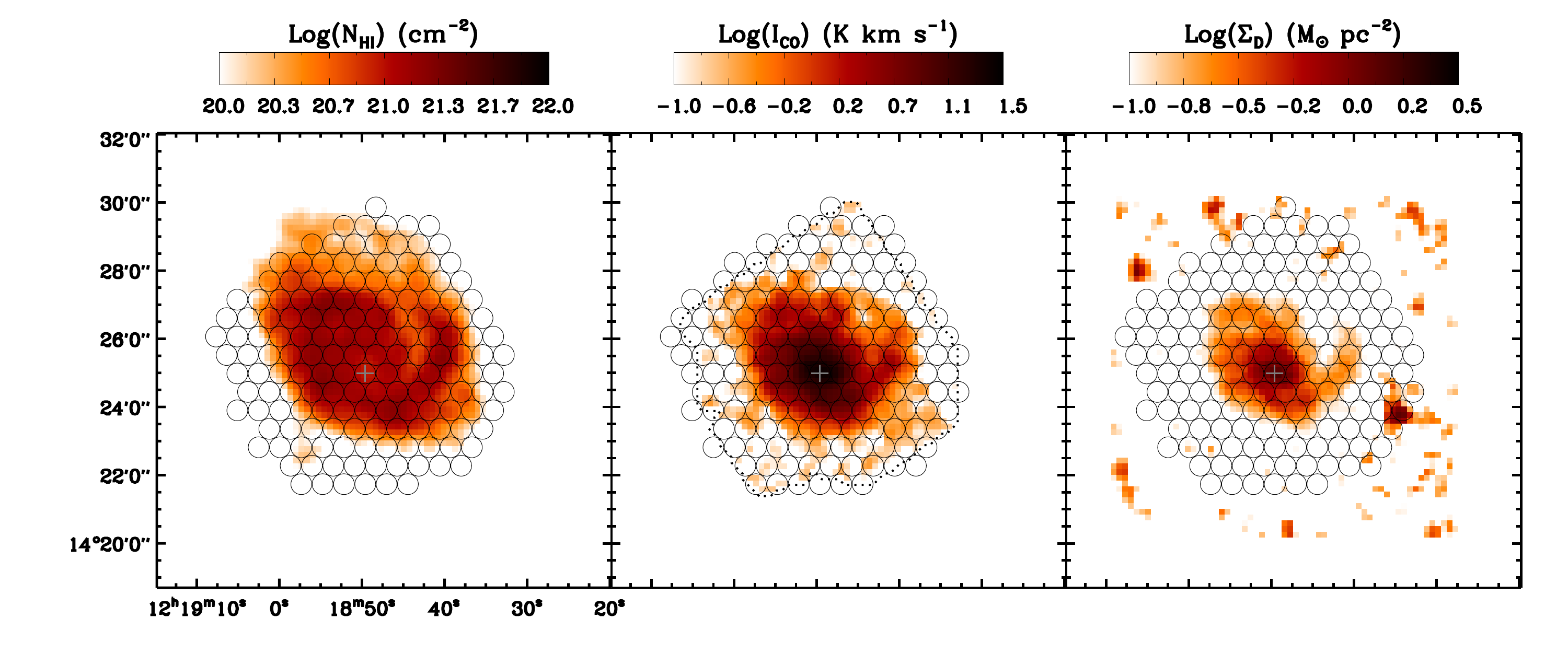}{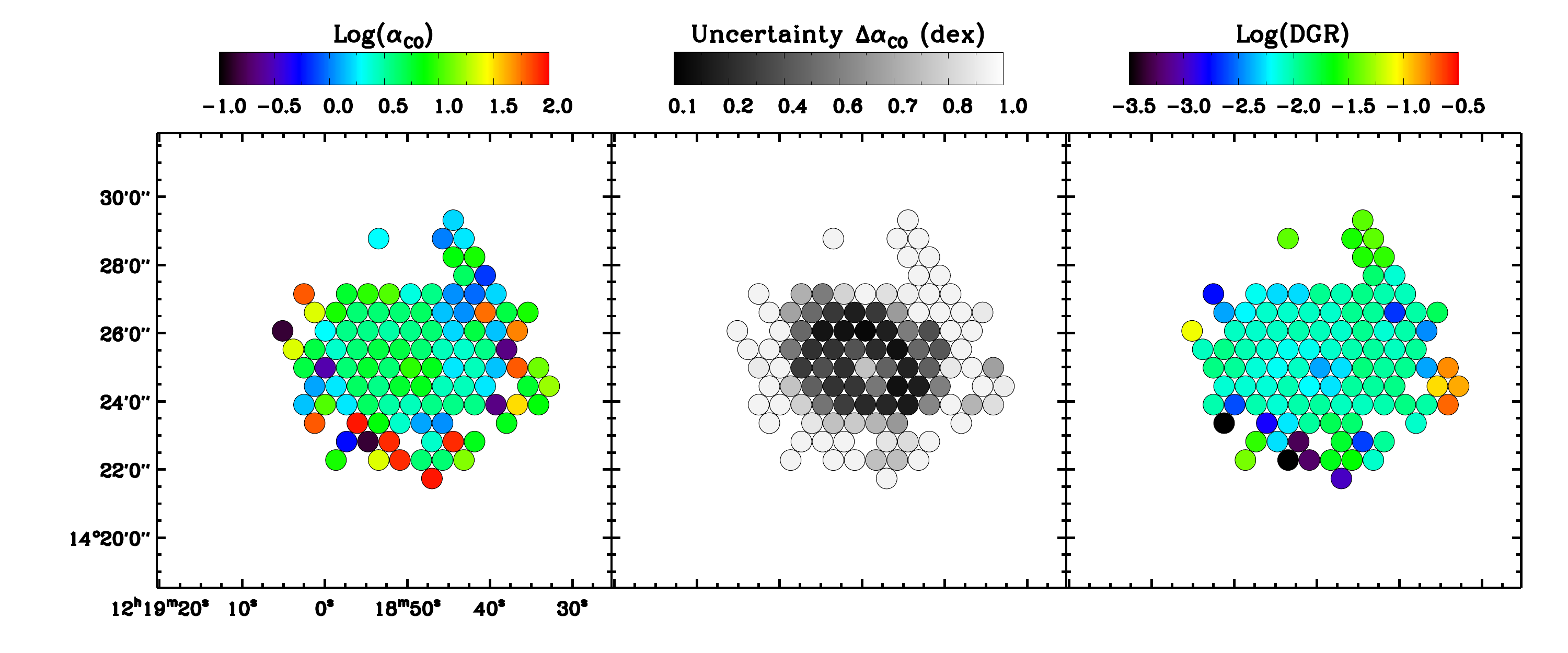}
\epsscale{1.0}
\plotone{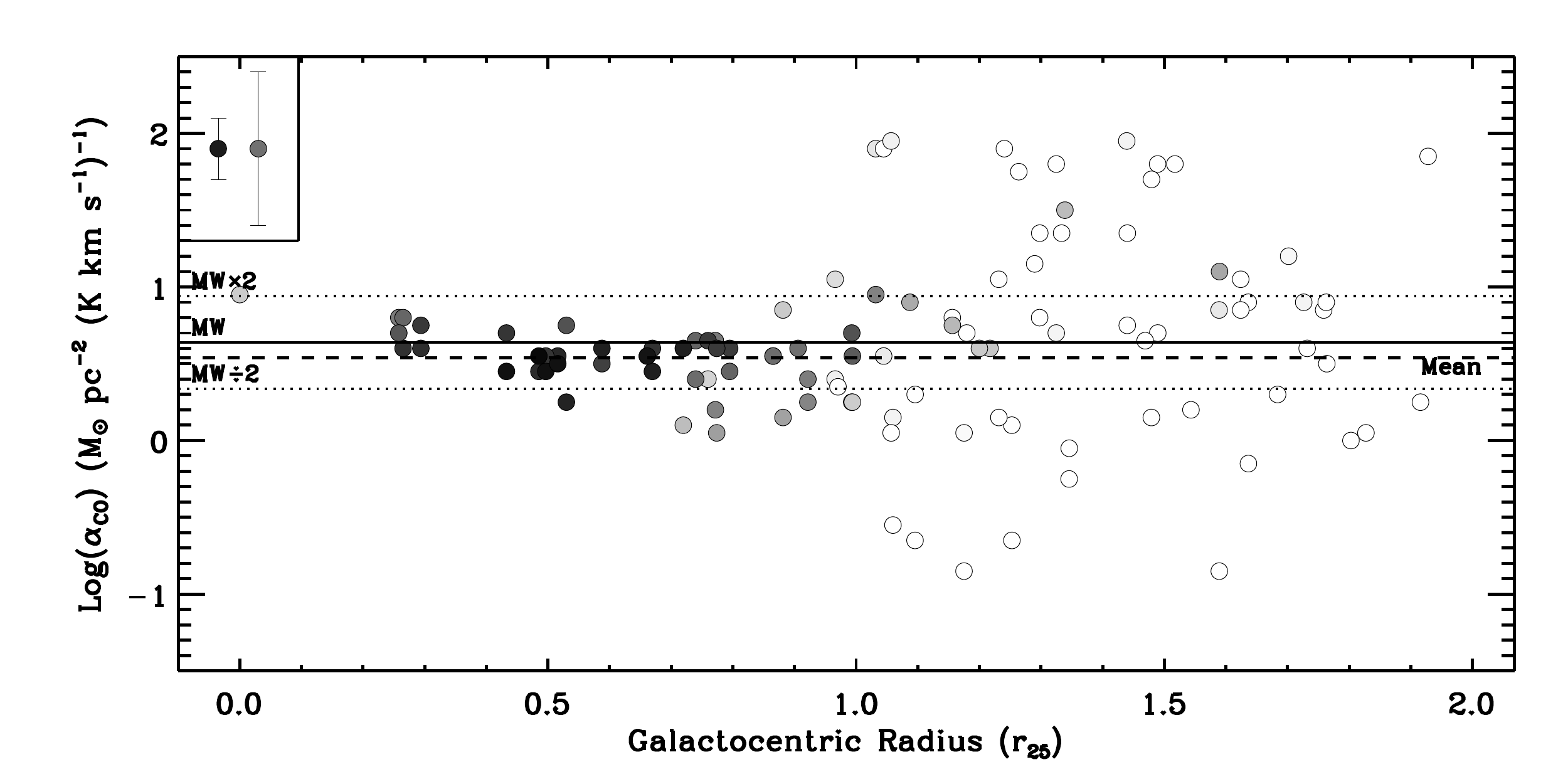}
\caption{Results for NGC 4254 (D = 14.4 Mpc; 1\arcsec = 70 pc).}
\label{fig:ngc4254_panel3}
\end{figure*}

\begin{figure*}
\centering
\epsscale{2.2}
\plottwo{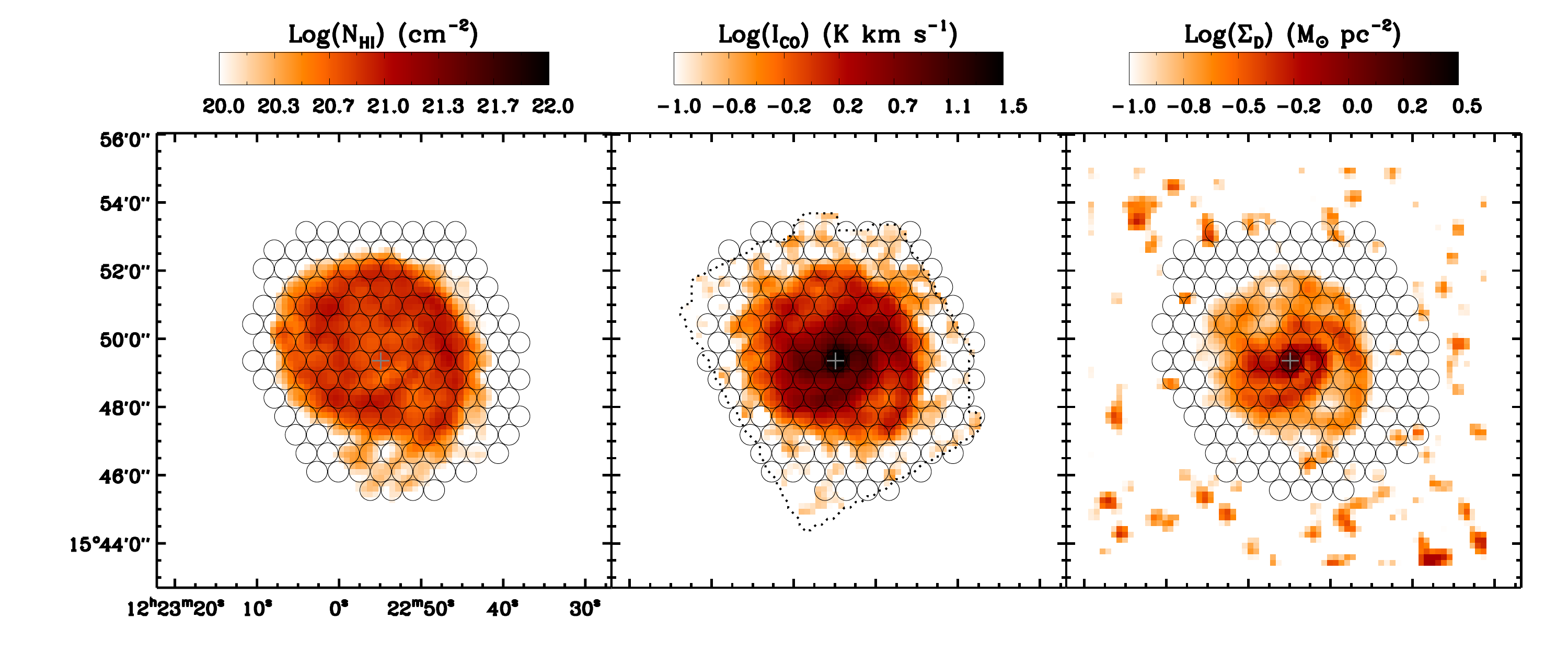}{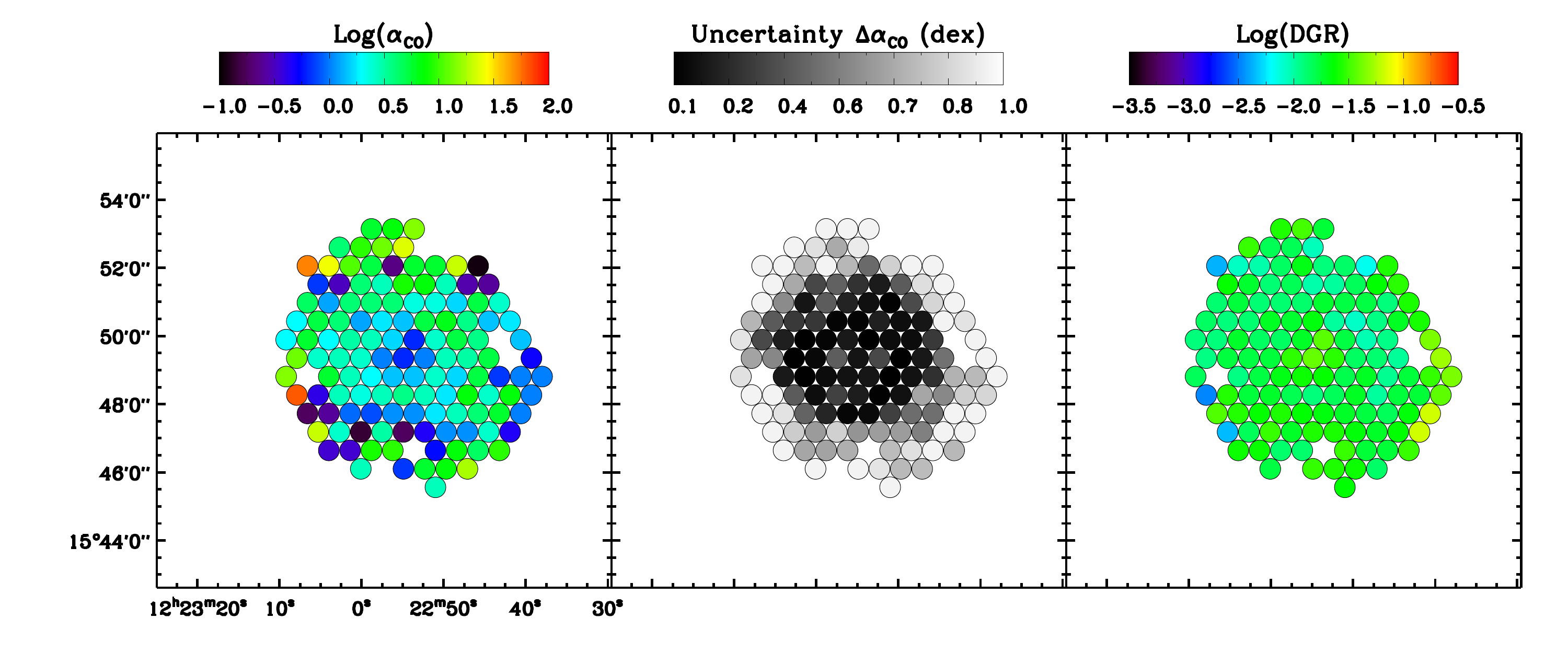}
\epsscale{1.0}
\plotone{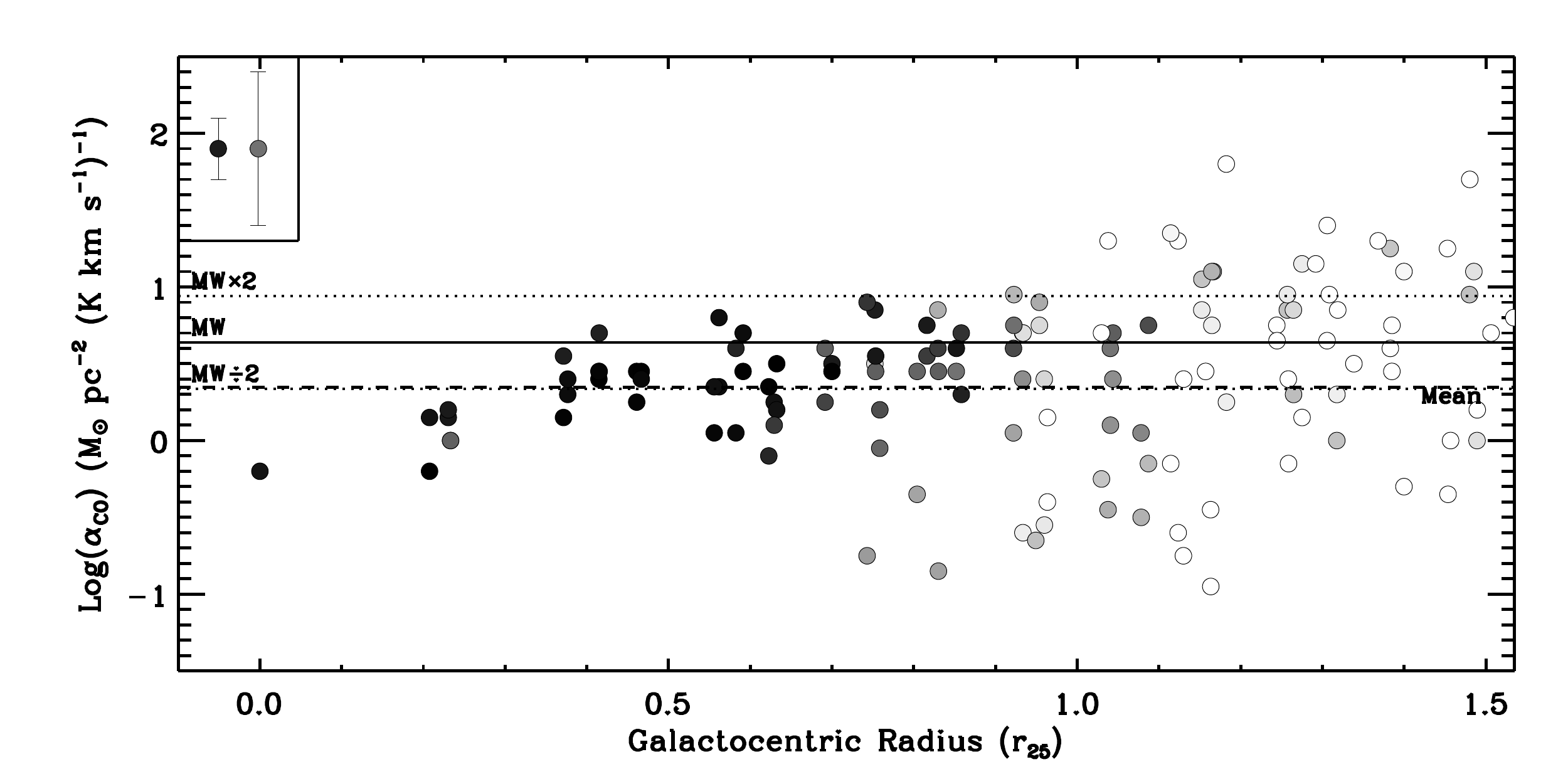}
\caption{Results for NGC 4321 (D = 14.3 Mpc; 1\arcsec = 69 pc).}
\label{fig:ngc4321_panel3}
\end{figure*}

\begin{figure*}
\centering
\epsscale{2.2}
\plottwo{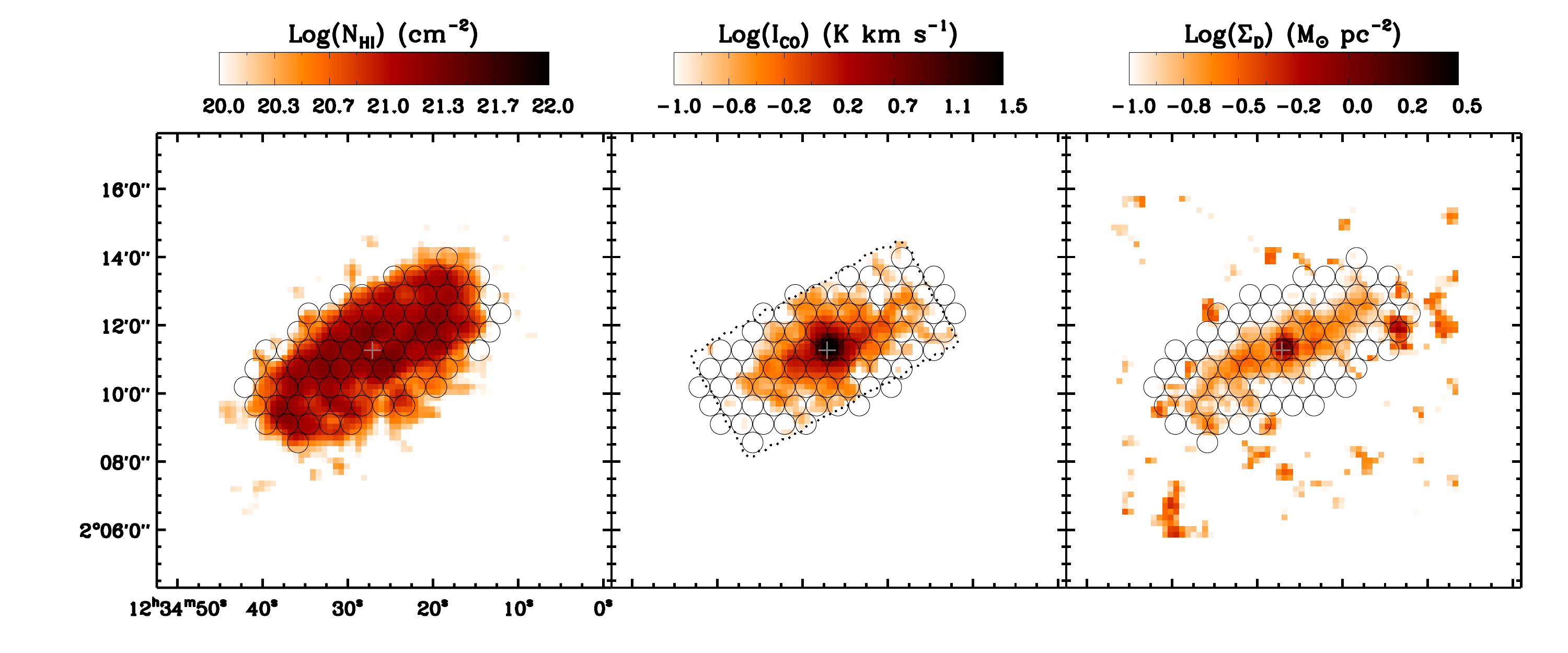}{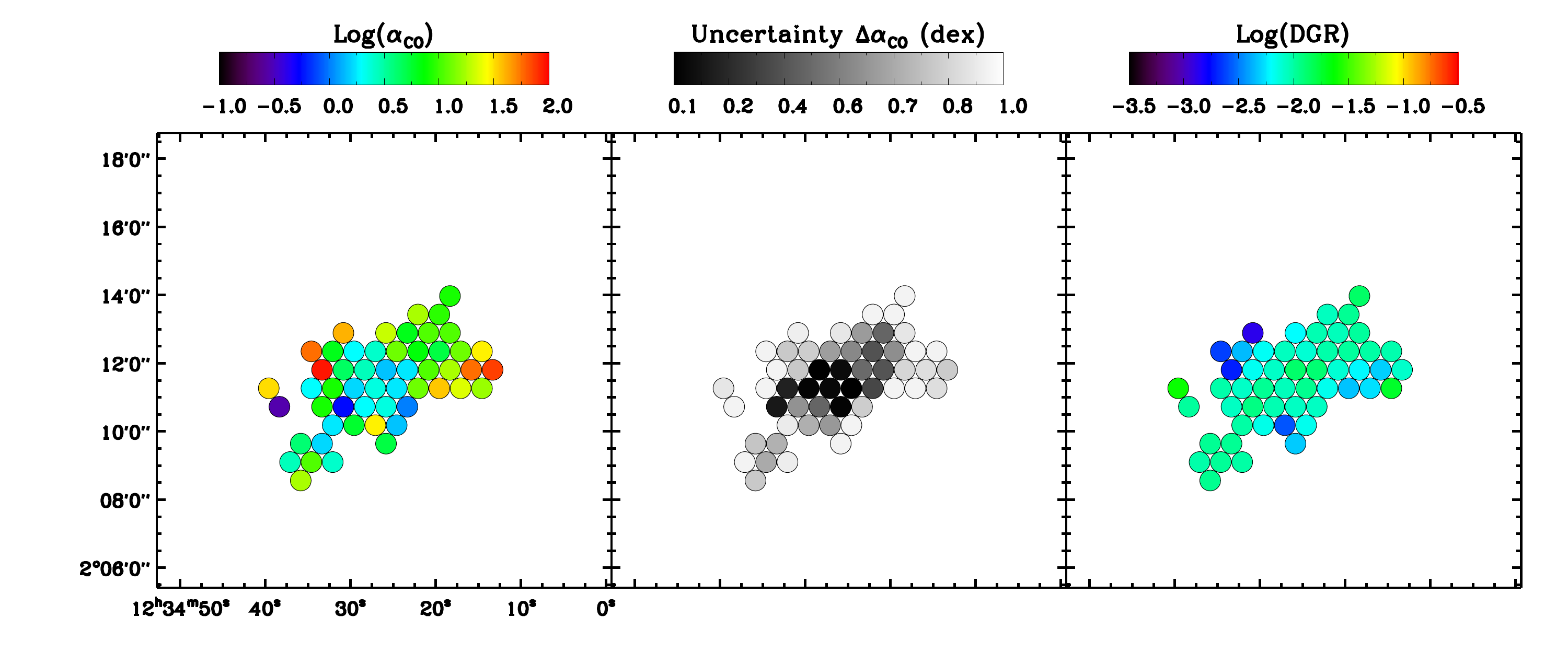}
\epsscale{1.0}
\plotone{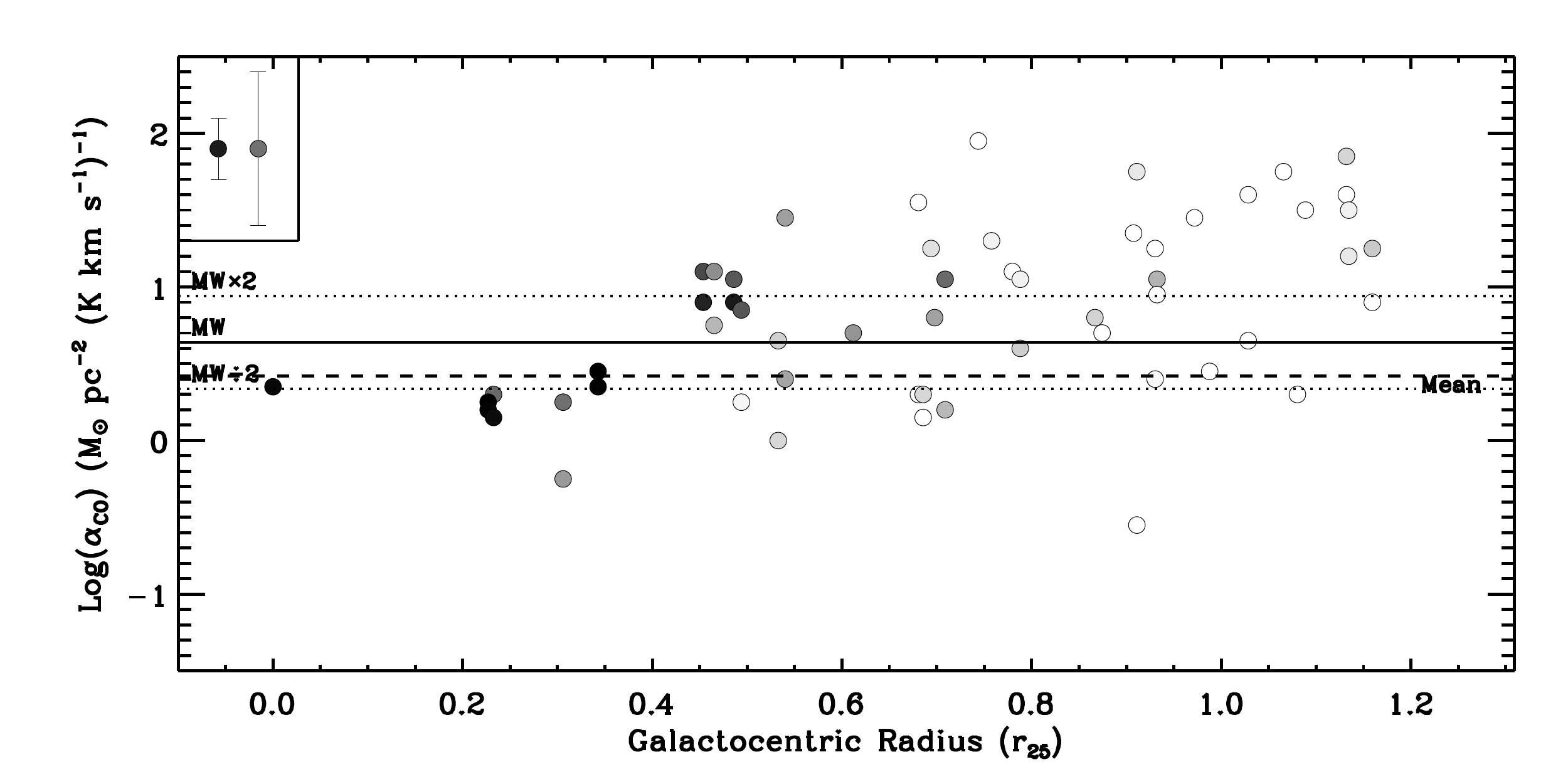}
\caption{Results for NGC 4536 (D = 14.5 Mpc; 1\arcsec = 70 pc).}
\label{fig:ngc4536_panel3}
\end{figure*}

\begin{figure*}
\centering
\epsscale{2.2}
\plottwo{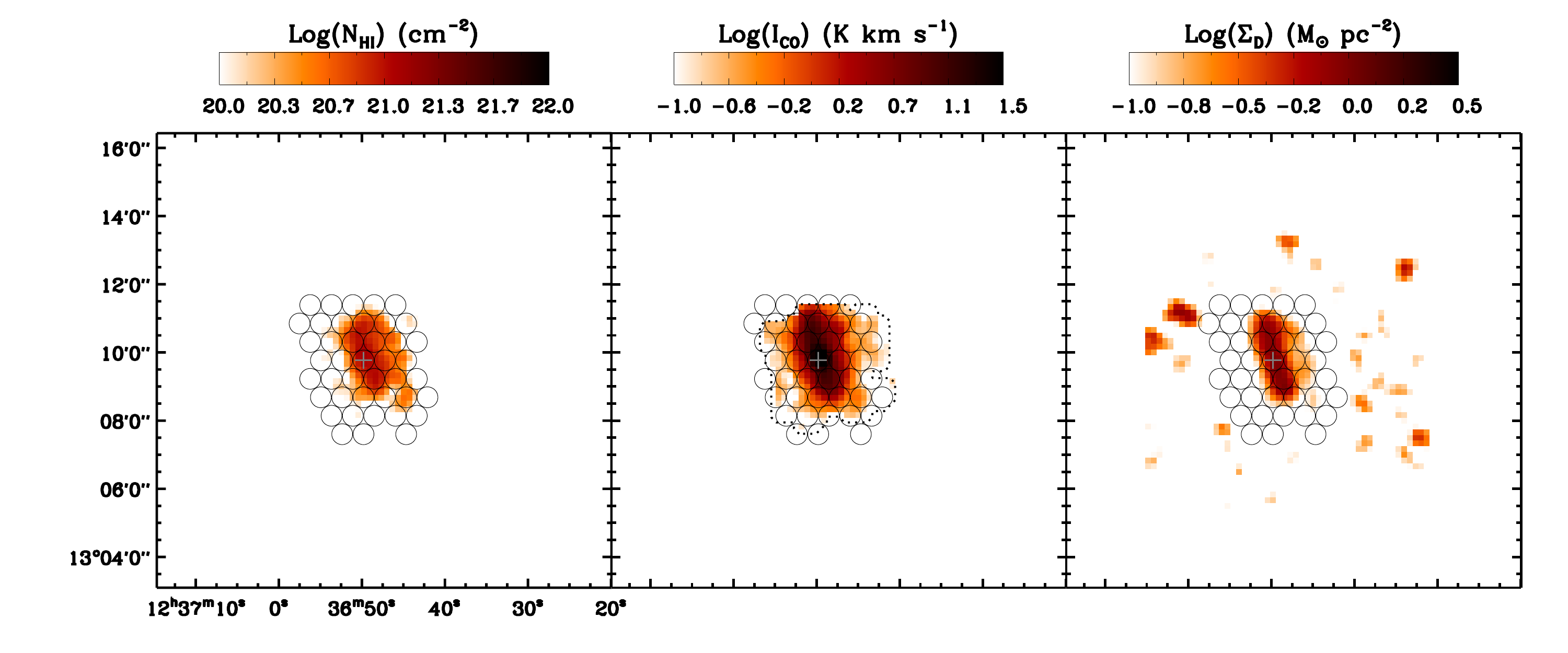}{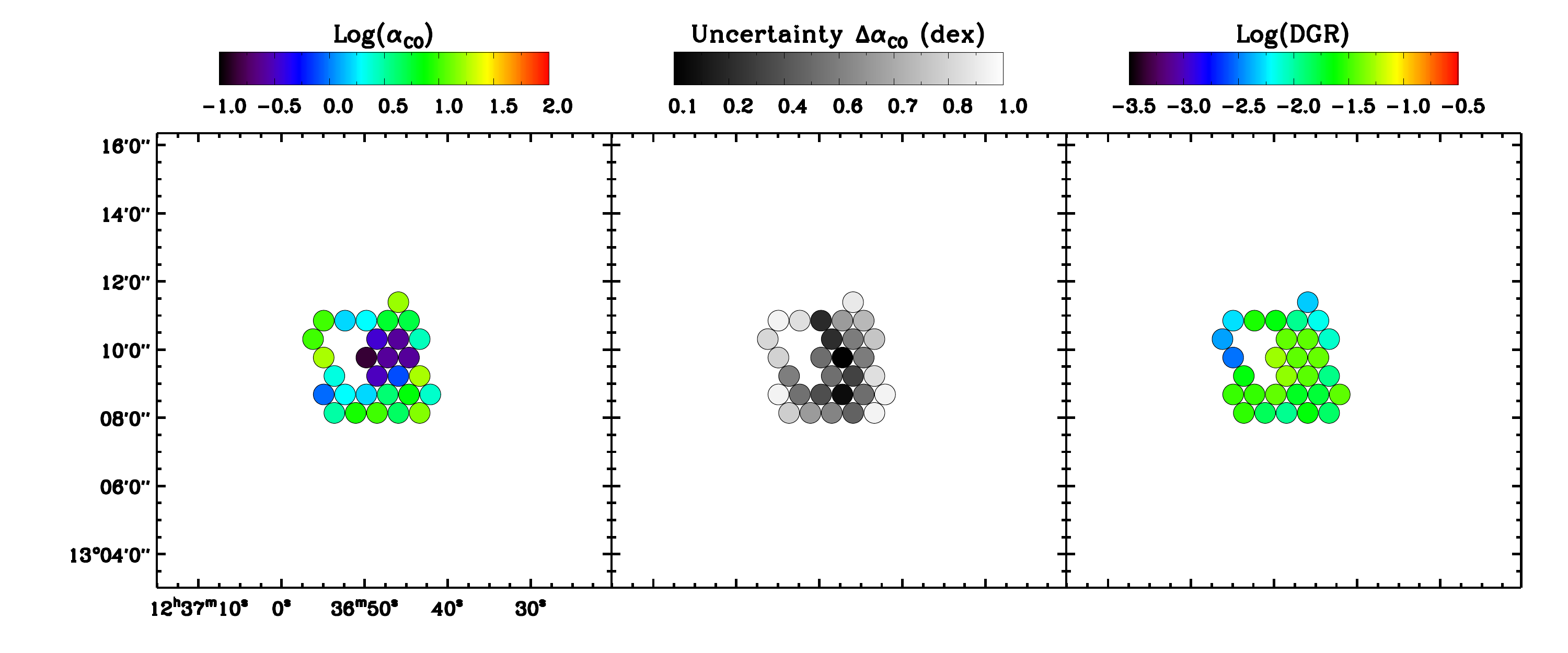}
\epsscale{1.0}
\plotone{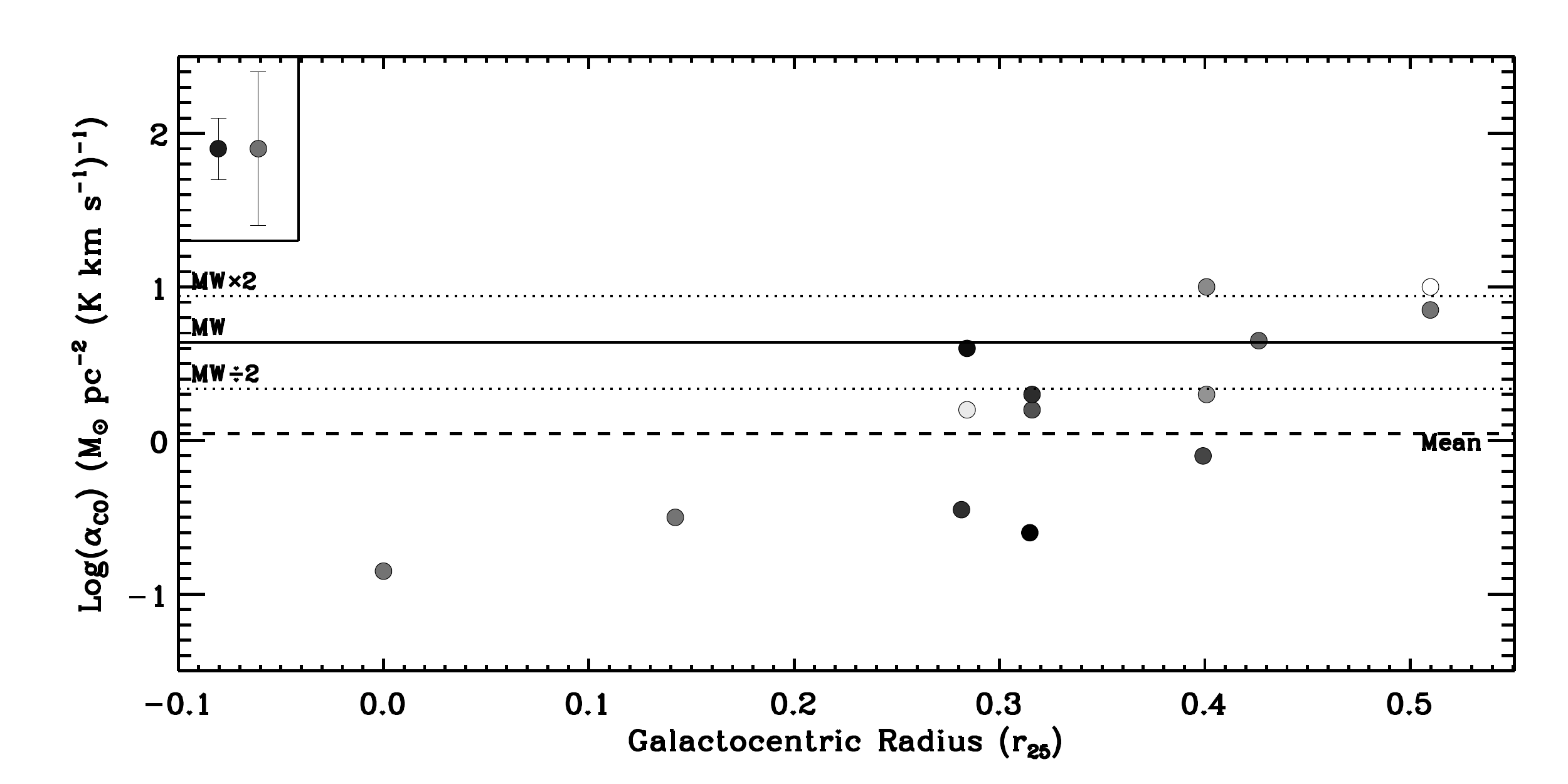}
\caption{Results for NGC 4569 (D = 9.9 Mpc; 1\arcsec = 48 pc).}
\label{fig:ngc4569_panel3}
\end{figure*}

\begin{figure*}
\centering
\epsscale{2.2}
\plottwo{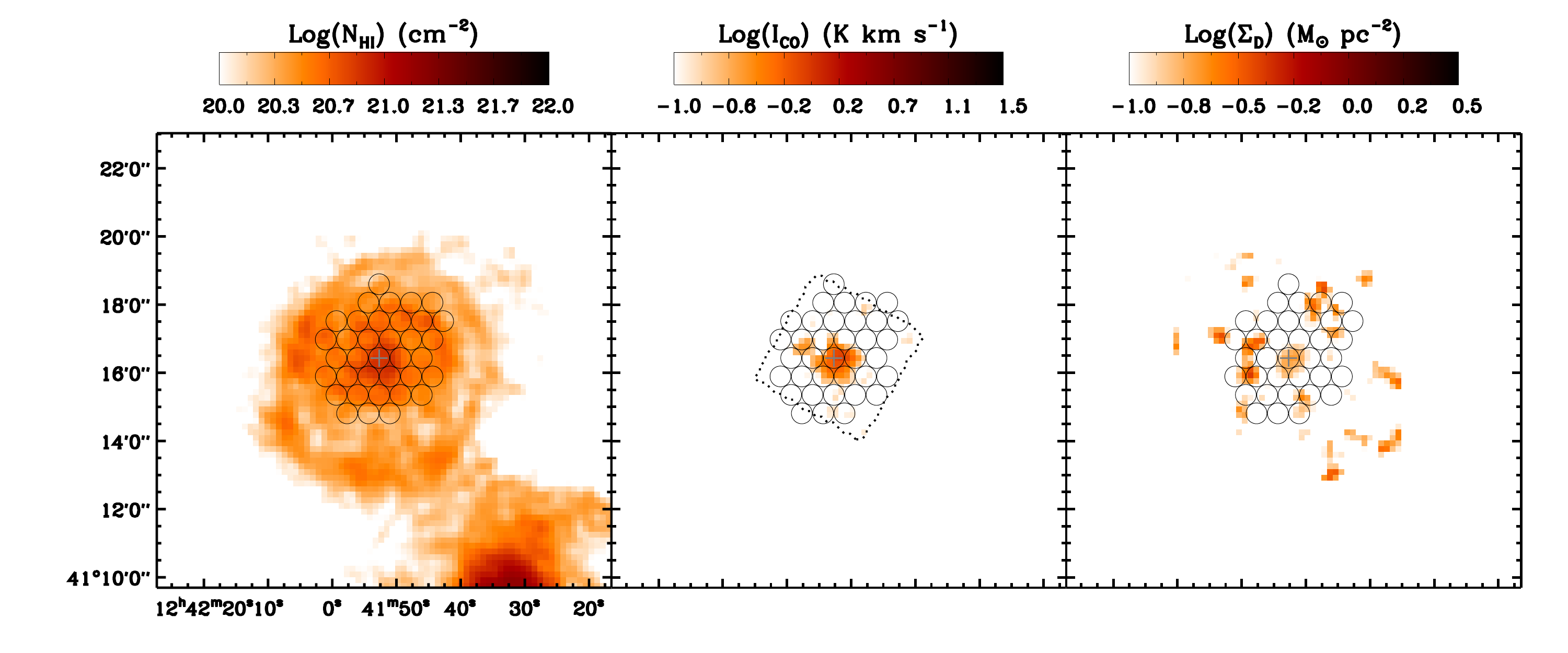}{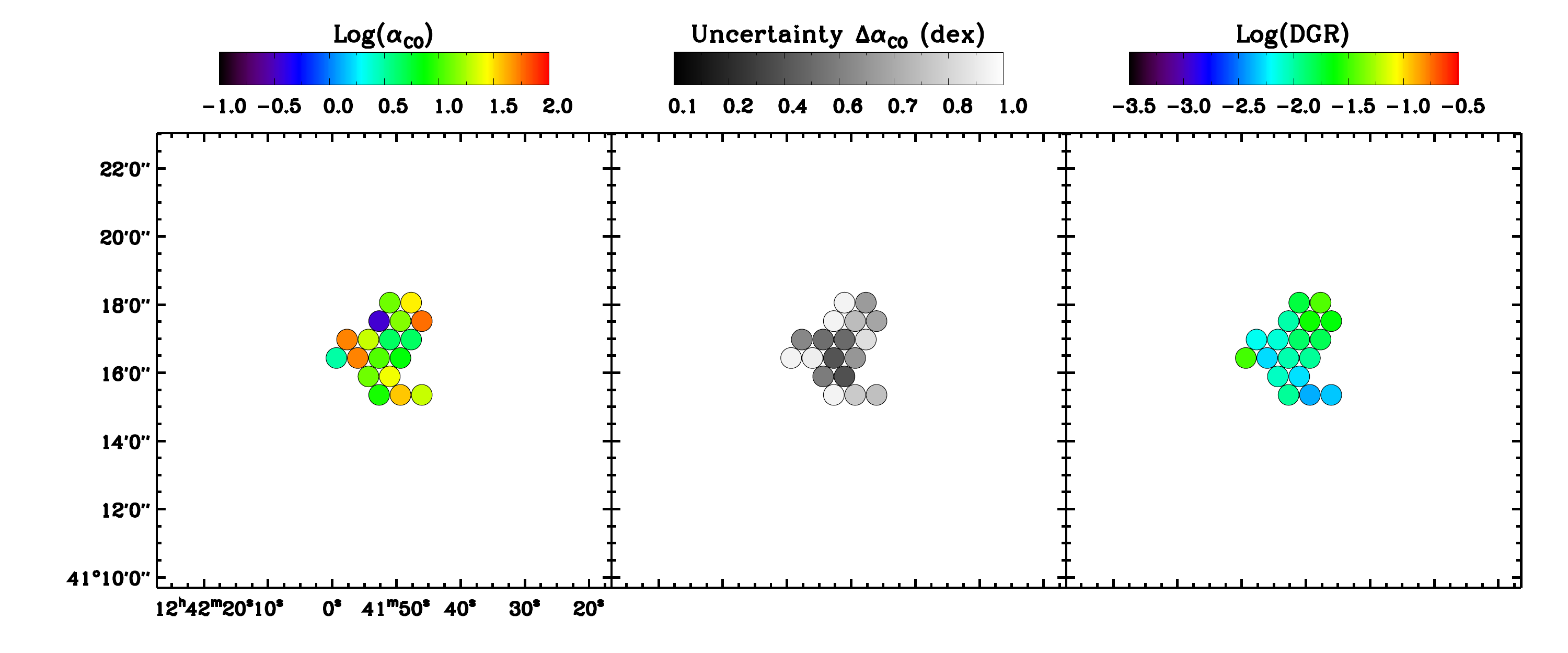}
\epsscale{1.0}
\plotone{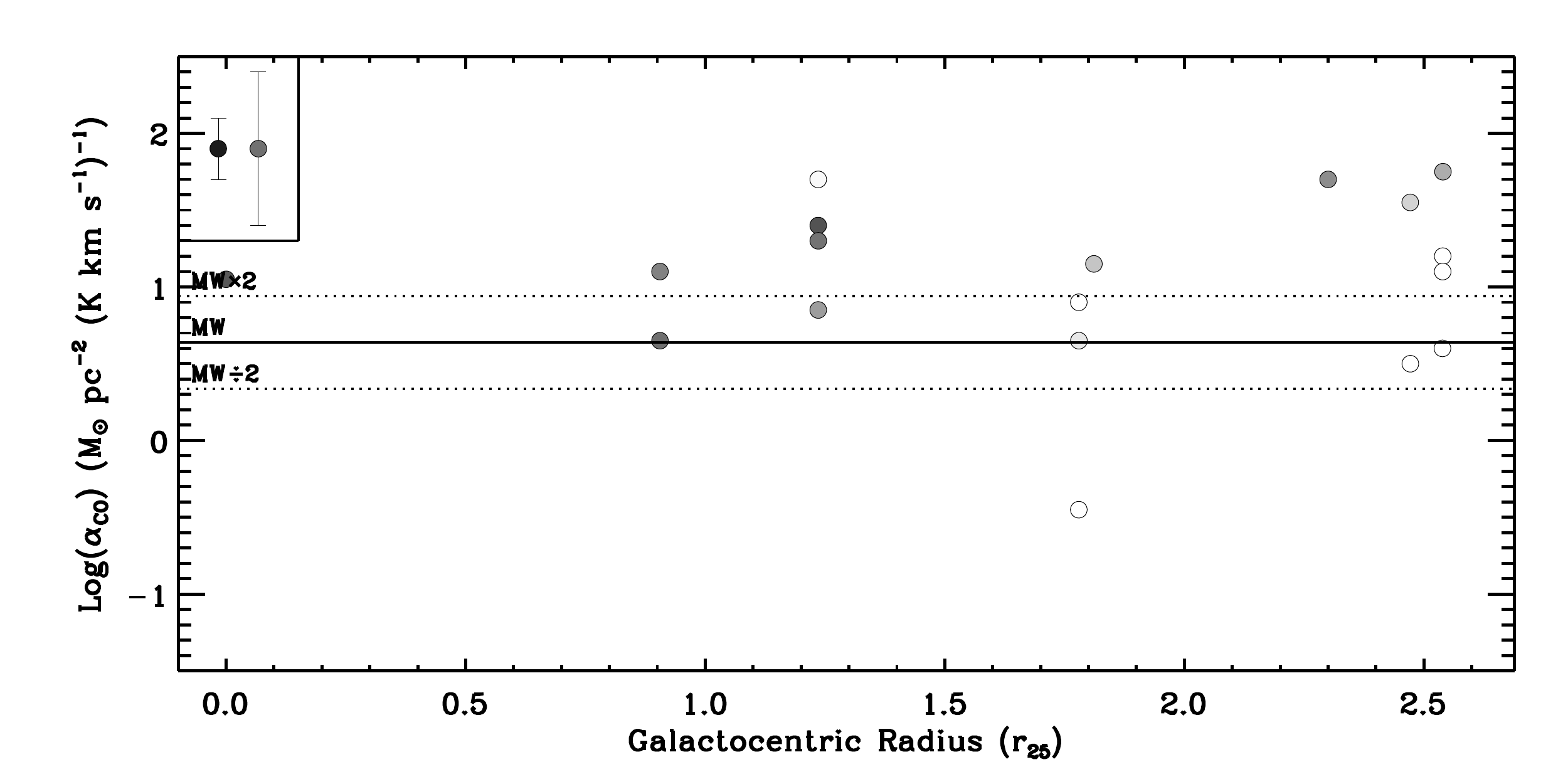}
\caption{Results for NGC 4625 (D = 9.3 Mpc; 1\arcsec = 45 pc).}
\label{fig:ngc4625_panel3}
\end{figure*}

\begin{figure*}
\centering
\epsscale{2.2}
\plottwo{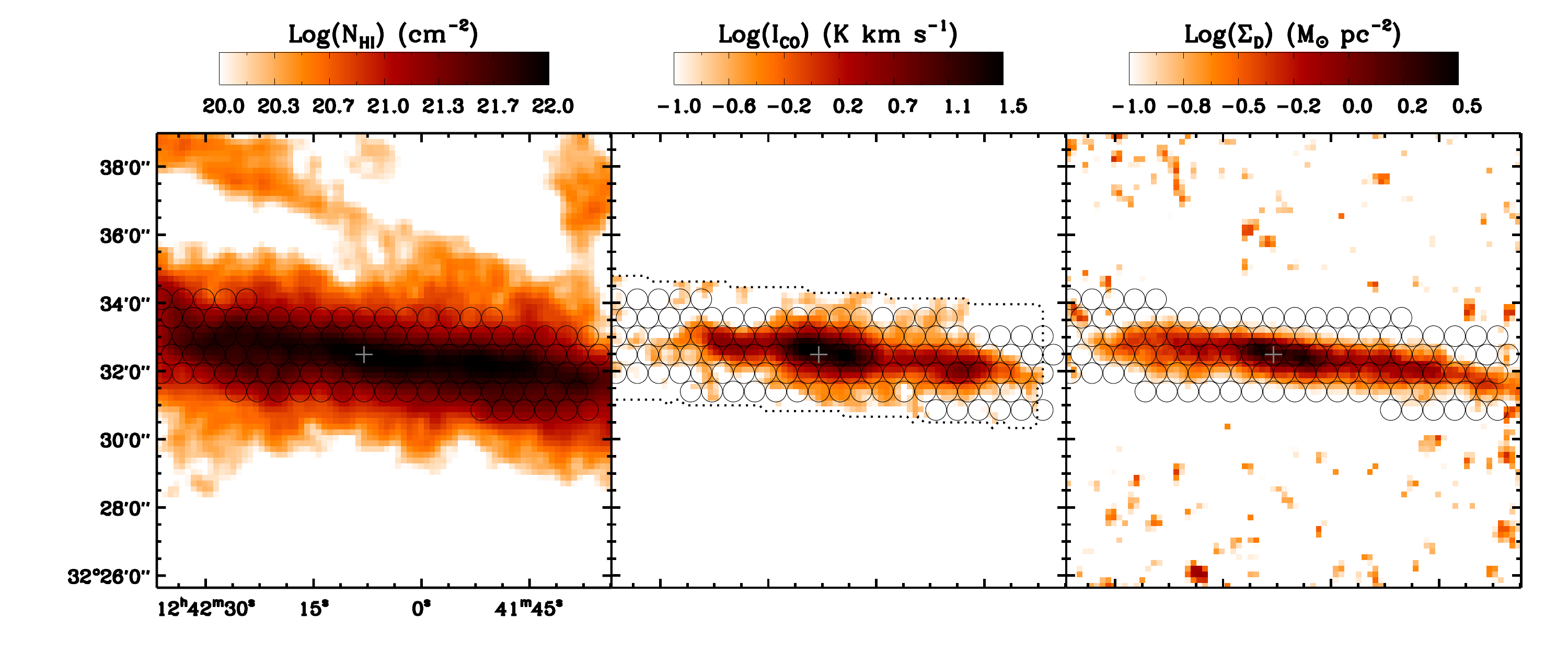}{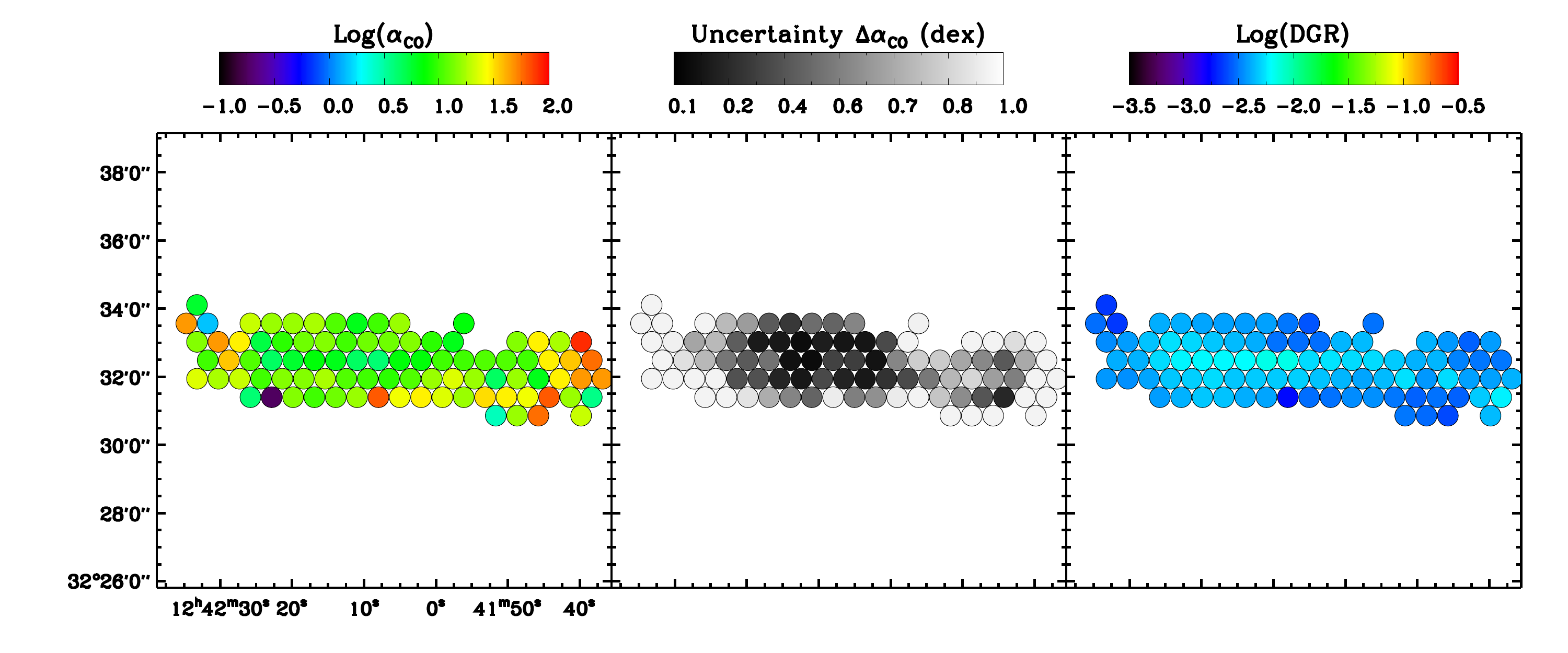}
\epsscale{1.0}
\plotone{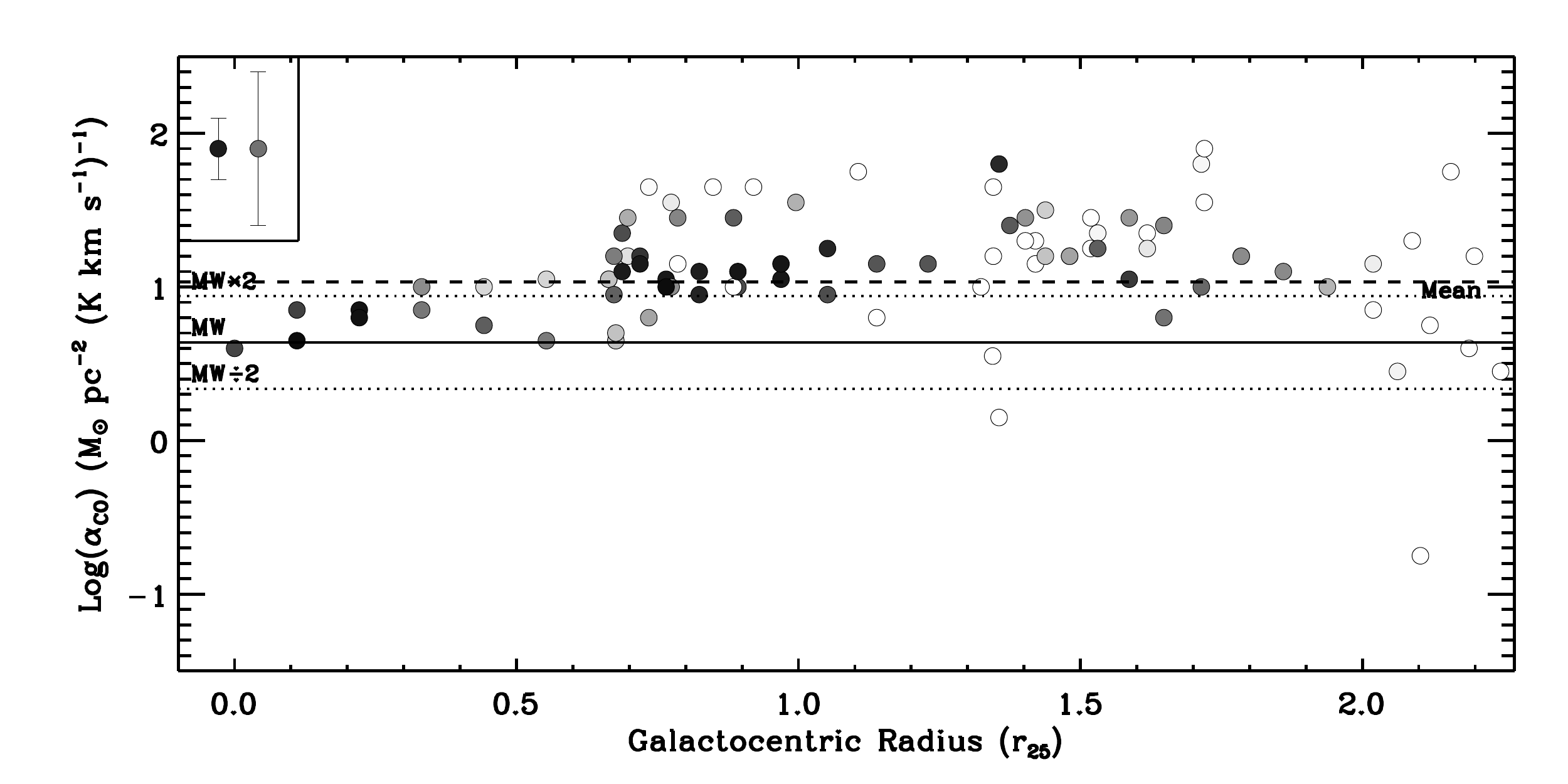}
\caption{Results for NGC 4631 (D = 7.6 Mpc; 1\arcsec = 37 pc).}
\label{fig:ngc4631_panel3}
\end{figure*}

\begin{figure*}
\centering
\epsscale{2.2}
\plottwo{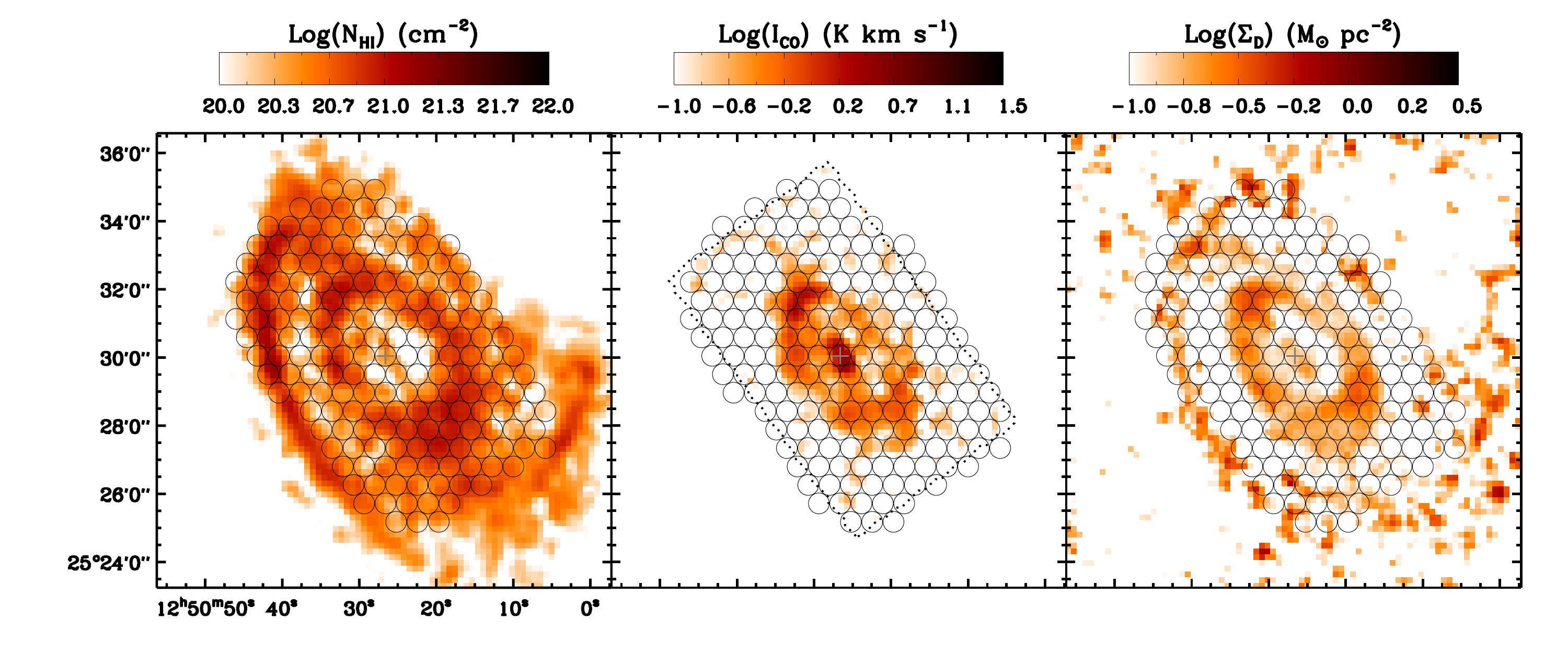}{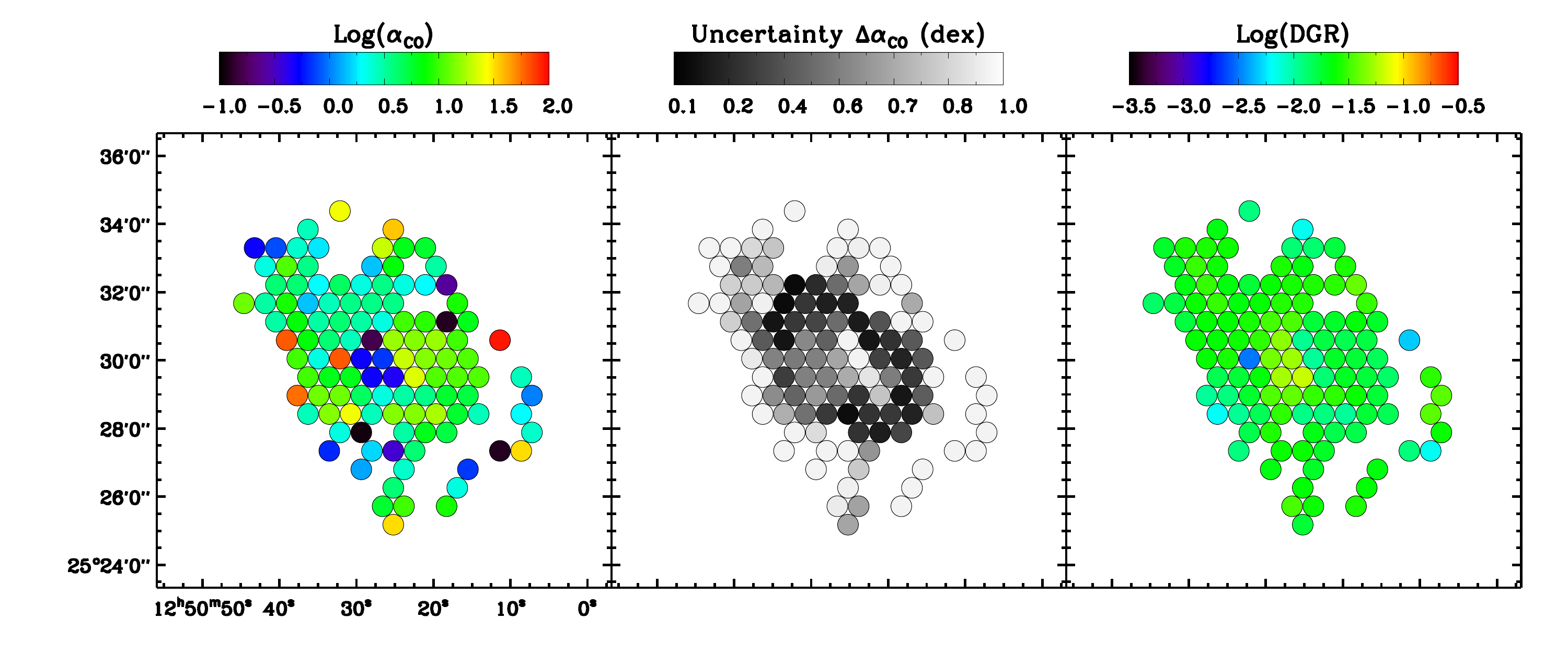}
\epsscale{1.0}
\plotone{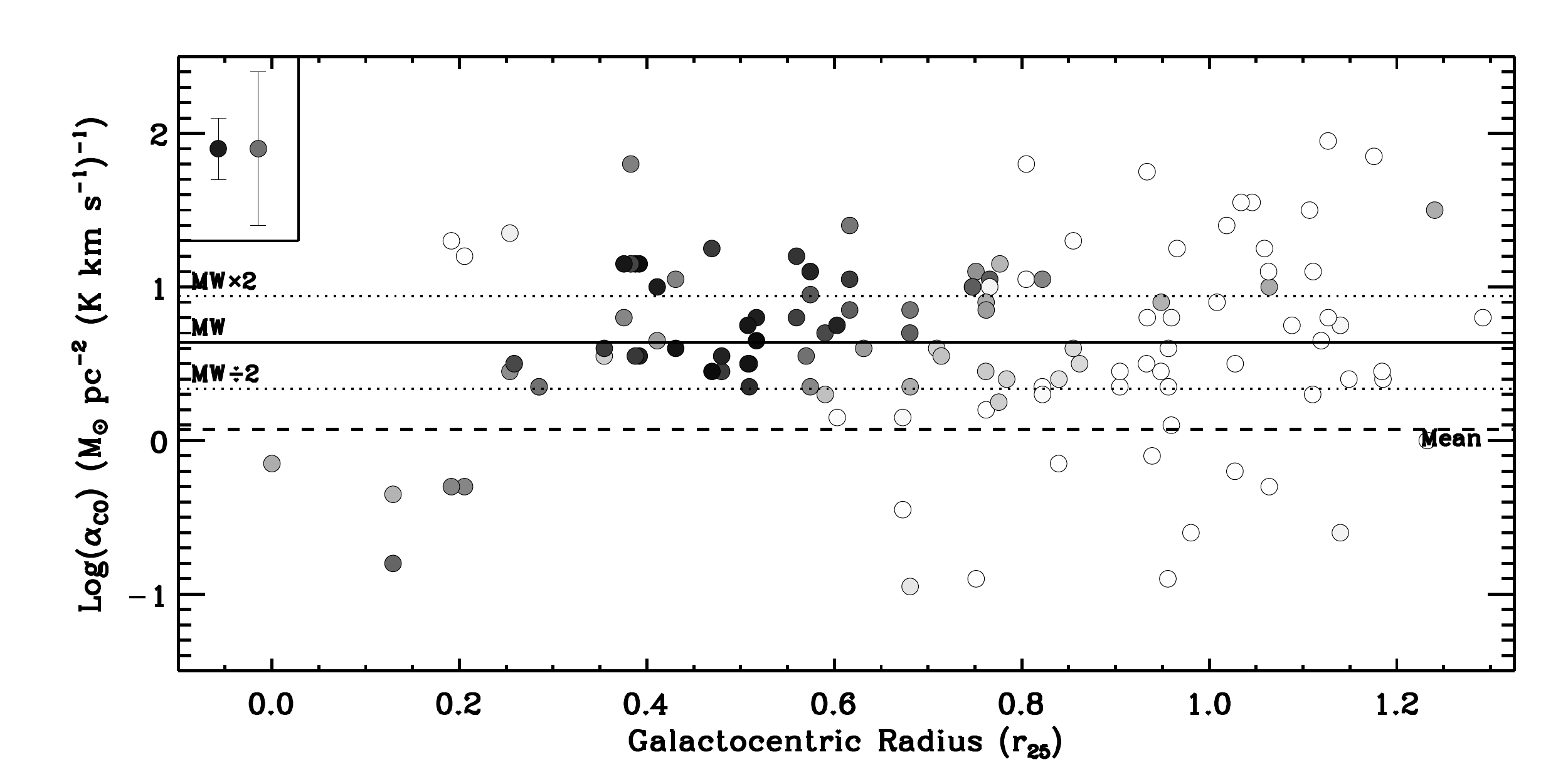}
\caption{Results for NGC 4725 (D = 11.9 Mpc; 1\arcsec = 58 pc).}
\label{fig:ngc4725_panel3}
\end{figure*}

\begin{figure*}
\centering
\epsscale{2.2}
\plottwo{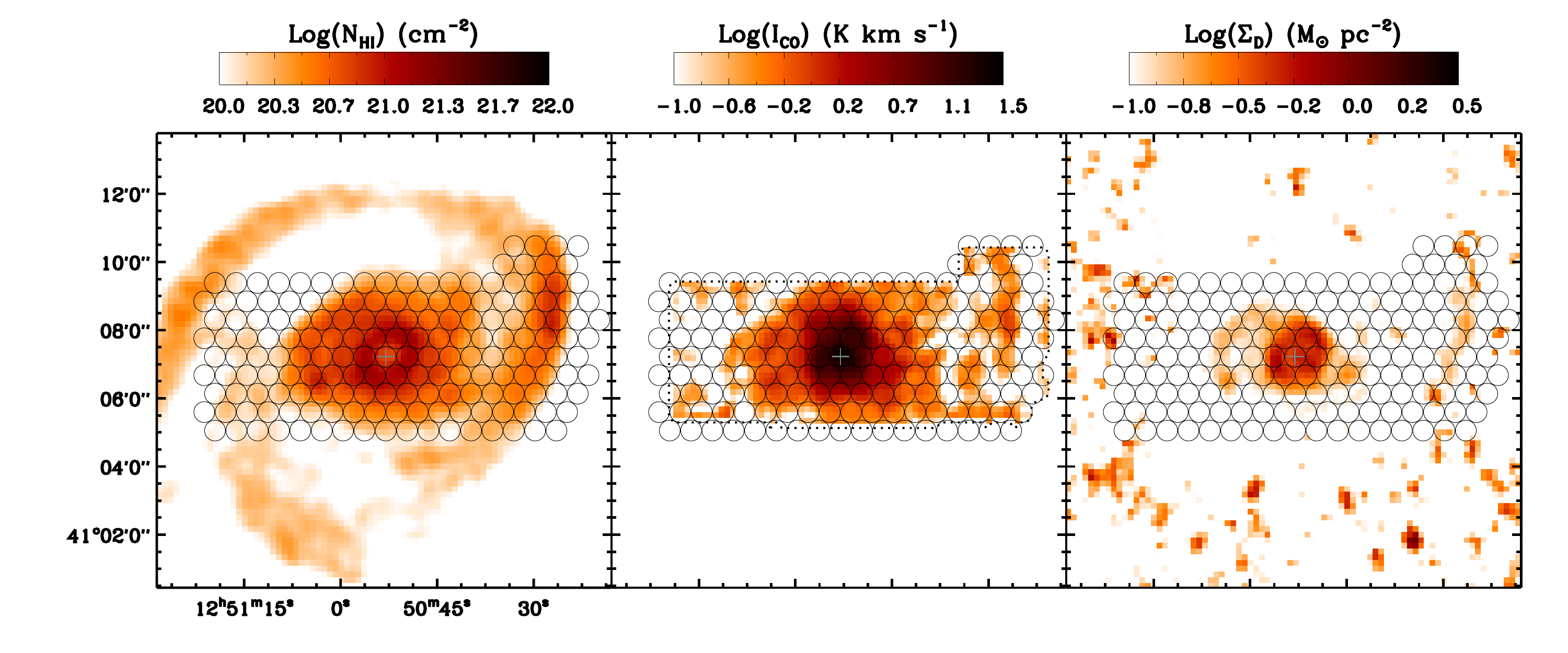}{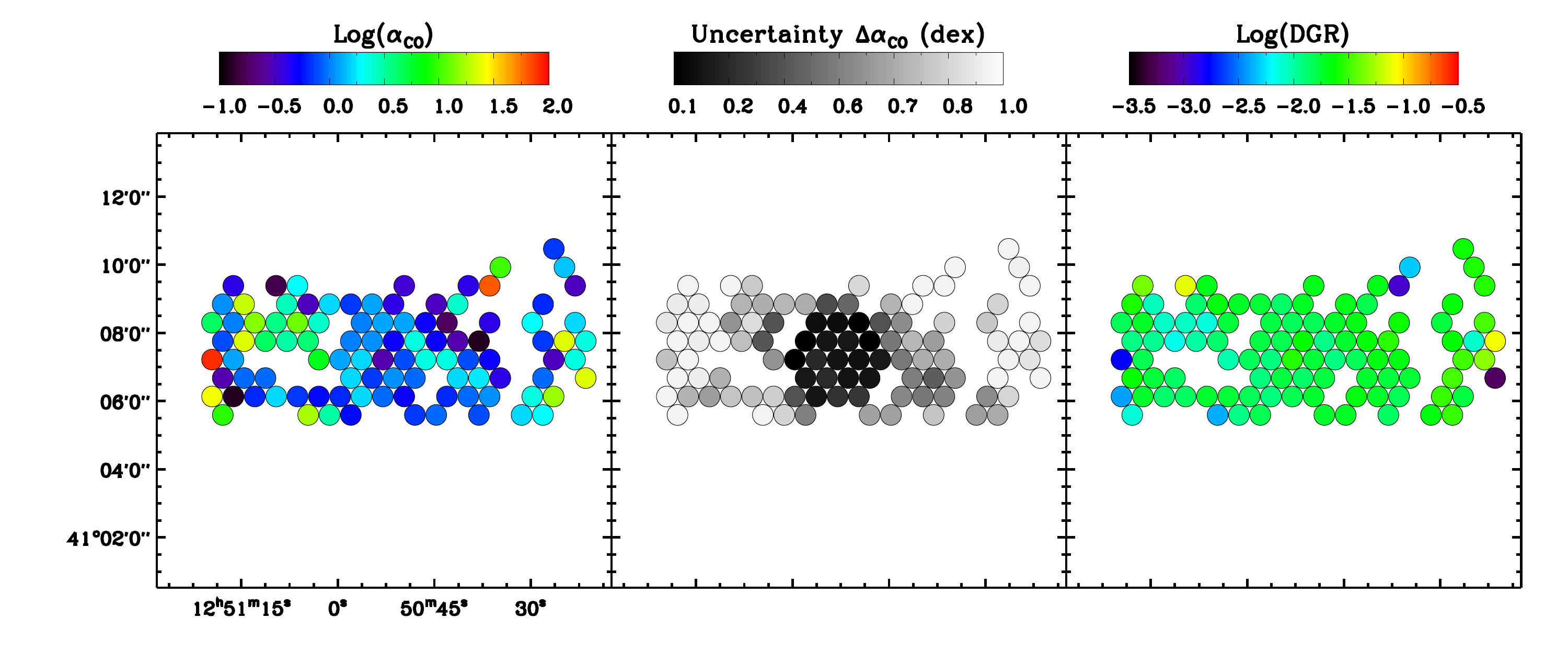}
\epsscale{1.0}
\plotone{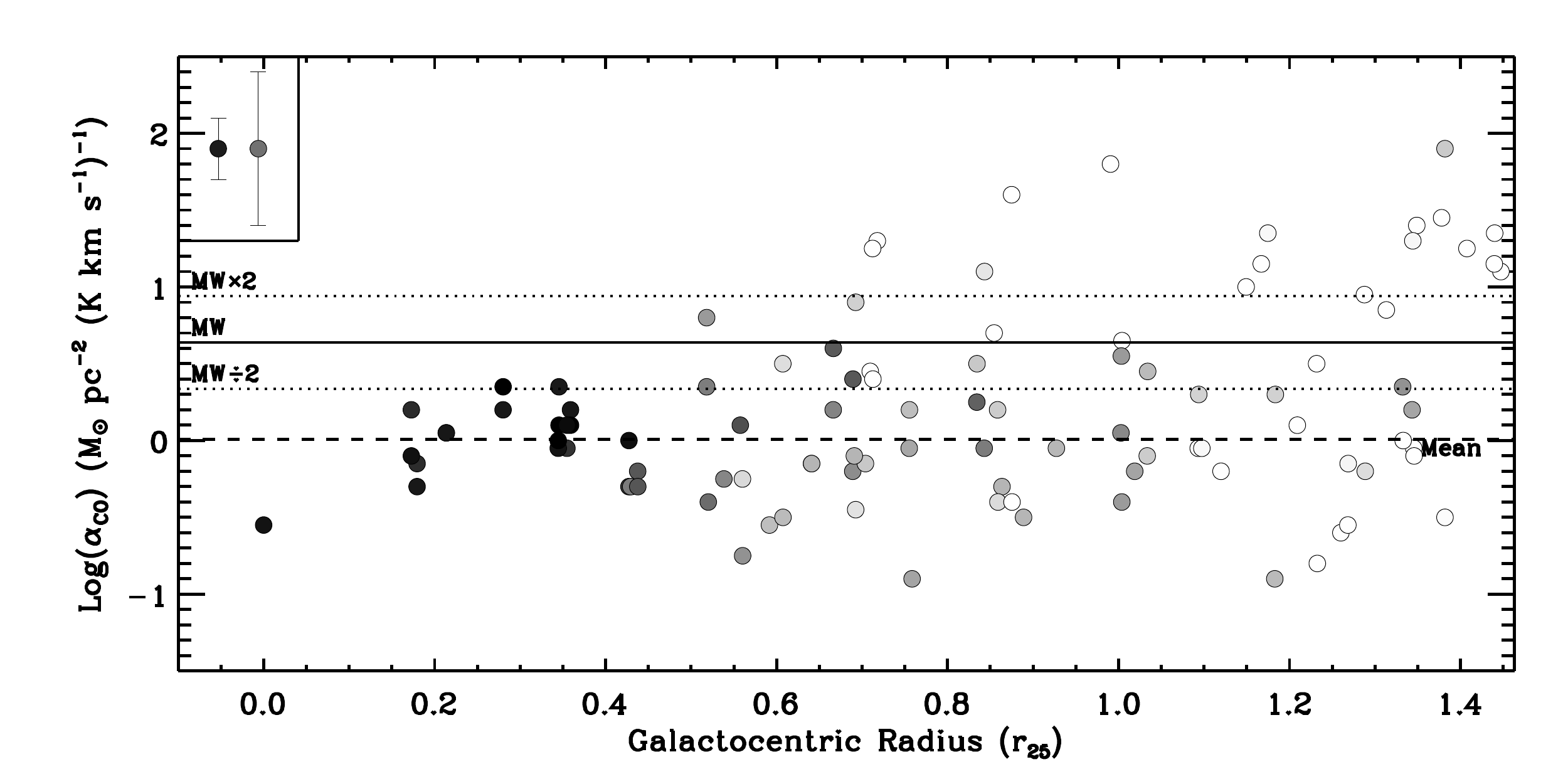}
\caption{Results for NGC 4736 (D = 4.7 Mpc; 1\arcsec = 23 pc).}
\label{fig:ngc4736_panel3}
\end{figure*}

\begin{figure*}
\centering
\epsscale{2.2}
\plottwo{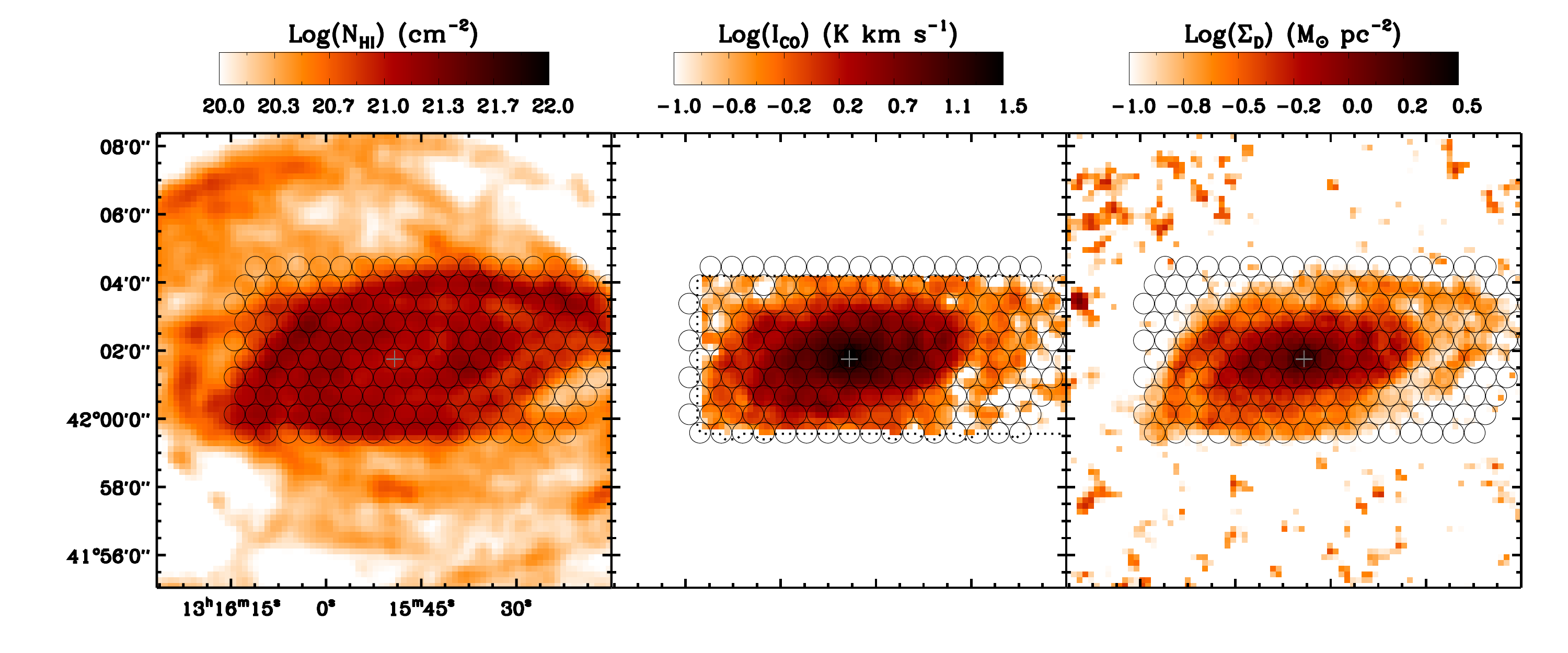}{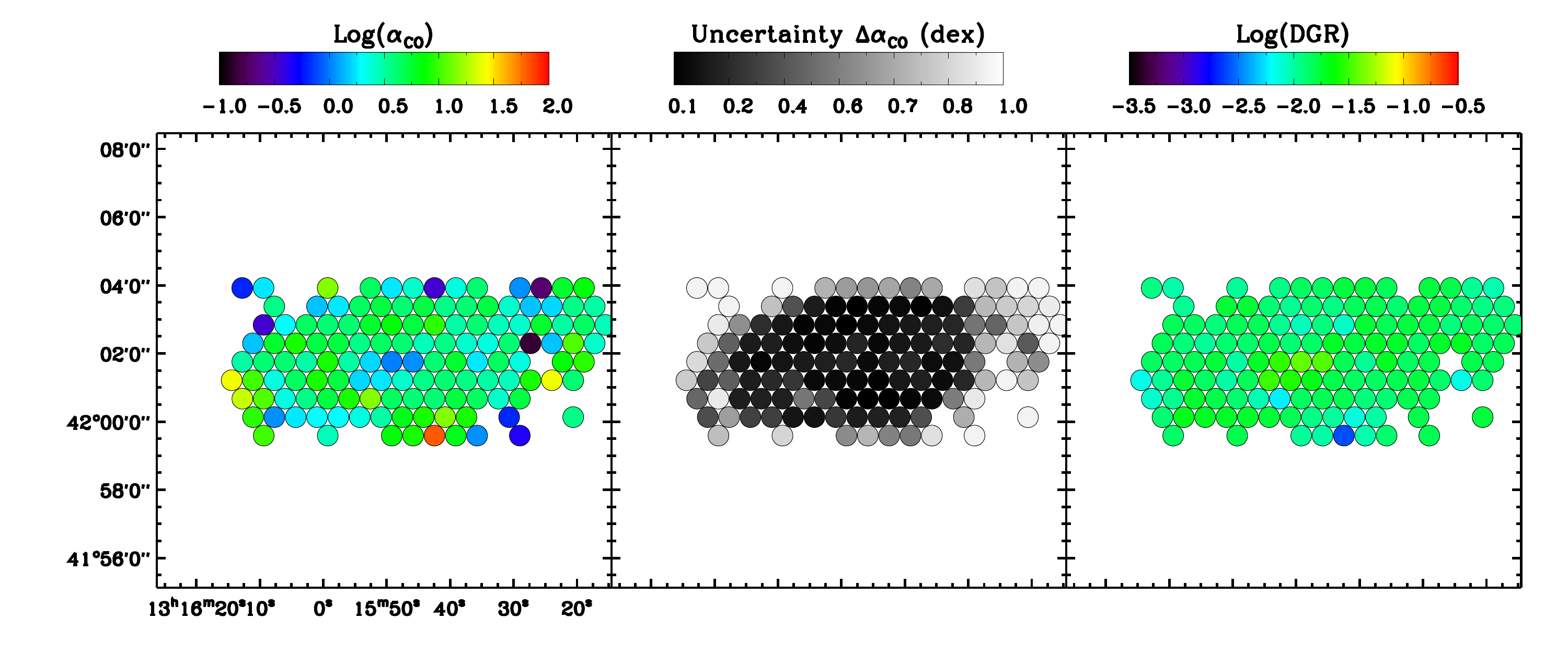}
\epsscale{1.0}
\plotone{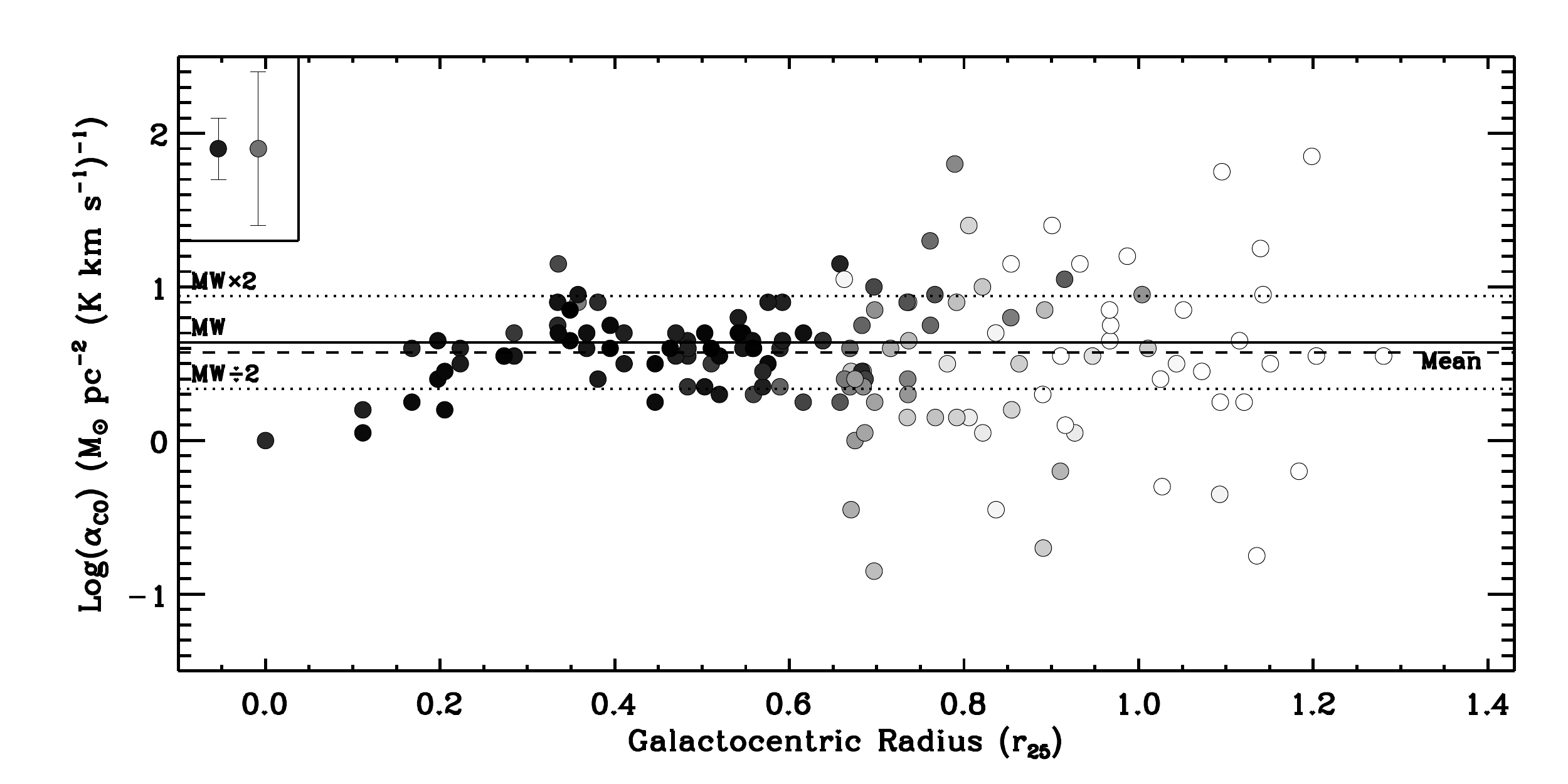}
\caption{Results for NGC 5055 (D = 7.9 Mpc; 1\arcsec = 38 pc).}
\label{fig:ngc5055_panel3}
\end{figure*}

\begin{figure*}
\centering
\epsscale{2.2}
\plottwo{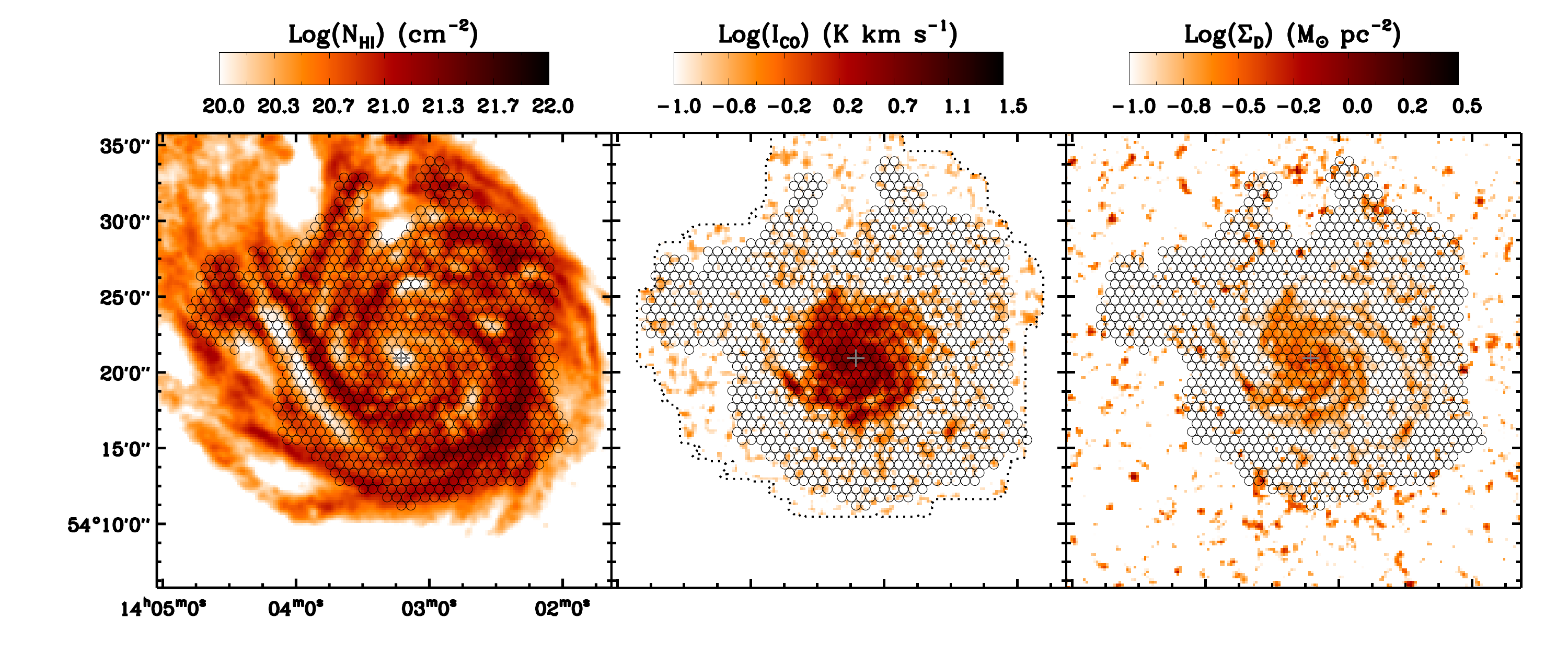}{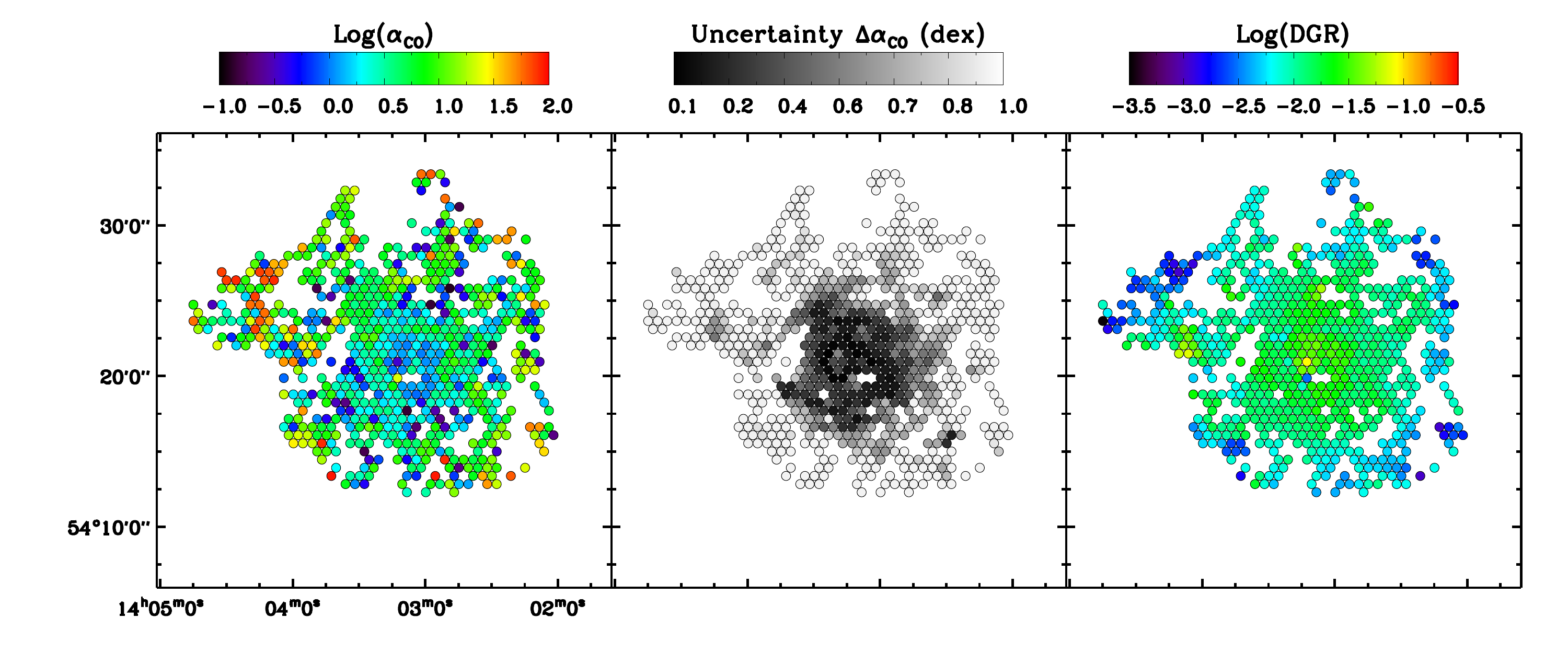}
\epsscale{1.0}
\plotone{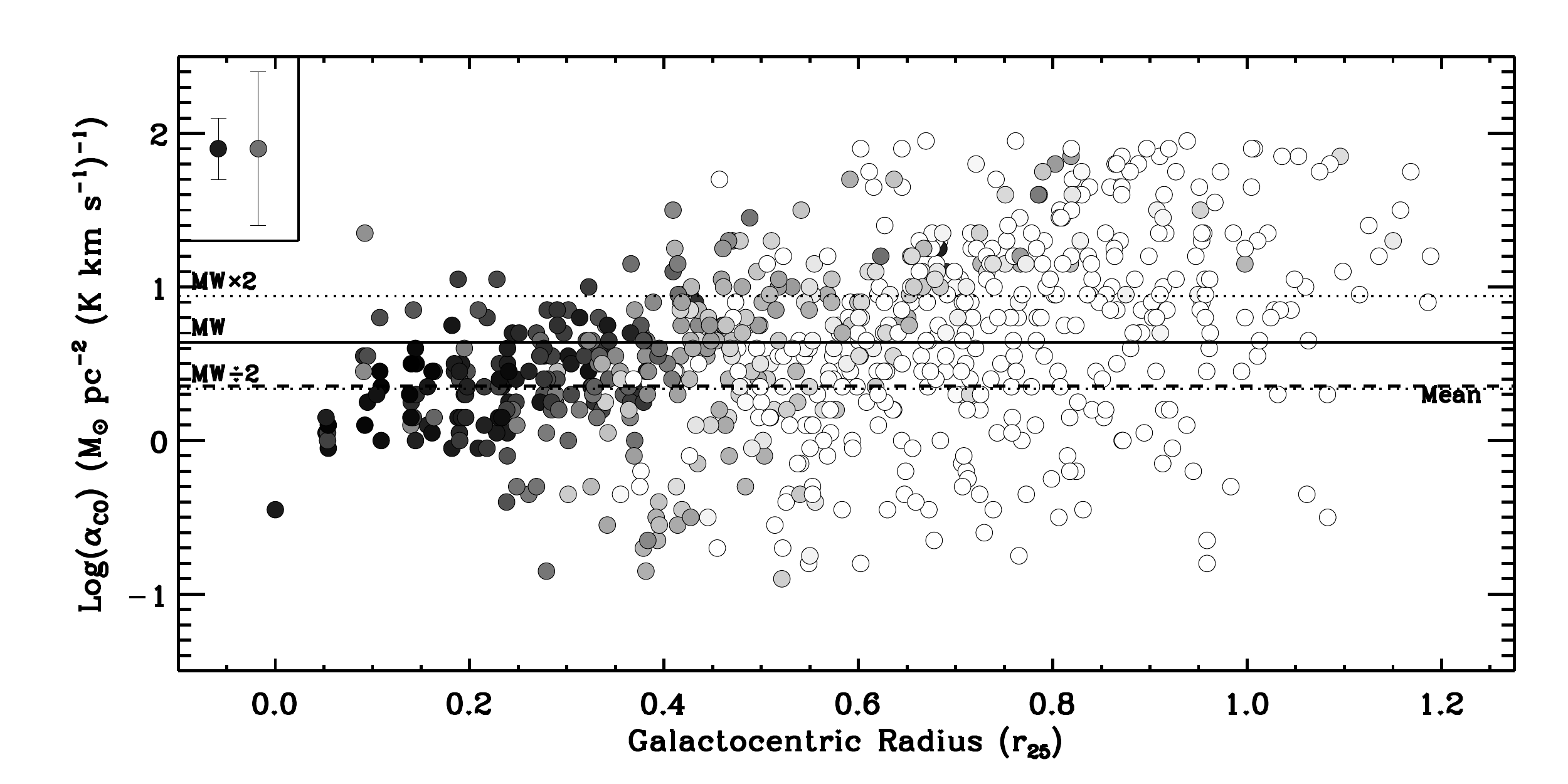}
\caption{Results for NGC 5457 (D = 6.7 Mpc; 1\arcsec = 32 pc).}
\label{fig:ngc5457_panel3}
\end{figure*}

\begin{figure*}
\centering
\epsscale{2.2}
\plottwo{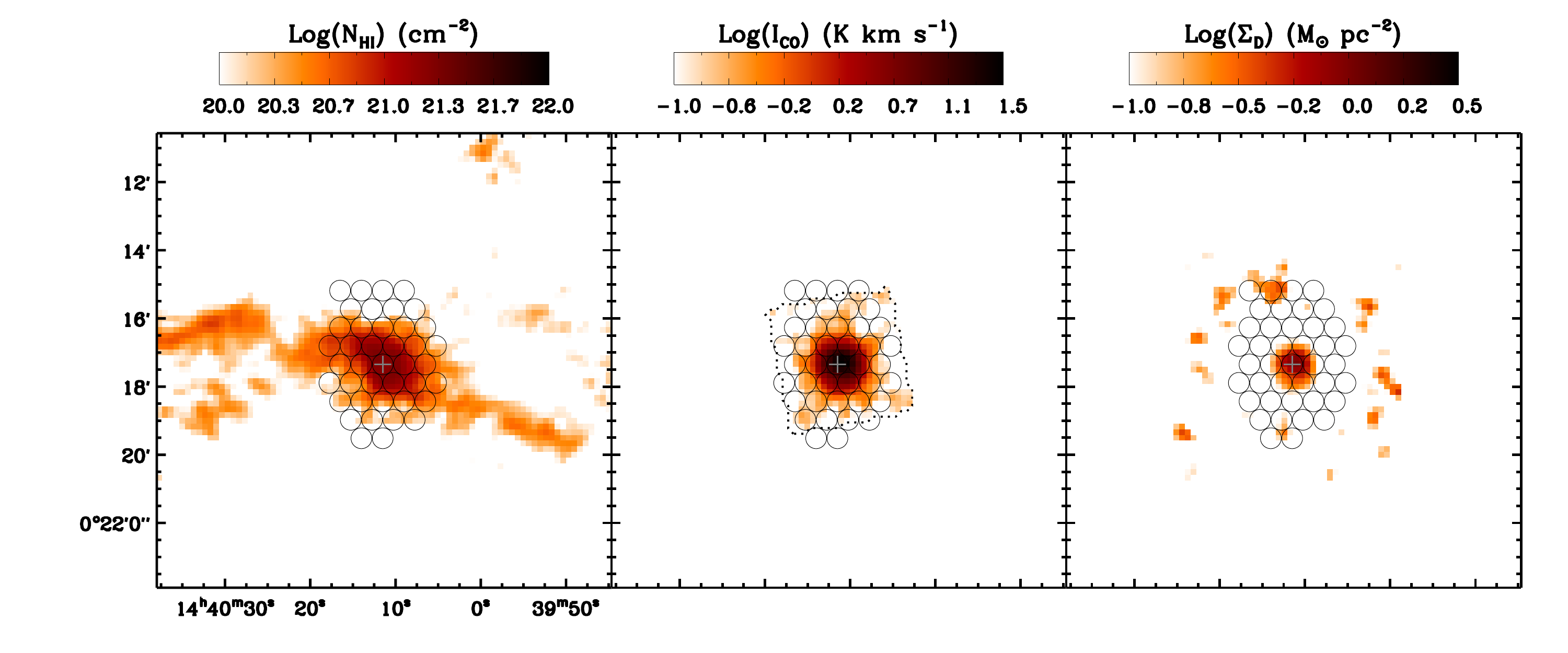}{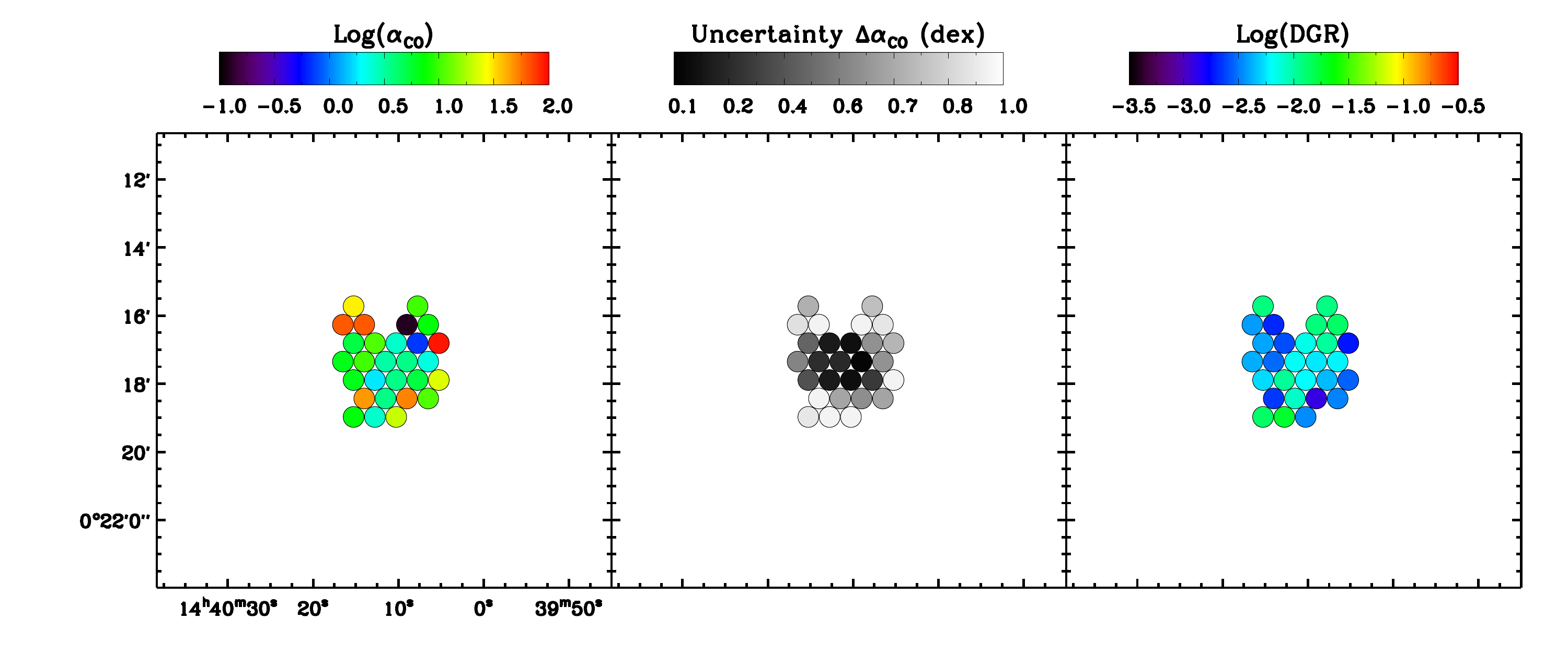}
\epsscale{1.0}
\plotone{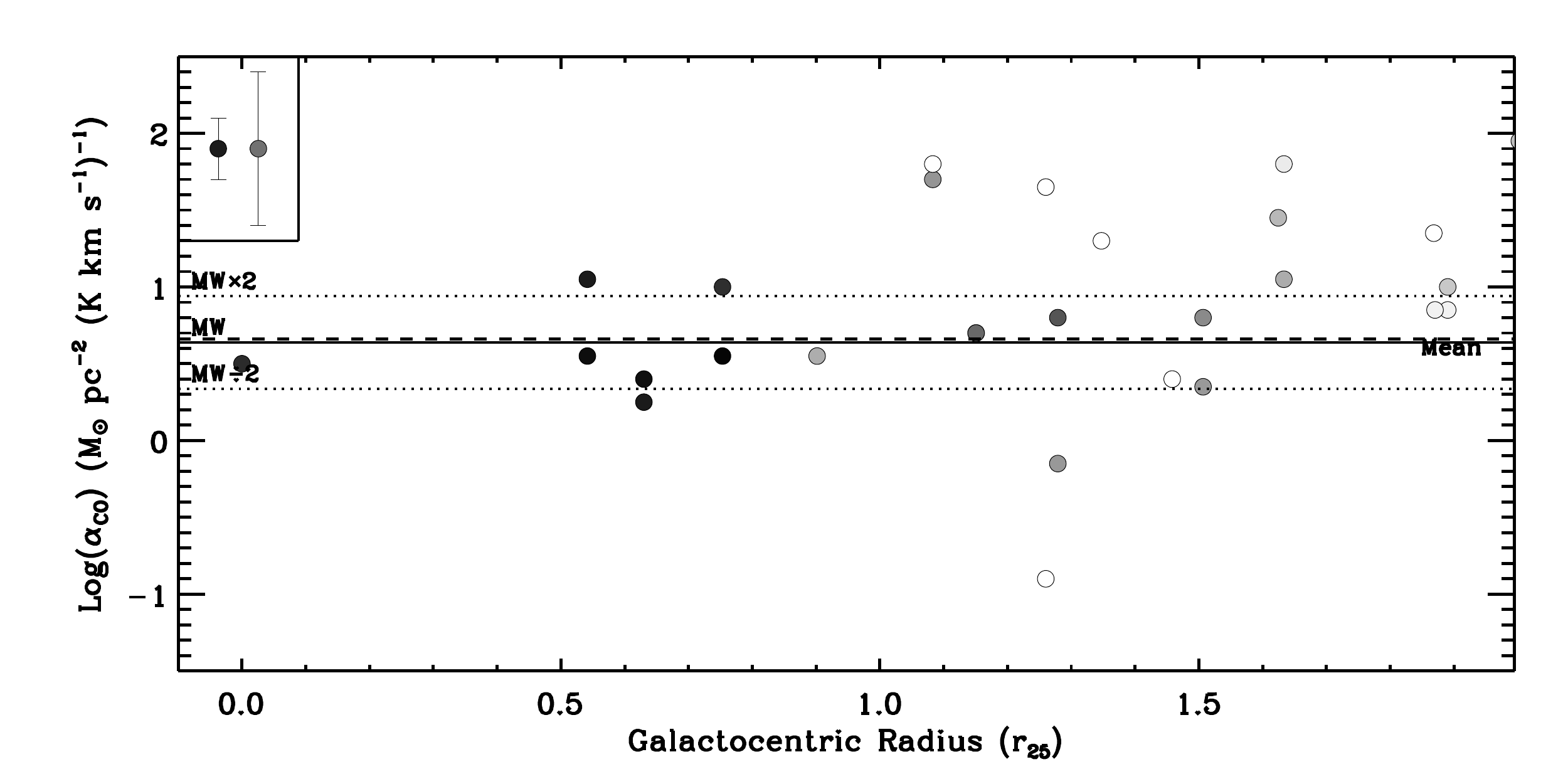}
\caption{Results for NGC 5713 (D = 21.4 Mpc; 1\arcsec = 104 pc).}
\label{fig:ngc5713_panel3}
\end{figure*}

\begin{figure*}
\centering
\epsscale{2.2}
\plottwo{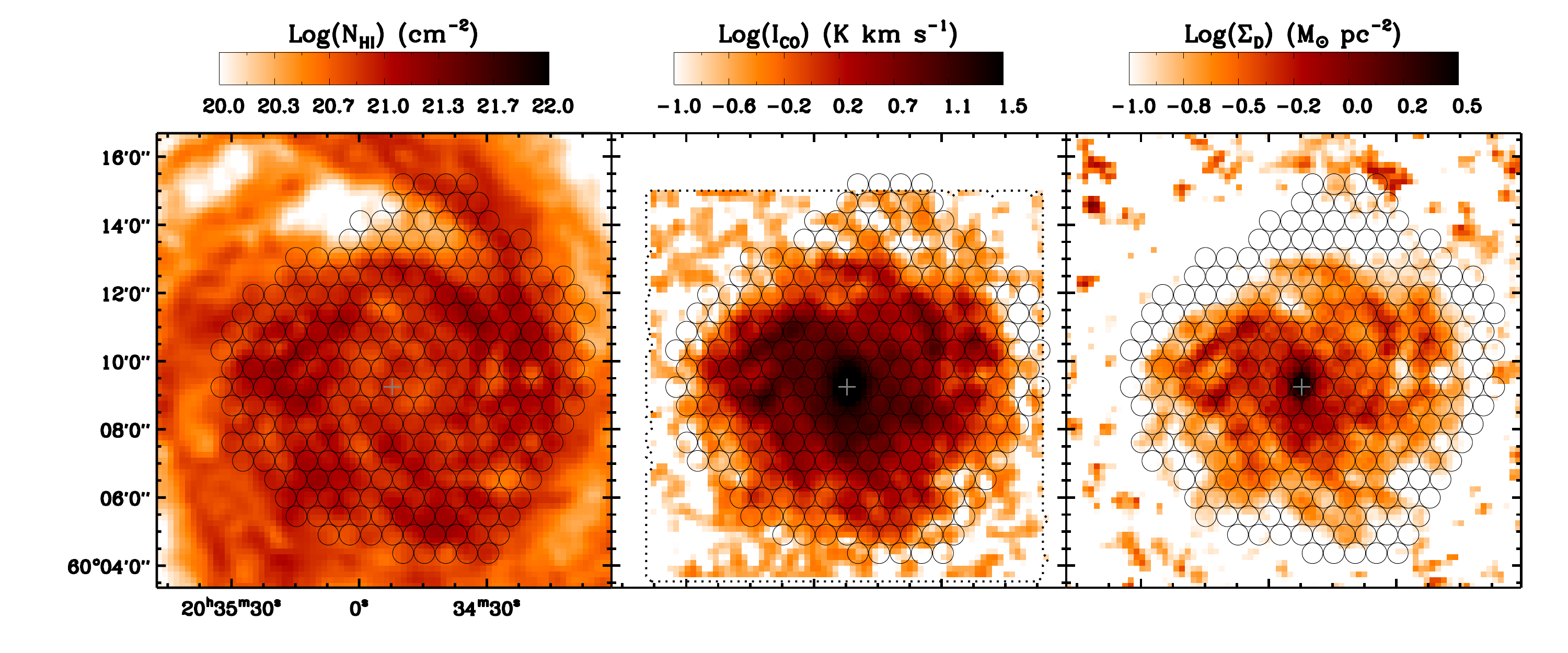}{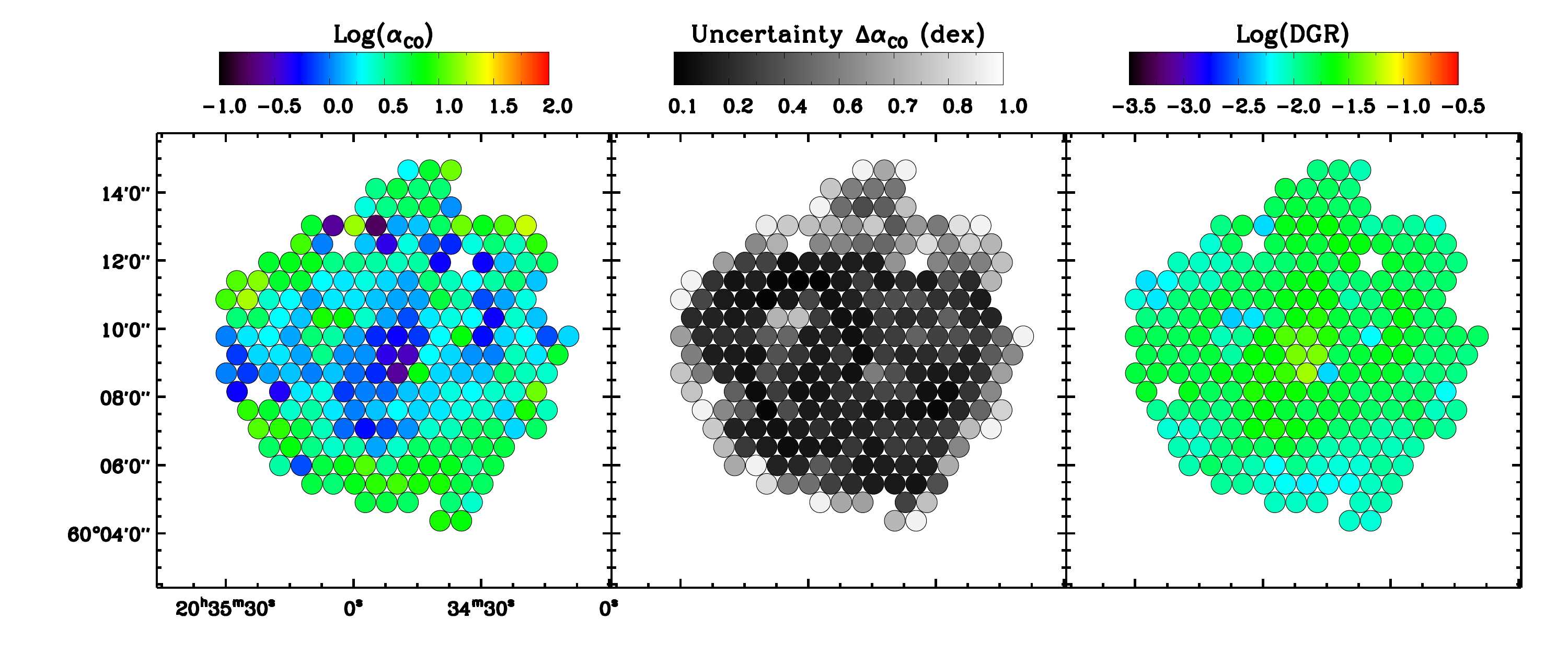}
\epsscale{1.0}
\plotone{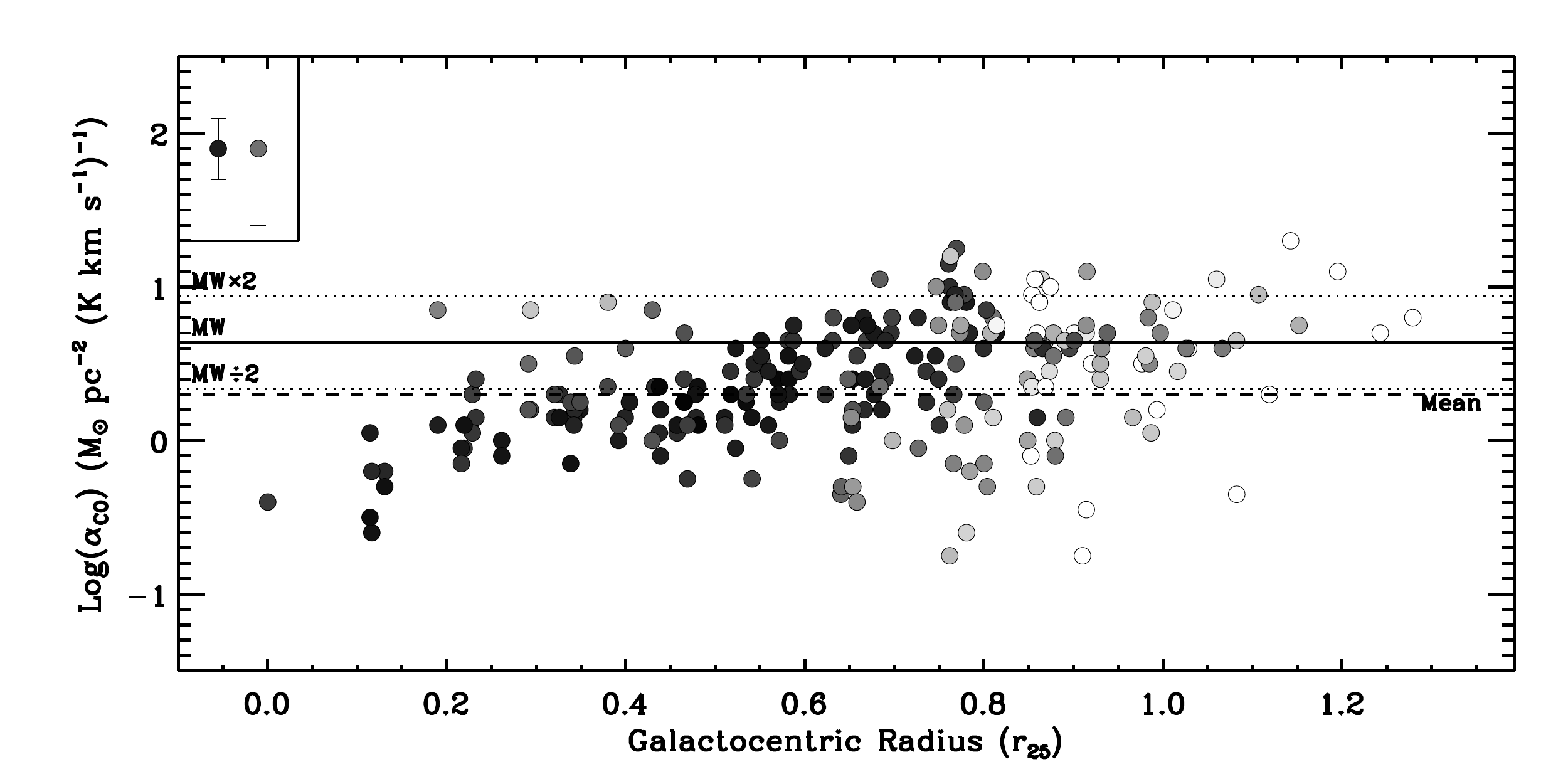}
\caption{Results for NGC 6946 (D = 6.8 Mpc; 1\arcsec = 33 pc).}
\label{fig:ngc6946_panel3}
\end{figure*}

\begin{figure*}
\centering
\epsscale{2.2}
\plottwo{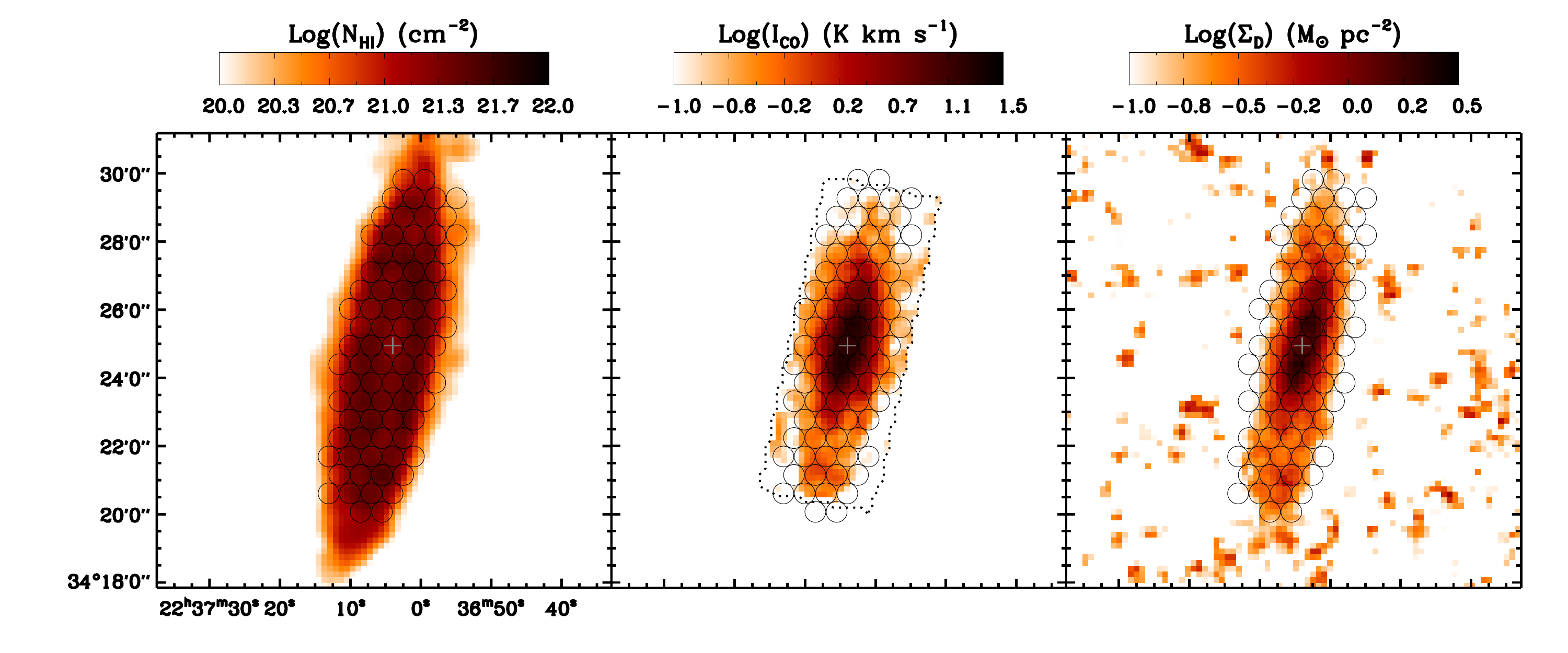}{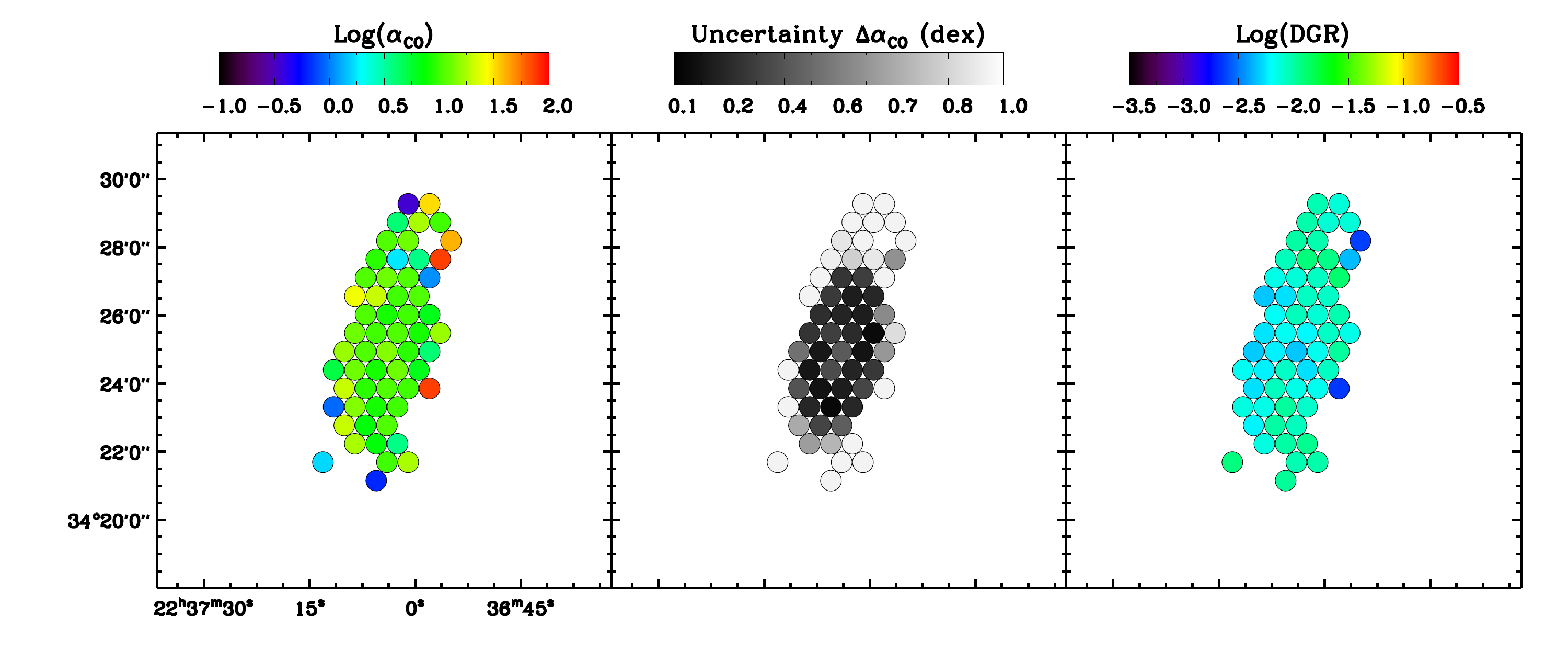}
\epsscale{1.0}
\plotone{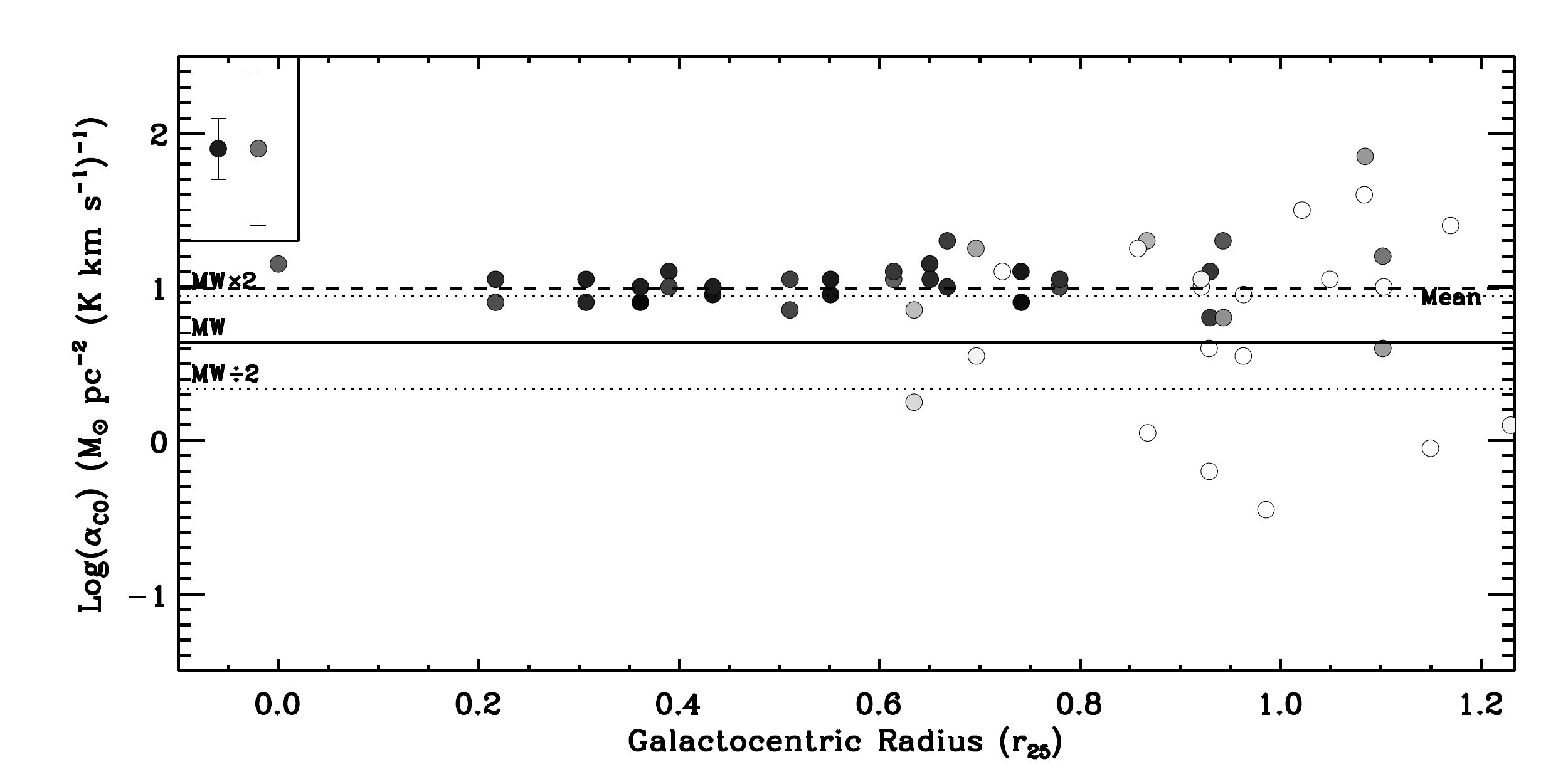}
\caption{Results for NGC 7331 (D = 14.5 Mpc; 1\arcsec = 70 pc).}
\label{fig:ngc7331_panel3}
\end{figure*}

In the following Appendix we present the results for individual solution pixels in all of the galaxies we have considered.  In the top panel of Figures~\ref{fig:ngc0337_panel3} through Figure~\ref{fig:ngc7331_panel3} we show the \hi, CO and \sigd\ maps at matched resolution.  The circles overlaid on these figures show the central position of each of the 37-point, hexagonal solution pixels in which we determine \aco.  In the middle panel of these same figures we show the \aco\ values on the left in color and the associated uncertainties in the right in gray scale. Solution pixels where the solution failed are omitted. Finally, in the bottom panels of the figures, we plot the measured \aco\ values as a function of galactocentric radius ($r_{25}$).  The color of the points in this plot reflects the uncertainties on the \aco\ values as shown in the gray scale color table in the middle panel.

% NGC 0337
\newpage

% NGC 0925
\newpage

% NGC 2841
\newpage

% NGC 2976
\newpage

% NGC 3077
\newpage

% NGC 3184
\newpage

% NGC 3198
\newpage

% NGC 3351
\newpage

% NGC 3521
\newpage

% NGC 3627
\newpage

% NGC 3938
\newpage

% NGC 4236
\newpage

% NGC 4254
\newpage

% NGC 4321
\newpage

% NGC 4536
\newpage

% NGC 4569
\newpage

% NGC 4625
\newpage

\clearpage

% NGC 4631
\newpage

% NGC 4725
\newpage

% NGC 4736
\newpage

% NGC 5055
\newpage

% NGC 5457
\newpage

% NGC 5713
\newpage

% NGC 6946
\newpage

% NGC 7331
\newpage

\end{document}